\documentclass[twocolumn,openany]{aa}
\usepackage{graphicx}
\usepackage{grffile}
\usepackage{txfonts}
\usepackage{newtxtext}
\usepackage{amsbsy}
\usepackage{wasysym}
\usepackage{natbib}
\usepackage{mathtools}
\usepackage{amsmath}
\usepackage{longtable}
\usepackage{pdflscape}
\usepackage{afterpage}
\usepackage[dvipsnames]{xcolor}
\usepackage[colorlinks=True,allcolors=blue]{hyperref}
\bibpunct{(}{)}{;}{a}{}{,}

\begin{document} 

\title{A global view on star formation: The GLOSTAR Galactic plane survey}
\subtitle{III.\ 6.7~GHz methanol maser survey in Cygnus X }

   \author{Gisela \ N.\ Ortiz-Le\'on \inst{1}
           \and
          Karl M.\ Menten\inst{1}
          \and
          Andreas Brunthaler\inst{1}
          \and
          Timea Csengeri\inst{2}
          \and
          James S.\,Urquhart\inst{3}
          \and
          Friedrich Wyrowski\inst{1}
          \and
          Yan Gong\inst{1}
          \and
          Michael R.\ Rugel\inst{1}
           \and
          Sergio A.\ Dzib\inst{1}
          \and
          Aiyuan Yang\inst{1}         
          \and
          Hans Nguyen\inst{1}
          \and 
          William D.\ Cotton\inst{4}
          \and
           Sac Nict\'e X.\ Medina\inst{1}
          \and
          Rohit Dokara\inst{1}
           \and
          Carsten K{\" o}nig\inst{1}
          \and
          Henrik Beuther\inst{5}
          \and
          Jagadheep D.\ Pandian\inst{6}
          \and
          Wolfgang Reich\inst{1}
          \and
          Nirupam Roy \inst{7}
          %
          }

   \institute{
              Max Planck Institut f\"ur Radioastronomie, Auf dem H\"ugel 69,  D-53121 Bonn, Germany
              \email{gortiz@mpifr-bonn.mpg.de}
              \and
              Laboratoire d'astrophysique de Bordeaux, Univ. Bordeaux, CNRS, B18N, all\'ee Geoffroy Saint-Hilaire, 33615 Pessac, France
               \and
               Centre for Astrophysics and Planetary Science, University of Kent, Canterbury, CT2 7NH, UK 
              \and
              National Radio Astronomy Observatory (NRAO), 520 Edgemont Road, Charlottesville, VA 22903, USA.
              \and
              Max Planck Institute for Astronomy, K\"onigstuhl 17, 69117, Heidelberg, Germany
              \and
             Department of Earth Space Sciences, Indian Institute of Space Science and Technology, Trivandrum 695547, India
             \and
             Department of Physics, Indian Institute of Science, Bengaluru, India 560012
               }


\titlerunning{GLOSTAR - Cygnus X} 
\abstract{
The Cygnus X complex is covered by the Global View of Star Formation in the Milky Way (GLOSTAR) survey, an unbiased radio-wavelength Galactic plane survey, in 4--8 GHz continuum radiation and several spectral lines. The GLOSTAR survey observed the 6.7~GHz transition of methanol (CH$_3$OH), an exclusive tracer of high-mass young stellar objects. Using the Very Large Array 
in both the B and D configurations, we  observed an area in Cygnus~X of 
$7^{\rm o}\times3^{\rm o}$
in size and simultaneously covered the methanol line and the continuum, allowing cross-registration.
We detected thirteen sources with Class~II methanol maser emission and one source with methanol absorption. Two methanol maser sources are newly detected; in addition, we found four new velocity components associated with known masers. Five masers are concentrated
in the DR21 ridge and W75N. 
We determined the characteristics of the detected masers 
and investigated the association with
infrared, (sub)millimeter, and radio continuum emission. All maser sources are associated with (sub)millimeter dust continuum  emission, which is consistent with the picture of masers tracing regions in an active stage of star formation. On the other hand, only five masers 
($38\pm17\%$) have radio continuum counterparts seen with GLOSTAR within $\sim$1$''$, testifying to their youth.
Comparing the distributions of the bolometric luminosity and the luminosity-to-mass ratio of cores that host 6.7~GHz methanol masers with those of the full core population, we identified lower limits 
$L_{\rm Bol}\sim200~L_\odot$ and $L_{\rm Bol}/M_{\rm core}\sim1~L_\odot~M^{-1}_\odot$ for a dust source to host maser emission.
}

\maketitle

\section{Introduction}\label{sec:intro}

The Cygnus X complex is a nearby (1.4~kpc\footnote{This distance comes from the average of the individual distances to several star-forming regions in Cygnus~X North. AFGL~2591, which is projected in the direction of Cygnus~X South, was found to be more distant, at 3.3~kpc. Whether Cygnus~X is a single connected region is still under debate.}; \citealt{Rygl2012}), $\sim$10$^{\rm o}$ wide region rich in molecular clouds that is undergoing active star formation and hosts a number of OB associations that testify to the prodigious star formation that has occurred over the past few million years. 
The vast richness of the present star formation activity in this region is revealed by a multitude of observations of centimeter-wavelength free-free radio emission, (sub)millimeter molecular line emission, dust continuum emission, and infrared-wavelength imaging (see \citealt{Reipurth2008} for a review). Numerous ($\sim$150) massive dense cores and young stellar objects have been identified \citep{Motte2007,Beerer2010,Bontemps2010,Cao2019}, as have around ten ultracompact (UC) HII regions \citep{Roy2011,Cao2019} and hundreds of more developed compact and extended HII regions \citep{Wendker1991}. Products of past star formation manifest themselves as  OB associations, with the 3--5 Myr old Cygnus OB2 association being the most prominent \citep{Wright2010}, as well as dozens of evolved objects, including Wolf–Rayet and carbon stars,  planetary nebulae, supernova remnants,
 and the famous X-ray binary Cygnus X 1 \citep{Uyaniker2001,Kraemer2010}. Figure 1 of \citet{Wendker1984} gives a schematic overview of Cygnus~X and its rich composition. 

\begin{figure*}[tb]
\begin{center}
 \includegraphics[width=0.8\textwidth,angle=0]{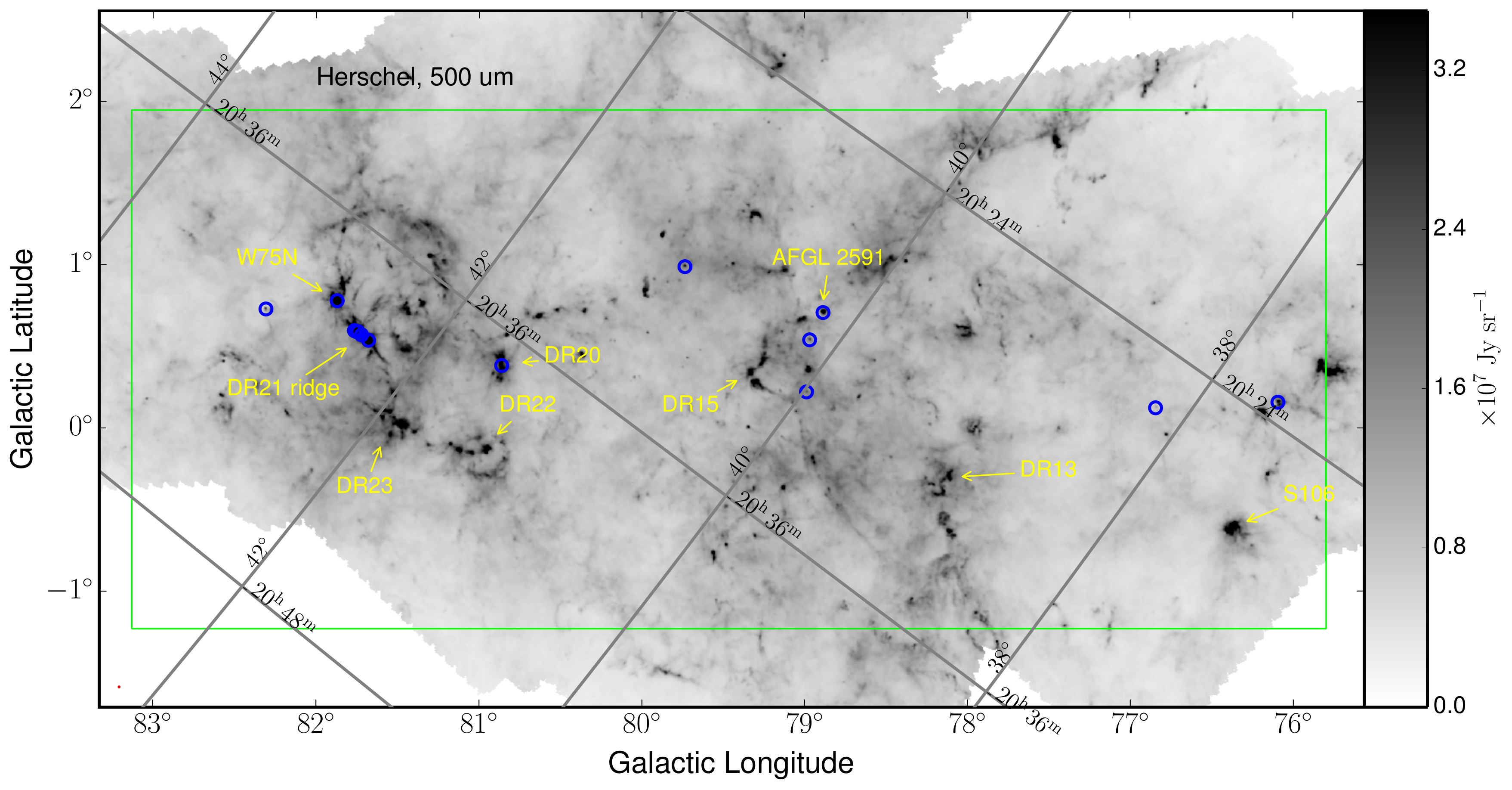}
\caption{Locations of methanol sources detected with the VLA  (blue circles), overlaid on a
Herschel SPIRE 500~$\mu$m 
continuum image of the Cygnus X region (grayscale). The mosaic was generated by \cite{Cao2019} from data obtained as part of the HOBYS \citep{Motte2010,Hennemann2014} 
and Hi-GAL projects \citep{Molinari2010}. The region outlined in green represents the area mapped with GLOSTAR. A few well-known star-forming regions and radio continuum sources \citep{DR1966} are labeled in yellow. 
}
\label{fig:allSource}
\end{center}
\end{figure*}

\begin{table*}
\caption{Summary of the VLA observations.}
\label{tab:obs} 
\centering 
\begin{tabular}{c c c c c c c c c c c}  
\hline\hline  
Observing Date &Observing Date &   Galactic coverage   & Beam size\tablefootmark{a}; PA   \\ 
  D-conf.            &   B-conf.           &                                   & (arcsec$\times$arcsec; deg)            \\
\hline 
2014-08-16      &       2015-05-02      &       $76^{\rm o} < l < 77^{\rm o}$     ;       $-1.0^{\rm o} < b<+0.5^{\rm o}$ &       $14\times12;~-28^{\rm o}$     &       \\
2014-08-25      &       2015-05-08      &       $76^{\rm o} < l < 77^{\rm o}$     ;       $+0.5^{\rm o} < b< +2.0^{\rm o}$        &       $17\times11;~-43^{\rm o}$     &       \\
2014-08-27      &       2015-05-07      &       $77^{\rm o} < l < 78^{\rm o}$     ;       $-1.0^{\rm o} < b< +0.5^{\rm o}$        &       $15\times12;~-54^{\rm o}$     &       \\
2014-08-24      &       2015-05-06      &       $77^{\rm o} < l < 78^{\rm o}$     ;       $+0.5^{\rm o} < b< +2.0^{\rm o}$        &       $18\times11;~-45^{\rm o}$     &       \\
2014-08-22      &       2015-04-25      &       $78^{\rm o} < l < 79^{\rm o}$     ;       $-1.0^{\rm o} < b< +0.5^{\rm o}$        &       $14\times11;~-46^{\rm o}$     &       \\
2014-08-07      &       2015-04-04      &       $78^{\rm o} < l < 79^{\rm o}$     ;       $+0.5^{\rm o} < b< +2.0^{\rm o}$        &       $16\times12;~-27^{\rm o}$     &       \\
2014-08-16      &       2015-04-24      &       $79^{\rm o} < l < 80^{\rm o}$     ;       $-1.0^{\rm o} < b<+0.5^{\rm o}$         &      $15\times12;~-46^{\rm o}$     &       \\
2014-08-15      &       2015-04-26      &       $79^{\rm o} < l < 80^{\rm o}$     ;       $+0.5^{\rm o} < b< +2.0^{\rm o}$        &       $15\times12;~-42^{\rm o}$     &       \\
2014-08-11      &       2015-05-14      &       $80^{\rm o} < l < 81^{\rm o}$     ;       $-1.0^{\rm o} < b< +0.5^{\rm o}$        &       $15\times11;~-29^{\rm o}$     &       \\
2014-08-17      &       2015-05-03      &       $80^{\rm o} < l < 81^{\rm o}$     ;       $+0.5^{\rm o} < b< +2.0^{\rm o}$        &       $13\times11;~-27^{\rm o}$     &       \\
2014-08-06      &       2015-05-08      &       $81^{\rm o} < l < 82^{\rm o}$     ;       $-1.0^{\rm o} < b< +0.5^{\rm o}$        &       $16\times12;~-29^{\rm o}$     &       \\
2014-08-18      &       2015-05-10      &       $81^{\rm o} < l < 82^{\rm o}$     ;       $+0.5^{\rm o} < b< +2.0^{\rm o}$        &       $14\times11;~-26^{\rm o}$     &       \\
2014-08-26      &       2015-05-09      &       $82^{\rm o} < l < 83^{\rm o}$     ;       $-1.0^{\rm o} < b< +0.5^{\rm o}$        &       $14\times11;~-41^{\rm o}$     &       \\
2014-08-12      &       2015-05-09      &       $82^{\rm o} < l < 83^{\rm o}$     ;       $+0.5^{\rm o} < b< +2.0^{\rm o}$        &       $14\times11;~-38^{\rm o}$     &       \\
\hline 
\end{tabular}
\tablefoot{
\tablefoottext{a}{These values are for the methanol dirty cubes constructed for the D-array configuration data. }
}
\end{table*}

Due to its proximity (being one of the nearest massive star-forming regions) and large gas reservoir, Cygnus X serves as an excellent laboratory for studying the formation process of high-mass stars. The region extends roughly between $76^{\rm o}$ and $83^{\rm o}$ in Galactic longitude and  $-2^{\rm o}$ and $2^{\rm o}$ in latitude (see Fig. 1 in \citealt{Schneider2006}). With a total mass of $3-4\times10^{6}~M_\odot$ \citep{Schneider2006}, this complex is one of the most massive giant molecular clouds (GMCs) in the extended solar neighborhood.

The $5_1-6_0~A^+$ line of methanol at 6.7~GHz was first detected in the interstellar medium by \citet{Menten1991ApJ380L}. The observed complex line shapes, high intensities, and association with star-forming regions immediately indicated maser emission -- in fact, after the 22.2 GHz transition from H$_2$O, it is the second strongest and second most frequently found interstellar maser line \citep[e.g.,][]{Caswell1995}.

The comprehensive Methanol Multibeam (MMB) survey in this transition that covered the Galactic plane from $l=186^{\rm o}$ to $l=60^{\rm o}$ found 972 maser sources \citep[see][]{Green2009,Green2010,Green2012,Green2017,Caswell2010,Caswell2011}.  Various studies have shown that 6.7 GHz methanol masers are exclusively associated with high-mass
star formation \citep{Minier2003,Bourke2005,Ellingsen2006,Pandian2008,Xu2008,Pandian2010}. Most of these masers were found to have no radio continuum counterpart (at the few mJy sensitivity level of previous observational efforts) by \citet{Walsh1998}, who first suggested that most methanol masers are associated with massive young (proto)stellar objects (MYSOs) in their earliest evolutionary phases (i.e., before they excite a UC~HII region), 
which has been widely confirmed. The comprehensive studies of \cite{Urquhart2013, Urquhart2015} and \cite{Billington2019} found that 99\% of the 958 methanol masers identified by the MMB in the Galactic plane are also associated with submillimeter continuum radiation from dust detected by the $870~\mu$m
APEX Telescope Survey of the Galaxy \citep[ATLASGAL;][]{Schuller2009, Csengeri2014, Urquhart2014} and the $850~\mu$m JCMT Plane Survey \citep{Moore2015, Eden2017} conducted with the 15 meter James Clerk Maxwell Telescope (JCMT). Given that 99\% of sources have been associated with dust, it is probably safe to assume that all methanol masers are thus associated with compact dust clumps that are massive enough to harbor at least one  high-mass \mbox{(proto)star}, along with its associated lower-mass (proto)stellar cluster members. 
We note further that the 6.7 GHz line is a Class II methanol maser transition and thus, by definition, is radiatively pumped \citep{Batrla1987, Menten1991, Menten1991ApJ380L}. These masers thus naturally require a source emitting strongly in the mid-infrared regime in their close vicinity \citep{Sobolev1994, Cragg2005}, which gives a natural explanation for the above-discussed maser--MYSO association.

The Global View of Star Formation in the  Milky Way (GLOSTAR) project is a large radio survey aimed at mapping a significant part of the Galactic mid-plane to characterize the properties of sites with star formation in the earliest evolutionary stages. It takes advantage of the upgraded wideband capabilities of the Karl G.\ Jansky Very Large Array (VLA), which provides a vastly increased bandwidth and better sensitivity. An overview of the survey is given by \cite{BrunthalerPilot}, while the first catalog of continuum radio sources detected in our pilot region ($l=28^{\rm o}$ to $36^{\rm o}$ with $b = \pm1^{\rm o}$) is presented in \cite{Medina2019}.

Here we report on new VLA observations of 6.7~GHz methanol masers that were carried out in the course of our GLOSTAR coverage of the Cygnus X region, from $l=76^{\rm o}$ to $83^{\rm o}$ and $b=-1^{\rm o}$  to $+2^{\rm o}$ (see Fig.\, \ref{fig:allSource} for survey coverage). Primarily, this paper presents the properties of the detected methanol masers and their association with radio continuum emission. Since the maser positions are accurate at the subarcsecond level, we establish their associations with other tracers of star formation to investigate the relationship between methanol masers and massive star formation in the Cygnus X complex.
In Sect. \ref{sec:obs} we describe the observations and data analysis. Section \ref{sec:search} presents the method for source extraction and the detected methanol masers. In Sect. \ref{sec:results} we discuss the properties of the masers and their association with radio continuum, (sub)millimeter, and infrared emission. We also investigate the properties of the dust cores with which the masers are associated and comment on the possible nature of each maser source. Finally, in Sect. \ref{sec:discuss} we discuss the relationship between maser emission and dust continuum source properties.

\section{Observations and data reduction}\label{sec:obs}

The observations were carried out
as part of the GLOSTAR survey, which is described in detail in \cite{BrunthalerPilot}. 
Here we discuss the Cygnus X region portion of the survey.
To characterize the radio emission and various  tracers of star formation, we carried out simultaneous observations of C-band continuum emission (covering 4.2--5.2 and 6.5-7.5~GHz), the methanol (CH$_3$OH) line at 6.7~GHz, seven radio recombination lines (H96$\alpha$, H98$\alpha$,  H99$\alpha$,  H110$\alpha$, H112$\alpha$,  H113$\alpha$, and H114$\alpha$), and formaldehyde (H$_2$CO) transition at 4.8~GHz. Observations were taken in both the D  and B configurations to simultaneously achieve the best sensitivity for extended emission and excellent angular resolution (see Table\,\ref{tab:obs} for details). The D- and B-configuration data provide angular resolutions of $15''\times11''$ and $1\rlap.{''}5\times1\rlap.{''}0$, respectively. This  corresponds to approximate 
physical scales of 0.1~pc and 1700~au, respectively, at a distance of 1.4~kpc.  
Two 1 GHz wide base bands in full polarization mode were registered for the continuum. These were centered at 4.7 and 6.9~GHz and consisted of eight 128 MHz wide spectral windows each. 
The methanol maser line, which has a rest frequency of 6668.5192~MHz \citep{Breckenridge1995}, was observed with 8~MHz of bandwidth and 2048 channels, resulting in a  
channel spacing of 0.176~km~s$^{-1}$, and a total velocity coverage of 
360~km~s$^{-1}$.

Our observations, which covered almost the whole Cygnus X complex  (i.e., an area of $7^{\rm o}\times3^{\rm o}$), required a total of 28 epochs, which were scheduled as part of GLOSTAR under program ID 14A-420. Each epoch observed a $1^{\rm o}\times1.5^{\rm o}$ strip (see Table \ref{tab:obs}),~typically using 532 single pointings. Two 11-second scans were spent on each pointing for a total (on-source) integration time of $\sim$15 seconds. The spectral line data  from each epoch were calibrated independently  within the Common Astronomy Software Applications ({\tt CASA}) package using version v4.6.0 and a customized version of the VLA pipeline\footnote{\url{https://science.nrao.edu/facilities/vla/data-processing/pipeline}}. The continuum subtraction was performed on the visibility data using the {\tt CASA} task {\tt uvcontsub}, where we excluded channels expected to contain emission lines in order to fit the continuum.
Specifically, we excluded the local standard of rest (LSR)
velocity range from $\sim-30$ to $+20$~km~s$^{-1}$ for the methanol line, which covers the velocity range of the complex \citep{Schneider2006}. 
      
The calibrated visibilities were imaged using the CLEAN algorithm as implemented in the package {\tt tclean} within  {\tt CASA} version v5.4 with a pixel size of $2\rlap.{''}5$ and $0\rlap.{''}2$ for D- and B-configuration data, respectively, and a spectral resolution of 0.18 km~s$^{-1}$. Each $1^{\rm o}\times1.5^{\rm o}$ stripe requires 2800$\times$2800 pixels in D-configuration maps, while B-configuration maps require about 100 times more pixels. 
Due to limitations of the computer resources required to perform the deconvolution of such a large map over a large number of spectral channels, we only constructed dirty D-configuration images for the methanol line and performed the maser detection search on these dirty cubes (this is described in more detail in Sect. \ref{sec:search}). Dirty cubes of the methanol line were produced for the velocity range from $-200$ to $+100$~km~s$^{-1}$. Maps of the B-configuration data were only constructed, in a subsequent step, for smaller regions ($3\rlap{.}'4\times3\rlap{.}'4$) centered at the positions of the detected masers determined in the D-configuration cubes and using the data from all neighboring pointings. The synthesized beams are on average $15''\times11''$ at position angle (PA)=$-49^{\rm o}$ and $1\rlap.{''}5\times1\rlap.{''}0$ at PA=$-33^{\rm o}$ in the D and B configuration, respectively. The 1$\sigma$ rms noise measured in channels free of maser emission is, on average, 0.028~Jy~beam$^{-1}$ per 0.18~km~s$^{-1}$ channel for both configurations.   

To complement our study, we also analyzed GLOSTAR continuum images toward sites with maser emission.  
A full description of the GLOSTAR continuum data calibration and imaging is given in \cite{BrunthalerPilot}, while the full analysis of continuum images of Cygnus X will be presented in a forthcoming paper. Here, we briefly discuss the imaging strategy. The calibration and imaging of the continuum data was performed with the Obit package \citep{Cotton2008}. The 2 GHz bandwidth was first rearranged into nine frequency subbands, which were used to image each pointing individually. Then, for each frequency subband the pointings were combined into large individual mosaics to cover the entire observed area. Finally, we combined the different frequencies to obtain the image at the reference frequency, which has circular beams of 19$''$ and 1.5$''$ in the D  and B configuration, respectively. 
Continuum and methanol line maps from Effelsberg observations have also been obtained as part of the GLOSTAR survey (\citealt{BrunthalerPilot}, Rugel et al.\ in prep.) We note that continuum images were constructed for Effelsberg data, the VLA D configuration, the VLA B configuration, a combination of the VLA D and B (D+B) configurations, and a combination of the VLA D configuration and Effelsberg observations. The central frequency of these images is 5.8~GHz. Here, we only use B-configuration continuum maps to study the region of the investigated methanol maser positions and D+B maps of the region around DR21 (see Sect. \ref{sec:dr21}). Methanol line data from Effelsberg were also inspected to look for flux variations in the VLA-detected masers (Sect. \ref{sec:comments}). 
The noise in the continuum images is not uniform, but rather varies across the mapped region, and can be high around strong sources with complex or extended emission. We locally measured the noise in regions close to the maser locations, resulting in 1$\sigma$ values in the range from 0.056 to 0.43~mJy~beam$^{-1}$ for B-configuration images. For the D configuration, the 1$\sigma$ rms noise ranges from 0.10 to 2.6~mJy~beam$^{-1}$. The higher values measured in D-configuration data are due to bright extended emission, which is present across the Cygnus X region, and are resolved out by the array in the B configuration. The highest local rms noise occurs around the strong radio source, DR21, a compact HII region. 

\begin{table*}
\caption{Properties of methanol sources from D-configuration maps.}
\label{tab:jmfit-D} 
\centering 
{\footnotesize
\begin{tabular}{l c c c c c c c c c c}  
\hline\hline  
Maser   &  Name  & $\alpha$/$\Delta\alpha$  & $\delta$/$\Delta\delta$ & $V_{\rm LSR}$  & $S_{\nu,~\rm Peak}$        & $S_{\nu,~\rm Int.}$ & Common    \\
ID     &         & (h:m:s)/(arcsec) & ($^{\rm o}$:$'$:$''$)/(arcsec) & (km~s$^{-1}$)   &(Jy~beam$^{-1}$)  &   (Jy)    &    Name   \\
(1)    & (2)     & (3)              & (4)                            & (5)             & (6)              & (7)       &     (8)   \\ 
\hline
1 & G76.0932+0.1580 & 20:23:23.683 & +37:35:36.17 &  4.84 & 0.54 $\pm$ 0.02 & 0.55 $\pm$ 0.03 & -- \\ 
   &                 &      0.46      &      0.70   & -6.50 & 0.46 $\pm$ 0.02 & 0.53 $\pm$ 0.04 &     \\ 
   &                 &      1.52      &     -1.32   &  6.46 & 0.41 $\pm$ 0.02 & 0.49 $\pm$ 0.04 &     \\ 
2 & G76.8437+0.1233 & 20:25:43.758 & +38:11:13.26 & -5.42 & 4.61 $\pm$ 0.04 & 4.70 $\pm$ 0.08 & -- \\ 
3 & G78.8870+0.7087 & 20:29:24.947 & +40:11:19.81 & -7.04 & 0.77 $\pm$ 0.02 & 0.80 $\pm$ 0.04 & AFGL2591 \\ 
4 & G78.9690+0.5410 & 20:30:22.725 & +40:09:23.90 &  4.84 & 1.16 $\pm$ 0.03 & 1.22 $\pm$ 0.06 & IRAS~20286+3959 \\ 
5 & G78.9884+0.2211 & 20:31:47.286 & +39:59:00.52 & -68.60 & 1.87 $\pm$ 0.03 & 1.97 $\pm$ 0.06 & -- \\ 
   &                 &     -0.34      &     -0.64   & -66.26 & 0.29 $\pm$ 0.02 & 0.32 $\pm$ 0.04 &     \\ 
   &                 &     -0.72      &      1.28   & -55.46 & 0.21 $\pm$ 0.02 & 0.20 $\pm$ 0.03 &     \\ 
   &                 &     -0.00      &     -0.06   & -54.92 & 0.20 $\pm$ 0.02 & 0.18 $\pm$ 0.03 &     \\ 
   &                 &      1.66      &     -2.84   & -60.32 & 0.14 $\pm$ 0.02 & 0.15 $\pm$ 0.03 &     \\ 
6 & G79.7358+0.9904 & 20:30:50.699 & +41:02:27.43 & -5.60 & 23.21 $\pm$ 0.10 & 23.42 $\pm$ 0.17 & IRAS~20290+4052 \\ 
   &                 &      0.13      &     -0.10   & -3.98 & 5.03 $\pm$ 0.03 & 5.04 $\pm$ 0.06 &     \\ 
   &                 &      0.09      &     -0.07   & -3.08 & 4.28 $\pm$ 0.04 & 4.38 $\pm$ 0.06 &     \\ 
7 & G80.8617+0.3834 & 20:37:00.991 & +41:34:56.18 & -4.16 & 11.79 $\pm$ 0.07 & 11.89 $\pm$ 0.12 & DR20   \\ 
   &                 &      0.04      &      0.29   & -2.00 & 4.18 $\pm$ 0.03 & 4.21 $\pm$ 0.06 &     \\ 
   &                 &     -1.09      &     -0.58   & -11.00 & 0.90 $\pm$ 0.02 & 0.98 $\pm$ 0.04 &     \\ 
   &                 &     -0.12      &     -0.60   & -12.26 & 0.49 $\pm$ 0.01 & 0.51 $\pm$ 0.03 &     \\ 
   &                 &     -1.96      &     -0.05   &  1.06 & 0.35 $\pm$ 0.02 & 0.36 $\pm$ 0.03 &     \\ 
   &                 &      0.52      &      0.21   & -1.10 & 0.30 $\pm$ 0.02 & 0.33 $\pm$ 0.04 &     \\ 
8\tablefootmark{a}  & G81.6790+0.5378 & 20:39:01.165 & +42:19:33.34 & -1.10 & -0.70 $\pm$ 0.03 & -1.04 $\pm$ 0.07 & DR21 \\ 
9 & G81.7219+0.5711 & 20:39:01.066 & +42:22:48.89 & -2.72 & 5.11 $\pm$ 0.04 & 5.09 $\pm$ 0.06 & DR21(OH) \\ 
   &                 &      0.11      &     -0.02   & -3.08 & 3.79 $\pm$ 0.03 & 3.73 $\pm$ 0.06 &     \\ 
   &                 &     -0.29      &     -0.23   & -3.80 & 1.43 $\pm$ 0.03 & 1.50 $\pm$ 0.05 &     \\ 
   &                 &     -1.88      &     -0.78   &  8.98 & 0.23 $\pm$ 0.03 & 0.38 $\pm$ 0.07 &     \\ 
10 & G81.7444+0.5910 & 20:39:00.369 & +42:24:37.00 &  4.48 & 11.72 $\pm$ 0.21 & 12.32 $\pm$ 0.39 & W75S-FIR1 / \\ 
   &                 &      0.19      &     -0.00   &  3.76 & 8.68 $\pm$ 0.10 & 9.02 $\pm$ 0.18 &    DR21 B\tablefootmark{b}    \\ 
11 & G81.7523+0.5908 & 20:39:01.996 & +42:24:59.13 & -8.66 & 14.06 $\pm$ 0.07 & 14.37 $\pm$ 0.13 & W75S-FIR2 / \\ 
   &                 &     -0.06      &     -0.07   & -5.78 & 7.09 $\pm$ 0.05 & 7.30 $\pm$ 0.09 &   DR21 A\tablefootmark{b}   \\ 
   &                 &      0.03      &     -0.06   & -6.86 & 4.88 $\pm$ 0.05 & 5.03 $\pm$ 0.10 &     \\ 
   &                 &      0.24      &     -0.32   & -2.90 & 1.80 $\pm$ 0.05 & 1.92 $\pm$ 0.09 &     \\ 
   &                 &      0.27      &     -0.17   & -2.36 & 1.11 $\pm$ 0.04 & 1.14 $\pm$ 0.08 &     \\ 
12 & G81.7655+0.5972 & 20:39:02.945 & +42:25:50.93 & -1.28 & 3.18 $\pm$ 0.17 & 3.82 $\pm$ 0.33 & CygX-N53 \\ 
13 & G81.8713+0.7807 & 20:38:36.435 & +42:37:34.70 &  7.18 & 336.19 $\pm$ 1.65 & 334.63 $\pm$ 2.89 & W75N(B) \\ 
   &                 &     -0.16      &      0.39   &  4.66 & 286.99 $\pm$ 1.50 & 288.37 $\pm$ 2.65 &     \\ 
   &                 &     -0.13      &      0.46   &  4.12 & 235.08 $\pm$ 1.17 & 236.85 $\pm$ 2.06 &     \\ 
   &                 &     -0.27      &      0.34   &  5.74 & 120.86 $\pm$ 0.58 & 121.43 $\pm$ 1.02 &     \\ 
   &                 &     -0.26      &      0.38   &  5.20 & 109.74 $\pm$ 0.55 & 110.23 $\pm$ 0.97 &     \\ 
   &                 &     -0.12      &      0.45   &  3.40 & 92.67 $\pm$ 0.40 & 94.04 $\pm$ 0.71 &     \\ 
   &                 &     -0.14      &     -0.43   &  9.34 & 37.03 $\pm$ 0.16 & 37.43 $\pm$ 0.29 &     \\ 
14 & G82.3079+0.7296 & 20:40:16.642 & +42:56:29.12 & 10.42 & 20.67 $\pm$ 0.07 & 20.78 $\pm$ 0.13 & -- \\ 
\hline 
\end{tabular}
}
\tablefoot{Column 2 gives the GLOSTAR source name, constructed from Galactic coordinates; Cols. 3 and 4 are equatorial coordinates. We give the absolute position of the maser component with maximum intensity. For the other components, we list position offsets relative to the position of the strongest component. The position uncertainties are $\pm0\rlap.{''}4$ and $\pm0\rlap.{''}3$, in RA\ and Dec., respectively (see Sect. \ref{sec:search}). 
For each feature, Col. 5 gives the LSR radial velocity of the peak. Columns 6 and 7 are the peak and integrated fluxes at the channel corresponding to the peak of the feature. Column 8 gives the common source names used in the literature. For masers 6, 7, 9, 10, 11, and 13, \citet{Rygl2012} reported astrometric VLBI observations.
\tablefoottext{a}{Seen in absorption.}
\tablefoottext{b}{Alternative nomenclature used by \citet{Rygl2012}.}
}
\end{table*}

\begin{table*}
\caption{Properties of methanol sources from B-configuration maps.}
\label{tab:jmfit-B} 
\centering 
{\footnotesize
\begin{tabular}{l c c c c c c c c c c}  
\hline\hline  
Maser   &  Name  & $\alpha$/$\Delta\alpha$  & $\delta$/$\Delta\delta$ & $V_{\rm LSR}$  & $S_{\nu,~\rm Peak}$        & $S_{\nu,~\rm Int.}$ & Common    \\
ID     &         & (h:m:s)/(arcsec) & ($^{\rm o}$:$'$:$''$)/(arcsec) & (km~s$^{-1}$)   &(Jy~beam$^{-1}$)  &   (Jy)    &    Name   \\
(1)    & (2)     & (3)              & (4)                            & (5)             & (6)              & (7)       &     (8)   \\ 
\hline
1 & G76.0932+0.1580 & 20:23:23.7140 & +37:35:35.558 &  6.64 & 0.57 $\pm$ 0.01 & 0.59 $\pm$ 0.02 & -- \\ 
   &                 &     -0.103      &     0.055   &  4.84 & 0.38 $\pm$ 0.01 & 0.48 $\pm$ 0.01 &     \\ 
   &                 &     -0.281      &     -0.005   & -6.50 & 0.17 $\pm$ 0.01 & 0.18 $\pm$ 0.01 &     \\ 
2 & G76.8437+0.1233 & 20:25:43.7657 & +38:11:12.734 & -5.42 & 0.17 $\pm$ 0.00 & 0.17 $\pm$ 0.01 & -- \\ 
3 & G78.8870+0.7087 & 20:29:24.9436 & +40:11:19.626 & -7.04 & 0.10 $\pm$ 0.01 & 0.19 $\pm$ 0.03 & AFGL2591 \\ 
4 & G78.9690+0.5410 & 20:30:22.6762 & +40:09:23.348 &  4.84 & 1.25 $\pm$ 0.01 & 1.28 $\pm$ 0.02 & IRAS~20286+3959 \\ 
5 & G78.9884+0.2211 & 20:31:47.3082 & +39:58:59.903 & -68.96 & 0.25 $\pm$ 0.01 & 0.30 $\pm$ 0.01 & -- \\ 
   &                 &     -0.028      &     -0.103   & -66.26 & 0.21 $\pm$ 0.01 & 0.30 $\pm$ 0.02 &     \\ 
6 & G79.7358+0.9904 & 20:30:50.6680 & +41:02:27.403 & -5.42 & 22.96 $\pm$ 0.21 & 23.81 $\pm$ 0.39 & IRAS~20290+4052 \\ 
   &                 &     0.055      &     -0.010   & -3.98 & 5.88 $\pm$ 0.09 & 6.11 $\pm$ 0.15 &     \\ 
   &                 &     0.045      &     -0.039   & -3.26 & 4.97 $\pm$ 0.06 & 5.07 $\pm$ 0.12 &     \\ 
7 & G80.8617+0.3834 & 20:37:00.9571 & +41:34:55.606 & -3.98 & 4.61 $\pm$ 0.11 & 5.02 $\pm$ 0.21 & DR20 \\ 
   &                 &     -0.029      &     0.331   & -2.00 & 1.56 $\pm$ 0.06 & 1.77 $\pm$ 0.12 &     \\ 
9 & G81.7219+0.5711 & 20:39:01.0501 & +42:22:49.124 & -2.72 & 4.07 $\pm$ 0.05 & 4.30 $\pm$ 0.10 & DR21(OH) \\ 
   &                 &     0.202      &     0.048   & -3.08 & 2.72 $\pm$ 0.05 & 3.12 $\pm$ 0.10 &     \\ 
   &                 &     -0.325      &     -0.133   & -3.80 & 1.30 $\pm$ 0.03 & 1.43 $\pm$ 0.05 &     \\ 
   &                 &     -1.163      &     -0.226   &  8.98 & 1.22 $\pm$ 0.03 & 1.25 $\pm$ 0.06 &     \\ 
10 & G81.7444+0.5910 & 20:39:00.3722 & +42:24:37.089 &  4.48 & 17.18 $\pm$ 0.12 & 17.88 $\pm$ 0.22 & W75S-FIR1 \\ 
   &                 &     0.015      &     0.020   &  3.58 & 6.58 $\pm$ 0.08 & 6.61 $\pm$ 0.14 &     \\ 
11 & G81.7523+0.5908 & 20:39:01.9870 & +42:24:59.261 & -8.66 & 15.19 $\pm$ 0.10 & 15.67 $\pm$ 0.19 & W75S-FIR2 \\ 
   &                 &     -0.014      &     -0.140   & -5.78 & 7.93 $\pm$ 0.07 & 8.12 $\pm$ 0.13 &     \\ 
   &                 &     -0.036      &     -0.107   & -6.86 & 4.29 $\pm$ 0.04 & 4.37 $\pm$ 0.08 &     \\ 
   &                 &     0.123      &     -0.300   & -2.36 & 1.33 $\pm$ 0.03 & 1.45 $\pm$ 0.06 &     \\ 
   &                 &     0.074      &     -0.363   & -2.90 & 1.52 $\pm$ 0.03 & 1.70 $\pm$ 0.07 &     \\ 
12 & G81.7655+0.5972 & 20:39:02.9329 & +42:25:50.954 & -1.28 & 3.90 $\pm$ 0.05 & 4.09 $\pm$ 0.09 & CygX-N53 \\ 
13 & G81.8713+0.7807 & 20:38:36.4232 & +42:37:34.756 &  7.18 & 336.13 $\pm$ 0.96 & 346.76 $\pm$ 1.73 & W75N(B) \\ 
   &                 &     -0.176      &     0.425   &  4.66 & 324.83 $\pm$ 1.15 & 336.46 $\pm$ 2.08 &     \\ 
   &                 &     -0.170      &     0.444   &  4.48 & 238.51 $\pm$ 1.07 & 246.98 $\pm$ 1.95 &     \\ 
   &                 &     -0.261      &     0.346   &  5.74 & 126.12 $\pm$ 0.76 & 128.52 $\pm$ 1.37 &     \\ 
   &                 &     -0.241      &     0.352   &  5.20 & 114.47 $\pm$ 0.74 & 116.80 $\pm$ 1.33 &     \\ 
   &                 &     -0.100      &     0.480   &  3.40 & 88.83 $\pm$ 0.80 & 90.64 $\pm$ 1.43 &     \\ 
   &                 &     -0.122      &     -0.424   &  9.34 & 31.67 $\pm$ 0.61 & 35.80 $\pm$ 1.16 &     \\ 
14 & G82.3079+0.7296 & 20:40:16.6487 & +42:56:29.288 & 10.42 & 23.82 $\pm$ 0.18 & 24.19 $\pm$ 0.31 & -- \\ 
\hline 
\end{tabular}
}
\tablefoot{Column 2 gives source Galactic coordinates; Cols. 3 and 4 are equatorial coordinates. We give the absolute position of the maser component with maximum intensity. For the other components, we list position offsets relative to the position of the strongest component. The position uncertainties are $\pm0\rlap.{''}18$ and $\pm0\rlap.{''}13$, in RA\ and Dec., respectively (see Sect. \ref{sec:search}).
For each feature, Col. 5 gives the LSR radial velocity of the peak. Columns 6 and 7 are the peak and integrated fluxes at the channel corresponding to the peak of the feature. Column 8 gives the common source names used in the literature. 
}
\end{table*}

\begin{figure*}[!ht]
\begin{center}
 \includegraphics[width=0.95\textwidth,angle=0]{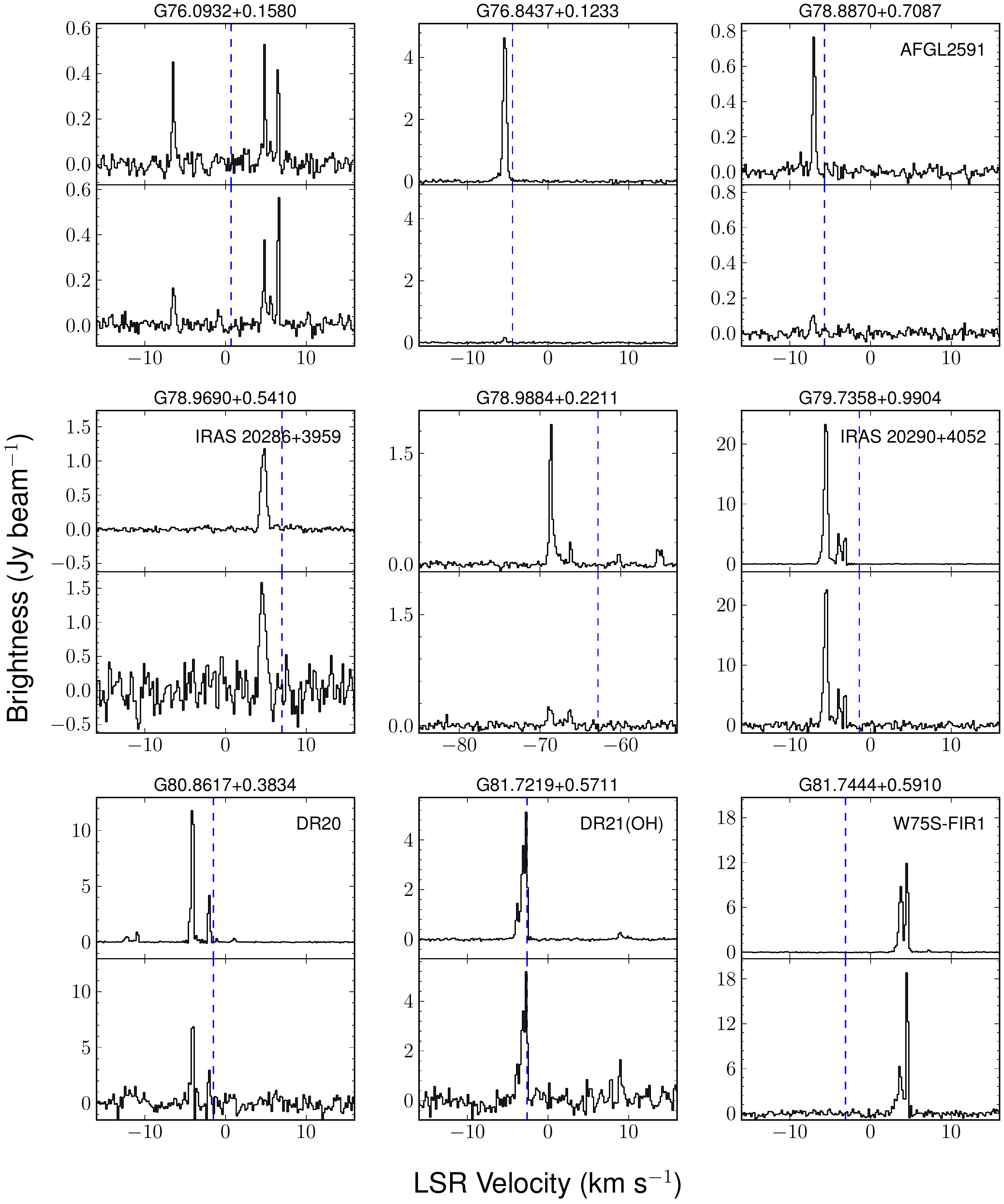} 
\caption{Observed spectra of the 13 sources with methanol maser emission and the one with methanol absorption. For every source, the top and bottom panels represent D-  and B-configuration data, respectively.  
These spectra were extracted at the peak pixel from the data cubes (see text).
The vertical lines show the systemic LSR velocities of the dense molecular gas of the star-forming regions that host the masers, which were taken 
from the literature (see Table \ref{tab:maserCtp}).
}
\label{fig:spectra}
\end{center}
\end{figure*}

\setcounter{figure}{0}
\renewcommand{\thefigure}{2}

\begin{figure*}[!ht]
\begin{center}
 \includegraphics[width=0.95\textwidth,angle=0]{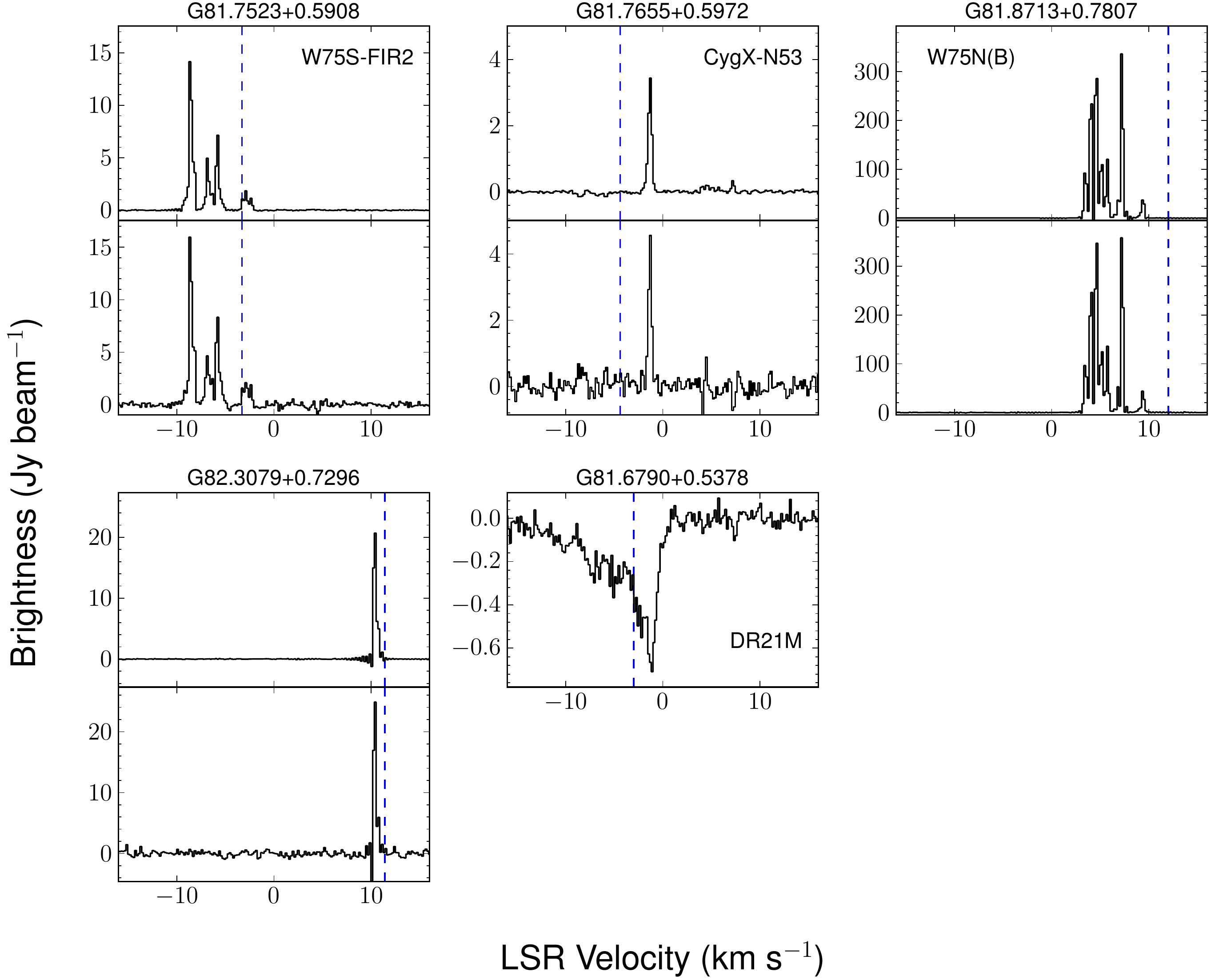}
\caption{{\it Continued.}}
\end{center}
\end{figure*}

\setcounter{figure}{2}
\renewcommand{\thefigure}{\arabic{figure}}

\section{Source detection}\label{sec:search}

 We searched for methanol line sources in the D-configuration dirty cubes using the source extraction code introduced in \cite{BrunthalerPilot}.
Briefly, this code scans  through the full data cubes to locate brightness peaks with a signal-to-noise ratio (S/N) above a fixed value; here we chose a threshold of S/N = 4, which was found to be a good balance between detecting weak sources and dismissing artifacts. For the calculation of the S/N,  the rms noise was measured in a small spatial box (typically with a size of 50$\times$50 pixel). Starting at the bottom-left corner of the image, a pixel within this box is considered to represent a potential source if the S/N of the emission remains above the S/N threshold in at least three consecutive spectral channels. After scanning the bandwidth for this spatial box, the search was repeated in the next adjacent spatial box until the full image ($\sim$1600$\times$2300~pixel) was scanned.  A few additional filters were applied to the candidate detection list in order to remove duplicate detections (both in position and velocity) and spurious detections caused by, for example, sidelobes. That is, we checked if the candidate detection was within 200 pixels of any previous detection. We kept the detection as a potential one when it appeared in the same channel range as the previous detection and had a higher S/N (in this case, we discarded the previous detection) or when it  appeared in a different channel range regardless of its S/N. 
A full description of the source finding algorithm will be given in an upcoming paper (Nguyen et al., in prep).  

Once a list of potential detections was produced by the extraction code, we imaged a region of $512\times512$~pixel ($1024\times1024$~pixel) or $\sim$21$'\times21'$ ($3\rlap{.}'4\times3\rlap{.}'4$) for the D configuration (B configuration) centered on the recorded brightness peak position. Additionally, we used the visibility data from all VLA pointings (observed at any date) that lay within the $21'\times21'$ region.  
The cubes were CLEANed 
to the expected rms by setting a threshold of 38~mJy.
This threshold corresponds to the expected 1$\sigma$ rms for a single field. The final sensitivities, however, are higher (cf.\ Sect. \ref{sec:obs}) because each field was covered by six neighboring fields.
D- and B-configuration data were imaged separately using a velocity spacing of 0.18 km~s$^{-1}$. The spectral extent of these cubes covers
LSR velocities from $-100$ to $+50$~km~s$^{-1}$. The CLEANed cubes produced for all potential source candidates were first scanned with our source extraction code and then inspected visually to search for
detections. Finally, we manually collected a list of channels that showed spectral peaks with a signal above 4$\sigma$. From this list we obtained the source positions and flux densities at individual channels by performing two-dimensional Gaussian fits to the brightness distribution with the {\tt CASA} task {\tt imfit}. 
For the Gaussian fits we selected a box, typically with spatial extents of $14\times14$~pixel and $20\times20$~pixel in the D  and B configuration, respectively ($\approx$1.3-1.7 times the beam), centered on the position of the brightest pixel. 

The statistical error in maser position is given by the astrometric uncertainty, ${\theta_{\rm res}}/ {(2\times{\rm S/N})}$, where $\theta_{\rm res}$ is the full width at half maximum (FWHM) size of the restoring beam and S/N is that of the source \citep{Thompson2017}. The D- and B-configuration maps of methanol have an average beam size of $13''$ and $1\rlap.{''}2$, respectively. For a maser detected at S/N=10, the formal (statistical) precision in position is $\approx 0\rlap.{''}7$ and $\approx 0\rlap.{''}06$ for the D  and B configurations, respectively. However, additional errors -- due to, for instance, uncompensated atmospheric delays that arise when applying the phase corrections determined for the calibrator sources to the Cygnus X data -- likely worsen the position accuracy. In order to estimate realistic position uncertainties, we compared our VLA positions with the positions determined from observations with the European Very-long-baseline interferometry (VLBI) Network (EVN; \citealt{Rygl2012}; see Appendix \ref{sec:supplm} for details). For the B configuration, the differences have rms values of $0\rlap.{''}15$ and $0\rlap.{''}12$ and mean values of $0\rlap.{''}09$ and $0\rlap.{''}05$ in RA\ and Dec., respectively. Thus, additional position errors of $\pm0\rlap.{''}18$ and $\pm0\rlap.{''}13$, respectively, need to be added to the formal errors. Similarly, from the differences between the VLA D configuration and EVN positions, we obtained systematic errors of  $\pm0\rlap.{''}4$ and $\pm0\rlap.{''}3$ in RA\ and Dec., respectively, for the D configuration.

\section{Results}\label{sec:results}

\subsection{Detected methanol masers}

We detected methanol maser emission toward 13 different locations and one case of methanol absorption (toward the continuum emission of the compact HII region DR21); two of these maser sources (G76.8437+0.1233 and G78.9884+0.2211) are newly detected.  The spatial distribution of these methanol sources 
is shown in Fig.\ \ref{fig:allSource}, overlaid on a 
mosaic image obtained at 500~$\mu$m with the Spectral and Photometric Imaging REceiver (SPIRE) aboard the Herschel Space Observatory  \citep{Cao2019} from data taken as part of the Herschel imaging survey of OB Young Stellar objects (HOBYS;  \citealt{Motte2010,Hennemann2014}) and the Herschel infrared Galactic Plane Survey (Hi-Gal; \citealt{Molinari2010}). Remarkably, several of the masers (five in total) are associated with the ridge of dense molecular material that extends from DR21 to W75N, which is the region with the highest degree of star formation activity in Cygnus X and home to numerous high-mass stars. 

Table \ref{tab:jmfit-D} lists the names, equatorial coordinates, peak LSR velocity, and peak and integrated fluxes of the main velocity components (or {``features''}) of the methanol sources detected in the D-configuration maps. Maser properties of the features as measured in the B-configuration
cubes are shown in Table \ref{tab:jmfit-B}, while properties and maps of the distribution of all detected maser spots above 4$\sigma$ are provided in the appendix (Table \ref{tab:catalog} and Fig. \ref{fig:spots}, respectively). Here,  ``spot'' refers to emission detected in a single velocity channel and ``feature'' to emission observed in contiguous velocity channels at nearly the same position. The features consist of several contributing spots.

Class II methanol masers are, in general, time variable, and the 8--9 month time span between our D- and B-configuration observations (see Table\,\ref{tab:obs}) allows us to check the variability in our sources. Individual maser components can spread by up to several arcseconds on the sky (see Fig. \ref{fig:spots} and also \citealt{Hu2016}), which is comparable to our B-configuration synthesized  beam; as such, to allow for a meaningful comparison, we 
convolved the B-configuration cubes to the D-configuration beam
and then, for every source, extracted each spectrum at the same pixel. 

In Fig.\,\ref{fig:spectra}, the D- and B-configuration spectra are compared with one another. We note that the noise in the B-configuration images convolved with the D-configuration beam is higher by a factor of $\approx$10. However, this is expected since the noise, in brightness temperature units, decreases with the square root of the beam area, and the brightness temperature is converted to flux density by multiplying with the beam area. This results in a linear increase in the noise with the beam size. Because of this, we do not show the  spectra from the convolved data cubes for the weakest lines (those with $S_{\rm peak}<1$~Jy~beam$^{-1}$) in Fig.\ \ref{fig:spectra}.
At our 0.18 km~s$^{-1}$ channel spacing, even narrow single velocity features are generally spectrally resolved. In 
a few cases, extremely narrow features are unresolved or only 
partially resolved. In the case of strong features, 
this leads to observable ``Gibbs ringing,'' which appears most 
pronounced in the G82.3079+0.7296 spectrum between 7 and 10 km~s$^{-1}$   but can also be seen in a few of the other 
maser spectra. The vertical lines show the systemic velocities of 
the molecular gas in the line of sight, which were taken from the literature.  
The LSR velocity of all but one of the maser features ranges from $-12$ to $+10$~km~s$^{-1}$. The exception is source G78.9884+0.2211, which has $V_{\rm LSR}\approx-70$~km~s$^{-1}$. In all cases, methanol-maser velocities are observed to be close, within $\lesssim7$~km~s$^{-1}$, to the systemic velocities.

\begin{table*}
\caption{Properties of the cores associated with methanol masers.}
\label{tab:coresProp} 
\centering 
\footnotesize
\begin{tabular}{l c c c c c c c c c c}  
\hline\hline 
 GLOSTAR  & Common  &  $\Delta V_{\rm D}$ & $\Delta V_{\rm B}$ & $L_{\rm maser,~D}$ & $L_{\rm maser,~B}$ & Size & $M_{\rm core}$ &  $T_{\rm core}$ & $L_{\rm FIR}$   & $L_{\rm Bol}$ \\ 
 Name &  Name & (km s$^{-1}$) & (km s$^{-1}$) & ($L_\odot$)   &    ($L_\odot$)  & ($''$) & ($M_\odot$) & (K) &  ($L_\odot$)    &  ($L_\odot$)         \\ 
 (1)      &         (2)      &     (3)         &       (4)      &       (5)  & (6) & (7) & (8) & (9) & (10)  & (11)   \\
\hline 
G76.0932+0.1580  &  -  &  1.3  &  1.6  &  5.8$\times 10^{-9}$  &  5.9$\times 10^{-9}$  &  18  &   27 $\pm$   3  &  20.4 $\pm$ 0.9  &    189  &  227  \\ 
G76.8437+0.1233  &  -  &  2.2  &  0.5  &  3.8$\times 10^{-8}$  &  1.0$\times 10^{-9}$  &  25  &   12 $\pm$   2  &  16.0 $\pm$ 0.7  &     19  &  26  \\ 
G78.8870+0.7087  &  AFGL2591  &  0.7  &  0.9  &  2.3$\times 10^{-8}$  &  2.4$\times 10^{-8}$  &  24  &  937 $\pm$  72  &  30.2 $\pm$ 1.9  &  67750  &  66262  \\ 
G78.9690+0.5410  &  IRAS 20286+3959  &  1.4  &  1.4  &  1.3$\times 10^{-8}$  &  1.4$\times 10^{-8}$  &  22  &   72 $\pm$   9  &  13.0 $\pm$ 0.4  &     33  &  42  \\ 
G78.9884+0.2211\tablefootmark{a}  &  -  &  3.6  &  0.7  &  1.8$\times 10^{-8}$  &  2.9$\times 10^{-9}$  &  25  &    8 $\pm$   1  &  21.3 $\pm$ 1.2  &     69  &  176  \\ 
G79.7358+0.9904  &  IRAS 20290+4052  &  3.4  &  3.2  &  2.4$\times 10^{-7}$  &  2.4$\times 10^{-7}$  &  19  &   42 $\pm$   4  &  17.7 $\pm$ 0.5  &    122  &  233  \\ 
G80.8617+0.3834  &  DR20  &  4.7  &  1.4  &  1.0$\times 10^{-7}$  &  3.7$\times 10^{-8}$  &  23  &  122 $\pm$  11  &  20.9 $\pm$ 0.8  &    972  &  1328  \\ 
G81.7219+0.5711  &  DR21(OH)  &  2.3  &  2.5  &  5.2$\times 10^{-8}$  &  4.9$\times 10^{-8}$  &  16  &  930 $\pm$ 140  &  20.9 $\pm$ 1.3  &   7344  &  7299  \\ 
G81.7444+0.5910  &  W75S-FIR1  &  2.2  &  2.2  &  1.3$\times 10^{-7}$  &  1.3$\times 10^{-7}$  &  11  &   74 $\pm$   5  &  21.9 $\pm$ 0.6  &    776  &  1487  \\ 
G81.7523+0.5908  &  W75S-FIR2  &  5.4  &  5.4  &  1.8$\times 10^{-7}$  &  2.0$\times 10^{-7}$  &  9  &  298 $\pm$  34  &  16.4 $\pm$ 0.6  &    552  &  955  \\ 
G81.7655+0.5972  &  CygX-N53  &  1.3  &  1.6  &  2.3$\times 10^{-8}$  &  2.2$\times 10^{-8}$  &  11  &  325 $\pm$  27  &  14.5 $\pm$ 0.3  &    294  &  221  \\ 
G81.8713+0.7807  &  W75N(B)  &  10.3  &  7.2  &  5.8$\times 10^{-6}$  &  5.8$\times 10^{-6}$  &  --  &  1020 $\pm$  91  &  27.3 $\pm$ 1.3  &  40588  &  32930  \\ 
G82.3079+0.7296  &  -  &  2.3  &  1.3  &  1.2$\times 10^{-7}$  &  1.3$\times 10^{-7}$  &  16  &   24 $\pm$   3  &  19.7 $\pm$ 0.8  &    139  &  183  \\ 
\hline 
\end{tabular}
\tablefoot{Columns 1 and 2 report the maser name, which is based on Galactic coordinates, and its common label; Cols. 3 and 4 are the velocity ranges of the maser emission in D- and B-configurations; Cols. 5 and 6 are the maser integrated luminosities; Col. 7 is the deconvolved size at 160~$\mu$m band, where the dust is optically thin. It corresponds to the  intensity-weighted diameter given by $\theta_D = 2 \sum\limits_{i} r_i A_i / \sum\limits_i A_i$, where $r_i$ is the distance of the $i$th pixel to the center of the source and $A_i$ is the pixel intensity \citep{Purcell2013}; Cols. 8, 9, 10, and 11 are the  mass, temperature, integrated FIR luminosity, and bolometric luminosity of the associated dust cores.
\tablefoottext{a}{The mass, FIR luminosity, and bolometric luminosity of this core must be rescaled to 347~$M_\odot$, 3093~$L_\odot$, and 6237~$L_\odot$, respectively, for a distance of 9.35~kpc (see Sect. \ref{sec:comments}). }
}
\end{table*}

We see small differences in the maser emission between D- and B-configuration data. In general, fewer maser features (i.e., emission observed in contiguous velocity channels at nearly the same position) are seen in the B-configuration images.
The peak fluxes are also weaker in the B-configuration data for five sources; for these sources we see variations in the range of 20--90\%, while the velocity profiles do not change significantly (Fig. \ref{fig:spectra}). This variability is further discussed for individual sources in Sect. \ref{sec:comments}. 

\begin{figure}[tb]
\begin{center}
\includegraphics[width=0.35\textwidth,angle=0]{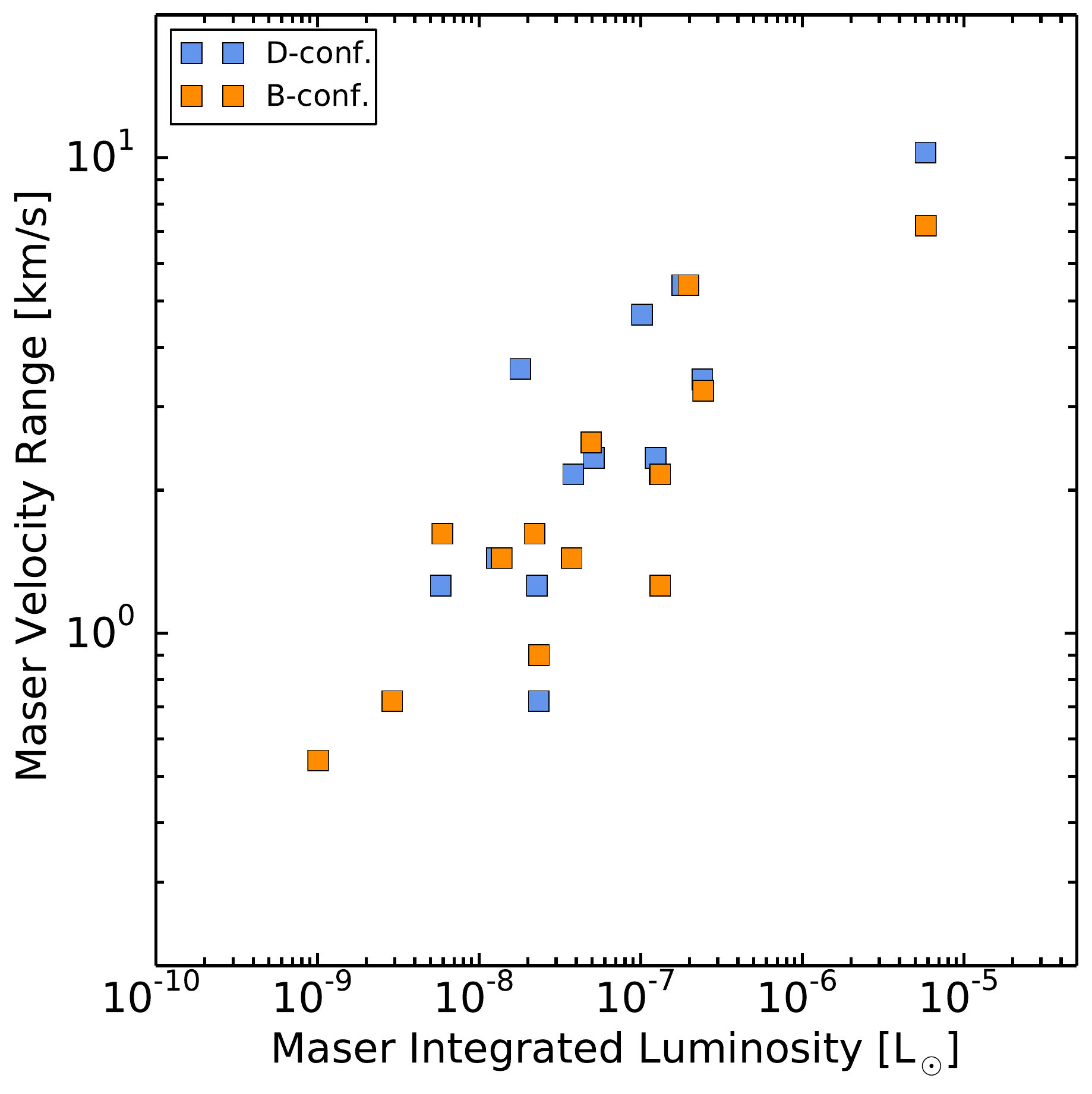}
\caption{Velocity range of detected maser emission in Cygnus X as a function of maser integrated luminosity. The colors indicate D- and B-array configurations (blue and orange, respectively).}
\label{fig:maser-vel-lum}
\end{center}
\end{figure}

Maser luminosities were estimated using
\begin{center}
\begin{equation}
L_{\rm maser} = 4\pi D^2 S_{\rm int} f/c, 
\end{equation}
\end{center}

\noindent where $S_{\rm int}$, in units of Jy~km~s$^{-1}$, is the maser flux integrated over the velocity range $\Delta V$ (i.e.,\ across the maser line with emission above 4$\sigma)$, $f=6668.5192$~MHz is the rest frequency of the $5_1-6_0~A^+$ methanol line, $c$ is the speed of light, and $D$ is the distance to the source. The term $f/c = \Delta f/\Delta v$ takes the change in frequency, $\Delta f$, into account, resulting from a change in velocity of $\Delta v=0.18$~km~s$^{-1}$. The velocity ranges as well as the resulting values of the maser luminosity are given in Table \ref{tab:coresProp}. For all sources (except AFGL~2591) we used a distance of $D=1.4$~kpc, which is the average distance of the Cygnus X complex measured from trigonometric parallaxes of several maser sources (\citealt{Rygl2012}; see also \citealt{Dzib2013}, who obtain a parallax measurement to a radio continuum source in the Cygnus~OB2 association). On the other hand, source AFGL~2591 is much farther away, at 3.33~kpc \citep{Rygl2012}, based also on maser parallax measurements. Figure \ref{fig:maser-vel-lum} shows the velocity range of the maser emission as a function of integrated maser luminosity.  
As expected, the maser velocity range (which can be used as a proxy for the line profile complexity) increases with maser luminosity. A similar relation was previously noticed by, for example,\ \cite{Breen2011} and \cite{Billington2019}. This may simply be the result of the fact that for a maser whose strongest component has a low flux density, any additional even weaker velocity components are below the S/N cutoff.

\begin{figure*}[tbh]
\begin{center}
\includegraphics[width=1.0\textwidth,angle=0]{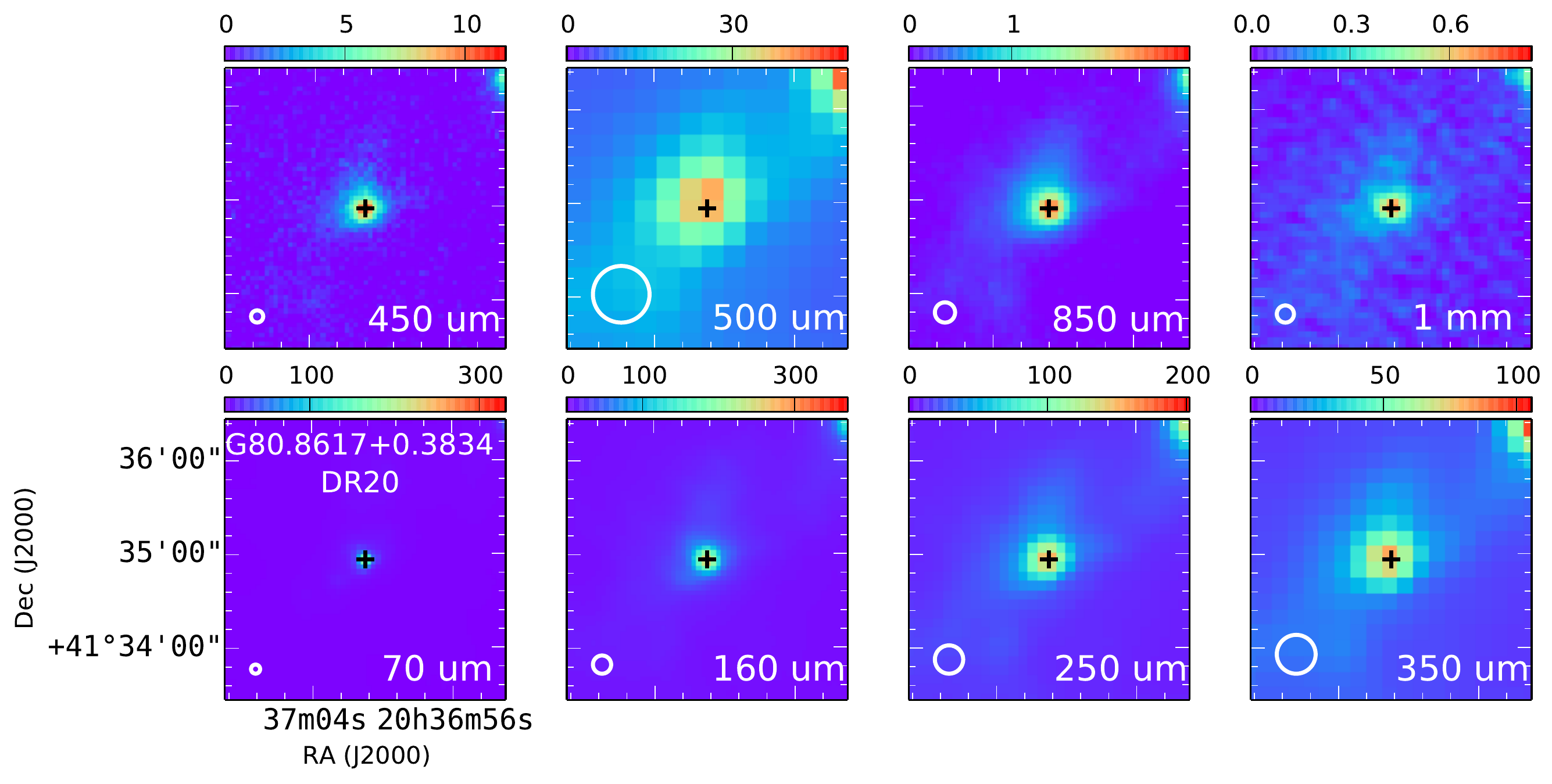} 
\caption{Dust continuum emission  images of DR20 from: {\it Herschel}  PACS (70 and 160~$\mu$m) and SPIRE (250, 350, and 500~$\mu$m); JCMT SCUBA-2 (450 and 850~$\mu$m); and either IRAM MAMBO/MAMBO-2 (1.2~mm) or CSO Bolocam (1.1~mm) instruments. 
The color scale is in units of Jy~beam$^{-1}$.
The VLA position of the detected methanol maser is indicated by the black cross. 
The galactic coordinate name of the GLOSTAR source is indicated at the top of the bottom-left panel.  The beams are shown at the bottom-left corner of each panel. Saturated pixels are shown in gray. 
The same figure for all sources is provided in Appendix \ref{sec:supplm}.
}
\label{fig:submm}
\end{center}
\end{figure*}

\begin{figure*}[t]
\begin{center}
 \includegraphics[width=0.7\textwidth,angle=0]{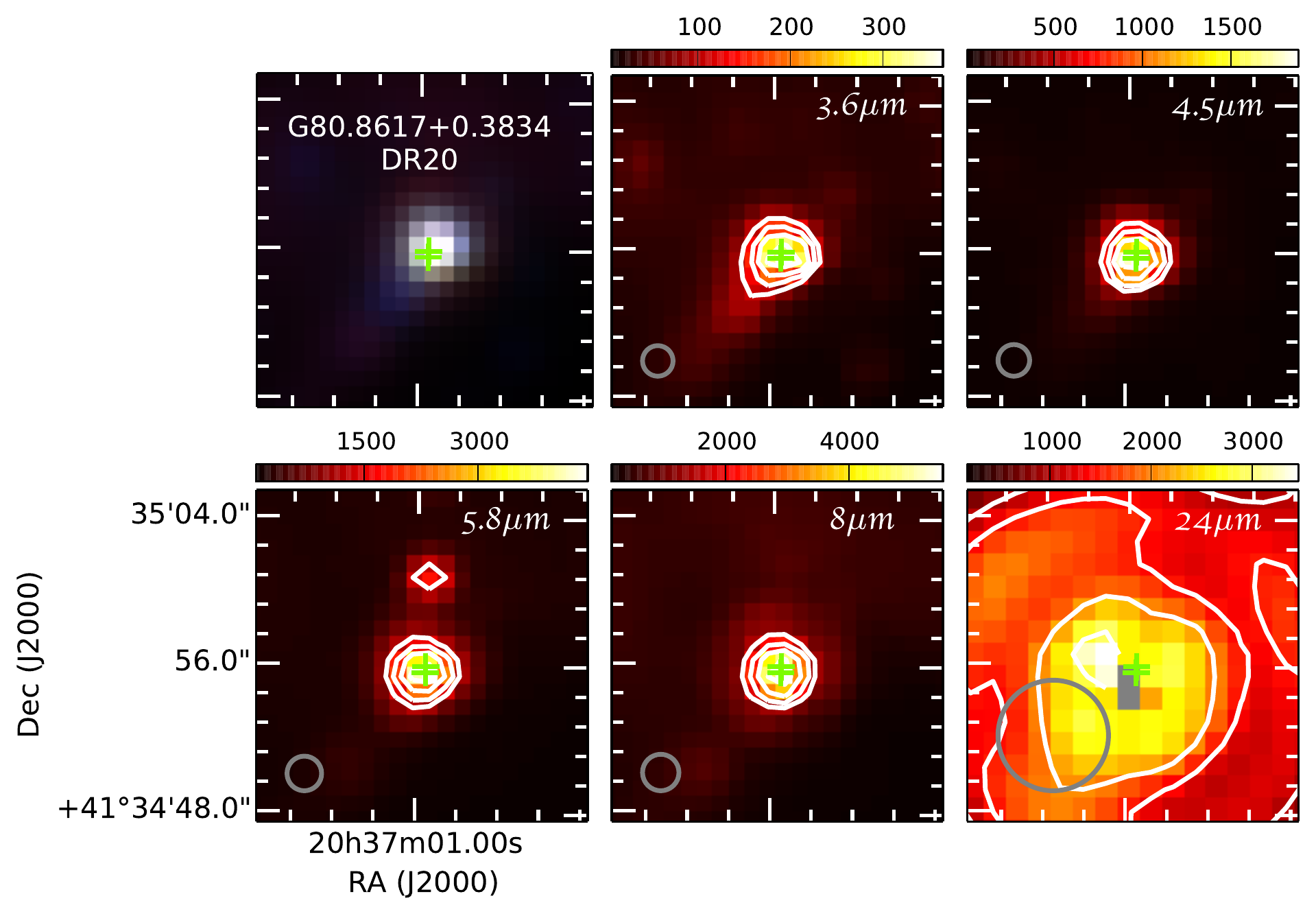} 
\caption{Infrared images (color scale and white contours) of the environment 
of the VLA-detected methanol masers toward DR20. 
The top-right panel shows a three-color map constructed using {\it Spitzer} 3.6~$\mu$m (blue), 4.5~$\mu$m (green), and 8~$\mu$m (red) images. The other four panels show infrared emission for each of the {\it Spitzer} bands. 
The $n$th white contour is at $\left({\sqrt{2}}\right)^{n}\times S_{\rm max} \times p$, where $S_{\rm max}$ is the maximum flux shown in the color bar (MJy~sr$^{-1}$) for each panel, $n$=0, 1, 2 ..., and $p$ is equal to 30\%.
The green crosses mark the positions of the maser features seen in B-array configuration maps. 
The Galactic 
coordinate name of the GLOSTAR source is indicated at the top of the left panel. The beams are shown at the bottom-left corner of each panel. Saturated pixels are shown in gray.
The same figure for all sources is provided in Appendix \ref{sec:supplm}.
}
\label{fig:infrared} 
\end{center}
\end{figure*}

\subsection{Association with millimeter-to-submillimeter emission}\label{sec:submm}

To compare the 6.7~GHz methanol maser activity with other tracers of star formation, we first compared the maser positions with those of the submillimeter continuum sources that pinpoint
the high density gas. For this, we used JCMT Submillimeter Common-User Bolometer Array 2 (SCUBA-2)
maps of Cygnus X at 450 and 850 micron \citep{Cao2019}.
We found that all detected masers have associated 850~$\mu$m  emission (Fig. \ref{fig:submm}) with a comparable beam size of 13$''$ (cf.\ Table 1 in  \citealt{Cao2019}). Thus, all our methanol masers are associated with dust continuum emission. This is consistent with the results of \cite{Urquhart2015} and \cite{Billington2019}, who found an almost ubiquitous association (99\%) between the 6.7~GHz methanol maser and dust continuum sources. In order to derive the properties (mass, temperature, and luminosities) 
of these dust cores, we fit the spectral energy distribution (SED) of the dust emission with a modified blackbody model,
\begin{center}
\begin{equation}\label{eq:sed}
F_\nu = \frac{\kappa_\nu B_\nu(T) M}{D^2},
\end{equation}
\end{center}
\noindent where $B_\nu(T)$ is the Planck function for the temperature, $T$, of the dust and $M$ is the core mass. 
Here, we adopted a dust mass opacity of $\kappa_\nu=\kappa_0 (\nu/\nu_0)^\beta$, where $\kappa_0=0.1~{\rm cm}^2~\rm{g}^{-1}$, $\nu_0=1$~THz, and $\beta=2$.
For the construction of the SED (see Fig.\ \ref{fig:sed-fit}), we used data from {\it Herschel} obtained as part of the HOBYS \citep{Motte2010,Hennemann2014} 
and Hi-GAL projects \citep{Molinari2010} in the Photodetector Array Camera and Spectrometer (PACS) 
70 and 160~$\mu$m and the Spectral and Photometric Imaging Receiver (SPIRE) 250, 350, and 500~$\mu$m bands\footnote{Data were retrieved from the Canadian Astronomy Data Centre.}, as well as JCMT SCUBA-2 data at 450 and 850~$\mu$m \citep{Cao2019}. In addition,  data at 1.1~mm are available from the Bolocam instrument installed at the 10.4 m Caltech Submillimeter Observatory, which have a resolution of 33$''$ and were taken as part of the Bolocam Galactic Plane Survey \citep[BGPS;][]{Aguirre2011,Rosolowsky2010}. At a higher angular resolution of 11$''$, 1.2~mm data are also available from the Max-Planck-Millimeter-Bolometer (MAMBO/MAMBO-2) 
instruments installed at the Institut de Radioastronomie Millim\'etrique (IRAM) 
telescope \citep{Motte2007}. Not all dust  cores associated with the methanol sources were covered by the observations of \cite{Motte2007}. Thus, we used 1.2~mm IRAM data for eight sources and 1.1~mm Bolocam data for six sources. Figure \ref{fig:submm} displays the images of (sub)millimeter continuum data for the GLOSTAR 6.7~GHz methanol masers. The images are $3'\times3'$ in size and are centered at the positions of the masers.

 In the SED we also show infrared data from the Infrared Array Camera (IRAC) 3.6, 4.5, 5.8, and 8~$\mu$m and the Multiband Imaging Photometer for Spitzer (MIPS) 24~$\mu$m 
 bands taken as part of the Spitzer Legacy Survey of the Cygnus~X Complex \citep{Beerer2010}. IRAC aperture photometry and MIPS point spread function (PSF) photometry were retrieved from the NASA/IPAC Infrared Science Archive (IRSA\footnote{\url{https://irsa.ipac.caltech.edu/data/SPITZER/Cygnus-X/}}). The corresponding images can be seen in Fig. \ref{fig:infrared}, where we present both three-color infrared images and single band images of $0\rlap.{'}7\times0\rlap.{'}7$ in size around the location of the masers.

\begin{table*}
\caption{Measured properties of radio continuum sources detected in GLOSTAR B-array maps close to the methanol line sources. }
\label{tab:contB} 
\centering 
\begin{tabular}{l c c c c c c c c c c}  
\hline\hline  
Name &   $l$             &     $b$         &  $\alpha$& $\delta$ & $S_{\rm Peak}$\tablefootmark{a}  & SNR & Offset\tablefootmark{b}  \\ 
           & ($^{\rm o}$) &($^{\rm o}$) & (J2000)  &   (J2000)&(mJy~beam$^{-1}$) &          & ($''$)   \\ 
    (1)  &         (2)        &     (3)          &     (4)      &       (5)    &      (6)                &    (7)  & (8)       \\
\hline 
G78.8870+0.7087 & 78.886737 & 0.708870 & 20h29m24.85s & +40d11m19.7s & 1.69 $\pm$ 0.37 & 4.5 & 1.2 \\ 
G80.8617+0.3834 & 80.861469 & 0.383462 & 20h37m00.91s & +41d34m56.0s & 0.79 $\pm$ 0.13 & 6.0 & 0.6 \\ 
G81.6790+0.5378\tablefootmark{c} & 81.678758 & 0.537054 & 20h39m01.30s & +42d19m31.5s & 1210  $\pm$  4\tablefootmark{d} & 333 & 2.4 \\ 
G81.7219+0.5711 & 81.721609 & 0.571246 & 20h39m00.95s & +42d22m48.8s & 3.05 $\pm$ 0.40 & 7.6 & 0.3 \\ 
G81.7444+0.5910 & 81.744319 & 0.591176 & 20h39m00.30s & +42d24m37.4s & 1.69 $\pm$ 0.22 & 7.8 & 0.8 \\ 
G81.8713+0.7807 & 81.871218 & 0.780755 & 20h38m36.39s & +42d37m34.9s & 4.44 $\pm$ 0.08 & 58.2 & 0.4 \\ 
G76.0932+0.1580 & -- & -- &    --    &    --    & $< 0.16 $ &    --    \\
G76.8437+0.1233 & -- & -- &    --    &    --    & $< 0.18 $ &    --    \\
G78.9690+0.5410 & -- & -- &    --    &    --    & $< 0.20 $ &    --    \\
G78.9884+0.2211 & -- & -- &    --    &    --    & $< 0.33 $ &    --    \\
G79.7358+0.9904 & -- & -- &    --    &    --    & $< 0.22 $ &    --    \\
G81.7523+0.5908 & -- & -- &    --    &    --    & $< 0.63 $ &    --    \\
G81.7655+0.5972 & -- & -- &    --    &    --    & $< 0.45 $ &    --    \\
G82.3079+0.7296 & -- & -- &    --    &    --    & $< 0.19 $ &    --    \\
\hline 
\end{tabular}
\tablefoot{The beam size of the continuum maps is $1\rlap.{''}5\times1\rlap.{''}5$.
\tablefoottext{a}{A $3\sigma$ upper limit is given for the cases where no radio continuum emission is detected around
the position of methanol masers. }
\tablefoottext{b}{Offset in position of the radio continuum source from the methanol maser intensity peak.}
\tablefoottext{c}{Methanol line is seen in absorption.}
\tablefoottext{d}{This flux was measured in the map made for the combination of D- and B-configuration data.}
}
\end{table*}

\begin{table*}
\caption{Properties of the detected methanol line sources.}
\label{tab:maserCtp} 
\centering 
\footnotesize
\begin{tabular}{l c c c c c c c c c c}  
\hline\hline  
ID   & GLOSTAR Name & SIMBAD or Common Name & Possible  & 1.2~mm                 & 6-GHz      & $V_{\rm LSR}$\tablefootmark{b}  & $V_{\rm LSR}$ \\ 
     &              &                       &  Nature   & core\tablefootmark{a}  & continuum  & (km~s$^{-1}$)  & Ref.          \\ 
 (1) &      (2)     &    (3)                &    (4)    &        (5)             &     (6)    &  (7)           & (8)           \\
\hline 
1  &  G76.0932+0.1580  &   J202323.73+373535.16\tablefootmark{d}  &  pre-UC HII  &  -  &  -  &  0.7  &  1  \\ 
2  &  G76.8437+0.1233  &   J202543.79+381112.98\tablefootmark{d}  &  pre-UC HII  &  -  &  -  &  -4.4  &  1  \\ 
3  &  G78.8870+0.7087  &  AFGL2591  &  UC HII region  &  S26  &  yes  &  -5.7  &  2  \\ 
4  &  G78.9690+0.5410  &  IRAS 20286+3959  &  pre-UC HII  &  -  &  -  &  7.0  &  3  \\ 
5  &  G78.9884+0.2211  &   J203147.25+395900.33\tablefootmark{d}  &  pre-UC HII  &  -  &  -  &  -62.8  &  4  \\ 
6  &  G79.7358+0.9904  &  IRAS 20290+4052  &  pre-UC HII  &  -  &  -  &  -1.4  &  5  \\ 
7  &  G80.8617+0.3834  &  DR20, IRAS 20352+4124  &  UC or HC HII region  &  N14  &  yes  &  -1.5  &  6  \\ 
8\tablefootmark{c}  &  G81.6790+0.5378  &  DR21  &  UC HII region  &  N46  &  yes  &  -3.0  &  7  \\ 
9  &  G81.7219+0.5711  &  DR21(OH)  &  UC HII region  &  N44  &  yes  &  -2.6  &  3  \\ 
10  &  G81.7444+0.5910  &  W75S-FIR1, DR21~B  &  UC or HC HII region  &  N43  &  yes  &  -3.1  &  7  \\ 
11  &  G81.7523+0.5908  &  W75S-FIR2, DR21~A  &  high-luminosity IR protostellar core  &  N51  &  -  &  -3.3  &  7  \\ 
12  &  G81.7655+0.5972  &  CygX-N53  &  massive IR-quiet protostellar core  &  N53  &  -  &  -4.4  &  8  \\ 
13  &  G81.8713+0.7807  &  W75N(B)  &  UC HII region  &  N30  &  yes  &  12.0  &  9  \\ 
14  &  G82.3079+0.7296  &   J204016.75+425629.49\tablefootmark{d}  &  pre-UC HII  &  -  &  -  &  11.4  &  1  \\ 
\hline 
\end{tabular}
\tablefoot{ 
\tablefoottext{a}{From \cite{Motte2007}.}
\tablefoottext{b}{Systemic LSR velocity of molecular gas.}
\tablefoottext{c}{Seen in absorption.}
\tablefoottext{d}{From \cite{Kryukova2014}.}
References: 1. \cite{Schneider2006}; 2. \cite{vanderTak1999}; 3. \cite{Keown2019}; 4. \cite{Yamagishi2018ApJS}; 5.
\cite{Bronfman1996}; 6. \cite{Pipenbrink1988}; 7. this work; 8. \cite{Duarte-Cabral2014}; 9. \cite{Gottschalk2012}. 
}
\end{table*}

The mass and luminosity of five maser-associated cores are not available in the literature, while for the other eight these properties have been measured, for example by \cite{Cao2019}. We thus derived these properties from the SED fitting not only for the five missing cores, but -- for consistency and to avoid systematic deviations introduced by different extraction techniques -- for all cores that have associated methanol maser emission detected by GLOSTAR. When fitting the SED, we note that the point at 70~$\mu$m shows, in almost all cases, significant excess emission to the fit since emission at this wavelength traces warm dust that cannot be reproduced with a single-temperature model \citep{Cao2019,Koenig2017}.
The 70 and 24~$\mu$m data points can be reproduced if we add a second, warmer component to the SED fit (see Fig.\ \ref{fig:sed-fit}). We thus carried out a two-component fit for sources for which 70 and 24~$\mu$m data are both available. We fit Eq. \ref{eq:sed} to 24--70 $\mu$m and 160--1000 $\mu$m data for the warm and cold component, respectively. The fit was done with the Python routine {\tt curve\_fit} from the {\tt scipy} package. The fluxes from all images that are saturated at the location of the masers (cf.\ Fig. \ref{fig:submm-appendix} and \ref{fig:infrared2-appendix}) are not shown in the SED and were thus not used in the fitting. Specifically, saturation occurs for G78.8870+0.7087\footnote{The image of this source has issues in the 3.6, 4.5, 5.8, and 8 $\mu$m bands as well. Thus, the fluxes from these bands are not shown in the SED.} (24 and 250~$\mu$m), G80.8617+0.3834 (24~$\mu$m), G81.7219+0.5711 (250 and 350~$\mu$m), G81.7444+0.5910 (24~$\mu$m), G81.7523+0.5908 (24~$\mu$m), and G81.8713+0.7807 (24, 160, 250, and 350~$\mu$m). 
In the case of G78.8870+0.7087 and G81.8713+0.7807, for which we have a limited number of data points due to saturation, the 70 $\mu$m point was included in the fit of the cold component, and the data were weighted with their 20\% flux errors.
The IRAC data points between 3.6 and 8~$\mu$m -- likely originating from a third, hot and inner component -- were not used in the SED fitting.

Finally, far-infrared (FIR) luminosities were obtained by integrating the SED of the cold component over wavelength,
\begin{equation}
L_{\rm FIR}=4 \pi D^2 \int_{0}^{+\infty} S_\lambda d\lambda.    
\end{equation}

To obtain bolometric luminosities from the SED, we should also have integrated the emission from the near- and mid-infrared; however, these data are not available for all our sources (cf.\ Fig. \ref{fig:infrared2-appendix}). Instead, we estimated the bolometric luminosities of the cores by using the empirical relation between luminosity and 70~$\mu$m flux found by \cite{Dunham2008}, 
\begin{center}
\begin{equation}
L_{\rm bol} =   3.3\times10^8 \left( \left( \frac{\nu F_{70~\mu{\rm m}}} {{\rm ergs~cm}^{-2}~{\rm s}^{-1}}\right)  \left( \frac{D}{140~{\rm pc}}\right)^2 \right)^{0.94}~L_\odot .
\end{equation}
\end{center}

The derived parameters for the cold components (masses, FIR luminosities, and bolometric luminosities) are listed in Table \ref{tab:coresProp}, and the SED and model fit are shown in Fig. \ref{fig:sed-fit}. We found that the two methods give luminosities that agree within 1$\sigma$ in eleven sources and within 3$\sigma$ in two sources. We now compare our results with those previously published in the literature.

For the core extraction, \cite{Cao2019} selected regions with column density $N_{\rm{H}_2} \geq 3.5\times10^{22}~{\rm cm}^{-2}$, which, for typical deconvolved sizes of 0.1~pc, corresponds to core masses above 23~$M_\odot$. Additional criteria were used by \cite{Cao2019} to reject cores with poorly extracted fluxes and/or large relative errors in mass and temperature estimation from the SED fitting, resulting in a minimum mass of 36~$M_\odot$ in their catalog.
We see that some methanol maser-associated cores not reported previously have masses close to the threshold of 23~$M_\odot$ adopted by \cite{Cao2019}, which explains why they were not reported by these authors. Although we follow a similar approach to \cite{Cao2019} to estimate core properties, here the fluxes were extracted individually at each wavelength using the {\tt radio flux} plug-in in {\tt DS9}\footnote{\url{https://www.extragalactic.info/\textasciitilde mjh/radio-flux.html}}. The integrated source flux was obtained over an aperture that delimits emission within a 30\% contour. 
The local background flux was estimated over an aperture farther out from the source and then subtracted from the source flux. 
We note that fluxes extracted in this way do not show large deviations from the model fit, unlike some fluxes extracted with the {\tt getsources} package used by \cite{Cao2019}, especially at 450~$\mu$m. The (sub)millimeter emission in the DR21 ridge is complex and extended, which probably affects the performance of background subtraction by {\tt getsources}. As a result, the masses derived by us agree with those from \cite{Cao2019} within $\sim$55\%.

When comparing the estimated masses with the masses obtained by \cite{Motte2007} for the eight overlapping cores from 1.2 mm data alone, after correcting for
the different distance and dust mass opacity used by these authors, we see a slightly better agreement, with a relative difference of $\sim$50\% on average. The properties derived above for the dust cores that are associated with a methanol maser source are used in Sect. \ref{sec:discuss} to investigate the relationship between methanol masers and massive star-forming dust cores. 

\begin{figure*}[ht]
\begin{center}
\includegraphics[width=0.38\textwidth,angle=0]{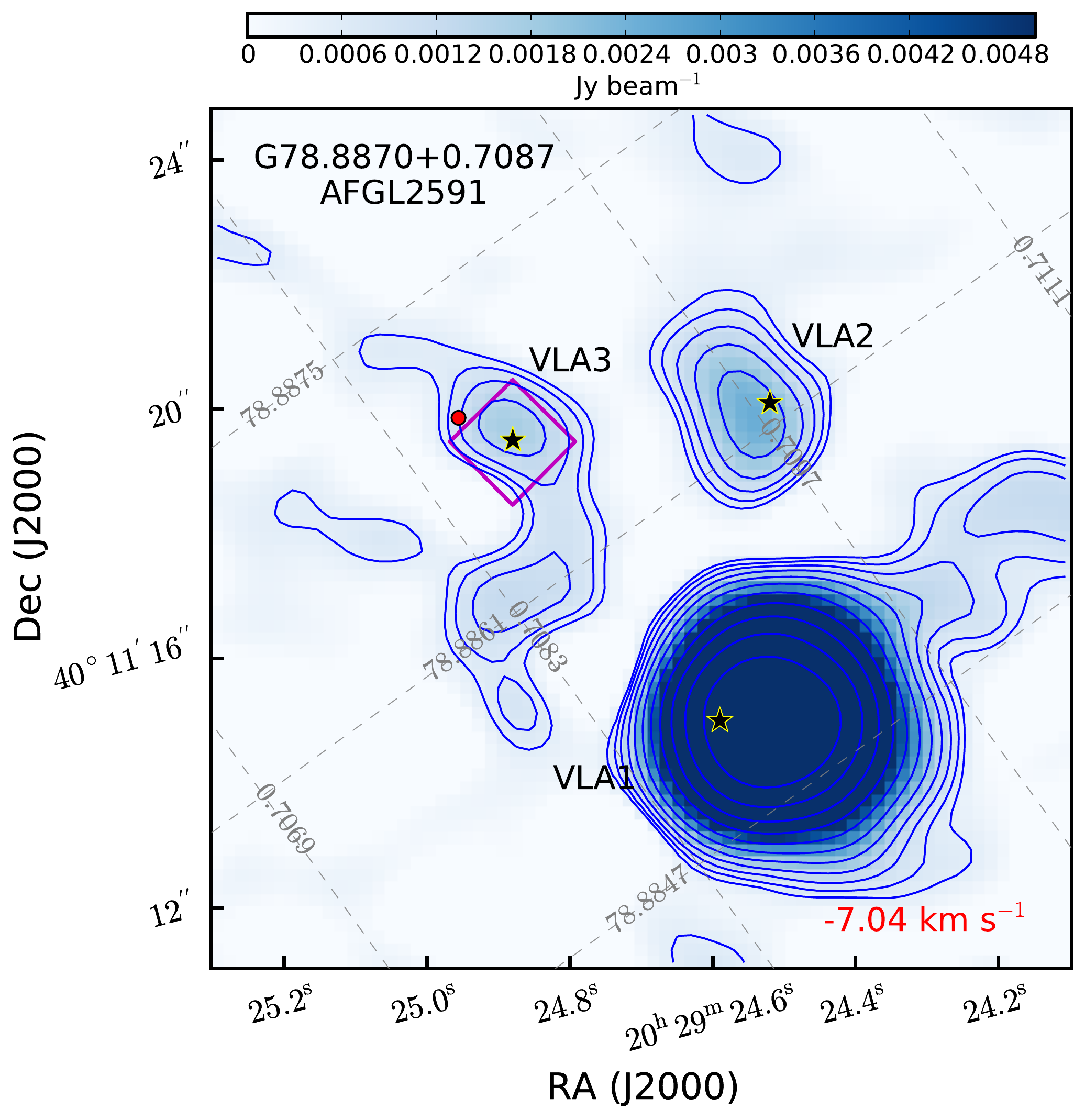}
\includegraphics[width=0.39\textwidth,angle=0]{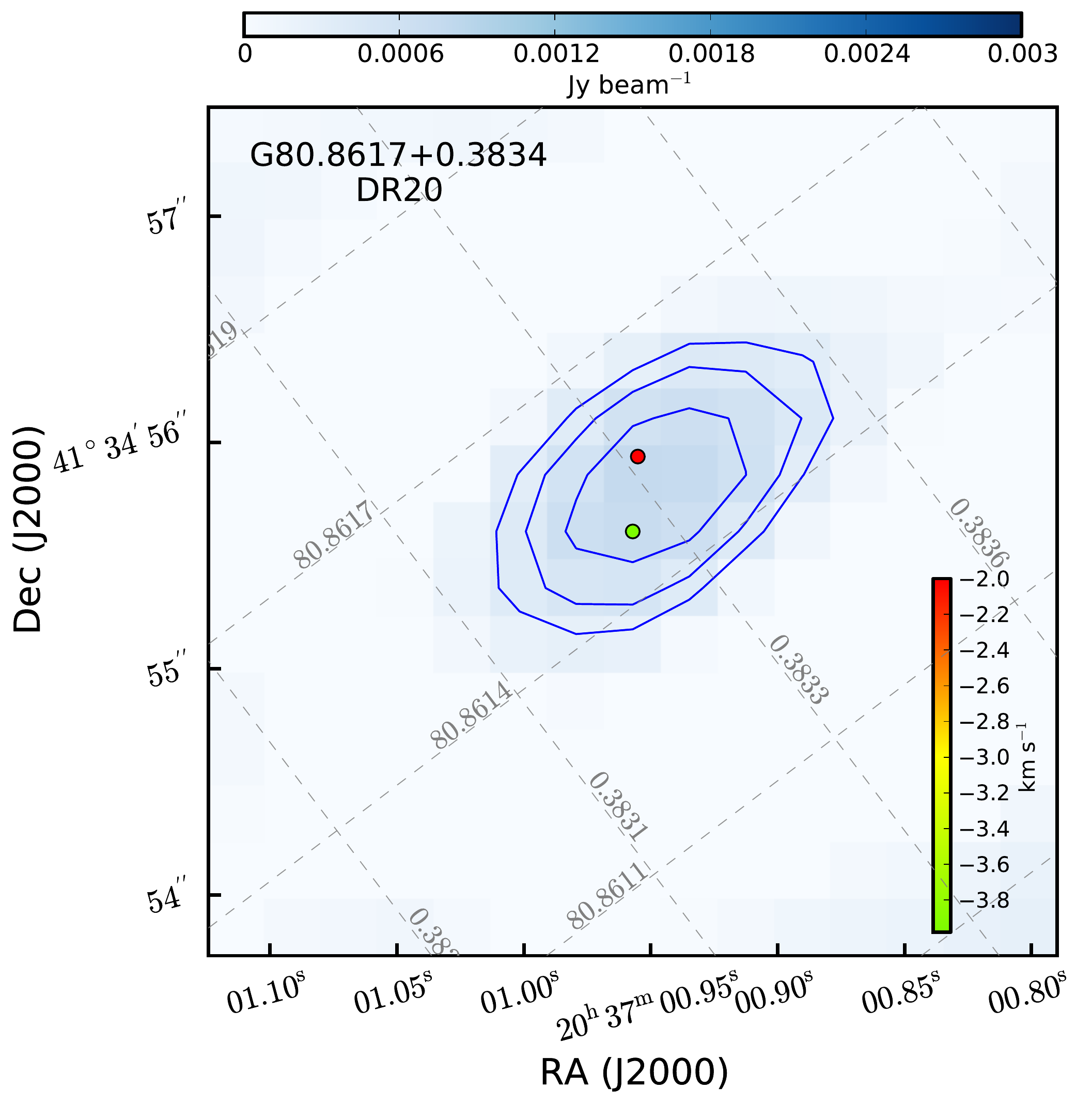}
\includegraphics[width=0.39\textwidth,angle=0]{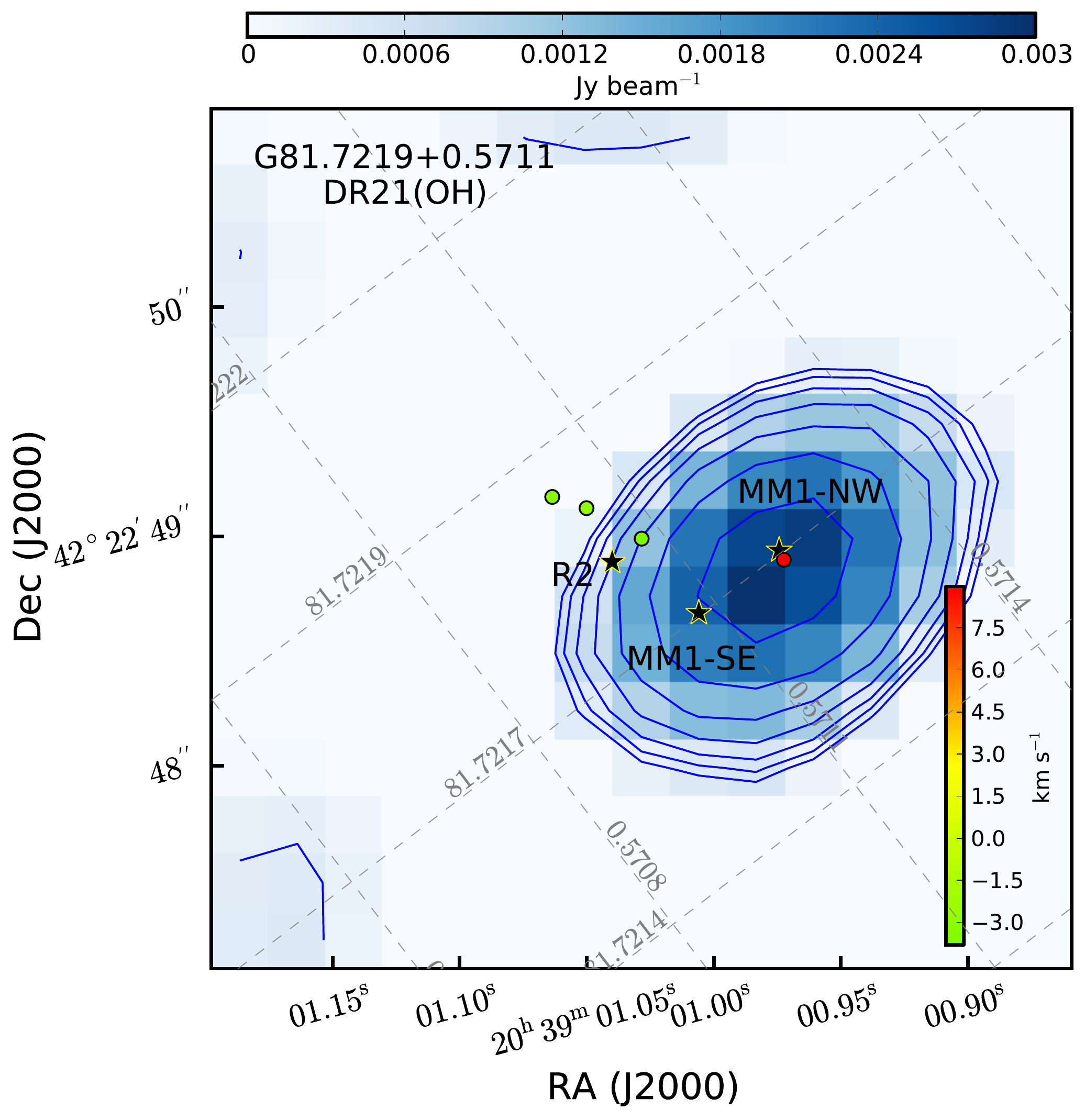}
\includegraphics[width=0.39\textwidth,angle=0]{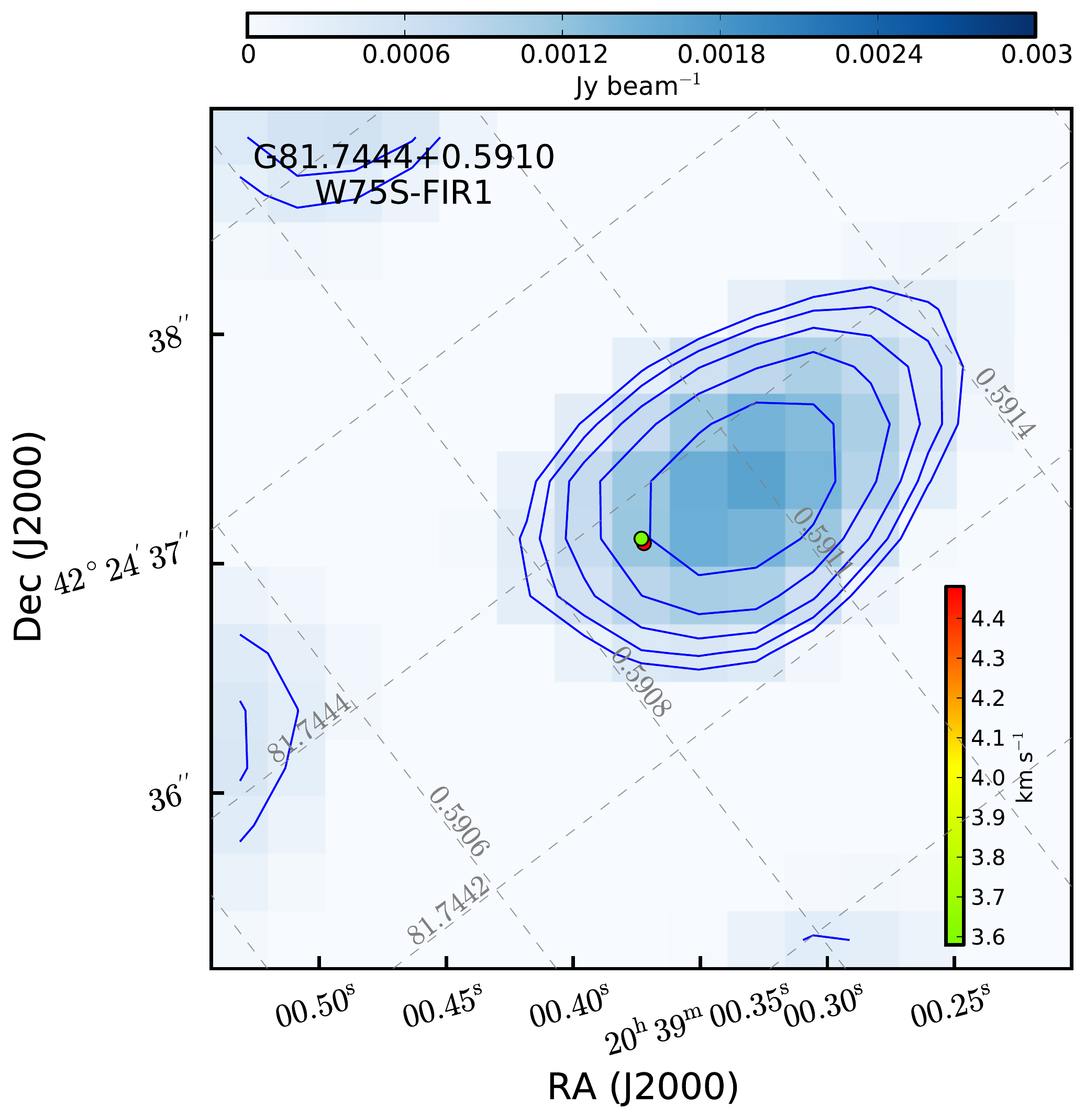}
\includegraphics[width=0.39\textwidth,angle=0]{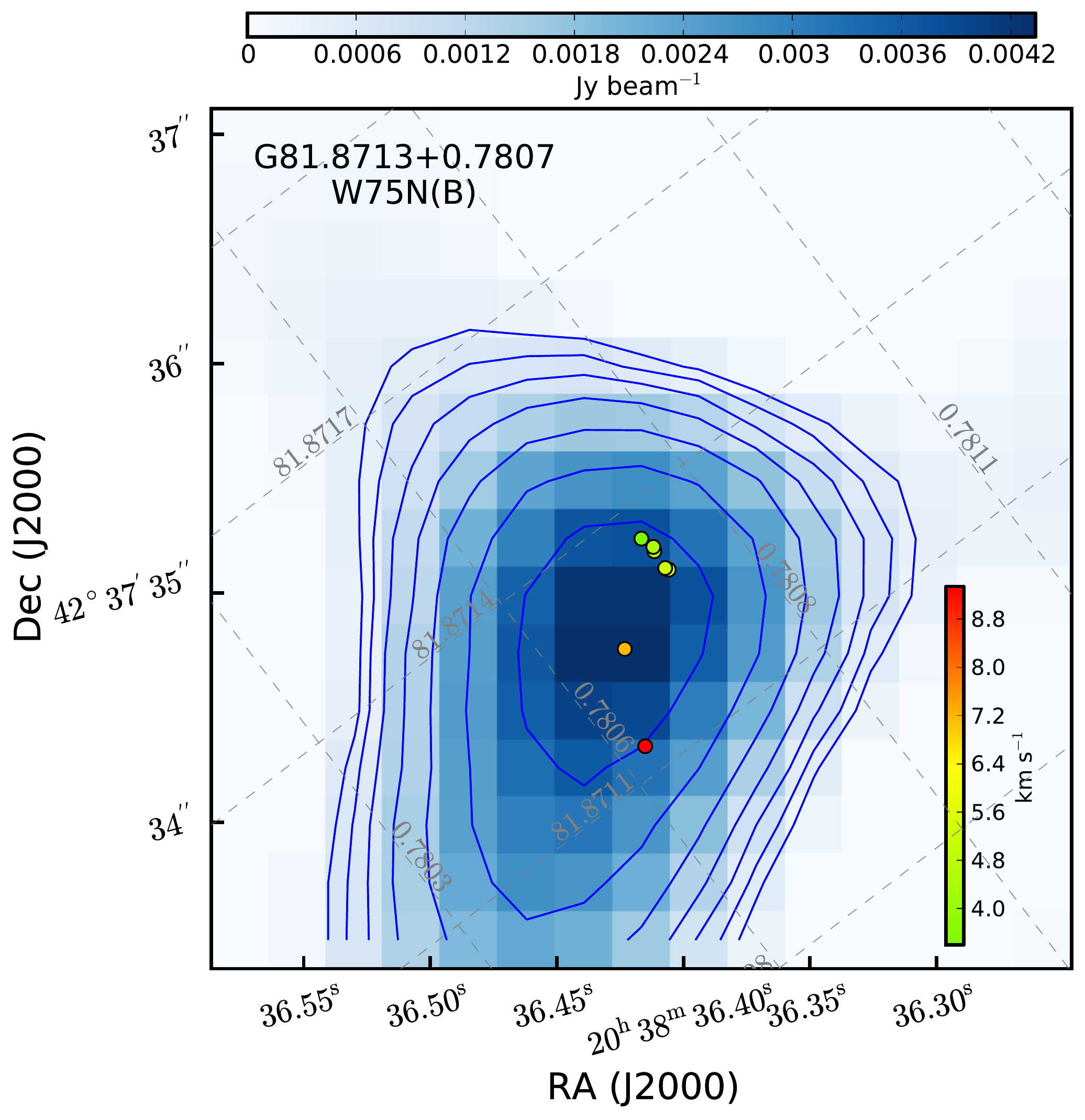} 
\caption{
Positions of maser features (filled circles) overlaid on continuum emission (blue scale and blue contours) for B-array data.
The masers are color coded by LSR velocity (color bar). When only one velocity component is detected, the velocity is 
given at the bottom of the panel. Names and positions of radio continuum
sources identified in previous observations are also marked (black stars). 
The $n$th blue contour is at $\left({\sqrt{2}}\right)^{n}\times S_{\rm max} \times p$, where $S_{\rm max}$ is the maximum continuum flux shown in the blue bar for each panel, 
$n$=0, 1, 2 ..., and $p$ is equal to 10\%.
The magenta rectangle in panel 1 depicts the approximate extent of a 1.3 mm continuum source detected with NOEMA \citep{Gieser2019}. The gray grid shows the Galactic coordinate system. 
The beam size of the radio continuum maps is $1\rlap.{''}5\times1\rlap.{''}5$.  
The maser position errors are $\approx0\rlap.{''}2$ (see Sect. \ref{sec:search}).
}
\label{fig:vla-cont}
\end{center}
\end{figure*}

\subsection{Comparison with other maser surveys}\label{sec:past-surveys}

\cite{Pestalozzi2005} published a catalog of all known 6.7~GHz methanol masers from both targeted and unbiased surveys in the Galactic plane. Their catalog contains six masers spread throughout the region observed by GLOSTAR in Cygnus X. Out of these six masers, four have a GLOSTAR counterpart within 1$'$, which is the average of the uncertainties in the maser coordinates of Pestalozzi's catalog. The sources from \cite{Pestalozzi2005} without counterparts are G78.62+0.98 and G80.85+0.43. Source G78.62+0.98 has a quoted peak flux of 3~Jy ($\sigma=1$~Jy) at $V_{\rm LSR} = -39.0$~km~s$^{-1}$. This velocity is very different from the velocities of our masers detected with GLOSTAR.  The source G80.85+0.43 has a quoted peak flux of 4.3~Jy at $V_{\rm LSR} = -4.1$~km~s$^{-1}$. This source is $2\rlap.{'}9$ away from a strong maser detected by GLOSTAR (G80.8617+0.3834/DR20), which shows a peak flux of 12~Jy ($\sigma=0.12$~Jy) at $V_{\rm LSR} = -4.16$~km~s$^{-1}$ and is not reported by \cite{Pestalozzi2005}. The two masers G80.85+0.43 and G80.8617+0.3834/DR20 could correspond to the same source given that the \cite{Pestalozzi2005} catalog collects positions obtained with a variety of single dish observations and hence different positional accuracy. They report an uncertainty of up to 3$'$ for their Galactic 
coordinates that would suggest that these two sources are the same. We created cleaned
images centered at the position of sources G80.85+0.43 and G78.62+0.98 and did not see emission. These masers are not detected in our Effelsberg maps either. 

Another targeted survey of this methanol maser line was conducted by \cite{Hu2016} with the VLA in C configuration. They selected all known methanol masers visible from the northern hemisphere and with peak fluxes above 2~Jy. The eight maser sources listed in this catalog within $76^{\rm o} <l< 83^{\rm o}$ and $-1^{\rm o} <b<+2^{\rm o}$ were also detected in our survey.
More recently, \cite{Yang2019} performed a new targeted survey in the whole Galaxy with the 65 m Shanghai Tianma Radio Telescope. They observed targets from the all-sky Wide Field Infrared Survey Explorer (WISE) point source catalog that could be associated with 6.7~GHz methanol masers based on their WISE magnitudes. We detected their three new masers and the five already known masers with WISE bright emission listed in \cite{Yang2019}. The only maser from this catalog not detected by us, likely due to variability, is G81.794+0.911. However, we do not see this maser in recent Effelsberg data either. In summary, out of thirteen GLOSTAR masers, four were previously reported by both \cite{Pestalozzi2005} and \cite{Hu2016}, five by both \cite{Hu2016} and \cite{Yang2019}, three others by \cite{Hu2016}, and three others by \cite{Yang2019}. Two maser sources (G76.8437+0.1233 and G78.9884+0.2211) are newly detected by our observations. These two sources display high flux variations between the two observed epochs, which may explain why they were missed by previous surveys.
Names, positions, and fluxes of all velocity components detected in the surveys by \cite{Hu2016} and \cite{Yang2019} are listed for reference in Table \ref{tab:hu16} of the appendix. 

\subsection{Comments on individual sources and their possible nature}\label{sec:comments}

In this section we give a short description of each 6.7~GHz methanol source detected in Cygnus X  (here, sources are referred to by their GLOSTAR identifier, GLL.llll+/-BB.bbbb, and also by their most common names if they are known sources). In order to discuss the association between these sources and their radio continuum emission, we show in Fig. \ref{fig:vla-cont} the positions of maser components determined from the  B-configuration images, overlaid on GLOSTAR radio continuum emission that is also from B-configuration data.  
Maser positions measured from D-configuration images are
less precise as compared with B-configuration positions, and thus they do not provide additional information on the distribution of maser spots. For this reason, we do not show D-configuration positions in Fig. \ref{fig:vla-cont}.

A clear association between maser (colored circles) and continuum emission (blue scale) and within a radius of $\sim$1$''$ is seen in five cases. The coordinates, peak fluxes, and the S/N of the detected radio continuum sources as measured with \texttt{BLOBCAT} \citep{Hales2012,Medina2019} are given in Table \ref{tab:contB}. A $3\sigma$ upper limit is given in the table when no radio continuum emission is detected.

\begin{figure*}[t]
\begin{center}
\includegraphics[width=0.95\textwidth,angle=0]{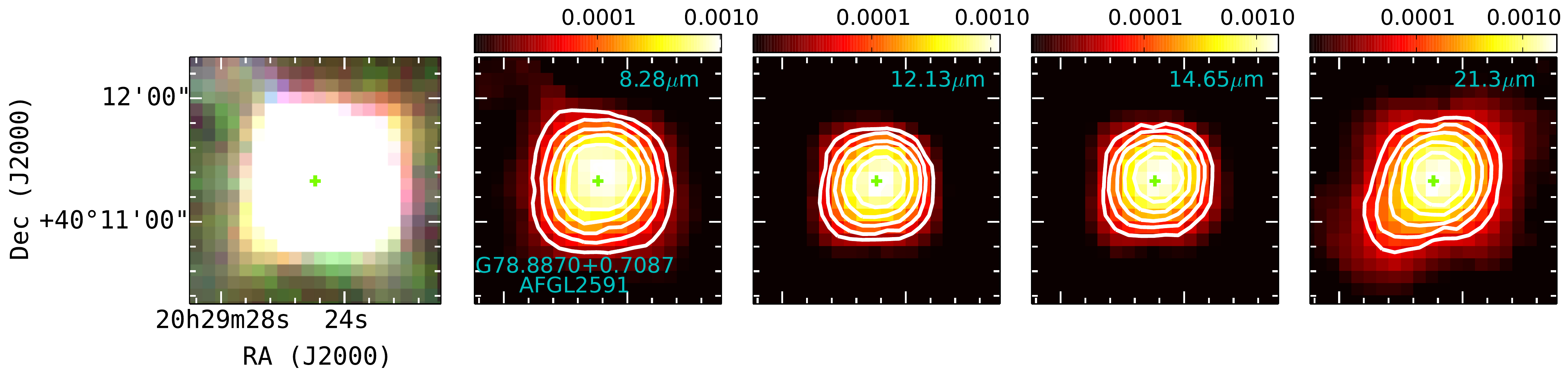} 
\includegraphics[width=0.95\textwidth,angle=0]{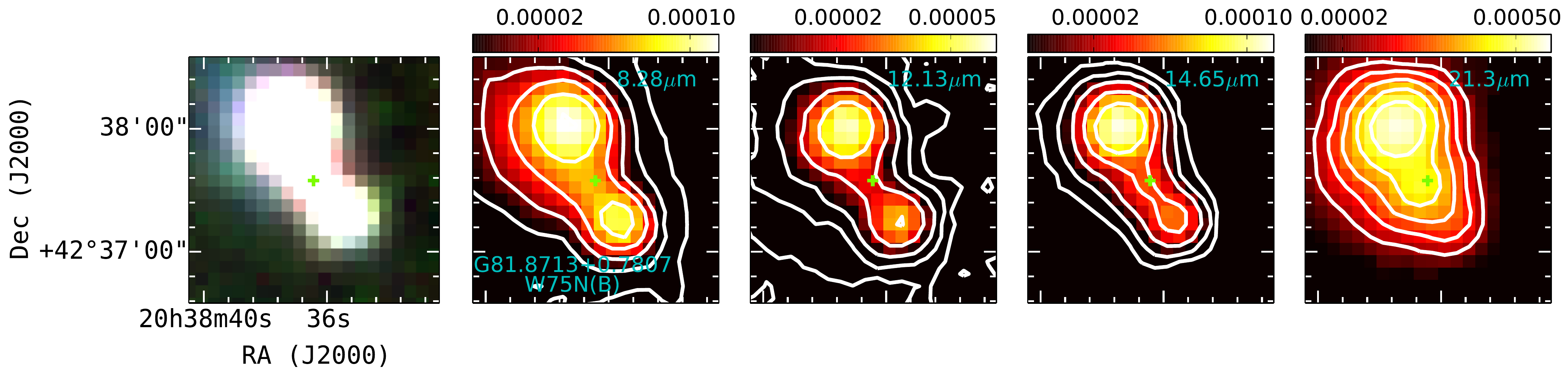} 
\caption{MSX infrared images (color scale and contours) of the environment around the location of G78.8870+0.7087 and G81.8713+0.7807. 
The {\it Spitzer} and WISE images of these two sources are saturated in at least one band. The first panel in each row shows a three-color map constructed using MSX 8.28~$\mu$m (blue), 12.13~$\mu$m (green), and 14.65~$\mu$m (red) images. 
The other four panels show infrared emission for each of the four MSX bands. 
The $n$th white contour is at $\left({\sqrt{2}}\right)^{n}\times S_{\rm max} \times p$, where $S_{\rm max}$ is the maximum flux shown in the color bar (W~m$^{-2}$~sr$^{-1}$) for each panel, $n$=0, 2, 4 ..., and $p$ is equal to 3\%.
The green crosses mark the positions
of the masers as measured in D-array configuration maps.
The Galactic
names of the GLOSTAR sources are indicated at the bottom of the second image in each row.
}
\label{fig:msx}
\end{center}
\end{figure*}

{\bf G78.8870+0.7087/AFGL2591}.  
This maser is close ($\sim1''$; cf.\ Table \ref{tab:contB}) to the VLA3 hot core \citep{Campbell1984,Johnston2013,Gieser2019} in the well-known high-mass star-forming region AFGL2591. VLA3 is also detected in the GLOSTAR B-configuration map (see panel 1 of Fig. \ref{fig:vla-cont}).   
A spectral index of $0.5\pm0.02$ between 8.3 and 43~GHz was estimated by \cite{Johnston2013} for VLA3.
Extended bright infrared emission at 3.6--24~$\mu$m is also present in the environment of the maser (see Figs. \ref{fig:infrared2-appendix} and \ref{fig:msx}). However, this infrared source is saturated in the {\it Spitzer} images. \cite{Motte2007} identified a dust core in their 1.2~mm observations whose peak position is closest to VLA3, and this dust core was recognized as a UC~HII region. More recently, \cite{Gieser2019} resolved the millimeter emission toward VLA3 with higher angular observations taken with the NOrthern Extended Millimeter Array (NOEMA) and revealed an extended, almost spherically symmetric envelope at $\sim 2''$ scales. We note that both the maser and the radio continuum emission lie within the extension of this source, which is approximately depicted by the magenta circle in panel 1 of Fig.\ \ref{fig:vla-cont}. The integrated flux of this maser in the peak channel changes from $0.80\pm0.04$~Jy in the D-configuration image to $0.29\pm0.04$~Jy in the B-configuration image due to potential variability.
The Effelsberg data, however, show an integrated flux of $\sim$0.7~Jy, which is consistent with the flux seen in the D-configuration map. Its 3.3 kpc maser parallax distance places AFGL 2591, like Cygnus X, in the local arm \citep[see, e.g., Fig. 1 of][]{Reid2019}.

{\bf G80.8617+0.3834/DR20}. We detected two additional features in the D-configuration spectrum, at $-12.3$ and 1~km~s$^{-1}$, which were not seen  in the spectrum from \cite{Hu2016} and \cite{Yang2019}. We note that these two components and the component at $-$11~km~s$^{-1}$ are not detected in the B-configuration data, likely due to variability. We see a weak source ($S_{\rm peak}$ = 0.79~mJy~beam$^{-1}$) in the B-array radio continuum map coincident with the location of the maser (panel 2 of Fig. \ref{fig:vla-cont}). 
The infrared environment around the maser shows slightly extended 3.6--8~$\mu$m emission (Fig. \ref{fig:infrared2-appendix}). The source at 24~$\mu$m is saturated. This object is recognized as a high-luminosity infrared protostellar core \citep{Motte2007}.

{\bf G81.7219+0.5711/DR21(OH)}. Compared to the spectrum from \cite{Hu2016}, we detected an additional weak feature at 9~km~s$^{-1}$. A bright infrared peak is seen at 3.6--24~$\mu$m close to the location of the maser (Fig. \ref{fig:infrared2-appendix}); however, its position does not exactly coincide with the maser intensity peak, showing a small offset of $\sim$1$''$. 
This maser is associated with the well-known high-mass star-forming region DR21(OH). Its position coincides with the main core in DR21(OH), which is a massive infrared-quiet (<10~Jy at 21~$\mu$m) protostar in an early evolutionary stage \citep{Motte2007}. We see a weak source in the B-array radio continuum map close to the location of the masers (panel 3 of Fig. \ref{fig:vla-cont}). The strongest maser component is offset by $1\rlap.{''}1$ with respect to the peak position of the continuum source; however, the component at $V_{\rm LSR}=9$~km~s$^{-1}$ coincides in position with the continuum source.
The continuum source has a peak flux density of $S_{\rm peak}=3$~mJy~beam$^{-1}$. In a previous work, \cite{Araya2009} conducted radio continuum observations at 3.6, 1.3, and 0.7~cm toward DR21(OH) at subarcsecond angular resolution. They detected a cluster of radio sources; their source MM1-NW ($S_{\rm peak}=1.14\pm0.02$~mJy~beam$^{-1}$ at 8.46~GHz) 
is the closest to the source we detect in our GLOSTAR B-array continuum data. Source MM1-NW is also coincident with the maser at $V_{\rm LSR}=9$~km~s$^{-1}$, while source R2 is the closest to the masers at smaller velocities (see panel 3 of Fig. \ref{fig:vla-cont}). \cite{Araya2009} suggested that the two brightest sources detected toward the molecular core (MM1-NW and MM1-SE) trace radio emission from shock-ionized gas in a jet, while the radio emission from source R2 was proposed to be gyrosynchrotron radiation from a low-mass star. We cannot resolve these individual continuum components in the GLOSTAR data, which prevents us from estimating spectral indices of the individual radio sources to confirm or reject the conclusion by \cite{Araya2009}. 
Based on their higher angular resolution images,  \cite{Araya2009} measured a spectral index of 0.8$\pm$0.2 between 8.4 and 43~GHz for both MM1-NW and MM1-SE, and 0.1$\pm$0.2 for R2. Thus, given the correspondence in the positions measured by these authors with the maser positions, it is tempting
to suggest that MM1-NW and R2 are UC~HII regions.

{\bf G81.7444+0.5910/W75S-FIR1}.  
We see a weak source in the  B-array radio continuum map very close (at $0\rlap.{''}8$) to the location of the maser position (panel 4 of Fig. \ref{fig:vla-cont}). The continuum source has a peak flux density of $S_{\rm peak}=1.7$~mJy. 
The maser has associated compact emission at 3.6--8~$\mu$m (Fig. \ref{fig:infrared2-appendix}). At 24~$\mu$m the source is saturated, but we see some extended emission surrounding the maser. The source with which this maser is associated is the high-luminosity ($\gtrsim10^3~L_\odot$) infrared object W75S~FIR1 \citep{Harvey1986}, a well-known massive embedded protostar \citep{Chandler1993,Motte2007}.

{\bf G81.8713+0.7807/W75N(B)}. 
We see that the masers trace a velocity gradient of a few km~s$^{-1}$ in the north-south direction across a scale of $\sim$2$''$. We detect continuum emission close to this maser in both D-array and B-array VLA maps. The maser position is within $0\rlap.{''}4$ of a radio continuum peak ($S_{\rm peak}=4.4$~mJy) seen in the B-array map (panel 5 of Fig. \ref{fig:vla-cont}). 
The radio continuum peak corresponds to VLA 1, a source that was found to have a jet-like appearance at 1.3~cm \citep{Torrelles1997}. Recently, \cite{RodriguezKamenetzky2020} conducted multifrequency VLA observations and concluded that VLA1 consists of a thermal radio jet surrounded by a hyper-compact (HC) HII region. The infrared environment toward the maser shows three peaks in the 3.6--8~$\mu$m maps (Fig. \ref{fig:infrared2-appendix}). At 24~$\mu$m, the source is saturated. The maser seems to be located between two infrared peaks seen in the 5.8~$\mu$m map, but its position coincides with an infrared peak we see at 21.3~$\mu$m in the Midcourse Space Experiment (MSX)  map (see the bottom panel of Fig. \ref{fig:msx}). 
The maser is found in the star-forming region W75N(B), which is mainly composed of two UC~HII regions \citep{Hunter1994}, namely W75N(Ba) and W75N(Bb) (also known as VLA1 and VLA3, respectively, in the nomenclature of \citealt{Torrelles1997}).

The following maser sources do not have counterparts in either our VLA radio continuum maps or previous radio continuum surveys (within a matching radius of 10$''$).

{\bf G76.0932+0.1580}. \cite{Yang2019} reported on the velocity components detected at 4.8 and 6.4~km~s$^{-1}$ (see Table \ref{tab:hu16}). Here, we detect an additional component at $-$6.5~km~s$^{-1}$ in the D-array VLA spectrum. The maser is associated with a compact weak source seen at 4.5-24~$\mu$m (Fig. \ref{fig:infrared2-appendix}).
The infrared source has been classified as a protostar candidate by \cite{Kryukova2014} and shows a rising infrared SED. 

{\bf G76.8437+0.1233}. This source is a newly detected 6.7~GHz maser. The maser is associated with compact bright emission at 4.5-24~$\mu$m (Fig. \ref{fig:infrared}).  
It also appears in the catalog of protostar candidates by \cite{Kryukova2014} as a rising SED protostar.  One single strong velocity component with an integrated flux in the peak channel of $4.7\pm0.08$~Jy is detected in the D-configuration image taken in August 2014.  The flux declines to $0.17\pm0.01$~Jy in the B-configuration image taken in May 2015. The maser is undetected in Effelsberg observations taken in August 2019 with a 3$\sigma$ upper limit of 0.36~Jy~beam$^{-1}$. Such variability in the 6.7~GHz emission may explain why the maser has not been reported in previous surveys of methanol masers.
  
{\bf G78.9690+0.5410/IRAS 20286+3959}. This maser was recently detected by \cite{Yang2019}, with a peak velocity and peak flux very consistent with our VLA detection. The maser is associated with slightly extended emission at 3.6--24~$\mu$m (Fig. \ref{fig:infrared2-appendix}). 
It has also been classified as a rising infrared SED protostar \citep{Kryukova2014}.

{\bf G78.9884+0.2211}. This 6.7~GHz maser has not been reported before. The maser LSR velocity ($V_{\rm peak}=-69$~km~s$^{-1}$) is completely different from the velocities of the other masers detected in Cygnus X, which are within $-$15 to 15 km~s$^{-1}$; therefore, this maser is probably at a different distance. The maser has five velocity components, with the strongest component having an integrated flux density at the peak channel of $1.97\pm0.06$~Jy in the D-configuration image. This declines to $0.32\pm0.02$~Jy in the B-configuration image, while the Effelsberg spectrum shows only one velocity component with an integrated flux of $\sim$0.6~Jy. The second strongest component does not change in flux between the two VLA epochs, while  the other three velocity components are not seen in the B-configuration map nor in the Effelsberg data. 

We inspected CO ($J=1-0$) data\footnote{\url{https://cygnus45.github.io/}} taken with the 45 m Nobeyama telescope by \citet{Yamagishi2018ApJS}. The spectrum extracted from the maser position shows two peaks, one at $V_{\rm LSR}=9.2$~km~s$^{-1}$ and one at $V_{\rm LSR}=-62.8$~km~s$^{-1}$, the latter of which is close
to the methanol maser velocity. The emission at this velocity is distributed across a region of $\approx2'$, while emission around $V_{\rm LSR}=+9.2$~km~s$^{-1}$ is more diffuse and extended. We calculated the distance to the $V_{\rm LSR}=-62.8$~km~s$^{-1}$ component using the Bayesian distance estimator developed by \cite{Reid2016}\footnote{\url{http://bessel.vlbi-astrometry.org/bayesian}}. The estimator yields a distance of $9.35\pm0.53$~kpc with an integrated probability of 100\% association with the Outer Arm. At a similar Galactic
longitude there are two  maser stars in the Outer Arm with parallaxes that are consistent with this distance  \citep[G075.29+01.32 and G073.65+00.19;][]{Sanna_2011,Reid2019}. 
Certainly, the CO-traced  $V_{\rm LSR}=-62.8$~km~s$^{-1}$ cloud associated with the methanol maser and likely also the $V_{\rm LSR}=+9.2$~km~s$^{-1}$ cloud are not part of the Cygnus X region,
but are rather unrelated clouds lying on the same line of sight by chance. 

Assuming that G78.9884+0.2211 is at 9.35~kpc, the mass, FIR luminosity, and bolometric luminosity must be rescaled to $347 M_\odot$, $3093 L_\odot$, and, $6237 L_\odot$, respectively. Given the uncertainty in the distance, we do not include this maser source in the analysis presented in Sect. \ref{sec:discuss}.

\cite{Wendker1970} reported on radio continuum emission at 2.695~GHz, but at $>8''$ from the maser location. There are several bright infrared peaks seen at 3.6--24~$\mu$m in the vicinity of the maser surrounded by extended emission (Fig. \ref{fig:infrared2-appendix}). The maser position coincides with one infrared peak. This infrared source also shows a rising SED \citep{Kryukova2014}.

{\bf G79.7358+0.9904/IRAS 20290+4052}.  
This maser is associated with bright, somewhat extended 3.6--24~$\mu$m emission (Fig. \ref{fig:infrared2-appendix}). 
\cite{Kryukova2014} classified the infrared source as a protostar candidate with a rising SED. The infrared source was also recognized early on as a UC~HII region based on their FIR colors \citep{Bronfman1996,vanderWalt1996}, which meet the criteria for UC~HII regions proposed by \cite{Wood1989}. However, as already pointed out, no radio continuum emission has been detected toward this source, which suggests that this object may be a precursor of a UC HII region.  

{\bf G81.7523+0.5908/W75S-FIR2}. 
The infrared environment toward this maser shows bright and slightly extended 3.6--8~$\mu$m emission (Fig. \ref{fig:infrared2-appendix}). The source is saturated at 24~$\mu$m. The maser position coincides with the location of object W75S~FIR2 \citep{Harvey1986}, a high-luminosity infrared protostellar core \citep{Motte2007}. 

{\bf G81.7655+0.5972/CygX-N53}.  
The emission seen in the D-array spectrum close to 7~km~s$^{-1}$ is from sidelobes by the strong nearby maser G81.8713+0.7807/W75N(B). The 3.6--8~$\mu$m emission seen at the location of the maser (Fig. \ref{fig:infrared2-appendix}) is remarkably weak (measured flux density at the location of the maser is 1--11~mJy~beam$^{-1}$, respectively). 
The associated molecular core has been recognized as a massive infrared-quiet protostellar core \citep{Motte2007}. Observations at 1.3~mm with the IRAM Plateau de Bure interferometer  revealed that the core is broken up into four smaller fragments \citep{Bontemps2010}. The maser position coincides with the strongest fragment, CygX-N53~MM1, which is a self-gravitating protostellar object and a good candidate to be a precursor of an OB star \citep{Bontemps2010}. Among the sample of 17 massive infrared-quiet protostellar cores identified by \cite{Motte2007}, CygX-N53 and DR21(OH) are the only ones that show maser emission. However, the infrared emission from CygX-N53 is about an order of magnitude weaker than that of DR21(OH).
The sample of \cite{Motte2007} comprises other infrared-quiet protostars ($S$<10~Jy at 21~$\mu$m) with similar properties (mass, size, density) as CygX-N53, also located in the DR21 ridge; thus, it is unclear which conditions favor the onset of maser emission in CygX-N53 and not in the other cores that have similar physical properties.

{\bf G82.3079+0.7296}. This maser was recently detected by \cite{Yang2019}. They reported a peak flux of 58.4~Jy, which is more than two times higher than the flux we detected with the VLA. The maser is associated with extended $3.6-24$~$\mu$m emission (Fig. \ref{fig:infrared2-appendix}). The infrared source has been classified as a rising SED protostar \citep{Kryukova2014}.

\clearpage

\begin{figure}[t]
\begin{center}
\includegraphics[width=0.4\textwidth,angle=0]{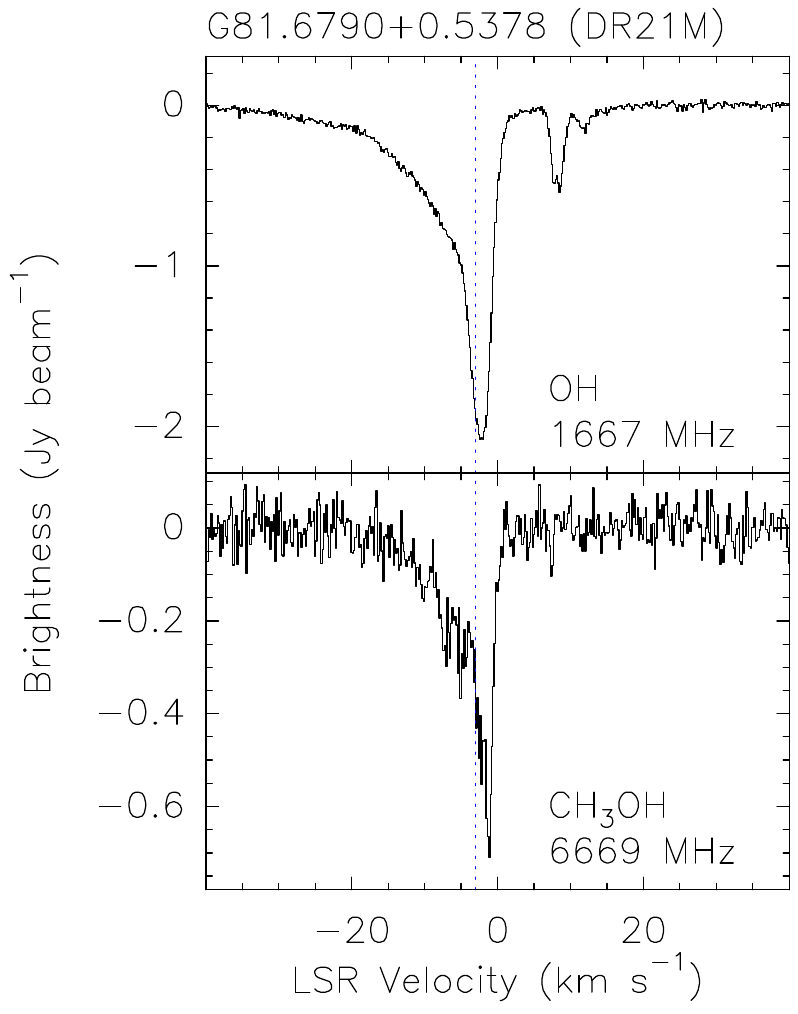} 
\caption{Absorption in the 6.7 GHz methanol line (lower panel) and the 1667 MHz OH line (upper panel) observed toward DR21~M. The dashed line indicates the systemic LSR velocity.
}
\label{fig:ohch3oh}
\end{center}
\end{figure}

\begin{figure}[t]
\includegraphics[width=0.5\textwidth,angle=0]{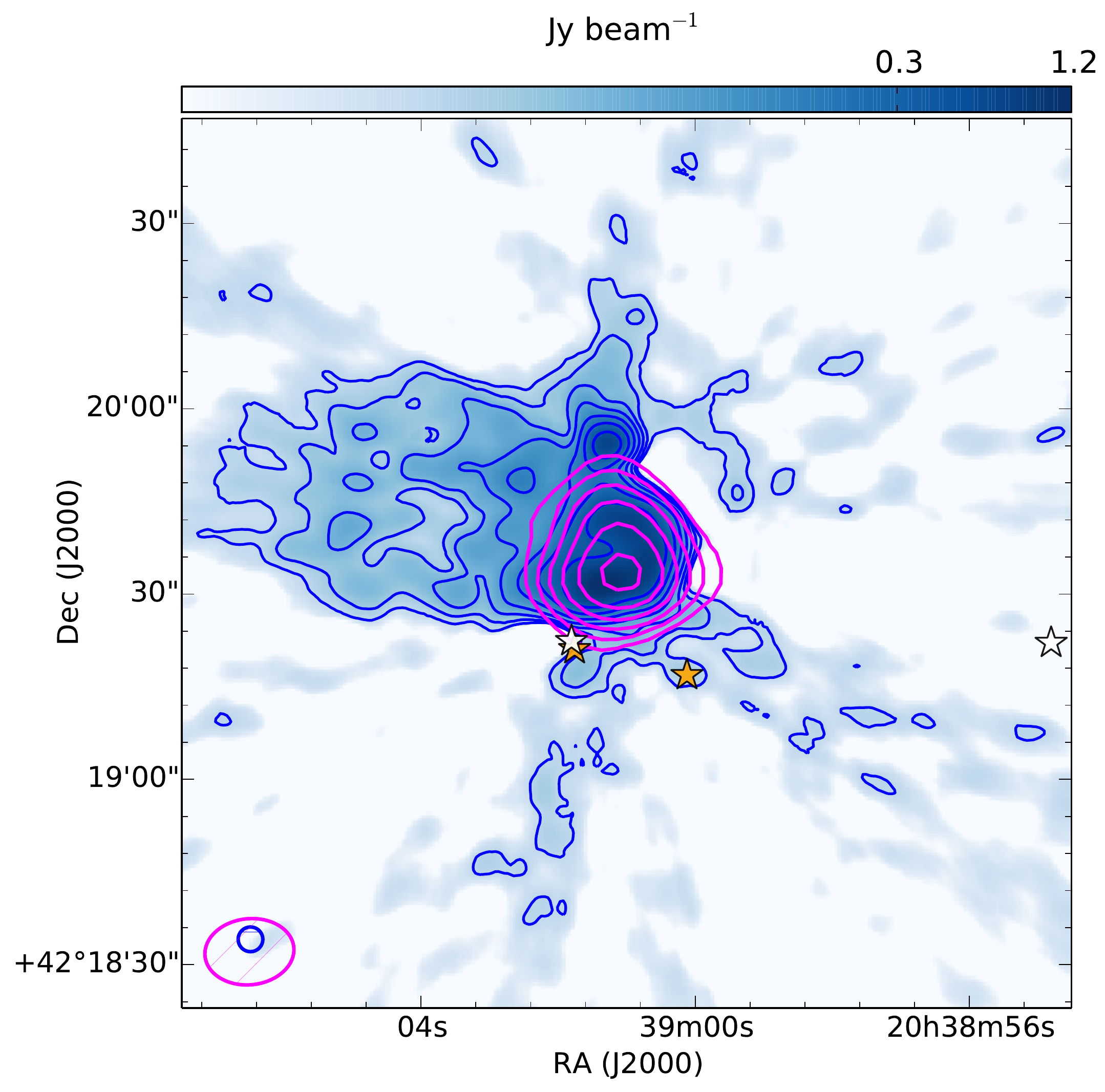} 
\caption{Methanol
absorption (magenta contours, integrated from $V_{\rm LSR}=-13.9$ to 0.5~km~s$^{-1}$) seen in D-array 
data, overlaid on extended continuum emission toward DR21 (blue scale and contours) as seen in our 
VLA D+B array configuration map. 
The $n$th contour is at $\left({\sqrt{2}}\right)^{n}\times S_{\rm max} \times p$, where: $S_{\rm max}=1.2$~Jy~beam$^{-1}$, 
$n$=1, 3, 5 ..., and $p$ is equal to 0.4\% for the radio continuum; and $n$=0, 1, 2 ..., $S_{\rm max}=-3.2$~Jy~beam$^{-1}$~km~s$^{-1}$, and $p$=16\% for the methanol absorption. 
The synthesized beams ($4''\times4''$ for 
the VLA D+B map and $14\rlap.{''}4\times10\rlap.{''}7$ at PA=$-$84.1$^{\rm o}$ for the D-configuration map) 
are shown at the bottom corner. The white stars mark the positions of Class I methanol masers observed in the 95 GHz $8_1-7_0~A^+$ line \citep{PlambeckMenten1990}. The orange stars are 44~GHz $7_0-6_1~A^+$ methanol masers \citep{Kurtz2004}.
}
\label{fig:vla-cont-2}
\end{figure}

\begin{figure}[tb]
\begin{center}
 \includegraphics[width=0.48\textwidth,angle=0]{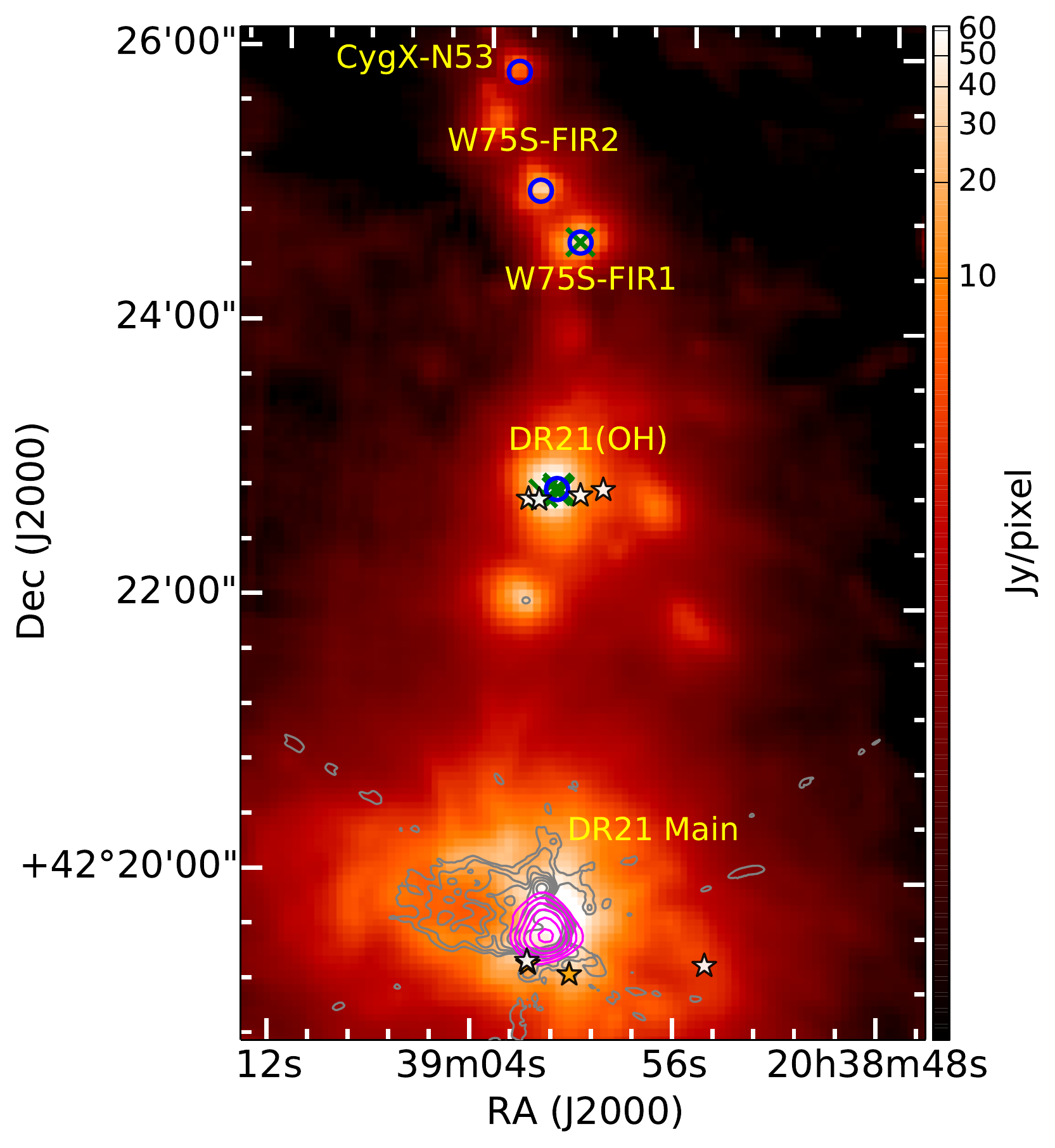}
\caption{Locations of 6.7 GHz Class II methanol masers detected with the VLA  (blue circles) toward the dense molecular ridge extending from DR 21 M in the south to DR 21 (OH) (and  beyond), overlaid on a $70\mu$m dust continuum image obtained with Herschel (red scale;  \citealt{Hennemann2012}).
 Gray contours represent the GLOSTAR combined B+D-array continuum emission.
The magenta contours show the velocity-integrated absorption of the 6.7 GHz methanol line. 
The $n$th contour is at $\left({\sqrt{2}}\right)^{n}\times S_{\rm max} \times p$, where: $S_{\rm max}=1.2$~Jy~beam$^{-1}$, $n$=1, 3, 5 ..., and $p$ is equal to 0.4\% for the continuum; and $n$=0, 1, 2 ..., $S_{\rm max}=-3.2$~Jy~beam$^{-1}$~km~s$^{-1}$, and $p$=16\% for the methanol absorption. 
The white stars and green Xs mark, respectively, the positions of Class I methanol masers observed in the 95 GHz $8_1-7_0~A^+$ line \citep{PlambeckMenten1990} and the 18 cm OH masers associated with DR21(OH) and W75S-FIR 1 \citep{Norris1982,Argon2000}. The orange stars represent 44~GHz Class I methanol masers \citep{Kurtz2004}.
}
\label{fig:dr21-ridge}
\end{center}
\end{figure}

\subsection{Methanol absorption toward DR21}\label{sec:dr21}

The region surrounding the prominent thermal radio source DR21 \citep{DR1966} hosts a compact (about $30''\times30''$) and massive molecular core, which is located  near the southern end of a long molecular cloud ridge oriented in the north-south direction and referred to as the DR21 ridge \citep[see, e.g.,][]{Reipurth2008,Hennemann2012}. 
This core harbors a group of five compact HII regions that are surrounded by a diffuse halo of ionized gas with a cometary morphology, 
with the most prominent compact component, DR21~M, at the head of the ``comet'' \citep{Harris1973,Roelfsema1989,Cyganowski2003}.

The DR21 M HII region itself is surrounded by a dense molecular envelope with a photodissociation region (PDR) as an interface \citep{Ossenkopf2010}. The more compact and denser envelopes of UC~HII regions often host Class II CH$_3$OH and OH masers, with W3(OH) the archetypal example \citep{Menten1992}. In contrast, the more developed envelope of DR21 M does not host any masers, but exhibits absorption features of OH in the hyperfine transitions of the rotational ground and excited states against the strong continuum emission of the compact HII region \citep[see][and references therein]{Jones1994}. As shown in Fig. \ref{fig:ohch3oh}, we detect absorption in the 6.7~GHz methanol line with a shape similar to that of the 1667 MHz OH line (taken from \citealt{Koley2020}). While the OH line extends to much lower velocities than that of CH$_3$OH, this may be a result of the poorer S/N of the latter. Both lines show their deepest  absorption at about $V_{\rm LSR}=-1~{\rm km~s}^{-1}$, close to the systemic velocity of DR21~M, $\approx 2$--3~km~s$^{-1}$. Figure \ref{fig:vla-cont-2} shows that the methanol absorption is seen only against the strongest continuum from DR21~M, although this may also be caused by our limited S/N toward regions with weaker extended continuum emission.

We note
that the OH transition, but not the CH$_3$OH transition, also shows absorption at velocities between +6 and +12 km~s$^{-1}$. This represents low density material from a second extended molecular cloud, sometimes called the ``W75 N 9 km~s$^{-1}$ cloud,'' that covers the part of the W75 region that is  kinematically distinct from the cloud, which is also extended, at the LSR velocity  of DR21 \citep{Dickel78}.
Emission or absorption from this cloud is only observed from transitions with a low critical density, that is, ground-state or near ground-state lines \citep[see][]{Koley2020}.

We searched the D-configuration data cubes for other sources of methanol absorption, specifically toward the DR21 ridge, but did not detect absorption at any other location. In particular, we did not find absorption toward the Class I methanol maser sources in the vicinity of DR21(OH) and DR 21 (see Fig. \ref{fig:dr21-ridge}). As explained by \citet{Menten1991ApJ380L} and  \citet{Leurini2016}, the same (purely collisional) pumping process that inverts the energy levels of the $J_{K=0}-(J-1)_{K=1}~A^+$ Class I maser lines of $A$-type CH$_3$OH ($J\ge7$) in molecular outflows also ``anti-inverts'' those of the 6.7 GHz $5_1-6_0~A^+$ line,  causing enhanced absorption (overcooling) over more extended regions; this has actually been observed with single dish telescopes \citep{Menten1991ApJ380L}\footnote{Analogously, for $E$-type CH$_3$OH, the
 $J_{k=-1}-(J-1)_{k=0}~E$ lines ($J\ge4$) are masing, whereas the 12.1 GHz 
 $2_0-3_{-1}~E$ line is overcooled.}. Apparently, in our data of the DR21 region, absorption signals confined to compact Class I maser spots are not strong enough to be detectable at our sensitivity.
 In Appendix \ref{sec:dr21ch3oh} we discuss constraints on the methanol abundance in the DR21~M molecular core derived from our absorption spectrum.

\section{Discussion}\label{sec:discuss}

\subsection{Association with radio continuum emission}

One important aspect to establish an evolutionary sequence model of methanol masers in star-forming regions is the coincidence of masers and UC HII regions. \cite{Walsh1998} proposed that methanol masers are present before the formation of a UC HII region around a massive star and persist until the destruction of methanol as the UC HII region evolves.
Our high angular resolution observations can probe, for the first time, the coincidence of masers and  UC HII regions (by means of their radio continuum emission) in the Cygnus~X complex down to scales of $\sim$1700~au. 
While the GLOSTAR observations are not ultra deep (our sensitivity limit is 0.06~mJy~beam$^{-1}$), they are more sensitive than most previous efforts and, notably, have resulted in the detection of weak (a few mJy) continuum emission toward five of our maser sources.

We found that five masers without a radio continuum counterpart are associated with infrared sources that show a rising SED between 4.5 and 24~$\mu$m \citep{Kryukova2014}, suggesting that they host high-mass protostellar objects. Since they do not have associated radio continuum, these sources may be in a younger evolutionary state compared to the UC~HII region (pre-UC~HII). Three masers are associated with known UC~HII regions identified previously in the literature (AFGL2591, \citealt{Motte2007}; IRAS 20290+4052, \citealt{Bronfman1996}; and W75N(B), \citealt{Hunter1994}). However, we found that G79.7358+0.9904/IRAS~20290+4052 does not show radio continuum emission at a 3$\sigma$ level of 0.22~mJy; thus, it could be in a pre-UC~HII state. DR21 is also associated with a UC~HII region; however, it does not have maser emission but rather absorption in the methanol line. Three other masers are found to be coincident (within $\sim$1$''$) with sources of compact radio continuum emission; therefore, they  are likely UC HII or HC~HII regions. We have indicated in Col. (4) of Table \ref{tab:maserCtp} whether the masers are associated with a pre-UC~HII or a UC or HC~HII 
region. In total, five dust cores host both a UC or HC~HII
region and maser emission\footnote{ 
We found that the number of expected background sources inside our matching radius is well below one source
(see Appendix \ref{sec:background}). This implies a very low probability for the association between masers and continuum on the line of sight being just by chance.}, and all of them have $L_{\rm Bol}>1000~L_\odot$. This implies an association rate between masers and HII regions of $38\pm17\%$, which is consistent within the error with that found in other studies \citep{Walsh1998,Beuther2002,Walsh2003,Hill2005,Hu2016,Billington2019}. However, we should note here that our sample is much smaller than the samples studied in these previous works, which is reflected by the large uncertainty in the association rate. As discussed in Sect. \ref{sec:intro}, many of the MYSOs that power a methanol maser could be in the HC~HII stage. Because of their small angular size ($\lesssim$0.03~pc; \citealt{Kurtz2005}), such sources are expected to have weak radio emission ($\sim$20~mJy at 1.4~kpc; \citealt{Thompson2016}) that, because of their high electron density ($\gtrsim$10$^6$~cm$^{-3}$; \citealt{Kurtz2005}), would be optically thick up to high radio frequency, implying a spectral index of $\approx$2. Unfortunately, the low flux levels of the continuum sources (which causes these sources to remain undetected at the individual subbands across the bandwidth; cf.\ Sect. \ref{sec:obs}) and the very limited frequency range covered (of only 2~GHz) preclude a meaningful determination of their spectral indices, which would have allowed us to unambiguously classify them as HC~HII regions. 

\begin{figure}[tbh]
\begin{center}
\includegraphics[width=0.35\textwidth,angle=0]{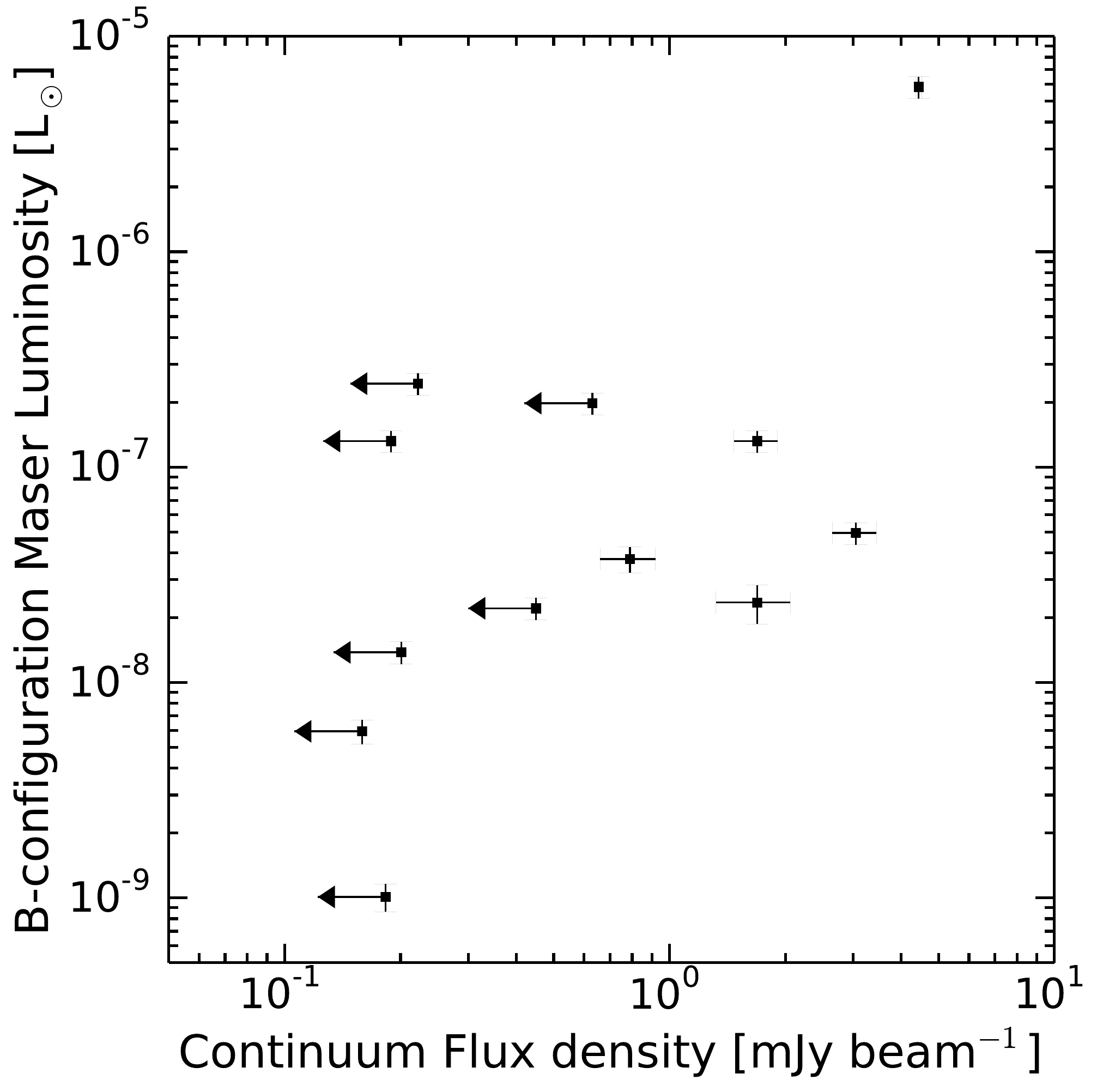}
\caption{Maser integrated luminosity against continuum flux density for the sample of masers detected by GLOSTAR in B configuration. The arrows indicate upper limits of continuum flux density.
}
\label{fig:maserLum-radio}
\end{center}
\end{figure}

In Fig. \ref{fig:maserLum-radio} we show the 6.7~GHz maser integrated luminosity against continuum flux density  for the five methanol masers that are associated with compact radio continuum emission as well as the 3$\sigma$ upper limits for the non-detections. 
We find no significant correlation between these values 
(a Spearman correlation test yields a coefficient of 
0.67 with a $p$-value $=0.22$).
If confirmed with more significant statistics, for instance by the GLOSTAR observations from the full Galactic plane, this would suggest
that the 6.7~GHz maser phenomenon is not directly linked to the ionizing radiation.

\begin{figure}[tbh]
\begin{center}
\includegraphics[width=0.3\textwidth,angle=0]{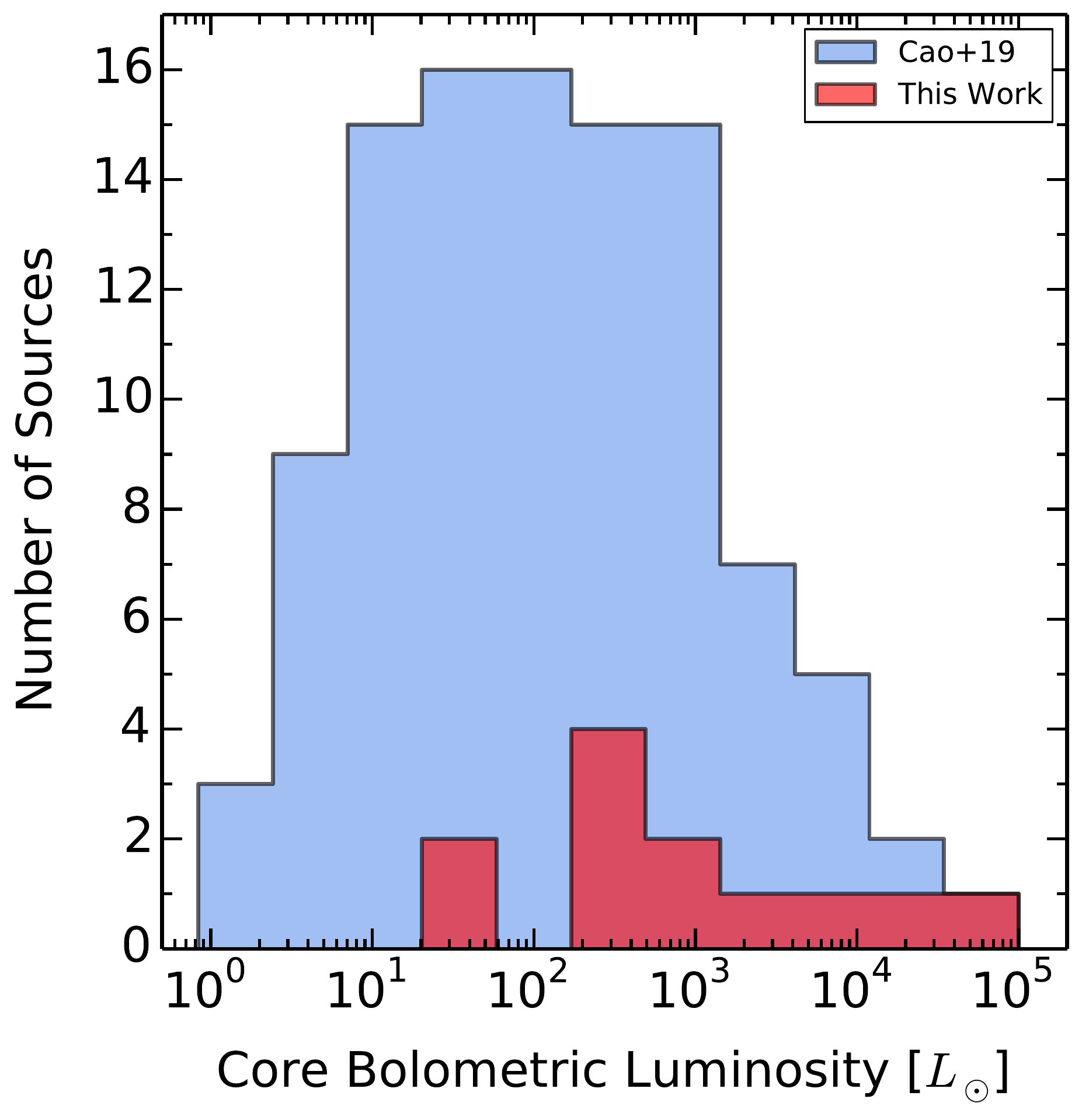}
\includegraphics[width=0.3\textwidth,angle=0]{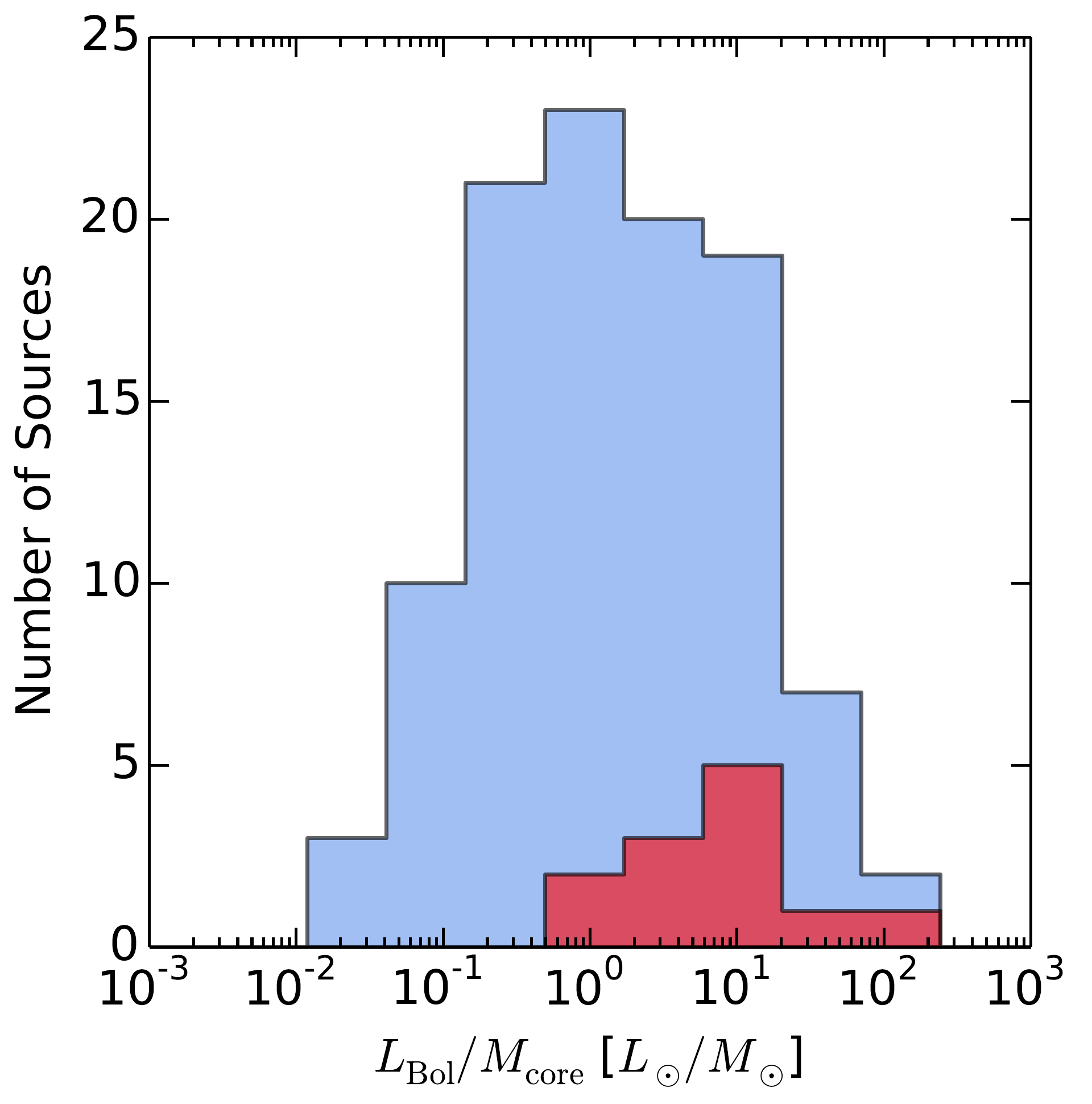}  
\caption{Distribution of core bolometric luminosities (top) and the $L_{\rm Bol}/M_{\rm core}$ ratio (bottom). 
In both panels the full sample of dust cores by \cite{Cao2019} is shown in blue. 
The distribution of cores with associated masers detected by GLOSTAR is shown in red. }
\label{fig:distributions}
\end{center}
\end{figure}

\begin{figure*}[tbh]
\begin{center}
\includegraphics[width=0.33\textwidth,angle=0]{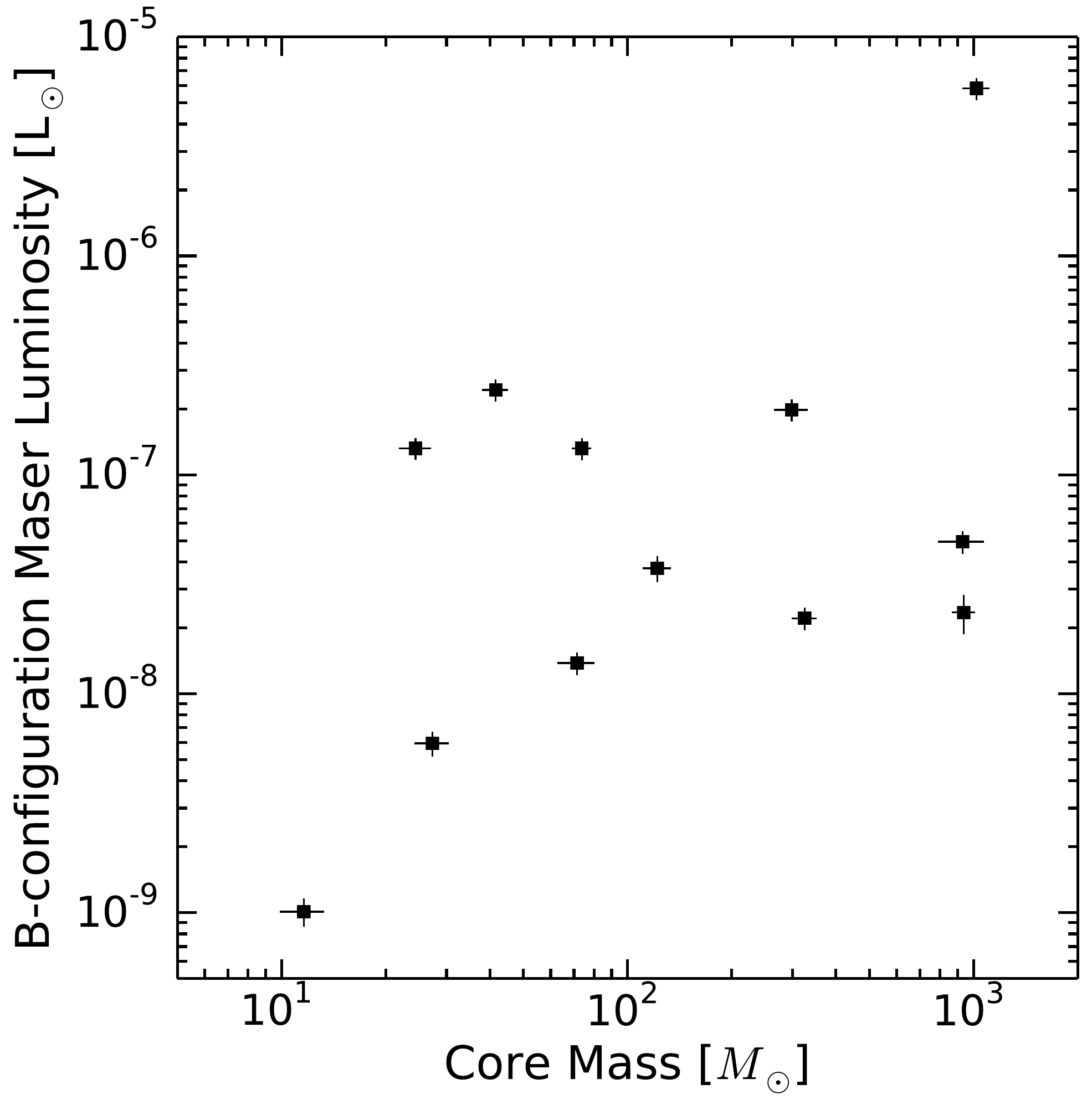}  
\includegraphics[width=0.33\textwidth,angle=0]{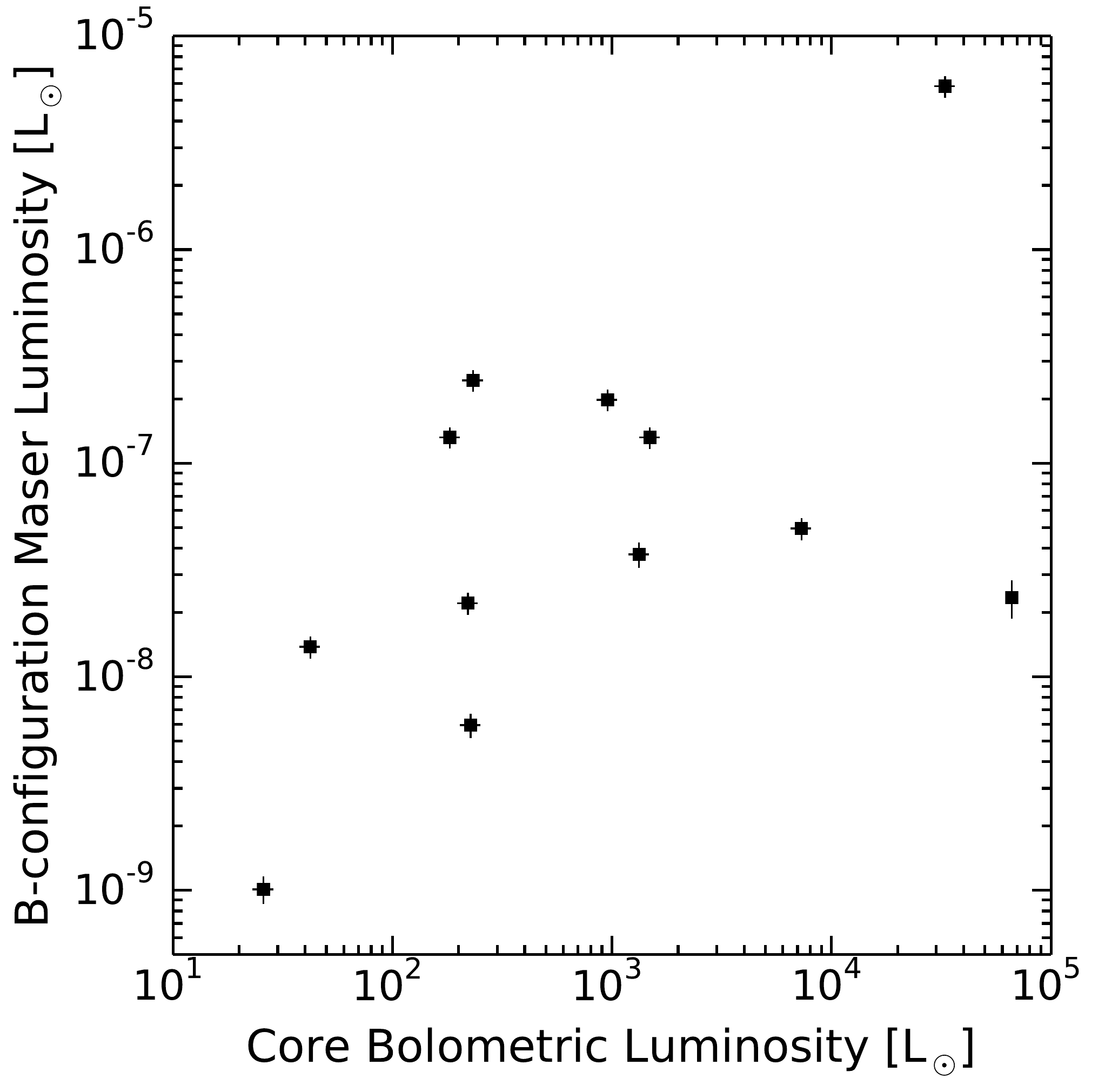}
\includegraphics[width=0.32\textwidth,angle=0]{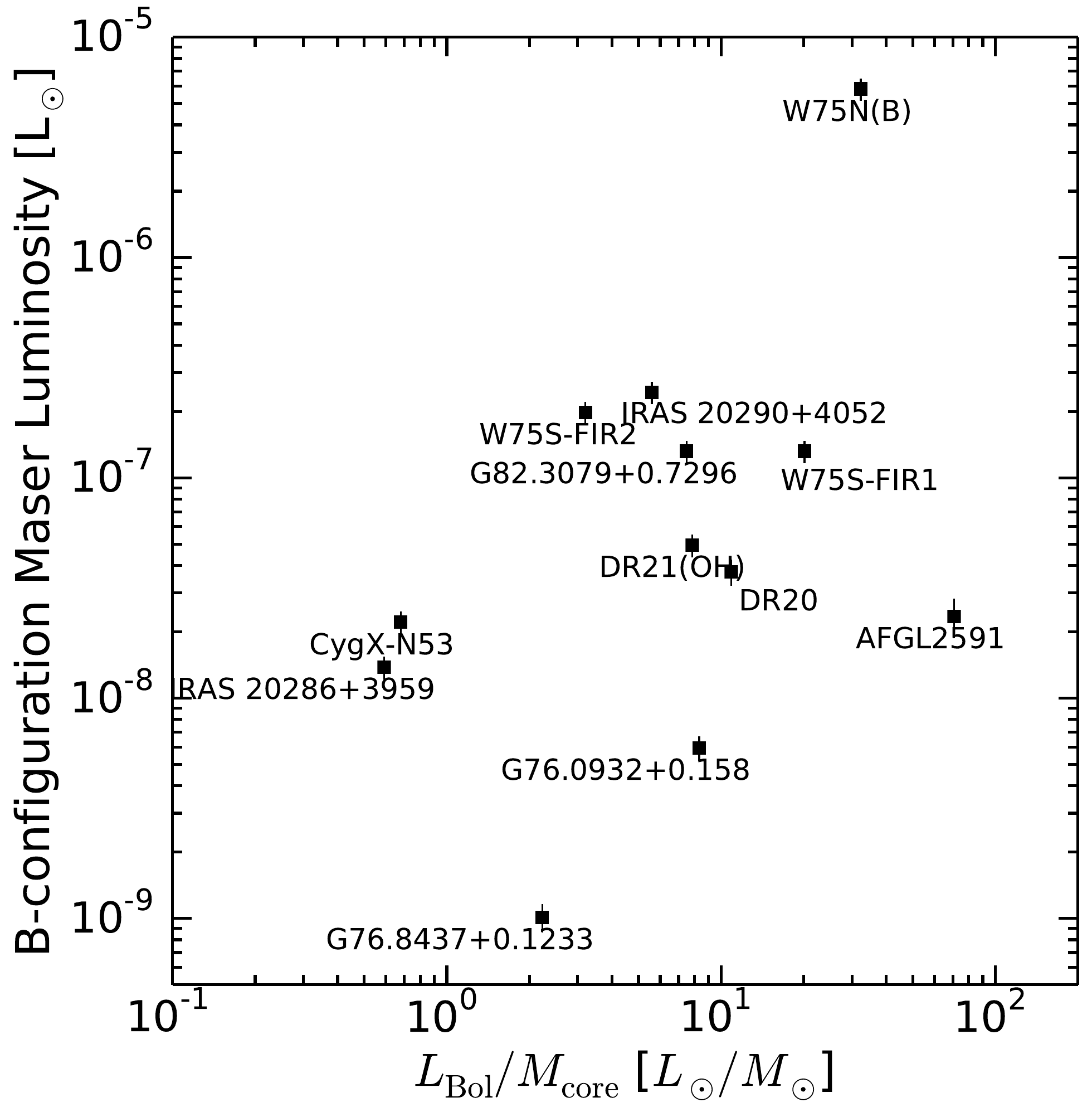}
\caption{Maser integrated luminosity measured in B-configuration GLOSTAR data as a function of core mass (left) and bolometric luminosity (middle). The right panel presents maser integrated luminosity against the $L_{\rm Bol}/M_{\rm core}$ ratio.
}
\label{fig:relations}
\end{center}
\end{figure*}

\subsection{Mass and luminosity correlation}

We estimated the properties of the dust cores that are associated with a methanol source (Sect. \ref{sec:submm}). In order to study the relationship between methanol masers and massive star-forming dust cores, in this section we investigate the physical properties of dust cores with and without associated masers as well as the correlations between the methanol masers and dust core properties.

The upper panel of Fig. \ref{fig:distributions} shows the distribution of bolometric luminosity for the sample of \cite{Cao2019} and the maser associated cores. Although the sample of cores with associated masers is small, it does seem to cover a narrower range in luminosity than the range occupied by the full sample of massive dense cores; however, it still covers a range of a few orders of magnitude (from $\sim10^{1.4}$ to $\sim10^{5}~L_\odot$). 
A similar result was reported by \cite{Billington2019} for the ATLASGAL clumps. They found that maser associated clumps have, on average, higher luminosities when compared to their full sample. Figure \ref{fig:distributions} also suggests a luminosity limit for maser emission of $\sim$200~$L_\odot$, which is lower than the limit of $10^3 L_\odot$ estimated by \cite{Bourke2005} and more similar to the limit found in \cite{Paulson2020} and \cite{Jones2020}.
The bottom panel of Fig. \ref{fig:distributions} shows the distribution of the bolometric luminosity-to-core-mass ratio, $L_{\rm Bol}/M_{\rm core}$,  for the sample of \cite{Cao2019} and the maser associated cores found with GLOSTAR. There seems to be a threshold of this parameter for maser emission since the minimum $L_{\rm Bol}/M_{\rm core}$ of a
source that is associated with maser emission is $\sim1~L_\odot~M^{-1}_\odot$. 
A similar value was suggested by \cite{Billington2019}.

Figure \ref{fig:relations} shows the correlations between methanol masers and dust core properties. The left and middle panels present maser integrated luminosity as a function of core mass and bolometric luminosity, respectively. We see in these figures that if we exclude the maser with the strongest luminosity, the correlation between these properties is nearly nonexistent.  
However, this may be due to the limited size of our sample. \cite{Billington2019} found a weak correlation between clump bolometric luminosity and maser luminosity for a much larger sample (958 dust clumps that host methanol masers). In another study, \cite{Paulson2020} also found a weak correlation between clump and maser luminosity for their sample of 320 MMB masers. These authors also noted that the relationship between clump and maser luminosity for the low-luminosity ($L<10^{-6} L_\odot$) maser population is consistent with no correlation. If this non-correlation is confirmed for a larger sample, it would suggest that, for low-luminosity masers, the maser luminosity is probably not mainly driven by core luminosity but by other factors, as suggested by \cite{Paulson2020}, such as gas density, gas temperature, and methanol fractional abundance, which are not explored in this paper. We note that the two new detected masers are associated with dust cores that have the lowest mass, $M_{\rm core}$, and also have the lowest luminosity, $L_{\rm maser,~B}$ (cf.\ Table \ref{tab:coresProp}). While previous maser studies toward Cygnus~X have selected the brightest sources \citep[e.g.,][]{Hu2016,Yang2019}, GLOSTAR is an unbiased survey and is also more sensitive than most of these previous efforts, allowing the detection of more new masers at the lower end of the maser luminosity distribution function. 

The luminosity-to-mass ($L/M$) ratio has been proposed to be an evolutionary indicator during the formation process of massive clumps (e.g., \citealt{Molinari2008}, see, however, \citealt{Ma2013}). The $L/M$ ratio increases with age as a result of the envelope mass falling into the forming star. Previous works have found a weak correlation between the $L/M$ ratio of maser associated clumps and maser luminosity \cite[e.g.,][]{Billington2019}, suggesting that higher-luminosity masers are associated with more evolved clumps. We investigated this relation for the maser host cores in Cygnus X. The right panel of Fig. \ref{fig:relations} shows maser integrated luminosity against the $L_{\rm Bol}/M_{\rm core}$ ratio. The correlation between these quantities is not significant (the Spearman correlation  coefficient is 0.3 with $p$-value =0.3),  
which can be attributed to the large scatter of the measurements and the small size of our sample. However, we note that the masers associated with UC or HC~HII
regions (cf.\ Table \ref{tab:maserCtp}) occupy the region in this diagram with high $L_{\rm Bol}/M_{\rm core}$ ratios, as expected for more evolved objects; specifically, they all lie above $L_{\rm Bol}/M_{\rm core}\approx8~L_\odot~M^{-1}_\odot$.

In summary, the results above suggest that there is a core bolometric luminosity threshold for the maser emission of $\sim$200~$L_\odot$, which is in agreement with that found in recent works that have studied a significant sample of maser host clumps in the Galactic plane but excluded the Cygnus X complex. For the minimum core mass, we found a value of $\approx$10~$M_\odot$
in our sample (cf.\ Table \ref{tab:coresProp}), which is likely sufficient to form a single high-mass star.
In a previous study, \cite{Billington2019} reported a minimum FWHM mass (mass within 50\%  of the 850 or 870 $\mu$m contour)
 of 17~M$_\odot$ for their distance-limited sample of cores that host methanol masers, 
while the minimum mass reported by \cite{Paulson2020} is $11~M_\odot$. 
The distribution of the luminosity-to-mass ratio of dust cores  that host maser emission is narrower than the distribution of the full core population, with methanol host cores having ratios between $\sim$1 and $10^2~L_\odot~M^{-1}_\odot$. 
This is consistent with that expected from the evolutionary sequence model of \cite{Breen2010} and \cite{Billington2019}, who suggest that the maser emission starts at a 
$L/M$ ratio of $10^{0.6}$~$L_\odot~M^{-1}_\odot$, followed by the maser emission and continuum emission from the HII region being simultaneously detectable at higher values of $L/M$, leading to the final decline of the maser emission due to the disruptive effects of the HII region at a
$L/M$ value of $10^{2.2}$~$L_\odot~M^{-1}_\odot$.
The threshold of the luminosity-to-mass ratio of $\sim$1~$L_\odot~M^{-1}_\odot$ is also consistent with what has been measured for a larger sample of star-forming regions in the Galactic plane \citep{Billington2019,Billington2020}. 

\section{Conclusions}

We have conducted the first unbiased survey of Class II methanol masers at 6.7~GHz in the Cygnus~X complex. A total of 13 masers were detected in the observed $7^{\rm o}\times3^{\rm o}$  area, for which we derive fluxes and integrated luminosities, as well as absolute positions of the maser features. In addition, methanol absorption is detected toward DR21 
against the compact continuum from the HII regions inside the DR21 core. 
We have established the associations between maser sources and radio continuum, (sub)millimeter, and infrared emission to within $\approx1''$. All masers are associated with dust continuum cores, and all of them are  close to or coincident with near-infrared or FIR peaks, as expected from the maser-MYSO relationship. Only a fraction of them ($38\pm17\%$) have radio continuum detected by our survey or in previous observations. These sources are more likely to be UC HII or HC~HII regions. 

By fitting the SED to the dust continuum emission, we have derived the properties of the cores that host maser emission and compared their distributions with those of the full population of dust cores in the Cygnus~X complex identified in the literature. We found thresholds for maser emission in luminosity and the luminosity-to-mass ratio of $\sim200~L_\odot$ and $\sim1~L_\odot~M^{-1}_\odot$, respectively. These values are in agreement with recent works that have investigated a larger sample of methanol maser hosts in the Galactic plane.

\begin{acknowledgements}
The authors are grateful to the anonymous referee, whose comments helped to improve this paper. G.N.O.-L. would like to thank Antonio Hernandez-G\'omez for a detailed reading of an early version of the manuscript and valuable suggestions.
We thank Yue Cao and Keping Qiu for providing mosaics maps of (sub)millimeter emission. We also thank Nicola Schneider for sharing her IRAM CO data. G.N.O.-L. acknowledges support from the von Humboldt Stiftung. T.Cs.\ has received financial support from the French State in the framework of the IdEx Universit\'e de Bordeaux Investments for the future Program. H.B.\ acknowledges  support from the European Research Council under the European Community's Horizon 2020 framework program (2014-2020) via the ERC Consolidator grant ``From Cloud to Star Formation (CSF)'' (project number 648505). H.B.\ further acknowledges support from the Deutsche Forschungsgemeinschaft (DFG) via Sonderforschungsbereich (SFB) 881 ``The Milky Way System'' (sub-project B1).
The National Radio Astronomy Observatory (NRAO) is operated by Associated Universities Inc., under cooperative agreement with the National Science  Foundation.
\end{acknowledgements}


\bibliographystyle{aa} 
\bibliography{ms} 

\appendix\label{sec:appendix}

\section{Supplementary figures and tables}\label{sec:supplm}

In this appendix, Figs. \ref{fig:submm-appendix} and \ref{fig:infrared2-appendix} show (sub)millimeter dust emission and {\it Spitzer} infrared emission in a region of $3{'}\times3{'}$ and $0\rlap.{'}7\times0\rlap.{'}7$, respectively, around maser positions. \\

Table \ref{tab:catalog} lists the catalog of maser spots detected above 4$\sigma$ in B-configuration data.\\

Figure \ref{fig:spots} displays the distribution of all methanol maser spots measured in the B-configuration images. For five of our sources, the 6.7 GHz methanol maser emission was imaged by \citet{Rygl2012} with the EVN with a resolution of a few milliarcseconds (mas), much better than our $1\rlap.{''}2$. A comparison between the EVN positions, which have absolute position uncertainties better than 1 mas, and even our higher resolution B-array VLA data is not straightforward for several reasons: Generally, in the VLA images several maser spots remain spatially unresolved, and our positions represent averages. Moreover, for several sources the EVN data show maser emission only over a fraction of the velocity range of the VLA data, leaving the determination of an emission centroid biased. 
Nevertheless, we note that for most sources the EVN-determined maser spots lie within the distributions of the VLA B-array spots (within tens of mas). For two sources, G81.7219+0.5711 [DR21(OH)] and G81.8713+0.78-7 [W75N(B)], we find offsets between the EVN and VLA positions of $\approx 0\rlap.{''}2 $ and $\approx 0\rlap.{''}15$, respectively. For DR21(OH), the significance of the offset is difficult to assess as many fewer spots (a total of three) appear in the EVN data than in the VLA data over a much narrower velocity range. Still, the offsets are smaller than or comparable to the position uncertainty discussed in Sect. \ref{sec:search}.

\clearpage
\onecolumn
\begin{longtable}{cccccccc}
\caption{\label{tab:catalog} Maser spot catalog for B-configuration data. } \\
\hline\hline
Name  & Ref.\ RA &  Ref.\ Dec. & $\Delta\alpha$ & $\Delta\delta$ & $V_{\rm LSR}$  &  $S_{\nu,~\rm Peak}$       & $S_{\nu,~\rm Int.}$   \\
      & (h:m:s) &  ($^{\rm o}$:$'$:$''$) & (arcsec) & (arcsec) & (km~s$^{-1}$)   &(Jy~beam$^{-1}$)  &   (Jy)       \\
(1)   & (2)     & (3)              & (4)              & (5)             & (6)              & (7)   & (8)      \\ 
\hline
\endfirsthead
\caption{Continued.}\\
\hline\hline
Name  & Ref.\ RA &  Ref.\ Dec. & $\Delta\alpha$ & $\Delta\delta$ & $V_{\rm LSR}$  & $S_{\nu,~\rm Peak}$        & $S_{\nu,~\rm Int.}$    \\
      & (h:m:s) &  ($^{\rm o}$:$'$:$''$) & (arcsec) & (arcsec) & (km~s$^{-1}$)   &(Jy~beam$^{-1}$)  &   (Jy)       \\
(1)   & (2)     & (3)              & (4)              & (5)             & (6)              & (7)   & (8)      \\ 
\hline
\endhead
\hline
\endfoot
\hline
\multicolumn{8}{l}{{\bf Notes.} The position uncertainties are
$\pm0\rlap.{''}18$ and $\pm0\rlap.{''}13$ 
in RA\ and Dec., respectively 
(see Sect. \ref{sec:search}).}
\endlastfoot
G76.0932+0.1580 & 20:23:23.7140 & +37:35:35.558 & 0.000 & 0.000 &  6.64 & 0.57 $\pm$ 0.01 & 0.59 $\pm$ 0.02 \\ 
G76.0932+0.1580 & 20:23:23.7140 & +37:35:35.558 & -0.089 & -0.024 &  6.46 & 0.38 $\pm$ 0.01 & 0.44 $\pm$ 0.02 \\ 
G76.0932+0.1580 & 20:23:23.7140 & +37:35:35.558 & -0.103 & 0.055 &  4.84 & 0.38 $\pm$ 0.01 & 0.48 $\pm$ 0.01 \\ 
G76.0932+0.1580 & 20:23:23.7140 & +37:35:35.558 & -0.281 & -0.005 & -6.50 & 0.17 $\pm$ 0.01 & 0.18 $\pm$ 0.01 \\ 
G76.0932+0.1580 & 20:23:23.7140 & +37:35:35.558 & -0.067 & -0.132 &  5.56 & 0.15 $\pm$ 0.01 & 0.12 $\pm$ 0.01 \\ 
G76.0932+0.1580 & 20:23:23.7140 & +37:35:35.558 & -0.202 & -0.126 & -6.32 & 0.14 $\pm$ 0.01 & 0.12 $\pm$ 0.01 \\ 
G76.0932+0.1580 & 20:23:23.7140 & +37:35:35.558 & 0.015 & -0.005 &  4.66 & 0.15 $\pm$ 0.00 & 0.18 $\pm$ 0.01 \\ 
G76.0932+0.1580 & 20:23:23.7140 & +37:35:35.558 & -0.143 & 0.135 &  5.02 & 0.10 $\pm$ 0.01 & 0.12 $\pm$ 0.01 \\ 
G76.0932+0.1580 & 20:23:23.7140 & +37:35:35.558 & -0.172 & -0.183 &  5.74 & 0.11 $\pm$ 0.01 & 0.11 $\pm$ 0.01 \\ 

G76.8437+0.1233 & 20:25:43.7657 & +38:11:12.734 & 0.000 & 0.000 & -5.42 & 0.17 $\pm$ 0.00 & 0.17 $\pm$ 0.01 \\ 
G76.8437+0.1233 & 20:25:43.7657 & +38:11:12.734 & 0.082 & 0.079 & -5.24 & 0.15 $\pm$ 0.00 & 0.18 $\pm$ 0.01 \\ 
G76.8437+0.1233 & 20:25:43.7657 & +38:11:12.734 & 0.110 & 0.156 & -5.60 & 0.07 $\pm$ 0.00 & 0.06 $\pm$ 0.01 \\ 

G78.8870+0.7087 & 20:29:24.9436 & +40:11:19.626 & 0.000 & 0.000 & -7.04 & 0.10 $\pm$ 0.01 & 0.19 $\pm$ 0.03 \\ 
G78.8870+0.7087 & 20:29:24.9436 & +40:11:19.626 & -0.054 & -0.160 & -7.22 & 0.09 $\pm$ 0.01 & 0.11 $\pm$ 0.01 \\ 
G78.8870+0.7087 & 20:29:24.9436 & +40:11:19.626 & 0.170 & -0.404 & -7.40 & 0.09 $\pm$ 0.01 & 0.09 $\pm$ 0.01 \\ 

G78.9690+0.5410 & 20:30:22.6762 & +40:09:23.348 & 0.000 & 0.000 &  4.84 & 1.25 $\pm$ 0.01 & 1.28 $\pm$ 0.02 \\ 
G78.9690+0.5410 & 20:30:22.6762 & +40:09:23.348 & -0.008 & -0.063 &  4.66 & 1.07 $\pm$ 0.00 & 1.08 $\pm$ 0.01 \\ 
G78.9690+0.5410 & 20:30:22.6762 & +40:09:23.348 & 0.035 & -0.014 &  4.48 & 0.95 $\pm$ 0.01 & 1.02 $\pm$ 0.01 \\ 
G78.9690+0.5410 & 20:30:22.6762 & +40:09:23.348 & 0.063 & -0.019 &  5.02 & 0.88 $\pm$ 0.01 & 0.94 $\pm$ 0.01 \\ 
G78.9690+0.5410 & 20:30:22.6762 & +40:09:23.348 & 0.025 & -0.131 &  4.30 & 0.57 $\pm$ 0.01 & 0.61 $\pm$ 0.01 \\ 
G78.9690+0.5410 & 20:30:22.6762 & +40:09:23.348 & 0.021 & 0.038 &  5.20 & 0.21 $\pm$ 0.01 & 0.21 $\pm$ 0.01 \\ 
G78.9690+0.5410 & 20:30:22.6762 & +40:09:23.348 & -0.098 & -0.019 &  4.12 & 0.12 $\pm$ 0.00 & 0.10 $\pm$ 0.00 \\ 
G78.9690+0.5410 & 20:30:22.6762 & +40:09:23.348 & 0.181 & -0.099 &  5.38 & 0.10 $\pm$ 0.00 & 0.09 $\pm$ 0.01 \\ 

G78.9884+0.2211 & 20:31:47.3082 & +39:58:59.903 & 0.000 & 0.000 & -68.96 & 0.25 $\pm$ 0.01 & 0.30 $\pm$ 0.01 \\ 
G78.9884+0.2211 & 20:31:47.3082 & +39:58:59.903 & -0.224 & 0.096 & -66.44 & 0.20 $\pm$ 0.00 & 0.16 $\pm$ 0.01 \\ 
G78.9884+0.2211 & 20:31:47.3082 & +39:58:59.903 & -0.028 & -0.103 & -66.26 & 0.21 $\pm$ 0.01 & 0.30 $\pm$ 0.02 \\ 
G78.9884+0.2211 & 20:31:47.3082 & +39:58:59.903 & -0.564 & 0.122 & -68.78 & 0.22 $\pm$ 0.01 & 0.34 $\pm$ 0.03 \\ 

G79.7358+0.9904 & 20:30:50.6680 & +41:02:27.403 & 0.000 & 0.000 & -5.42 & 22.96 $\pm$ 0.21 & 23.81 $\pm$ 0.39 \\ 
G79.7358+0.9904 & 20:30:50.6680 & +41:02:27.403 & -0.029 & -0.001 & -5.60 & 21.87 $\pm$ 0.24 & 22.88 $\pm$ 0.44 \\ 
G79.7358+0.9904 & 20:30:50.6680 & +41:02:27.403 & -0.028 & -0.009 & -5.78 & 11.84 $\pm$ 0.16 & 12.70 $\pm$ 0.30 \\ 
G79.7358+0.9904 & 20:30:50.6680 & +41:02:27.403 & -0.007 & -0.016 & -5.24 & 7.38 $\pm$ 0.09 & 7.85 $\pm$ 0.17 \\ 
G79.7358+0.9904 & 20:30:50.6680 & +41:02:27.403 & -0.047 & -0.001 & -5.96 & 5.62 $\pm$ 0.08 & 5.79 $\pm$ 0.15 \\ 
G79.7358+0.9904 & 20:30:50.6680 & +41:02:27.403 & 0.055 & -0.010 & -3.98 & 5.88 $\pm$ 0.09 & 6.11 $\pm$ 0.15 \\ 
G79.7358+0.9904 & 20:30:50.6680 & +41:02:27.403 & 0.045 & -0.039 & -3.26 & 4.97 $\pm$ 0.06 & 5.07 $\pm$ 0.12 \\ 
G79.7358+0.9904 & 20:30:50.6680 & +41:02:27.403 & 0.074 & -0.022 & -3.08 & 4.62 $\pm$ 0.09 & 4.81 $\pm$ 0.17 \\ 
G79.7358+0.9904 & 20:30:50.6680 & +41:02:27.403 & -0.010 & -0.015 & -3.80 & 2.86 $\pm$ 0.06 & 2.88 $\pm$ 0.11 \\ 
G79.7358+0.9904 & 20:30:50.6680 & +41:02:27.403 & 0.088 & -0.026 & -3.62 & 2.34 $\pm$ 0.05 & 2.41 $\pm$ 0.10 \\ 
G79.7358+0.9904 & 20:30:50.6680 & +41:02:27.403 & -0.013 & 0.004 & -6.14 & 2.26 $\pm$ 0.04 & 2.35 $\pm$ 0.08 \\ 
G79.7358+0.9904 & 20:30:50.6680 & +41:02:27.403 & 0.180 & 0.021 & -4.88 & 0.95 $\pm$ 0.04 & 1.03 $\pm$ 0.07 \\ 
G79.7358+0.9904 & 20:30:50.6680 & +41:02:27.403 & 0.018 & 0.020 & -6.32 & 0.90 $\pm$ 0.04 & 0.93 $\pm$ 0.07 \\ 
G79.7358+0.9904 & 20:30:50.6680 & +41:02:27.403 & 0.062 & 0.002 & -4.70 & 0.86 $\pm$ 0.03 & 0.82 $\pm$ 0.05 \\ 
G79.7358+0.9904 & 20:30:50.6680 & +41:02:27.403 & 0.012 & -0.054 & -4.16 & 0.68 $\pm$ 0.06 & 0.74 $\pm$ 0.10 \\ 
G79.7358+0.9904 & 20:30:50.6680 & +41:02:27.403 & -0.132 & -0.111 & -5.06 & 0.39 $\pm$ 0.05 & 0.55 $\pm$ 0.10 \\ 
G79.7358+0.9904 & 20:30:50.6680 & +41:02:27.403 & -0.250 & -0.031 & -4.52 & 0.37 $\pm$ 0.02 & 0.33 $\pm$ 0.03 \\ 
G79.7358+0.9904 & 20:30:50.6680 & +41:02:27.403 & -0.070 & 0.017 & -4.34 & 0.14 $\pm$ 0.01 & 0.10 $\pm$ 0.01 \\ 

G80.8617+0.3834 & 20:37:00.9571 & +41:34:55.606 & 0.000 & 0.000 & -3.98 & 4.61 $\pm$ 0.11 & 5.02 $\pm$ 0.21 \\ 
G80.8617+0.3834 & 20:37:00.9571 & +41:34:55.606 & -0.003 & 0.001 & -4.16 & 4.10 $\pm$ 0.12 & 4.55 $\pm$ 0.24 \\ 
G80.8617+0.3834 & 20:37:00.9571 & +41:34:55.606 & -0.029 & 0.331 & -2.00 & 1.56 $\pm$ 0.06 & 1.77 $\pm$ 0.12 \\ 
G80.8617+0.3834 & 20:37:00.9571 & +41:34:55.606 & 0.153 & -0.035 & -4.34 & 1.24 $\pm$ 0.07 & 1.42 $\pm$ 0.14 \\ 
G80.8617+0.3834 & 20:37:00.9571 & +41:34:55.606 & -0.000 & 0.320 & -2.18 & 0.89 $\pm$ 0.06 & 1.02 $\pm$ 0.11 \\ 
G80.8617+0.3834 & 20:37:00.9571 & +41:34:55.606 & 0.206 & -0.093 & -4.52 & 0.61 $\pm$ 0.06 & 0.81 $\pm$ 0.13 \\ 
G80.8617+0.3834 & 20:37:00.9571 & +41:34:55.606 & 0.603 & 0.044 & -3.44 & 0.14 $\pm$ 0.04 & 0.61 $\pm$ 0.20 \\ 
G80.8617+0.3834 & 20:37:00.9571 & +41:34:55.606 & -0.095 & 0.204 & -3.62 & 0.23 $\pm$ 0.02 & 0.26 $\pm$ 0.04 \\ 

G81.7219+0.5711 & 20:39:01.0501 & +42:22:49.124 & 0.000 & 0.000 & -2.72 & 4.07 $\pm$ 0.05 & 4.30 $\pm$ 0.10 \\ 
G81.7219+0.5711 & 20:39:01.0501 & +42:22:49.124 & 0.202 & 0.048 & -3.08 & 2.72 $\pm$ 0.05 & 3.12 $\pm$ 0.10 \\ 
G81.7219+0.5711 & 20:39:01.0501 & +42:22:49.124 & 0.216 & 0.044 & -3.26 & 2.19 $\pm$ 0.04 & 2.49 $\pm$ 0.07 \\ 
G81.7219+0.5711 & 20:39:01.0501 & +42:22:49.124 & 0.098 & 0.034 & -2.90 & 2.36 $\pm$ 0.04 & 2.54 $\pm$ 0.07 \\ 
G81.7219+0.5711 & 20:39:01.0501 & +42:22:49.124 & -0.325 & -0.133 & -3.80 & 1.30 $\pm$ 0.03 & 1.43 $\pm$ 0.05 \\ 
G81.7219+0.5711 & 20:39:01.0501 & +42:22:49.124 & -0.023 & 0.025 & -2.54 & 1.60 $\pm$ 0.04 & 1.79 $\pm$ 0.07 \\ 
G81.7219+0.5711 & 20:39:01.0501 & +42:22:49.124 & -1.163 & -0.226 &  8.98 & 1.22 $\pm$ 0.03 & 1.25 $\pm$ 0.06 \\ 
G81.7219+0.5711 & 20:39:01.0501 & +42:22:49.124 & 0.187 & 0.073 & -3.44 & 0.89 $\pm$ 0.02 & 1.03 $\pm$ 0.05 \\ 
G81.7219+0.5711 & 20:39:01.0501 & +42:22:49.124 & -0.433 & -0.126 & -3.98 & 0.88 $\pm$ 0.02 & 0.87 $\pm$ 0.04 \\ 
G81.7219+0.5711 & 20:39:01.0501 & +42:22:49.124 & -1.240 & -0.223 &  8.80 & 0.68 $\pm$ 0.03 & 0.70 $\pm$ 0.05 \\ 
G81.7219+0.5711 & 20:39:01.0501 & +42:22:49.124 & -0.320 & -0.141 & -3.62 & 0.55 $\pm$ 0.04 & 0.63 $\pm$ 0.07 \\ 
G81.7219+0.5711 & 20:39:01.0501 & +42:22:49.124 & -1.305 & -0.011 &  9.16 & 0.18 $\pm$ 0.02 & 0.33 $\pm$ 0.06 \\ 
G81.7219+0.5711 & 20:39:01.0501 & +42:22:49.124 & -0.467 & -0.105 & -4.16 & 0.15 $\pm$ 0.01 & 0.12 $\pm$ 0.01 \\ 
G81.7219+0.5711 & 20:39:01.0501 & +42:22:49.124 & -1.415 & -0.087 &  9.34 & 0.13 $\pm$ 0.01 & 0.13 $\pm$ 0.01 \\ 

G81.7444+0.5910 & 20:39:00.3722 & +42:24:37.089 & 0.000 & 0.000 &  4.48 & 17.18 $\pm$ 0.12 & 17.88 $\pm$ 0.22 \\ 
G81.7444+0.5910 & 20:39:00.3722 & +42:24:37.089 & 0.002 & 0.010 &  4.66 & 11.57 $\pm$ 0.09 & 11.89 $\pm$ 0.17 \\ 
G81.7444+0.5910 & 20:39:00.3722 & +42:24:37.089 & 0.015 & 0.020 &  3.58 & 6.58 $\pm$ 0.08 & 6.61 $\pm$ 0.14 \\ 
G81.7444+0.5910 & 20:39:00.3722 & +42:24:37.089 & 0.006 & 0.002 &  3.76 & 5.37 $\pm$ 0.07 & 5.42 $\pm$ 0.12 \\ 
G81.7444+0.5910 & 20:39:00.3722 & +42:24:37.089 & -0.025 & 0.004 &  3.94 & 2.89 $\pm$ 0.05 & 3.00 $\pm$ 0.09 \\ 
G81.7444+0.5910 & 20:39:00.3722 & +42:24:37.089 & -0.045 & 0.015 &  3.40 & 2.52 $\pm$ 0.04 & 2.56 $\pm$ 0.08 \\ 
G81.7444+0.5910 & 20:39:00.3722 & +42:24:37.089 & -0.027 & -0.039 &  4.30 & 2.00 $\pm$ 0.03 & 1.92 $\pm$ 0.06 \\ 
G81.7444+0.5910 & 20:39:00.3722 & +42:24:37.089 & -0.001 & 0.044 &  4.12 & 1.72 $\pm$ 0.04 & 1.73 $\pm$ 0.07 \\ 
G81.7444+0.5910 & 20:39:00.3722 & +42:24:37.089 & -0.074 & 0.049 &  3.22 & 1.50 $\pm$ 0.03 & 1.54 $\pm$ 0.06 \\ 
G81.7444+0.5910 & 20:39:00.3722 & +42:24:37.089 & -0.030 & 0.009 &  3.04 & 1.20 $\pm$ 0.03 & 1.25 $\pm$ 0.05 \\ 
G81.7444+0.5910 & 20:39:00.3722 & +42:24:37.089 & -0.005 & -0.089 &  5.02 & 0.27 $\pm$ 0.03 & 0.35 $\pm$ 0.05 \\ 
G81.7444+0.5910 & 20:39:00.3722 & +42:24:37.089 & 0.073 & -0.040 &  5.38 & 0.17 $\pm$ 0.01 & 0.12 $\pm$ 0.01 \\ 

G81.7523+0.5908 & 20:39:01.9870 & +42:24:59.261 & 0.000 & 0.000 & -8.66 & 15.19 $\pm$ 0.10 & 15.67 $\pm$ 0.19 \\ 
G81.7523+0.5908 & 20:39:01.9870 & +42:24:59.261 & 0.030 & 0.010 & -8.48 & 11.28 $\pm$ 0.09 & 11.62 $\pm$ 0.17 \\ 
G81.7523+0.5908 & 20:39:01.9870 & +42:24:59.261 & -0.014 & -0.140 & -5.78 & 7.93 $\pm$ 0.07 & 8.12 $\pm$ 0.13 \\ 
G81.7523+0.5908 & 20:39:01.9870 & +42:24:59.261 & 0.018 & -0.141 & -5.96 & 5.47 $\pm$ 0.06 & 5.63 $\pm$ 0.11 \\ 
G81.7523+0.5908 & 20:39:01.9870 & +42:24:59.261 & -0.036 & -0.107 & -6.86 & 4.29 $\pm$ 0.04 & 4.37 $\pm$ 0.08 \\ 
G81.7523+0.5908 & 20:39:01.9870 & +42:24:59.261 & 0.005 & 0.003 & -8.30 & 5.21 $\pm$ 0.06 & 5.52 $\pm$ 0.10 \\ 
G81.7523+0.5908 & 20:39:01.9870 & +42:24:59.261 & 0.020 & 0.021 & -8.12 & 4.44 $\pm$ 0.04 & 4.65 $\pm$ 0.08 \\ 
G81.7523+0.5908 & 20:39:01.9870 & +42:24:59.261 & 0.053 & -0.127 & -5.60 & 2.52 $\pm$ 0.04 & 2.65 $\pm$ 0.07 \\ 
G81.7523+0.5908 & 20:39:01.9870 & +42:24:59.261 & 0.009 & -0.110 & -6.68 & 2.56 $\pm$ 0.03 & 2.65 $\pm$ 0.05 \\ 
G81.7523+0.5908 & 20:39:01.9870 & +42:24:59.261 & -0.022 & -0.111 & -6.32 & 1.88 $\pm$ 0.03 & 1.97 $\pm$ 0.06 \\ 
G81.7523+0.5908 & 20:39:01.9870 & +42:24:59.261 & 0.002 & -0.094 & -7.04 & 1.97 $\pm$ 0.03 & 2.06 $\pm$ 0.05 \\ 
G81.7523+0.5908 & 20:39:01.9870 & +42:24:59.261 & -0.038 & -0.125 & -5.42 & 1.72 $\pm$ 0.03 & 1.83 $\pm$ 0.05 \\ 
G81.7523+0.5908 & 20:39:01.9870 & +42:24:59.261 & -0.020 & -0.020 & -8.84 & 1.81 $\pm$ 0.04 & 1.91 $\pm$ 0.07 \\ 
G81.7523+0.5908 & 20:39:01.9870 & +42:24:59.261 & -0.020 & -0.112 & -6.50 & 1.58 $\pm$ 0.03 & 1.69 $\pm$ 0.06 \\ 
G81.7523+0.5908 & 20:39:01.9870 & +42:24:59.261 & 0.123 & -0.300 & -2.36 & 1.33 $\pm$ 0.03 & 1.45 $\pm$ 0.06 \\ 
G81.7523+0.5908 & 20:39:01.9870 & +42:24:59.261 & 0.023 & 0.001 & -9.02 & 1.15 $\pm$ 0.02 & 1.17 $\pm$ 0.04 \\ 
G81.7523+0.5908 & 20:39:01.9870 & +42:24:59.261 & 0.074 & -0.363 & -2.90 & 1.52 $\pm$ 0.03 & 1.70 $\pm$ 0.07 \\ 
G81.7523+0.5908 & 20:39:01.9870 & +42:24:59.261 & -0.037 & -0.018 & -9.20 & 0.77 $\pm$ 0.02 & 0.89 $\pm$ 0.04 \\ 
G81.7523+0.5908 & 20:39:01.9870 & +42:24:59.261 & 0.035 & -0.124 & -5.24 & 0.66 $\pm$ 0.02 & 0.70 $\pm$ 0.04 \\ 
G81.7523+0.5908 & 20:39:01.9870 & +42:24:59.261 & -0.014 & -0.374 & -3.26 & 0.99 $\pm$ 0.04 & 1.19 $\pm$ 0.09 \\ 
G81.7523+0.5908 & 20:39:01.9870 & +42:24:59.261 & -0.055 & -0.090 & -7.22 & 0.50 $\pm$ 0.02 & 0.56 $\pm$ 0.04 \\ 
G81.7523+0.5908 & 20:39:01.9870 & +42:24:59.261 & -0.094 & -0.431 & -3.08 & 0.73 $\pm$ 0.04 & 0.86 $\pm$ 0.07 \\ 
G81.7523+0.5908 & 20:39:01.9870 & +42:24:59.261 & 0.110 & -0.365 & -2.72 & 0.78 $\pm$ 0.05 & 1.03 $\pm$ 0.10 \\ 
G81.7523+0.5908 & 20:39:01.9870 & +42:24:59.261 & 0.022 & -0.188 & -2.18 & 0.27 $\pm$ 0.02 & 0.34 $\pm$ 0.04 \\ 
G81.7523+0.5908 & 20:39:01.9870 & +42:24:59.261 & 0.113 & -0.051 & -6.14 & 0.28 $\pm$ 0.02 & 0.34 $\pm$ 0.04 \\ 
G81.7523+0.5908 & 20:39:01.9870 & +42:24:59.261 & 0.075 & 0.009 & -2.54 & 0.22 $\pm$ 0.04 & 0.52 $\pm$ 0.11 \\ 
G81.7523+0.5908 & 20:39:01.9870 & +42:24:59.261 & -0.106 & 0.155 & -9.38 & 0.12 $\pm$ 0.02 & 0.34 $\pm$ 0.08 \\ 
G81.7523+0.5908 & 20:39:01.9870 & +42:24:59.261 & -0.013 & -0.173 & -5.06 & 0.14 $\pm$ 0.02 & 0.19 $\pm$ 0.03 \\ 
G81.7523+0.5908 & 20:39:01.9870 & +42:24:59.261 & -0.062 & 0.090 & -7.40 & 0.10 $\pm$ 0.00 & 0.15 $\pm$ 0.01 \\ 
G81.7523+0.5908 & 20:39:01.9870 & +42:24:59.261 & 0.059 & 0.013 & -7.76 & 0.09 $\pm$ 0.00 & 0.11 $\pm$ 0.01 \\ 

G81.7655+0.5972 & 20:39:02.9329 & +42:25:50.954 & 0.000 & 0.000 & -1.28 & 3.90 $\pm$ 0.05 & 4.09 $\pm$ 0.09 \\ 
G81.7655+0.5972 & 20:39:02.9329 & +42:25:50.954 & -0.037 & 0.017 & -1.46 & 2.38 $\pm$ 0.03 & 2.49 $\pm$ 0.06 \\ 
G81.7655+0.5972 & 20:39:02.9329 & +42:25:50.954 & -0.009 & -0.003 & -1.10 & 1.32 $\pm$ 0.03 & 1.49 $\pm$ 0.06 \\ 
G81.7655+0.5972 & 20:39:02.9329 & +42:25:50.954 & 0.015 & 0.061 & -1.64 & 0.45 $\pm$ 0.04 & 0.70 $\pm$ 0.09 \\ 
G81.7655+0.5972 & 20:39:02.9329 & +42:25:50.954 & 0.171 & -0.057 & -1.82 & 0.16 $\pm$ 0.02 & 0.18 $\pm$ 0.03 \\ 
G81.7655+0.5972 & 20:39:02.9329 & +42:25:50.954 & -0.010 & -0.140 & -2.00 & 0.13 $\pm$ 0.01 & 0.16 $\pm$ 0.03 \\ 
G81.7655+0.5972 & 20:39:02.9329 & +42:25:50.954 & -0.359 & -0.088 & -0.74 & 0.09 $\pm$ 0.01 & 0.07 $\pm$ 0.01 \\ 
G81.7655+0.5972 & 20:39:02.9329 & +42:25:50.954 & -0.085 & 0.039 & -0.92 & 0.08 $\pm$ 0.01 & 0.07 $\pm$ 0.01 \\ 
G81.7655+0.5972 & 20:39:02.9329 & +42:25:50.954 & 0.154 & -0.025 & -2.18 & 0.08 $\pm$ 0.01 & 0.06 $\pm$ 0.01 \\ 

G81.8713+0.7807 & 20:38:36.4232 & +42:37:34.756 & 0.000 & 0.000 &  7.18 & 336.13 $\pm$ 0.96 & 346.76 $\pm$ 1.73 \\ 
G81.8713+0.7807 & 20:38:36.4232 & +42:37:34.756 & -0.176 & 0.425 &  4.66 & 324.83 $\pm$ 1.15 & 336.46 $\pm$ 2.08 \\ 
G81.8713+0.7807 & 20:38:36.4232 & +42:37:34.756 & -0.158 & 0.491 &  4.12 & 231.68 $\pm$ 0.89 & 238.50 $\pm$ 1.61 \\ 
G81.8713+0.7807 & 20:38:36.4232 & +42:37:34.756 & -0.170 & 0.444 &  4.48 & 238.51 $\pm$ 1.07 & 246.98 $\pm$ 1.95 \\ 
G81.8713+0.7807 & 20:38:36.4232 & +42:37:34.756 & -0.021 & 0.018 &  7.36 & 199.40 $\pm$ 0.95 & 206.21 $\pm$ 1.71 \\ 
G81.8713+0.7807 & 20:38:36.4232 & +42:37:34.756 & -0.124 & 0.470 &  3.94 & 181.99 $\pm$ 0.88 & 188.35 $\pm$ 1.59 \\ 
G81.8713+0.7807 & 20:38:36.4232 & +42:37:34.756 & -0.261 & 0.346 &  5.74 & 126.12 $\pm$ 0.76 & 128.52 $\pm$ 1.37 \\ 
G81.8713+0.7807 & 20:38:36.4232 & +42:37:34.756 & -0.241 & 0.352 &  5.20 & 114.47 $\pm$ 0.74 & 116.80 $\pm$ 1.33 \\ 
G81.8713+0.7807 & 20:38:36.4232 & +42:37:34.756 & -0.100 & 0.480 &  3.40 & 88.83 $\pm$ 0.80 & 90.64 $\pm$ 1.43 \\ 
G81.8713+0.7807 & 20:38:36.4232 & +42:37:34.756 & -0.233 & 0.349 &  5.02 & 89.76 $\pm$ 0.83 & 91.29 $\pm$ 1.49 \\ 
G81.8713+0.7807 & 20:38:36.4232 & +42:37:34.756 & -0.101 & 0.474 &  3.58 & 64.76 $\pm$ 0.74 & 66.49 $\pm$ 1.33 \\ 
G81.8713+0.7807 & 20:38:36.4232 & +42:37:34.756 & -0.022 & -0.138 &  6.82 & 41.99 $\pm$ 0.58 & 45.36 $\pm$ 1.08 \\ 
G81.8713+0.7807 & 20:38:36.4232 & +42:37:34.756 & -0.122 & -0.424 &  9.34 & 31.67 $\pm$ 0.61 & 35.80 $\pm$ 1.16 \\ 
G81.8713+0.7807 & 20:38:36.4232 & +42:37:34.756 & -0.309 & 0.381 &  5.56 & 35.79 $\pm$ 0.69 & 37.69 $\pm$ 1.26 \\ 
G81.8713+0.7807 & 20:38:36.4232 & +42:37:34.756 & -0.315 & 0.352 &  5.92 & 30.37 $\pm$ 0.61 & 31.77 $\pm$ 1.12 \\ 
G81.8713+0.7807 & 20:38:36.4232 & +42:37:34.756 & -0.193 & 0.414 &  4.84 & 45.54 $\pm$ 0.76 & 47.59 $\pm$ 1.38 \\ 
G81.8713+0.7807 & 20:38:36.4232 & +42:37:34.756 & -0.113 & -0.435 &  9.52 & 16.86 $\pm$ 0.52 & 20.40 $\pm$ 1.04 \\ 
G81.8713+0.7807 & 20:38:36.4232 & +42:37:34.756 & -0.229 & 0.347 &  5.38 & 25.80 $\pm$ 0.64 & 28.06 $\pm$ 1.19 \\ 
G81.8713+0.7807 & 20:38:36.4232 & +42:37:34.756 & -0.199 & 0.444 &  3.76 & 20.97 $\pm$ 0.46 & 22.90 $\pm$ 0.87 \\ 
G81.8713+0.7807 & 20:38:36.4232 & +42:37:34.756 & -0.076 & 0.000 &  7.72 & 6.08 $\pm$ 0.37 & 8.11 $\pm$ 0.79 \\ 
G81.8713+0.7807 & 20:38:36.4232 & +42:37:34.756 & -0.130 & 0.442 &  3.22 & 10.31 $\pm$ 0.44 & 12.30 $\pm$ 0.88 \\ 
G81.8713+0.7807 & 20:38:36.4232 & +42:37:34.756 & -0.172 & -0.060 &  6.46 & 2.47 $\pm$ 0.26 & 4.53 $\pm$ 0.70 \\ 
G81.8713+0.7807 & 20:38:36.4232 & +42:37:34.756 & 0.044 & -0.287 &  9.16 & 2.89 $\pm$ 0.32 & 6.33 $\pm$ 0.97 \\ 
G81.8713+0.7807 & 20:38:36.4232 & +42:37:34.756 & -0.025 & -0.349 &  8.80 & 0.93 $\pm$ 0.12 & 2.03 $\pm$ 0.36 \\ 
G81.8713+0.7807 & 20:38:36.4232 & +42:37:34.756 & -0.100 & -0.261 &  7.00 & 7.90 $\pm$ 0.62 & 9.60 $\pm$ 1.22 \\ 
G81.8713+0.7807 & 20:38:36.4232 & +42:37:34.756 & -0.099 & -0.366 &  8.98 & 2.15 $\pm$ 0.24 & 4.27 $\pm$ 0.69 \\ 
G81.8713+0.7807 & 20:38:36.4232 & +42:37:34.756 & -0.174 & 0.410 &  2.86 & 0.79 $\pm$ 0.10 & 1.88 $\pm$ 0.31 \\ 
G81.8713+0.7807 & 20:38:36.4232 & +42:37:34.756 & 0.130 & -0.086 &  9.70 & 0.90 $\pm$ 0.10 & 2.19 $\pm$ 0.35 \\ 
G81.8713+0.7807 & 20:38:36.4232 & +42:37:34.756 & -0.566 & 0.371 &  6.28 & 1.00 $\pm$ 0.10 & 1.46 $\pm$ 0.24 \\ 
G81.8713+0.7807 & 20:38:36.4232 & +42:37:34.756 & -0.426 & 0.400 &  4.30 & 7.97 $\pm$ 0.52 & 9.50 $\pm$ 1.04 \\ 
G81.8713+0.7807 & 20:38:36.4232 & +42:37:34.756 & -0.387 & -0.150 &  6.64 & 1.72 $\pm$ 0.25 & 3.76 $\pm$ 0.77 \\ 
G81.8713+0.7807 & 20:38:36.4232 & +42:37:34.756 & -0.021 & 0.108 &  8.08 & 0.60 $\pm$ 0.09 & 1.17 $\pm$ 0.26 \\ 
G81.8713+0.7807 & 20:38:36.4232 & +42:37:34.756 & -0.019 & -0.017 &  6.10 & 0.85 $\pm$ 0.15 & 1.95 $\pm$ 0.48 \\ 
G81.8713+0.7807 & 20:38:36.4232 & +42:37:34.756 & -0.053 & -0.448 &  8.62 & 0.52 $\pm$ 0.06 & 0.92 $\pm$ 0.17 \\ 
G81.8713+0.7807 & 20:38:36.4232 & +42:37:34.756 & 0.113 & -0.042 &  8.44 & 0.27 $\pm$ 0.05 & 0.71 $\pm$ 0.18 \\ 
G81.8713+0.7807 & 20:38:36.4232 & +42:37:34.756 & 0.081 & -0.498 &  9.88 & 0.27 $\pm$ 0.04 & 0.48 $\pm$ 0.10 \\ 
G81.8713+0.7807 & 20:38:36.4232 & +42:37:34.756 & 0.127 & 0.048 & 10.42 & 0.27 $\pm$ 0.04 & 0.69 $\pm$ 0.14 \\ 
G81.8713+0.7807 & 20:38:36.4232 & +42:37:34.756 & -0.213 & 0.344 &  2.50 & 0.26 $\pm$ 0.03 & 0.64 $\pm$ 0.11 \\ 
G81.8713+0.7807 & 20:38:36.4232 & +42:37:34.756 & 0.066 & -0.038 & 10.06 & 0.23 $\pm$ 0.03 & 0.67 $\pm$ 0.12 \\ 
G81.8713+0.7807 & 20:38:36.4232 & +42:37:34.756 & -0.574 & 0.603 &  2.68 & 0.14 $\pm$ 0.03 & 0.28 $\pm$ 0.07 \\ 

G82.3079+0.7296 & 20:40:16.6487 & +42:56:29.288 & 0.000 & 0.000 & 10.42 & 23.82 $\pm$ 0.18 & 24.19 $\pm$ 0.31 \\ 
G82.3079+0.7296 & 20:40:16.6487 & +42:56:29.288 & 0.027 & 0.006 & 10.24 & 15.28 $\pm$ 0.13 & 15.85 $\pm$ 0.23 \\ 
G82.3079+0.7296 & 20:40:16.6487 & +42:56:29.288 & 0.006 & 0.001 & 10.78 & 6.38 $\pm$ 0.07 & 6.46 $\pm$ 0.12 \\ 
G82.3079+0.7296 & 20:40:16.6487 & +42:56:29.288 & -0.079 & -0.048 & 10.60 & 4.47 $\pm$ 0.05 & 4.54 $\pm$ 0.09 \\ 
G82.3079+0.7296 & 20:40:16.6487 & +42:56:29.288 & 0.072 & 0.006 &  9.88 & 1.27 $\pm$ 0.03 & 1.30 $\pm$ 0.05 \\ 
G82.3079+0.7296 & 20:40:16.6487 & +42:56:29.288 & -0.042 & -0.021 & 11.14 & 1.04 $\pm$ 0.04 & 1.27 $\pm$ 0.07 \\ 
G82.3079+0.7296 & 20:40:16.6487 & +42:56:29.288 & 0.212 & -0.053 &  9.52 & 0.24 $\pm$ 0.03 & 0.39 $\pm$ 0.08 \\ 

\hline 
\end{longtable}
\clearpage

\twocolumn

\begin{figure*}[tbh]
\begin{center}
\includegraphics[width=0.8\textwidth,angle=0]{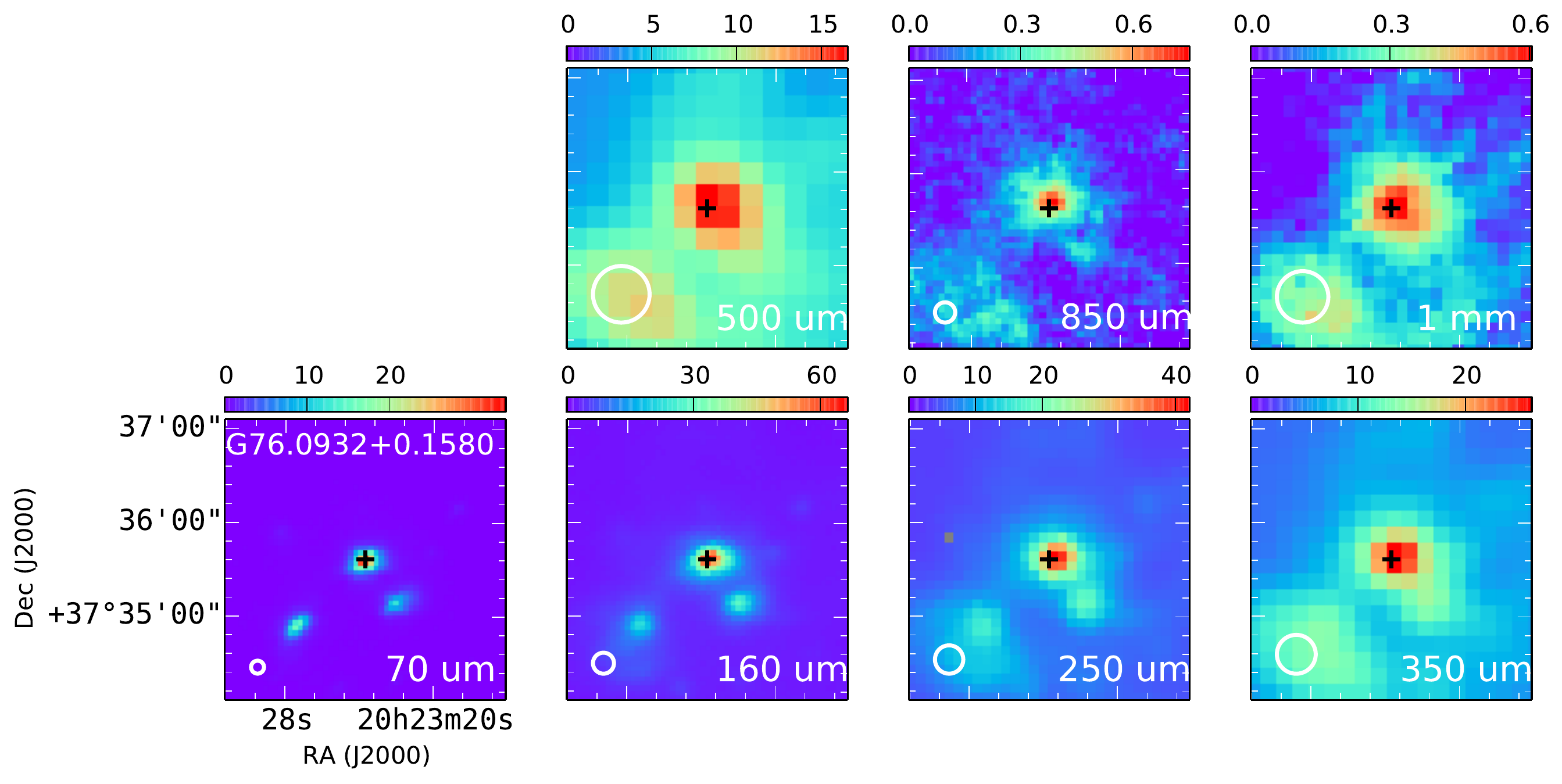} 
\includegraphics[width=0.8\textwidth,angle=0]{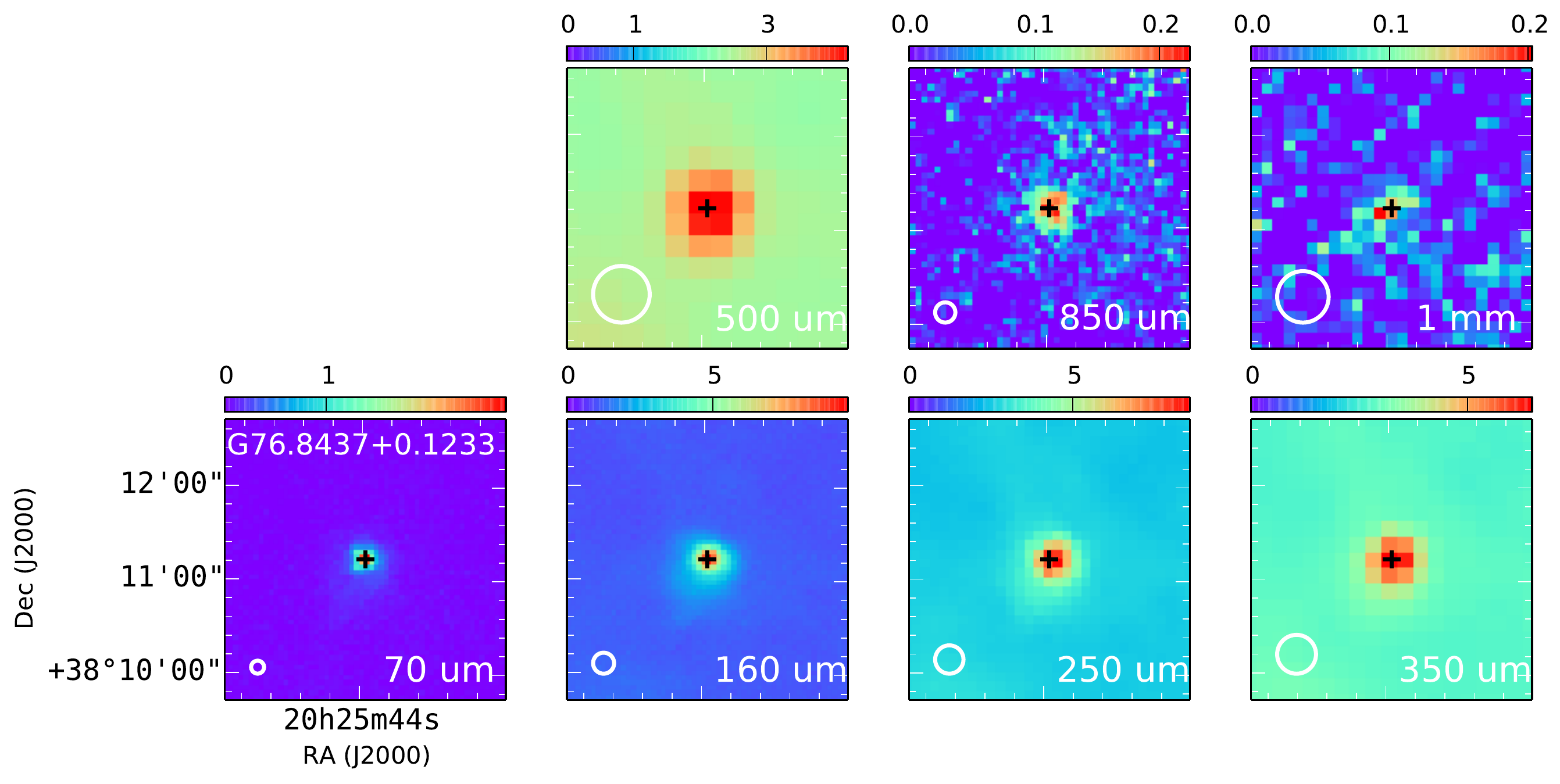} 
\includegraphics[width=0.8\textwidth,angle=0]{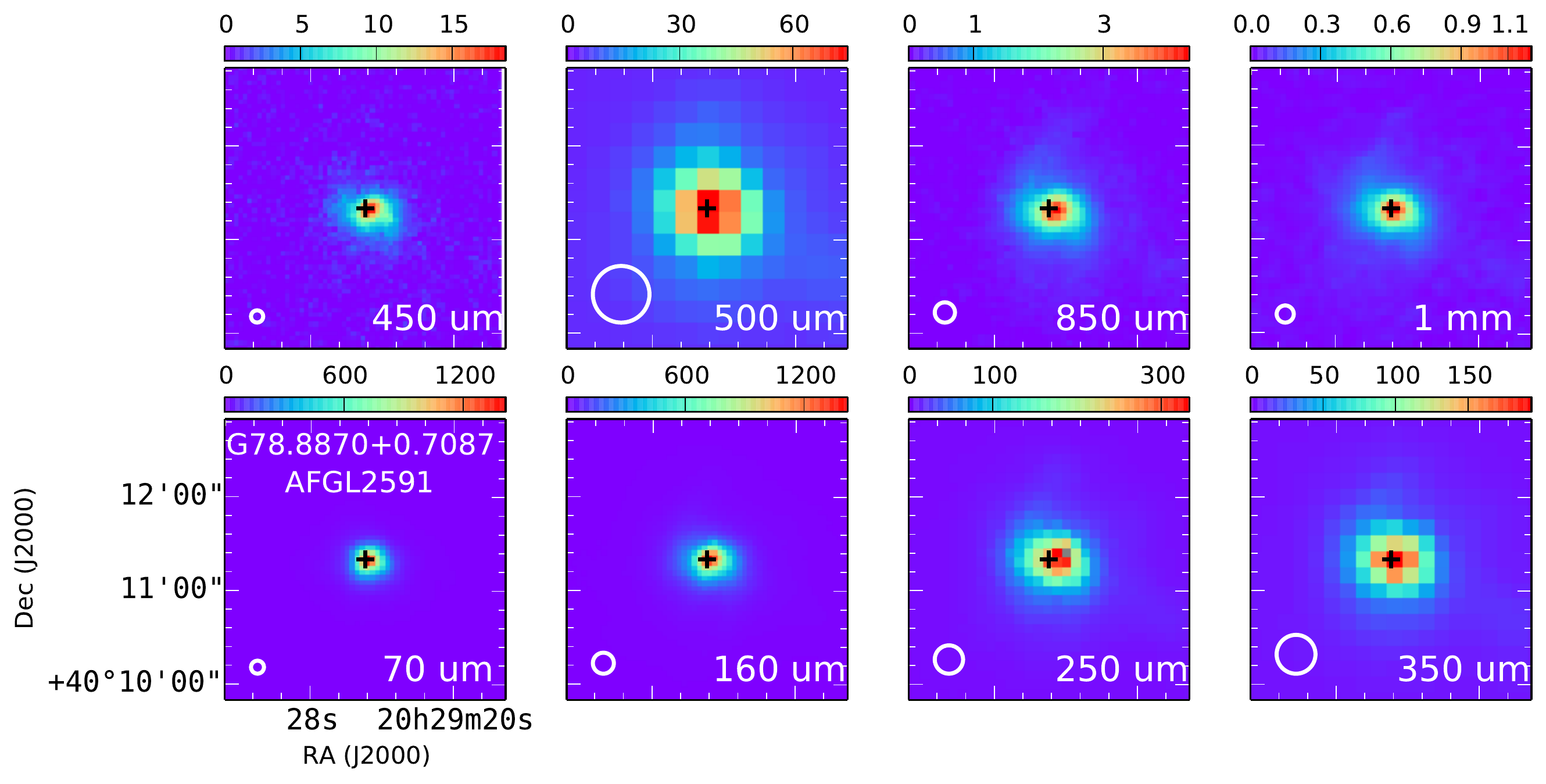}  
\caption{Dust continuum emission images from: {\it Herschel}  PACS (70 and 160~$\mu$m) and SPIRE (250, 350, and 500~$\mu$m); JCMT SCUBA-2 (450 and 850~$\mu$m); and either IRAM MAMBO/MAMBO-2 (1.2~mm) or CSO Bolocam (1.1~mm) instruments. 
The color scale is in units of Jy~beam$^{-1}$.
The location of the methanol masers detected with the VLA are indicated by the black crosses.
The Galactic
names of the GLOSTAR sources are indicated at the top of the bottom-left panels.
The beams are shown at the bottom-left corner of each panel. Saturated pixels are shown in gray. }
\label{fig:submm-appendix}
\end{center}
\end{figure*}

\setcounter{figure}{0}
\renewcommand{\thefigure}{A.1}

\begin{figure*}[tbh]
\begin{center}
\includegraphics[width=0.8\textwidth,angle=0]{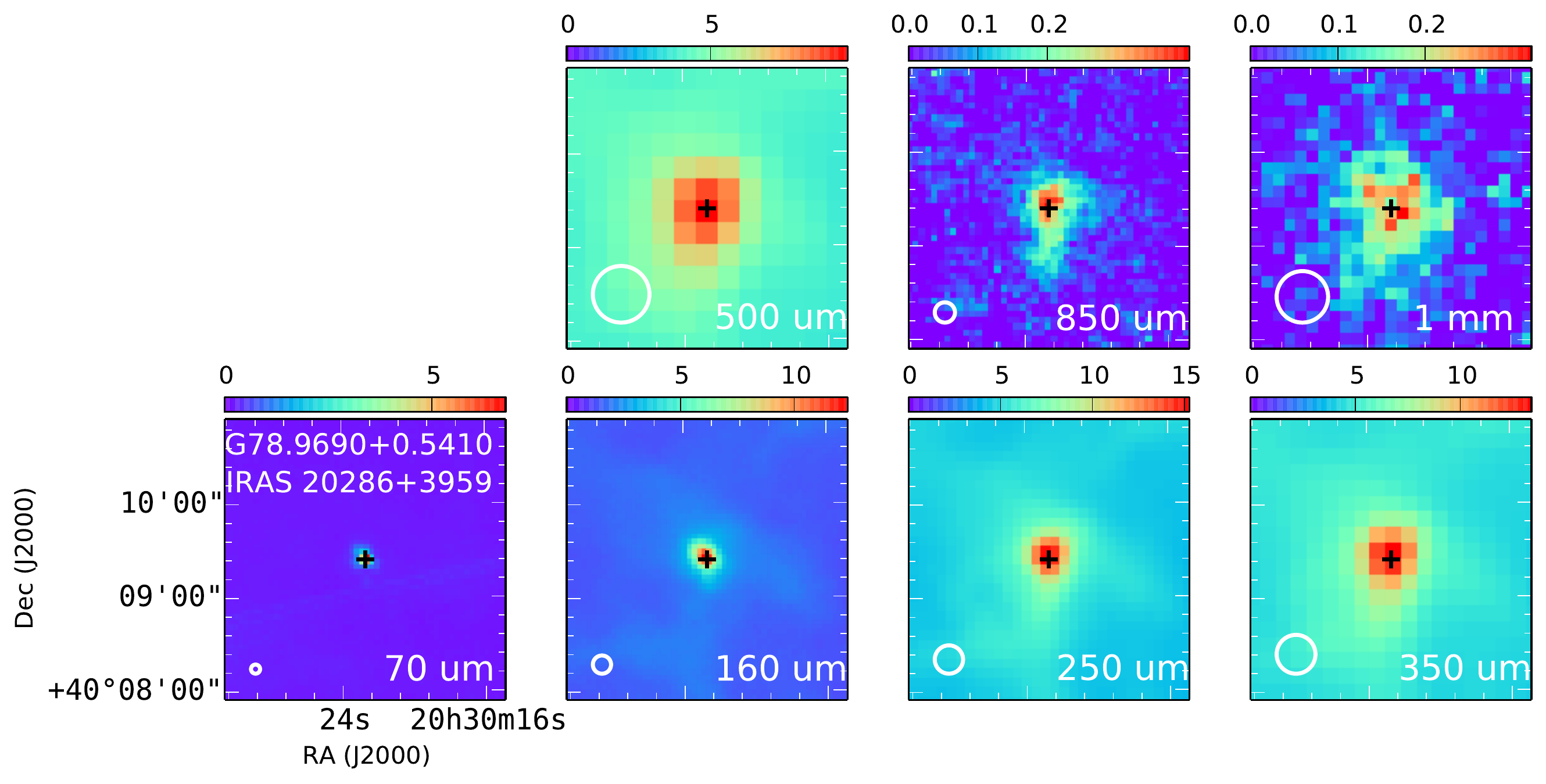} 
\includegraphics[width=0.8\textwidth,angle=0]{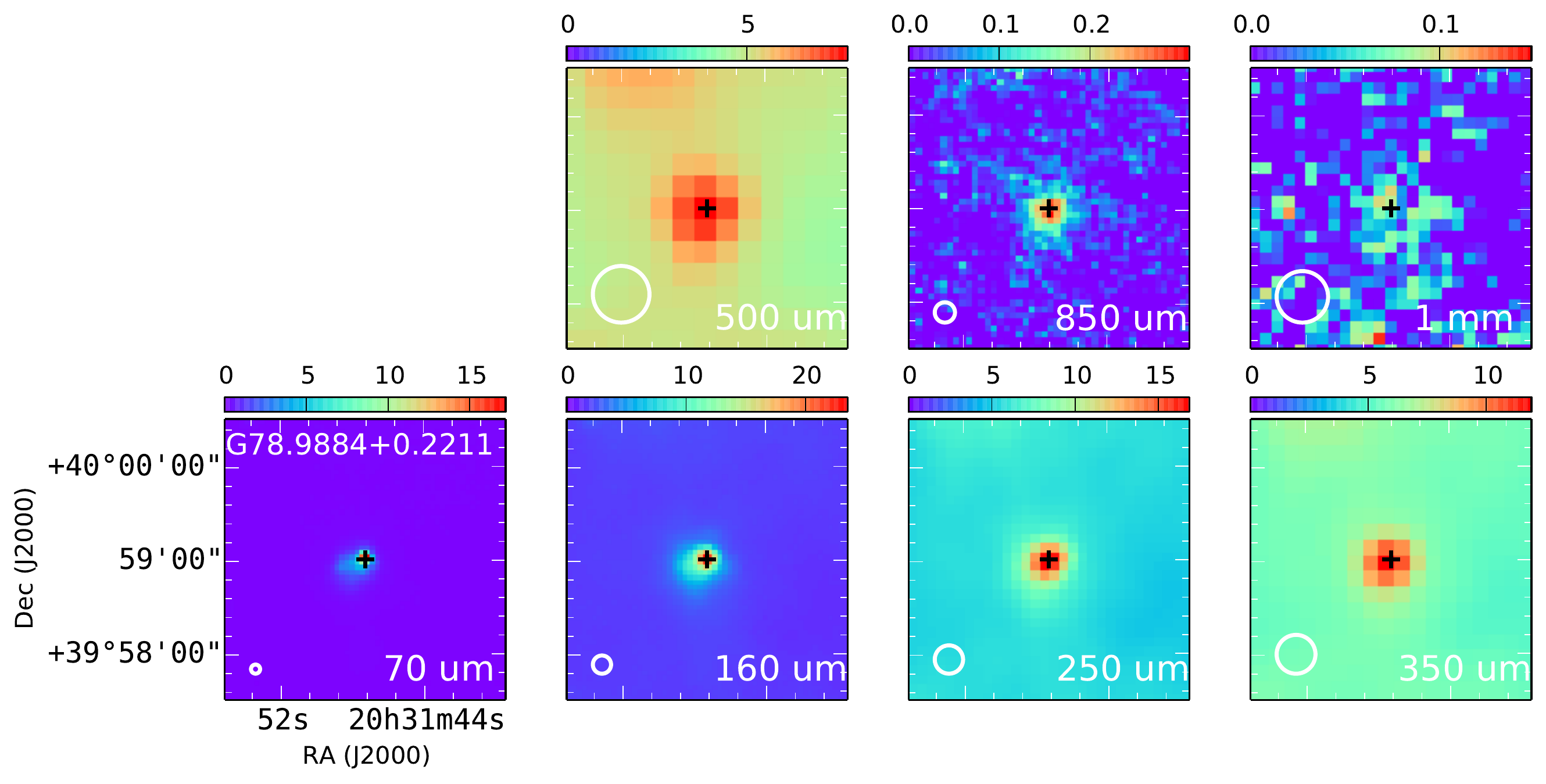}  
\includegraphics[width=0.8\textwidth,angle=0]{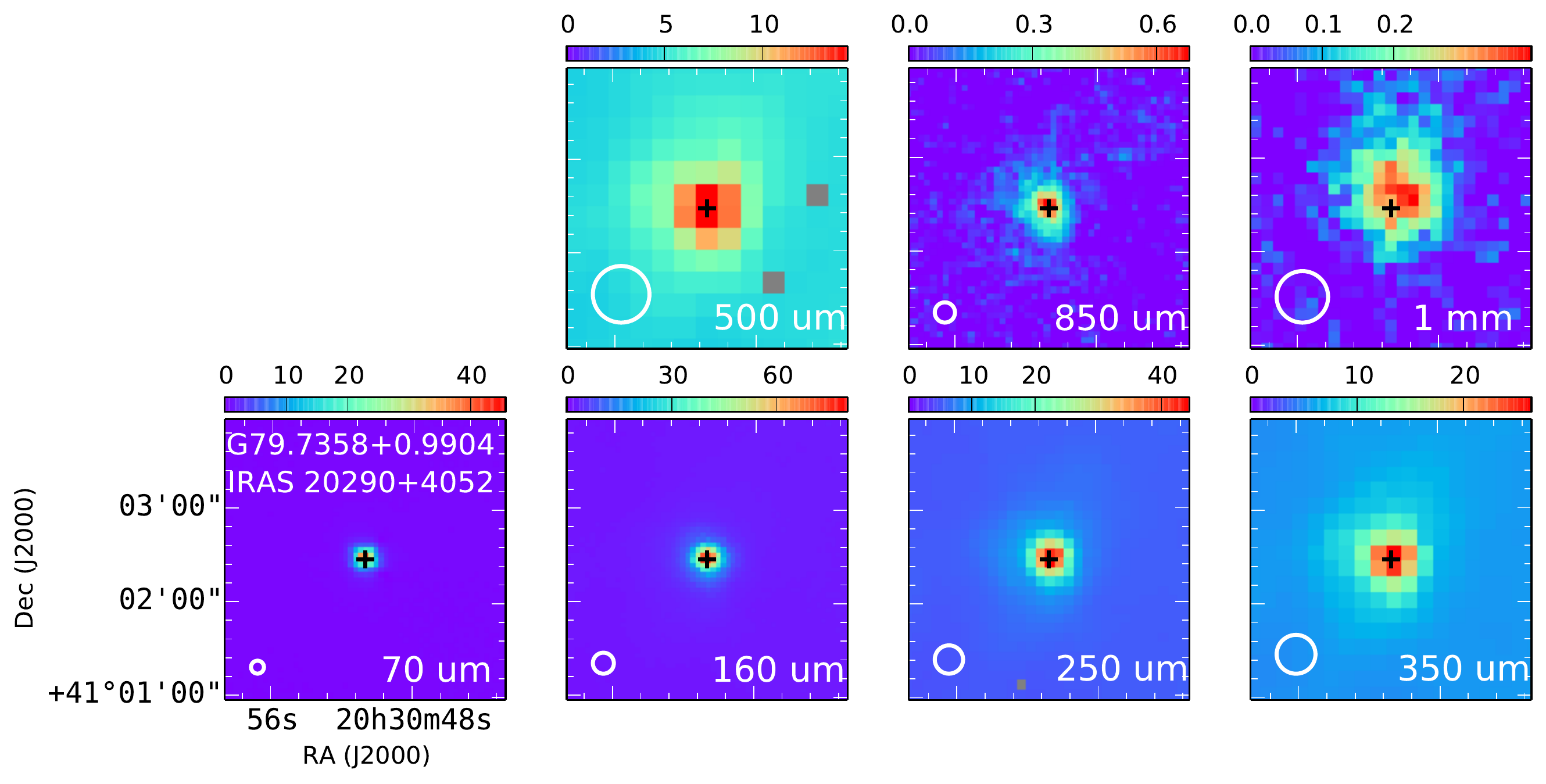}  
\caption{\it Continued.   }
\end{center}
\end{figure*}

\setcounter{figure}{0}
\renewcommand{\thefigure}{A.1}

\begin{figure*}[tbh]
\begin{center}
\includegraphics[width=0.8\textwidth,angle=0]{figures/herschel-3+808617+03834.pdf} 
\includegraphics[width=0.8\textwidth,angle=0]{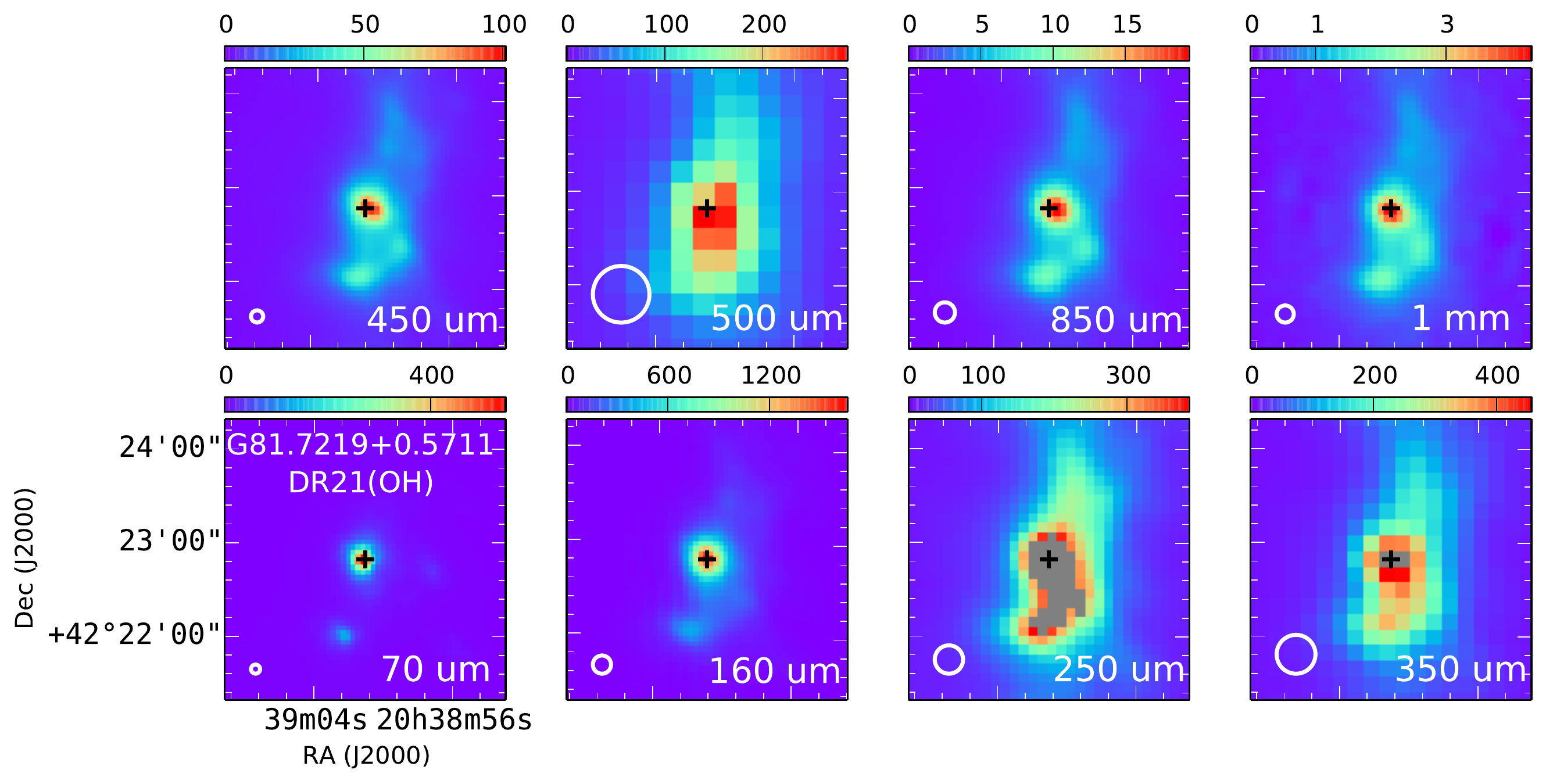}  
\includegraphics[width=0.8\textwidth,angle=0]{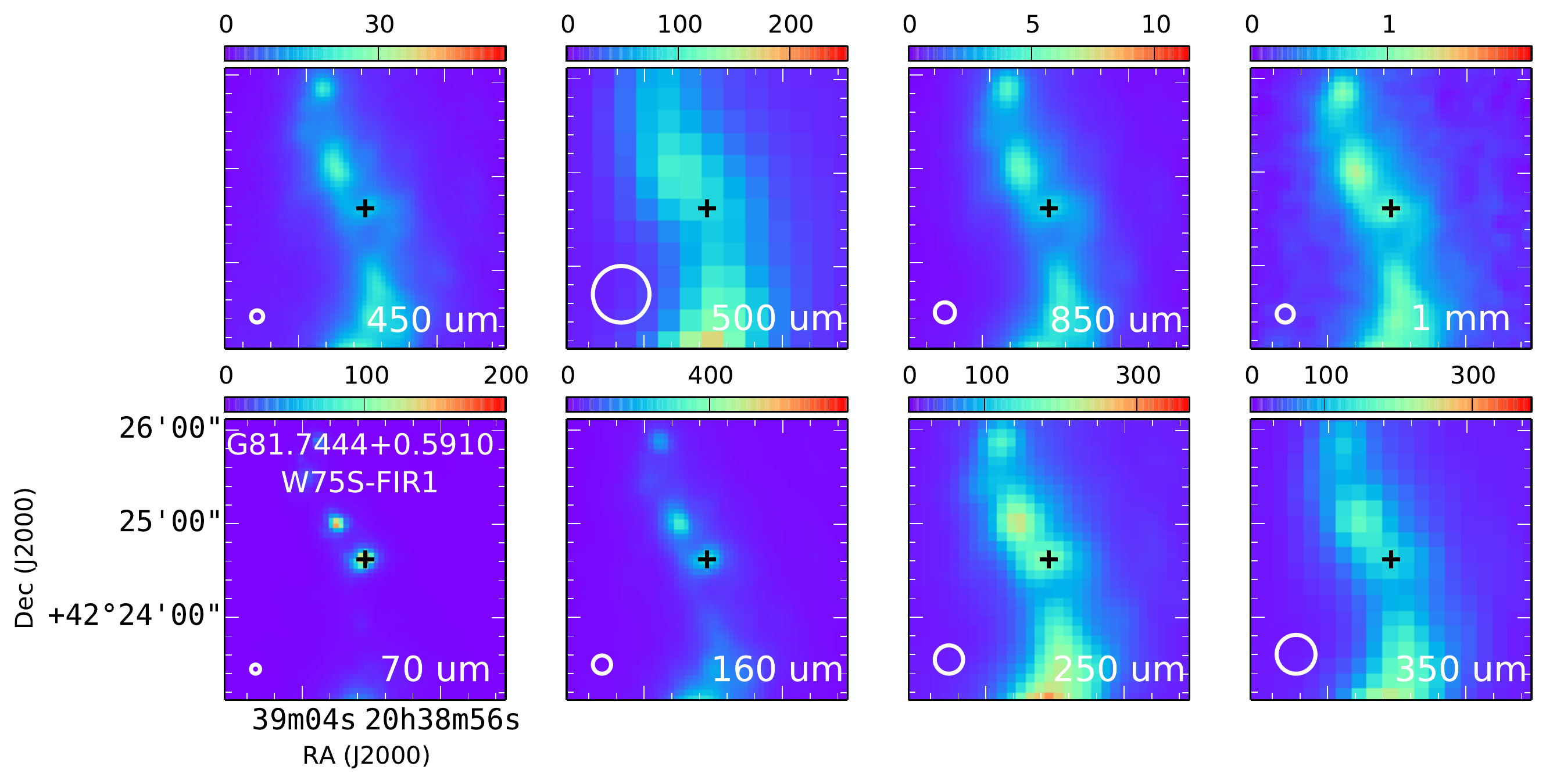} 
\caption{\it Continued.   }
\end{center}
\end{figure*}

\setcounter{figure}{0}
\renewcommand{\thefigure}{A.1}

\begin{figure*}[tbh]
\begin{center}
\includegraphics[width=0.8\textwidth,angle=0]{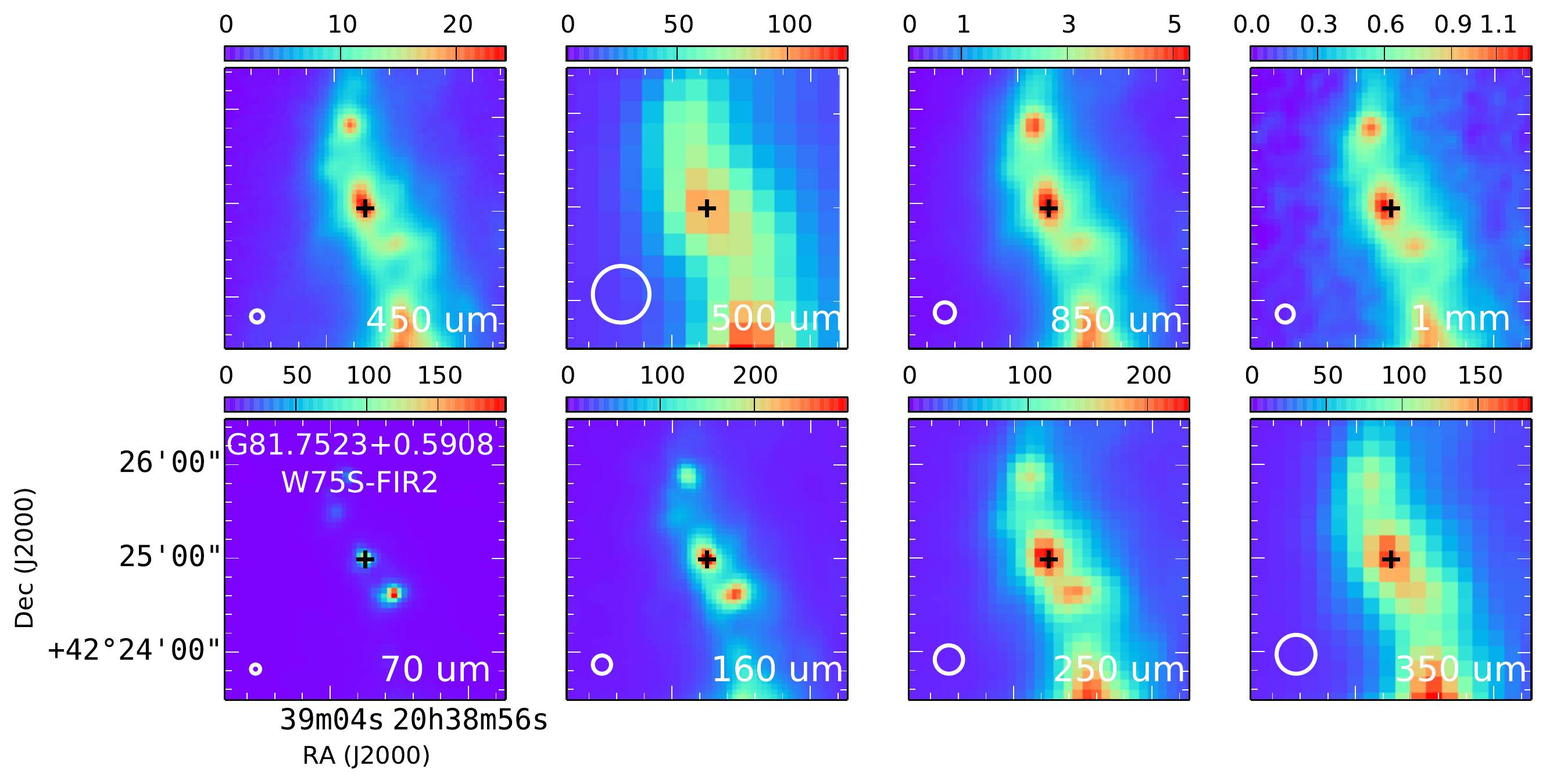}  
\includegraphics[width=0.8\textwidth,angle=0]{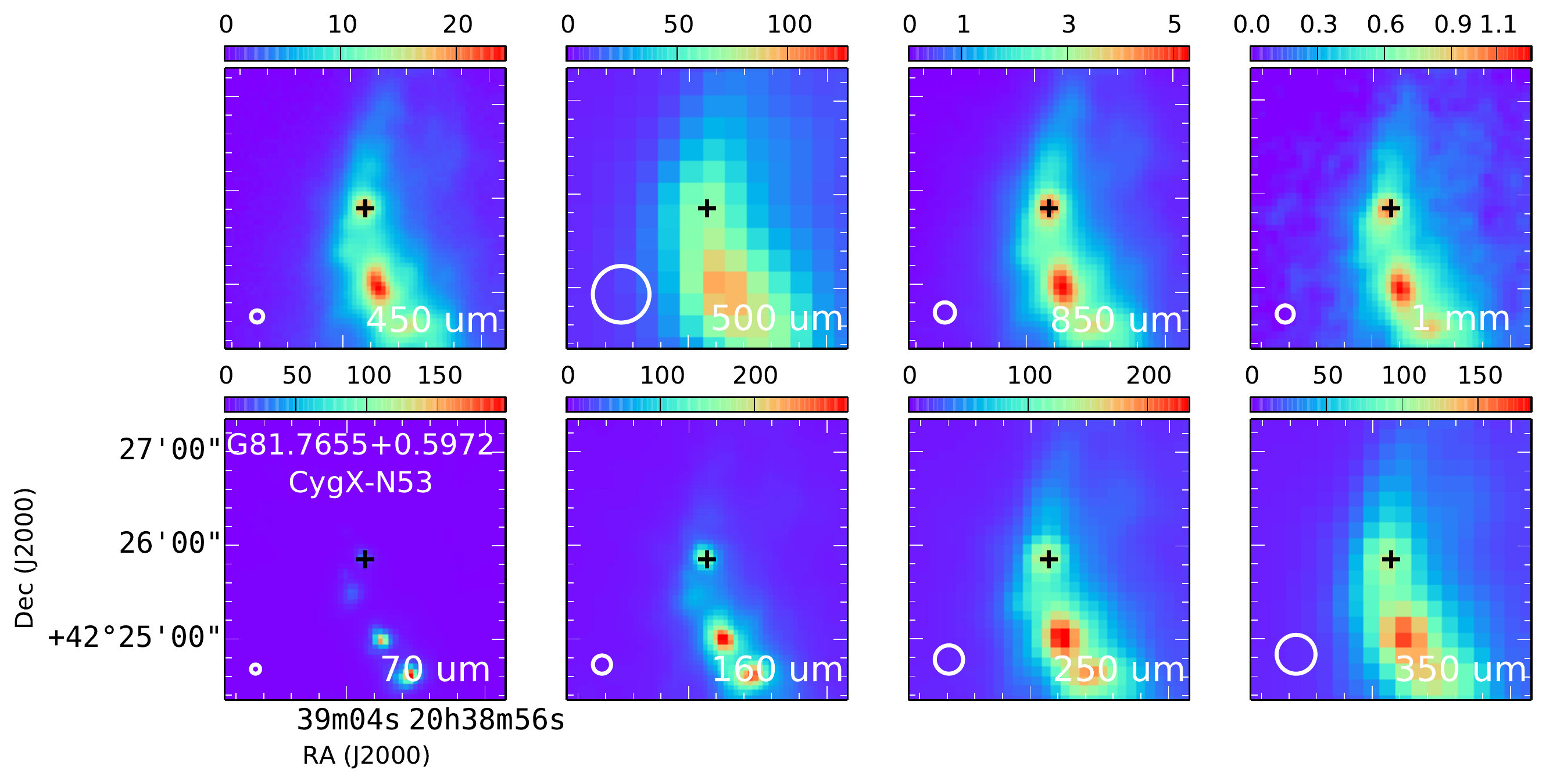} 
\includegraphics[width=0.8\textwidth,angle=0]{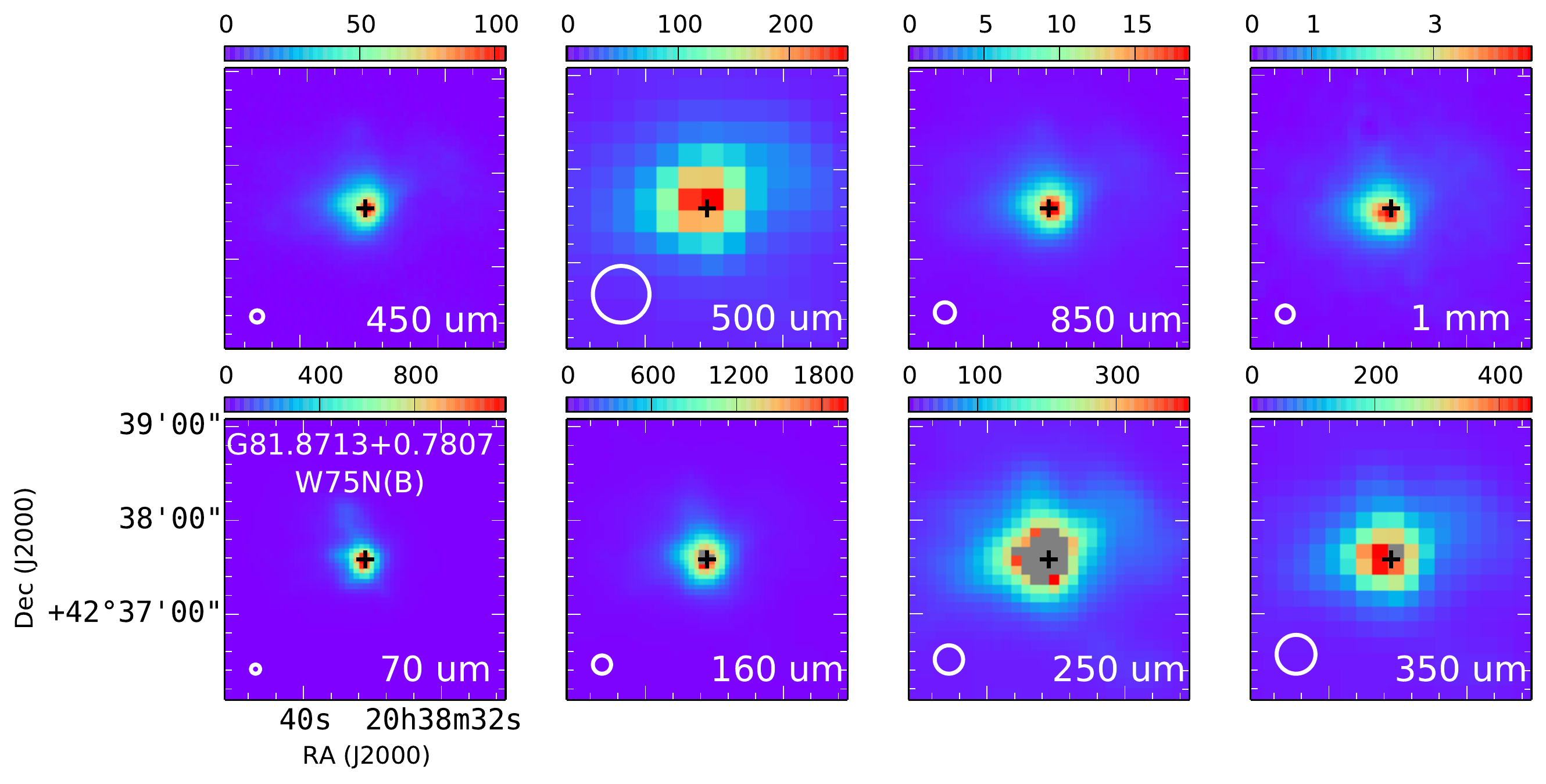} 
\caption{\it Continued.   }
\end{center}
\end{figure*}

\setcounter{figure}{0}
\renewcommand{\thefigure}{A.1}

\begin{figure*}[tbh]
\begin{center}
\includegraphics[width=0.8\textwidth,angle=0]{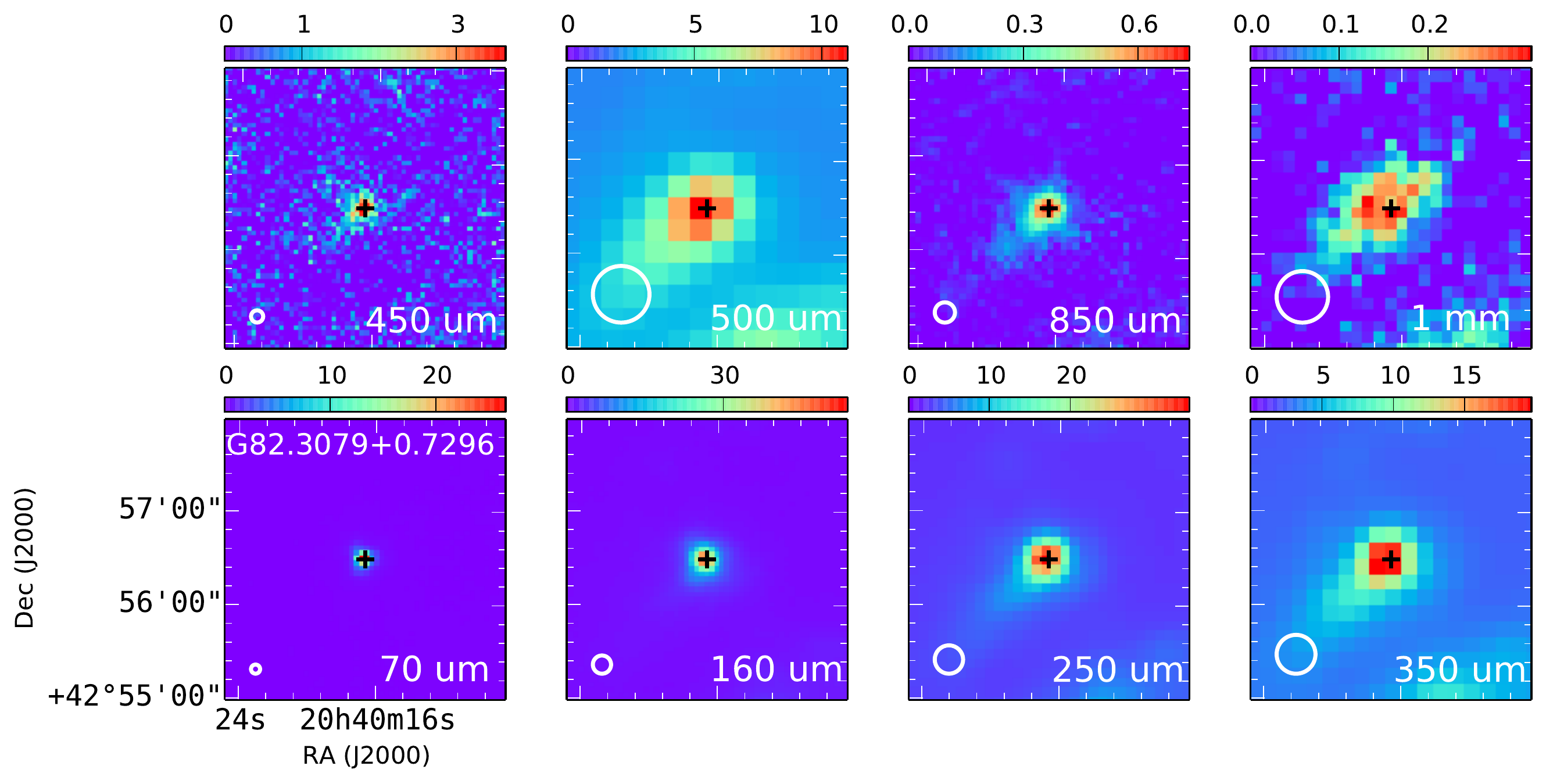} 
\caption{\it Continued.   }
\end{center}
\end{figure*}

\setcounter{figure}{2}
\renewcommand{\thefigure}{A.2}

\begin{figure*}[t]
\begin{center}
 \includegraphics[width=0.64\textwidth,angle=0]{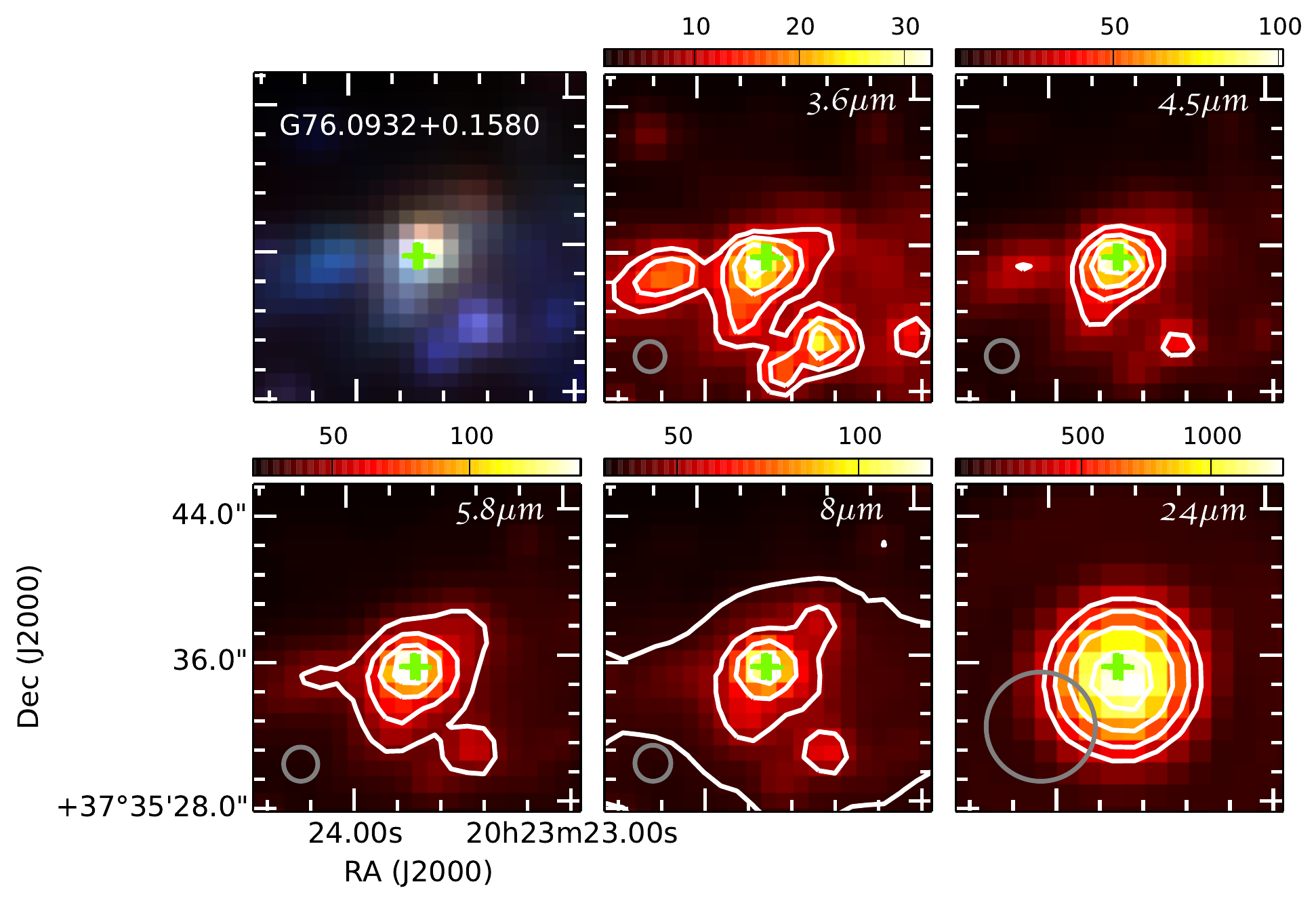} 
 \includegraphics[width=0.64\textwidth,angle=0]{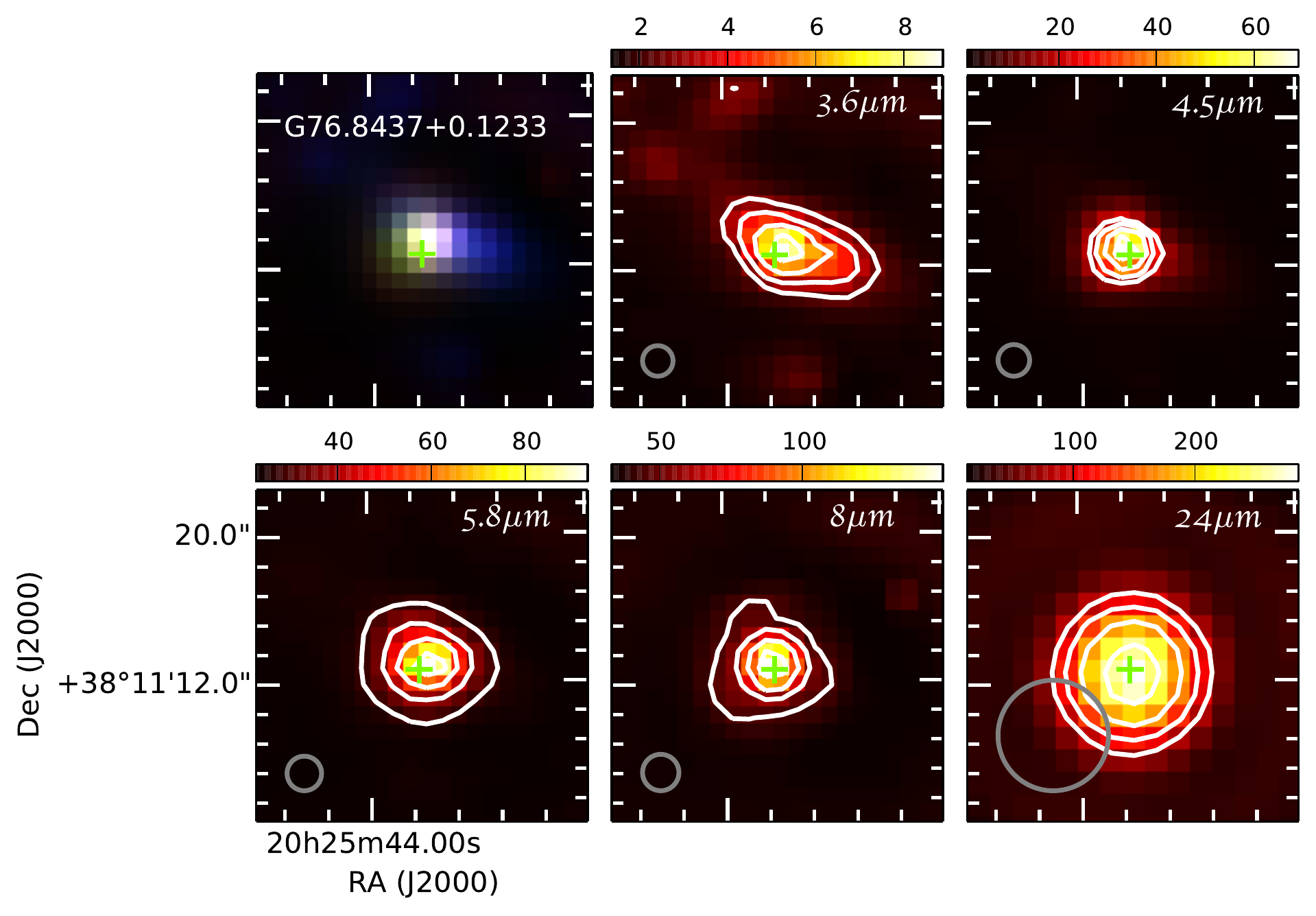} 
\caption{Infrared images (color scale and contours) of the environment around the location of the VLA-detected methanol masers. The first panel in each row shows a three-color map constructed using {\it Spitzer} 3.6~$\mu$m (blue), 4.5~$\mu$m (green), and 8~$\mu$m (red) images. The other four panels show infrared emission for each of the {\it Spitzer} bands. 
The $n$th white contour is at $\left({\sqrt{2}}\right)^{n}\times S_{\rm max} \times p$, where $S_{\rm max}$ is the maximum flux shown in the color bar (MJy~sr$^{-1}$) for each panel, $n$=0, 1, 2 ..., and $p$ is equal to 30\%.
The green crosses mark the positions of the maser features seen in B-array configuration maps. The Galactic
names of the GLOSTAR sources are indicated at the top of the left-most panels.
The beams are shown at the bottom-left corner of each panel. 
Saturated pixels are shown in gray. 
}
\label{fig:infrared2-appendix}
\end{center}
\end{figure*}

\setcounter{figure}{0}
\renewcommand{\thefigure}{A.2}

\begin{figure*}[t]
\begin{center} 
 \includegraphics[width=0.64\textwidth,angle=0]{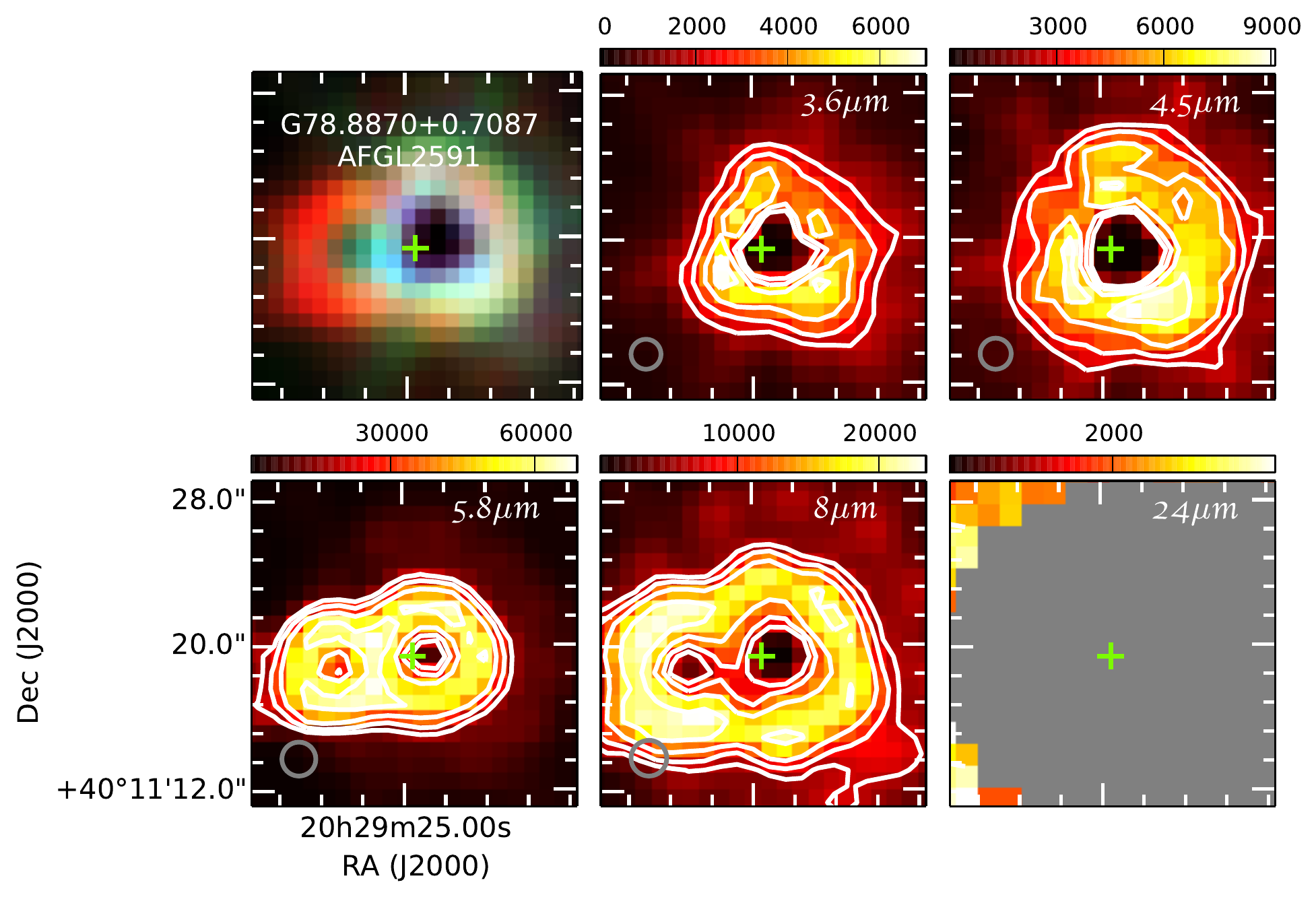} 
 \includegraphics[width=0.64\textwidth,angle=0]{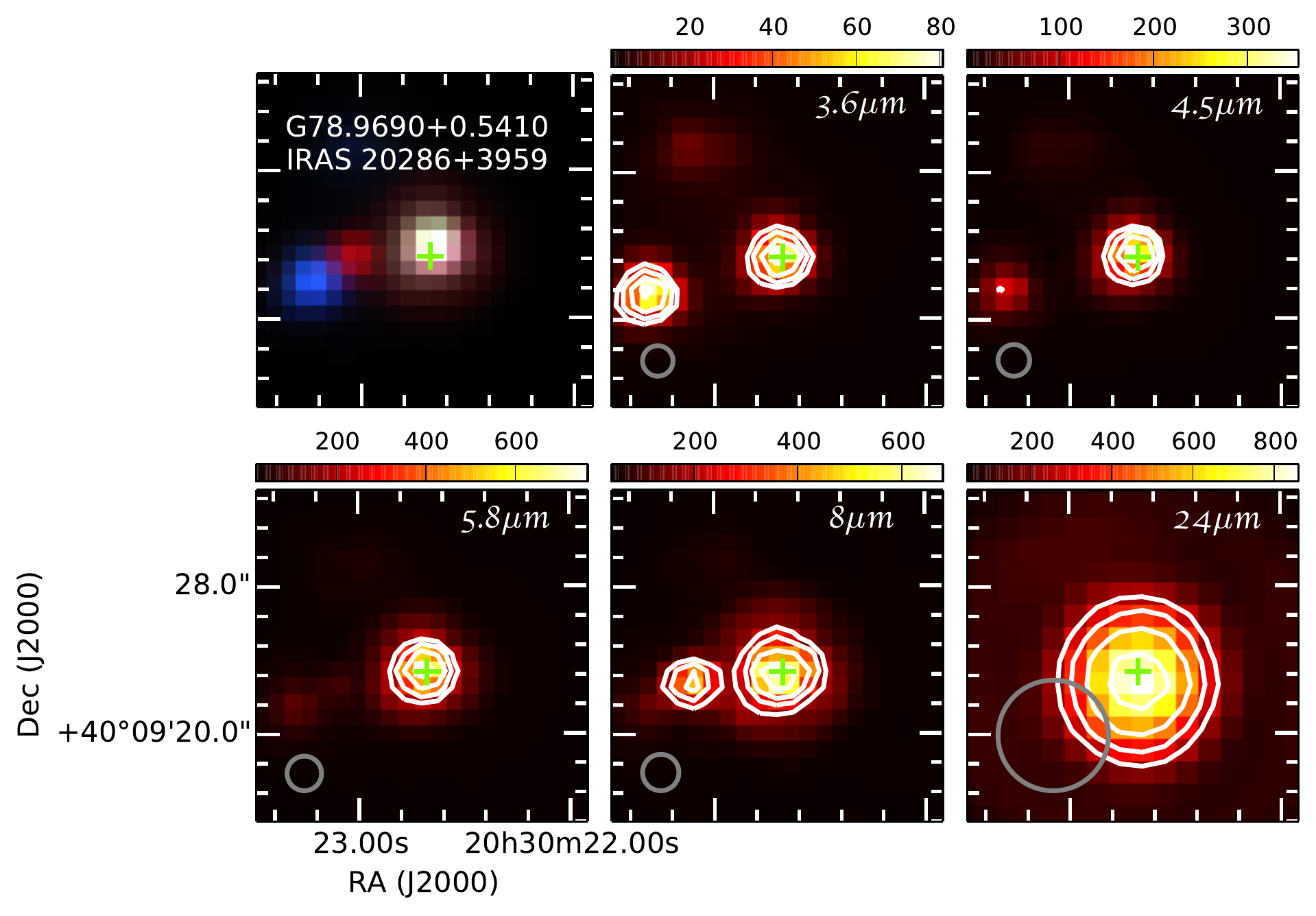} 
 \includegraphics[width=0.64\textwidth,angle=0]{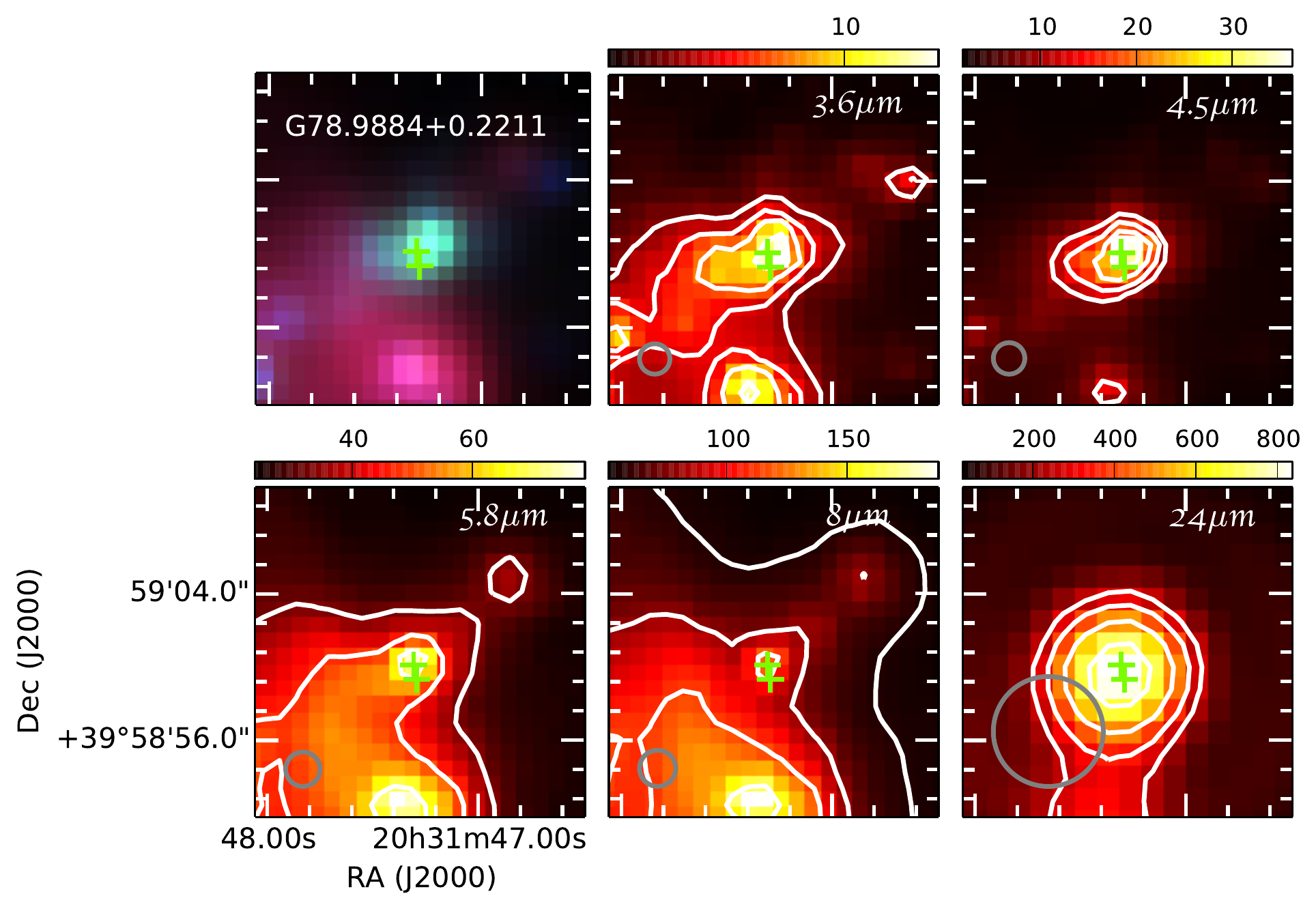} 
\caption{Continued.  }
\end{center}
\end{figure*}

\setcounter{figure}{0}
\renewcommand{\thefigure}{A.2}

\begin{figure*}[t]
\begin{center} 
 \includegraphics[width=0.64\textwidth,angle=0]{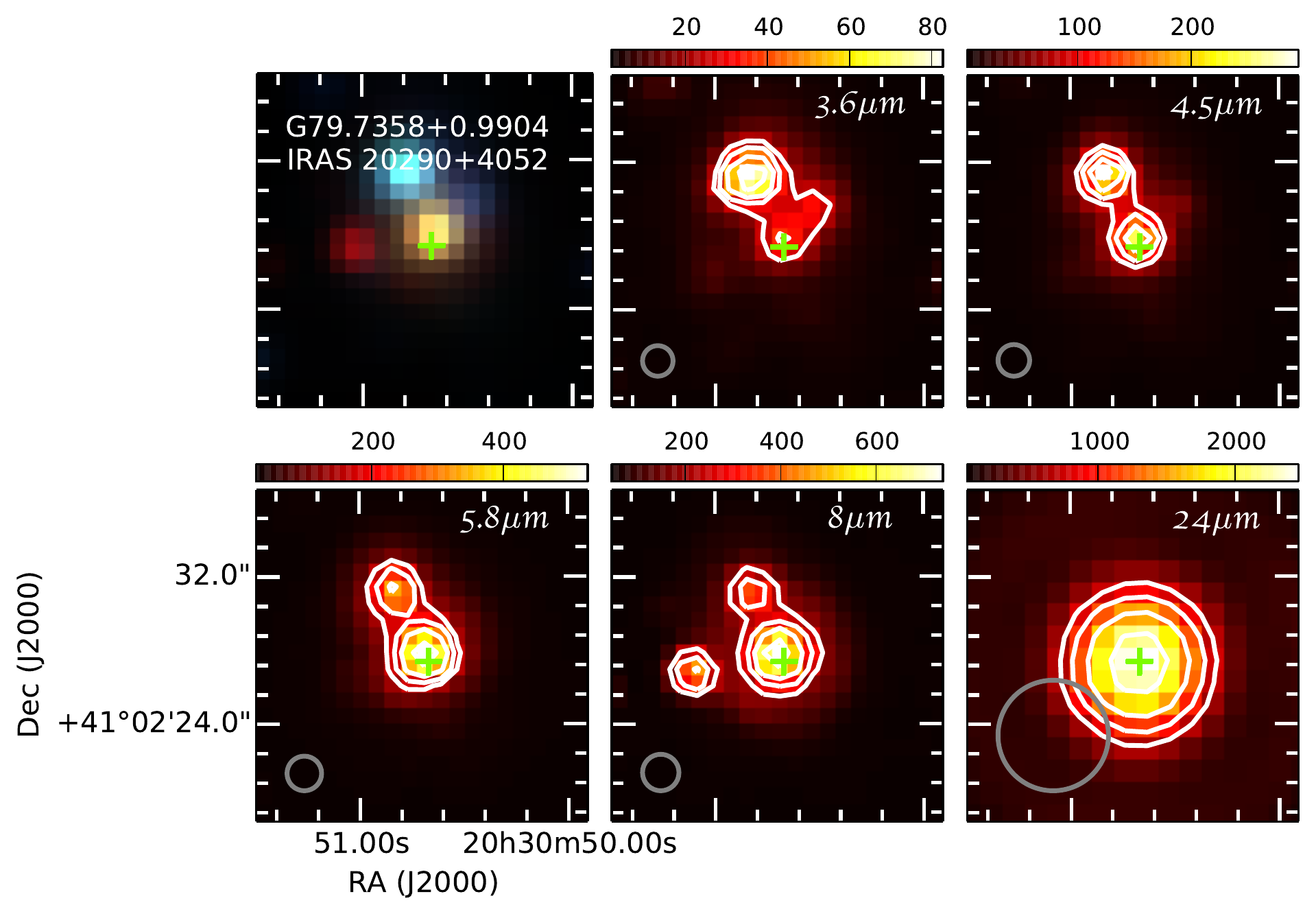}  
 \includegraphics[width=0.64\textwidth,angle=0]{figures/spitzer_l_80_8617_b_+00_3834-4bands.pdf}
 \includegraphics[width=0.64\textwidth,angle=0]{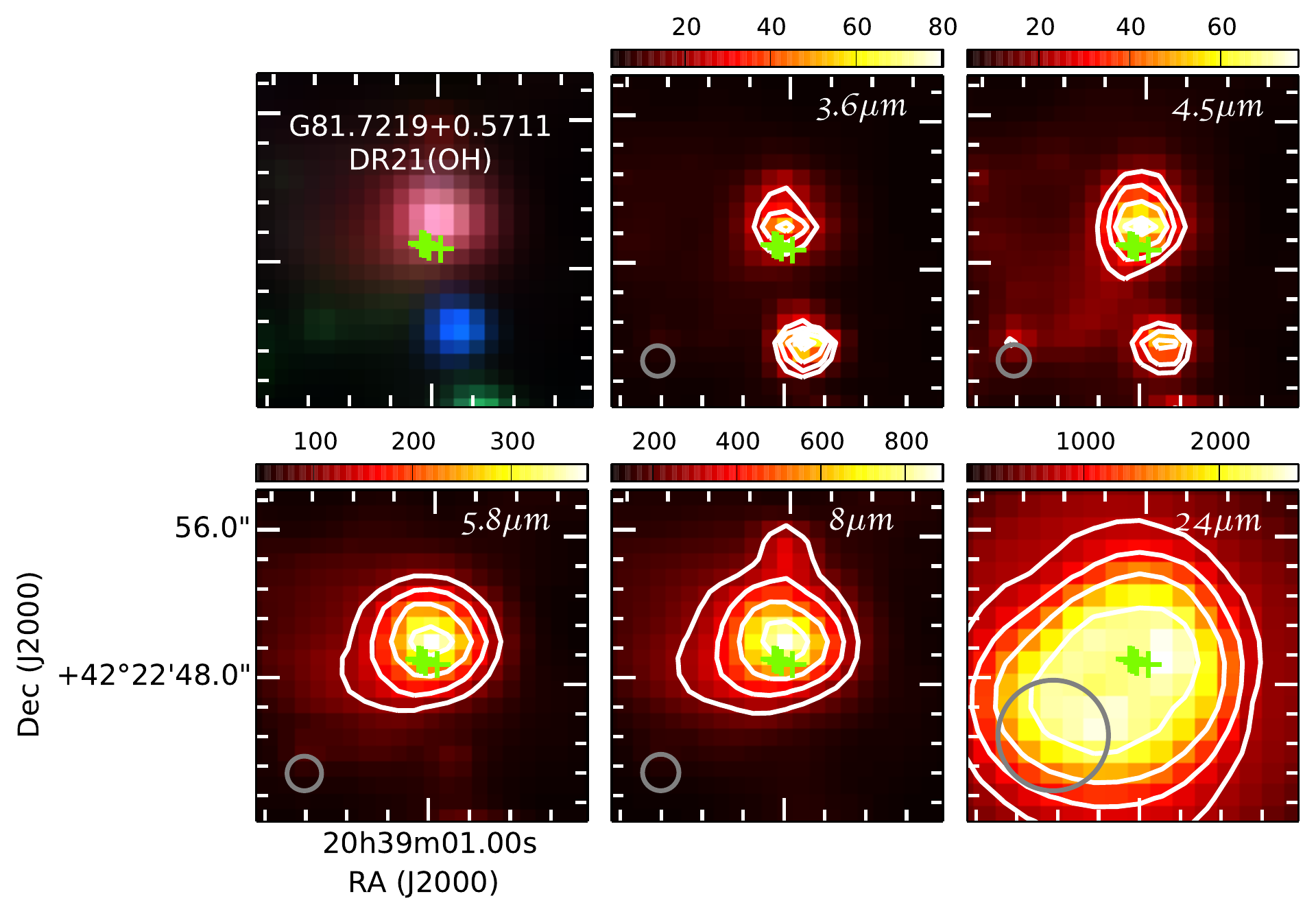} 
\caption{Continued.  }
\end{center}
\end{figure*}

\setcounter{figure}{0}
\renewcommand{\thefigure}{A.2}

\begin{figure*}[t]
\begin{center} 
 \includegraphics[width=0.64\textwidth,angle=0]{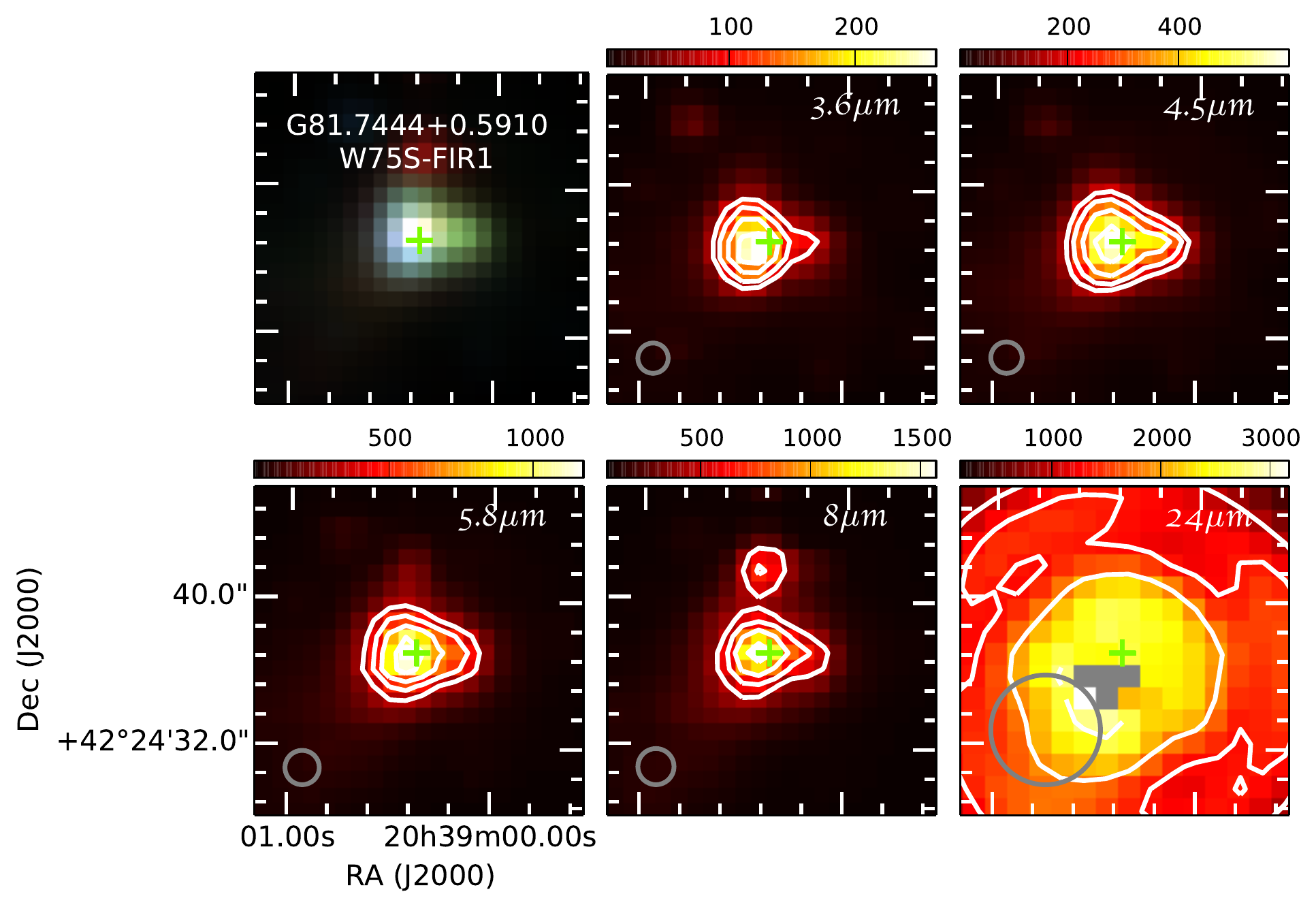}  
 \includegraphics[width=0.64\textwidth,angle=0]{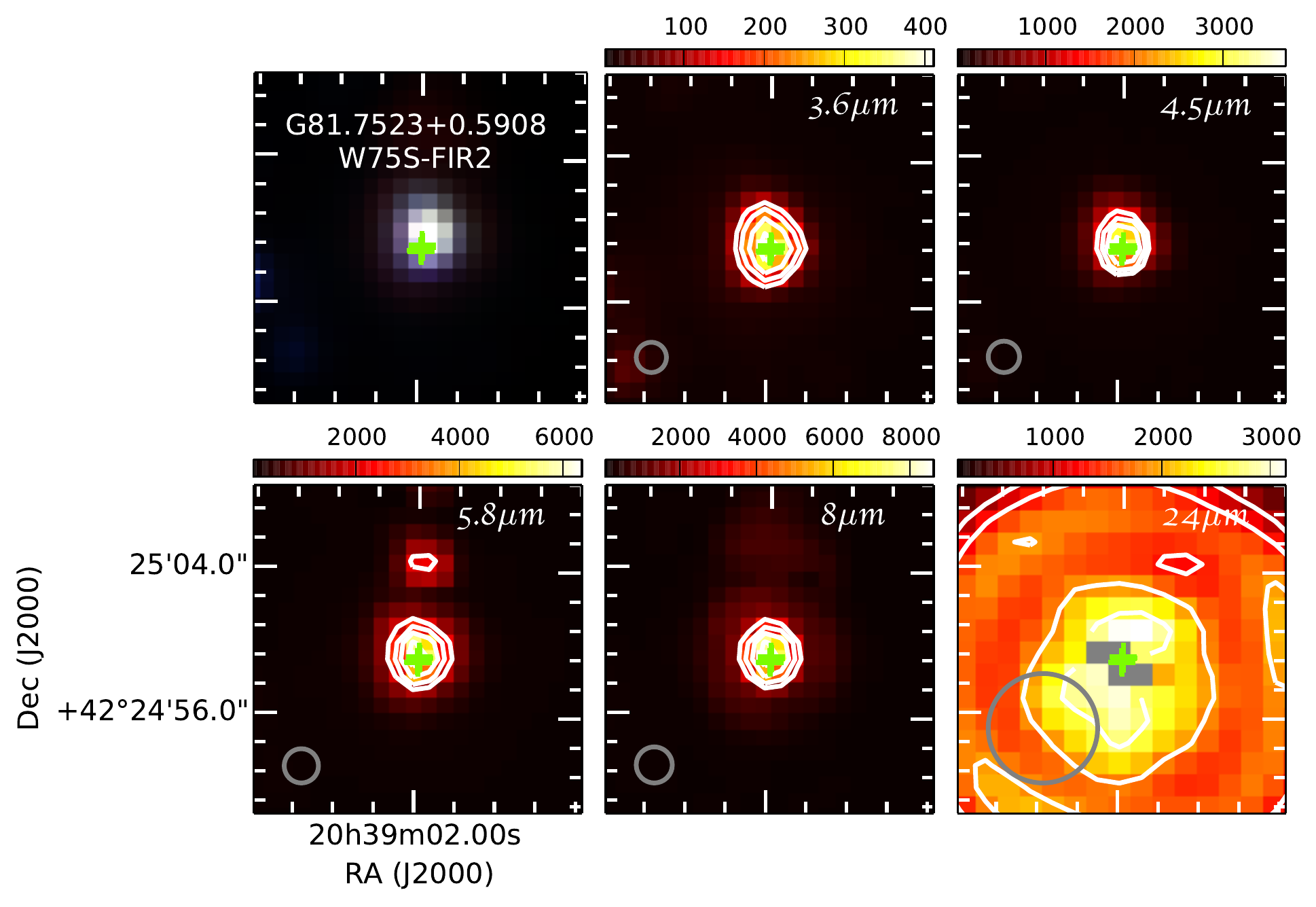} 
 \includegraphics[width=0.64\textwidth,angle=0]{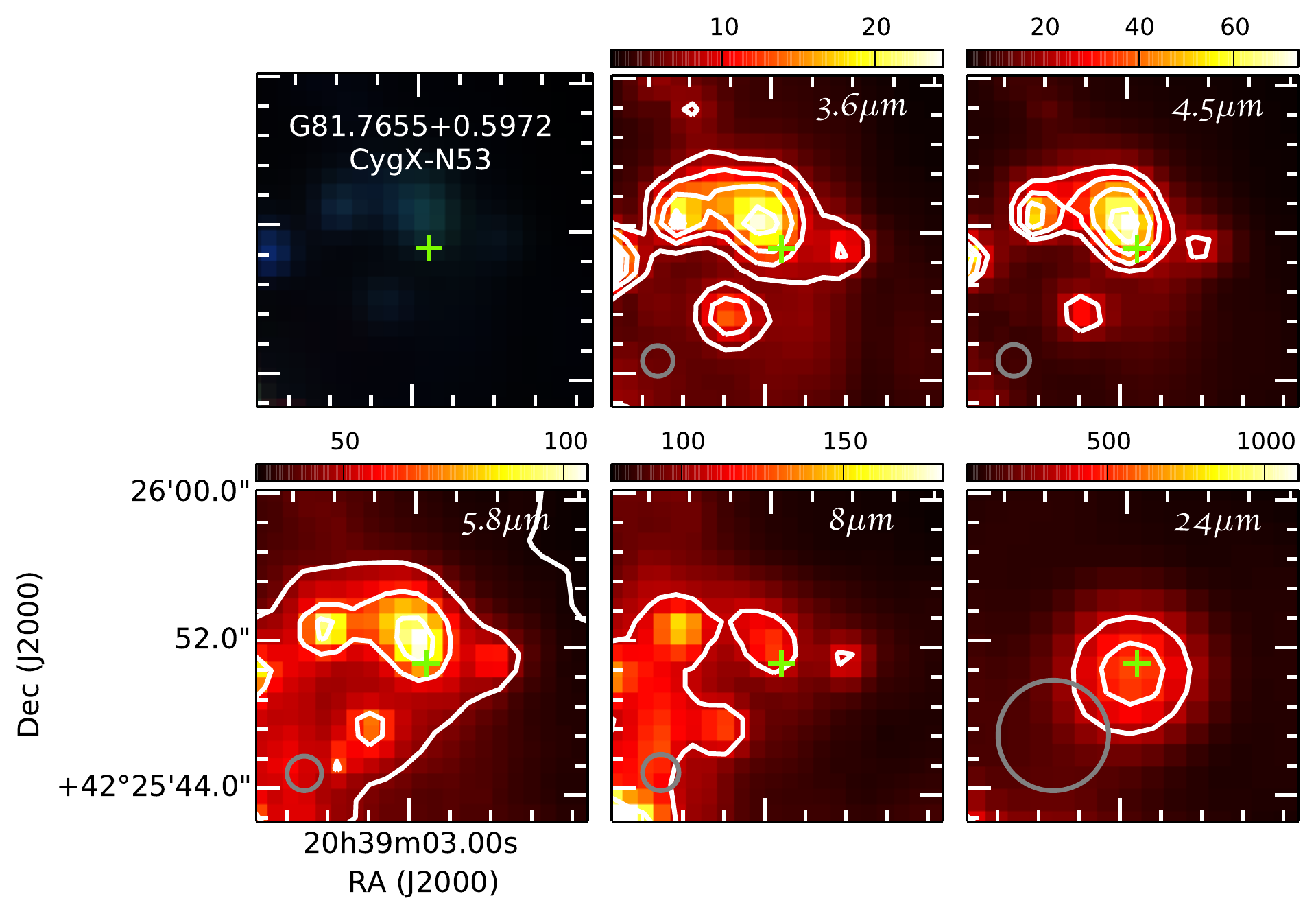} 
\caption{Continued.  }
\end{center}
\end{figure*}

\setcounter{figure}{0}
\renewcommand{\thefigure}{A.2}

\begin{figure*}[t]
\begin{center} 
 \includegraphics[width=0.64\textwidth,angle=0]{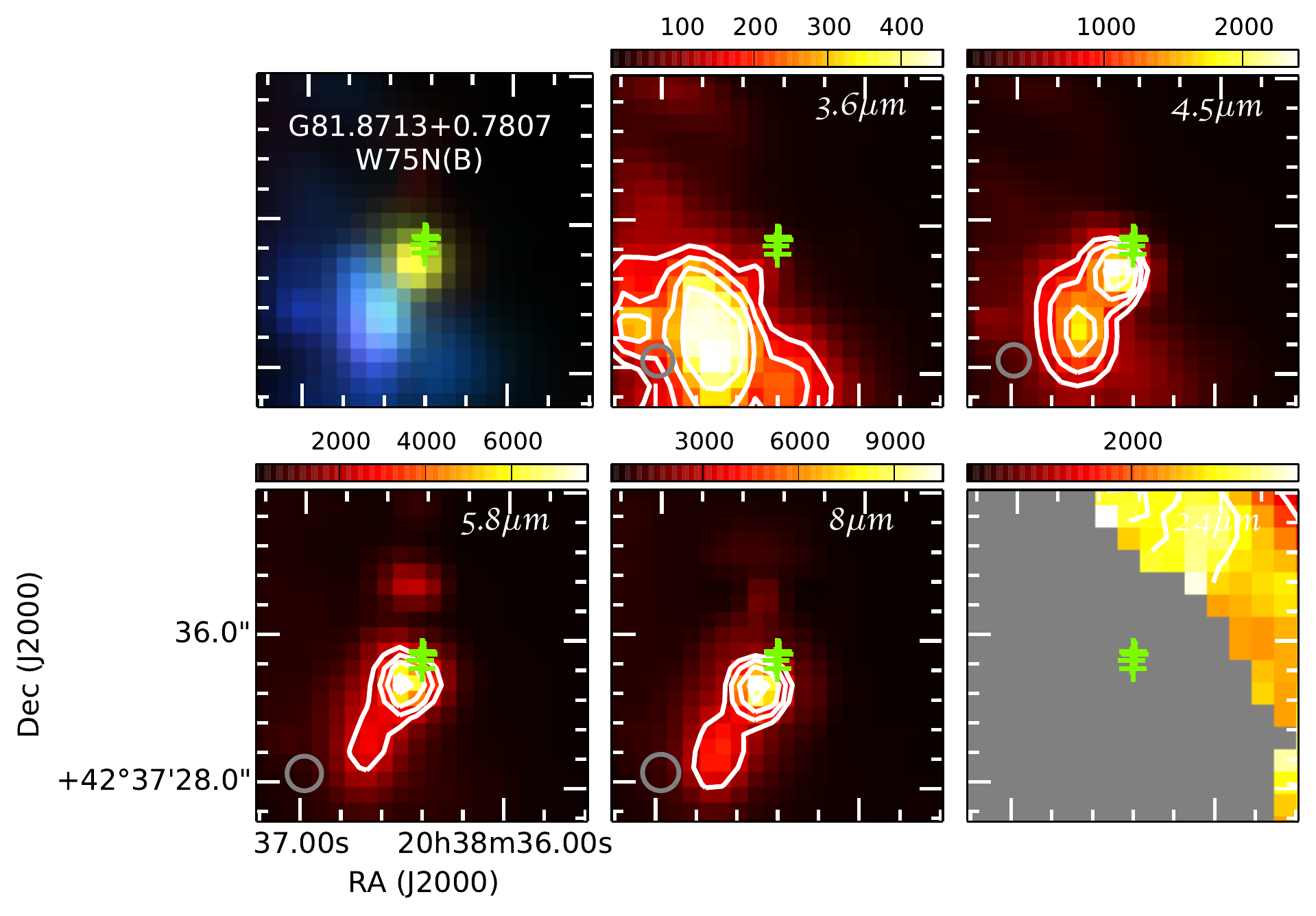} 
 \includegraphics[width=0.64\textwidth,angle=0]{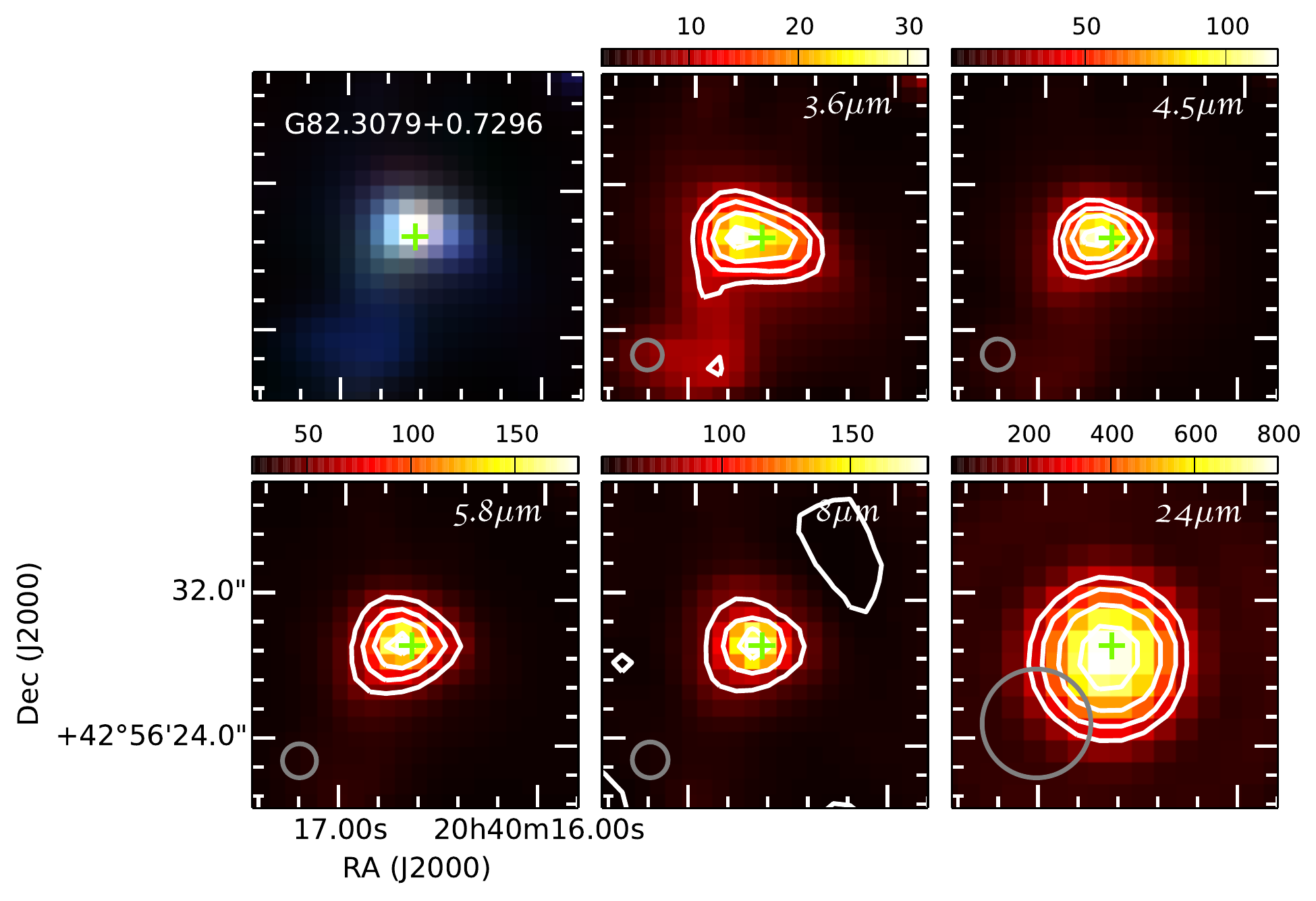}  
\caption{Continued.  }
\end{center}
\end{figure*}

\setcounter{figure}{3}
\renewcommand{\thefigure}{\arabic{figure}}
\renewcommand{\thefigure}{A.3}

\begin{figure*}[ht]
\begin{center}
\includegraphics[width=0.40\textwidth,angle=0]{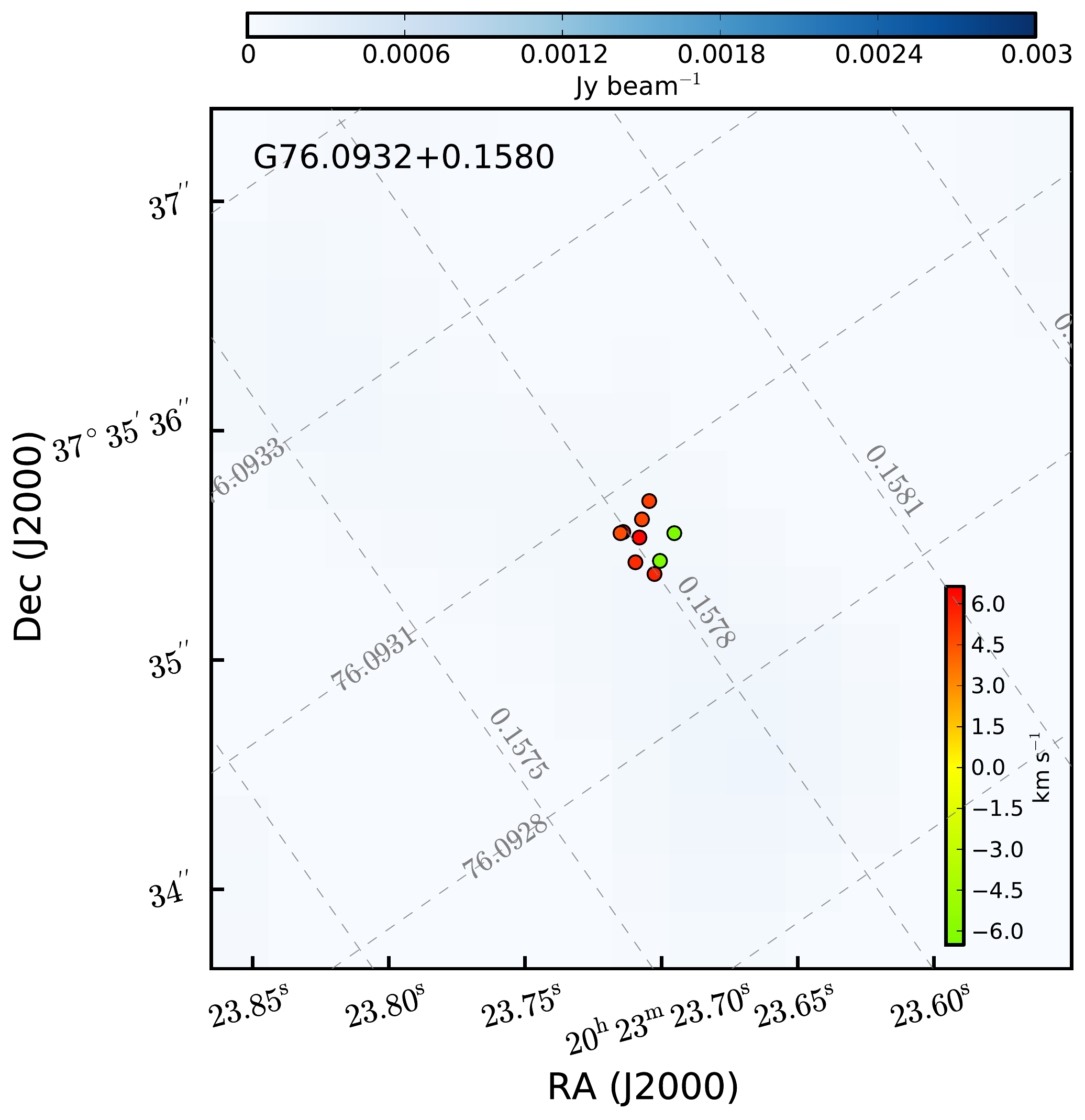} 
\includegraphics[width=0.40\textwidth,angle=0]{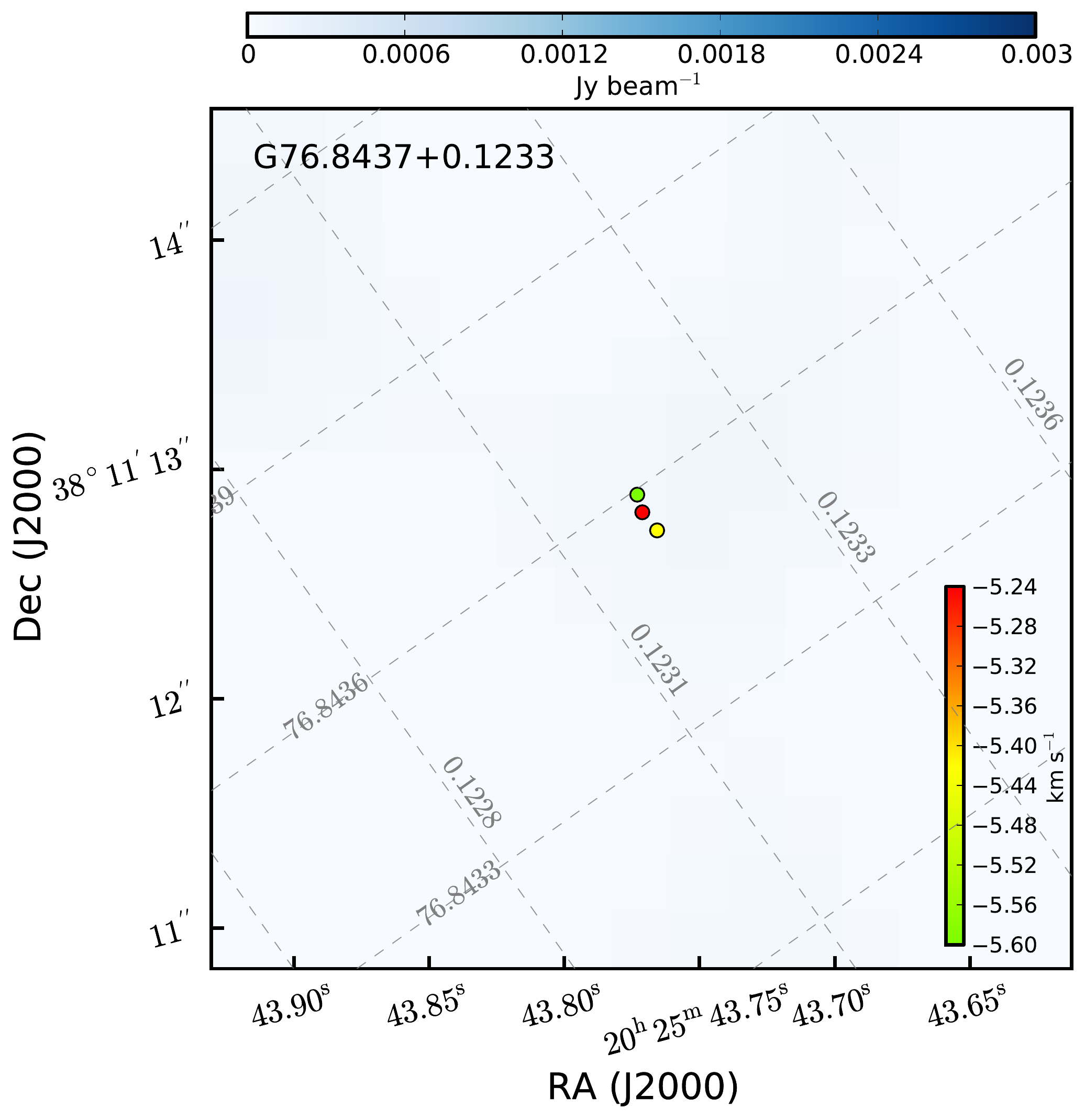}  
\includegraphics[width=0.40\textwidth,angle=0]{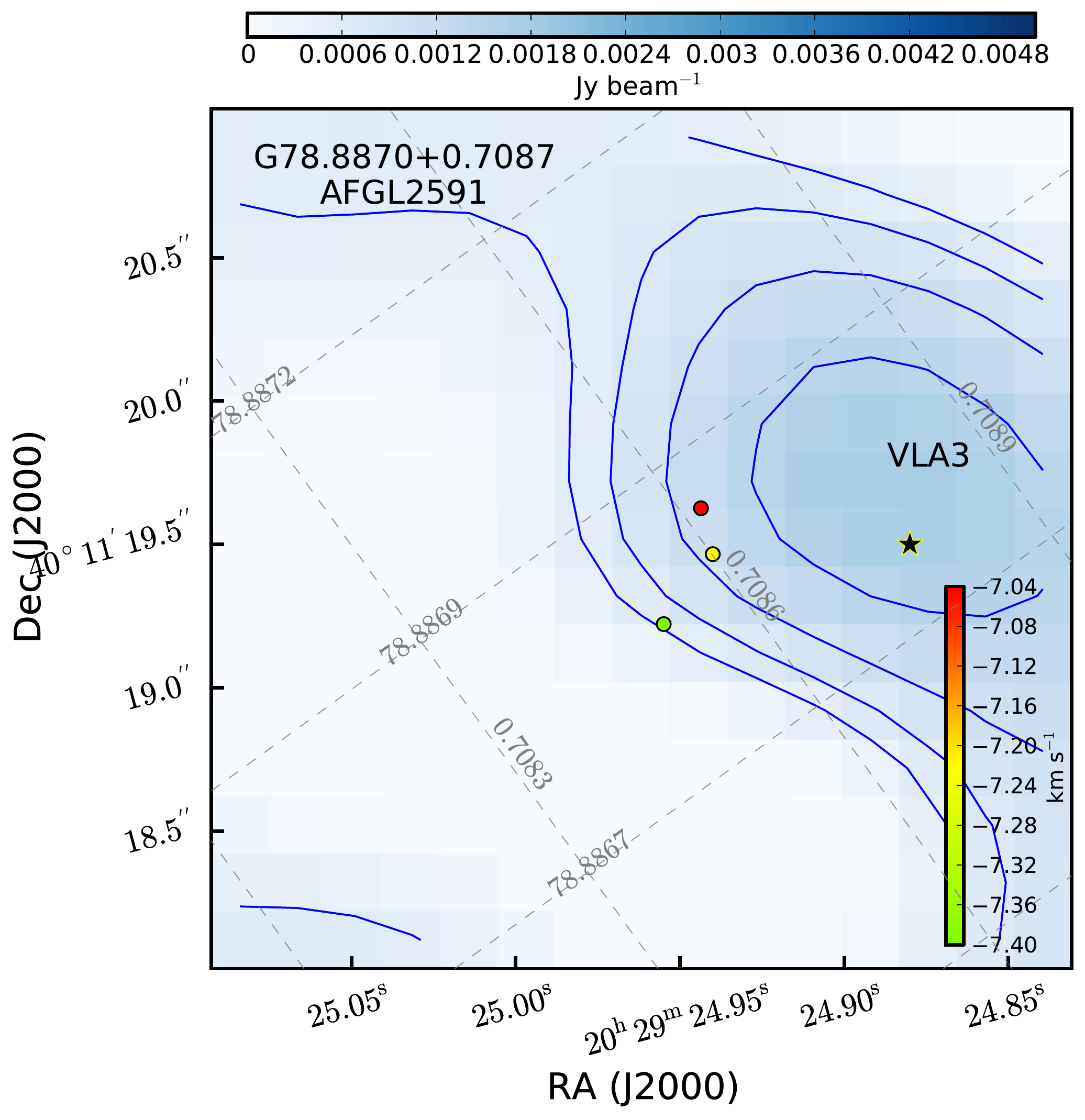}
\includegraphics[width=0.40\textwidth,angle=0]{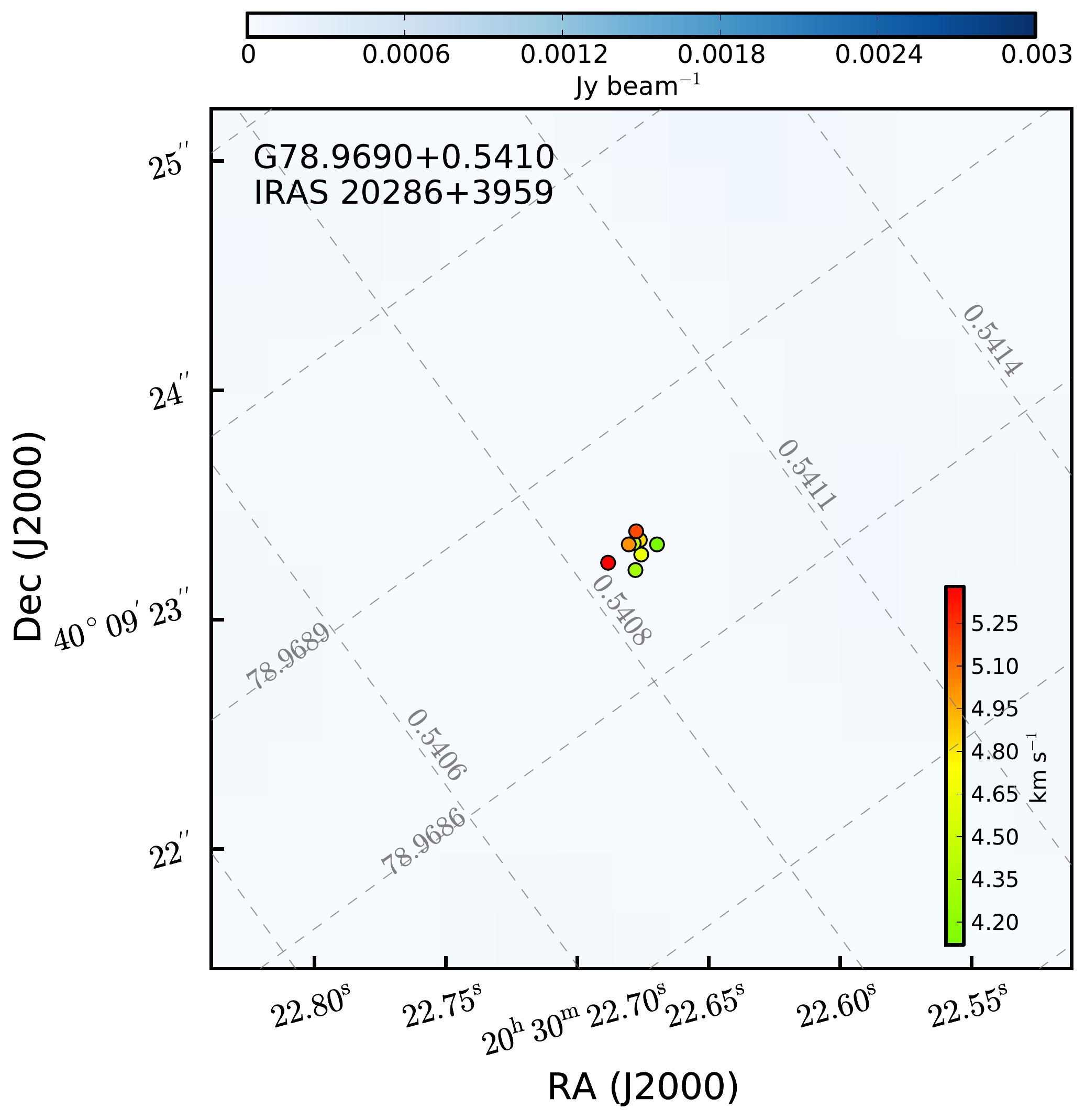}
\includegraphics[width=0.40\textwidth,angle=0]{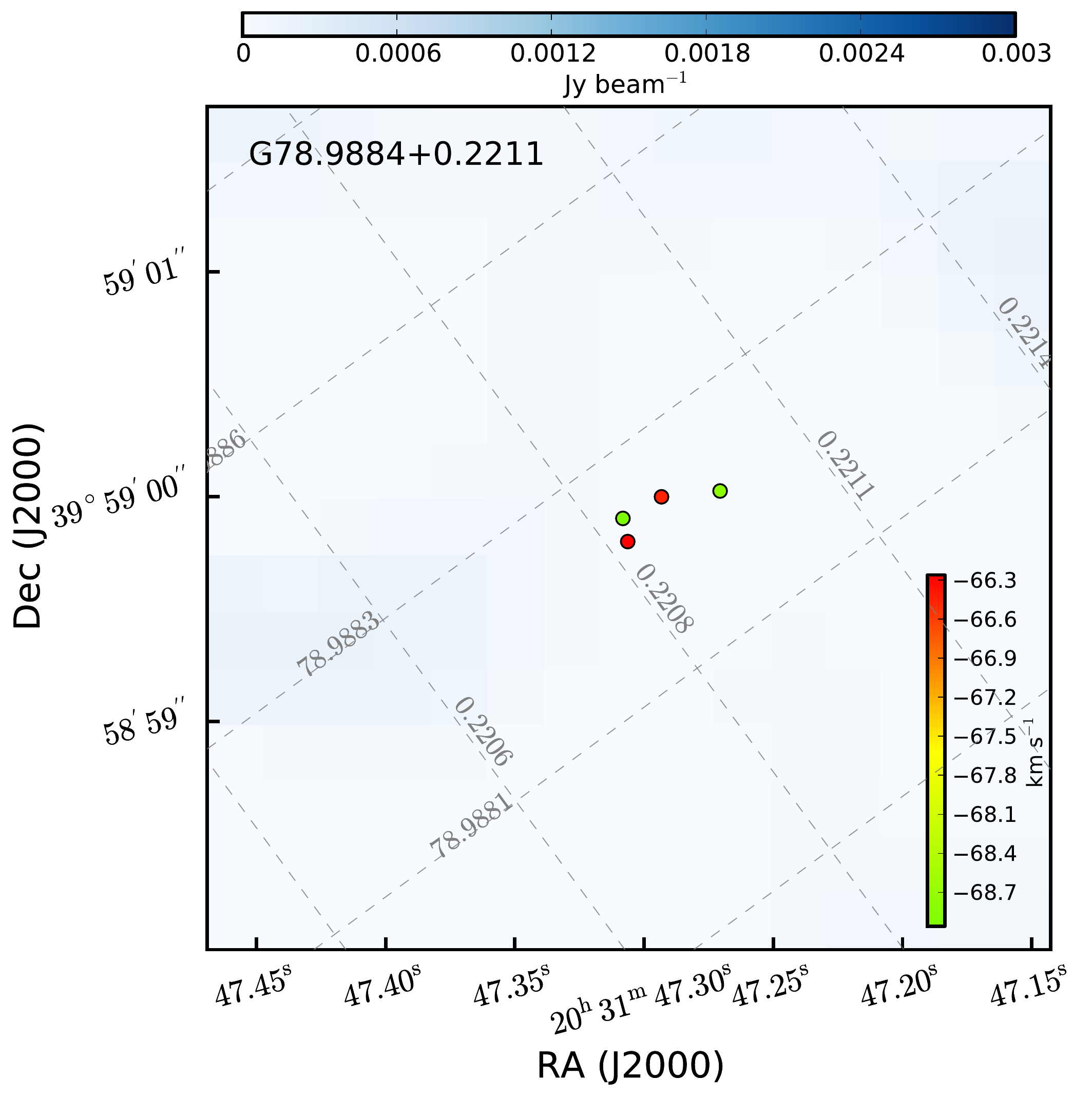} 
\includegraphics[width=0.40\textwidth,angle=0]{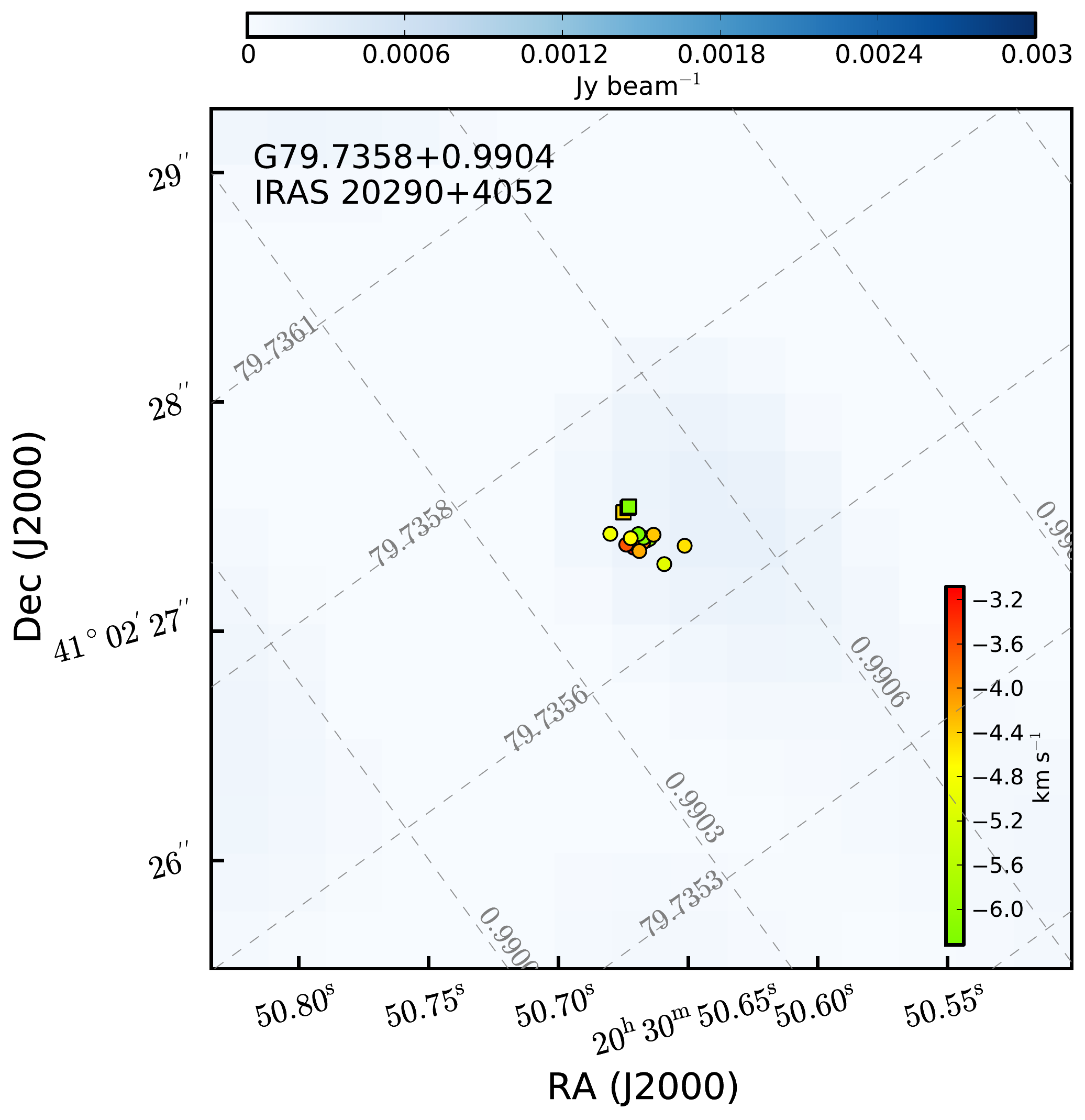} 
\caption{Distribution of all the maser spots (filled circles) contributing to the maser features for B-array data. The background image shows B-configuration continuum emission (blue scale and contours). 
The spots are color coded by LSR velocity (color bar). 
The $n$th blue contour is at $\left({\sqrt{2}}\right)^{n}\times S_{\rm max} \times p$, where $S_{\rm max}$ is the maximum continuum flux shown in the blue bar for each panel, 
$n$=0, 1, 2 ..., and $p$ is equal to 10\%. Positions of radio continuum sources identified in previous observations are marked with black stars along with their names. Maser spots detected with the EVN by \cite{Rygl2012} are shown as filled squares. 
The gray grid shows the Galactic coordinate system. The beam size of the radio continuum maps is $1\rlap.{''}5\times1\rlap.{''}5$. The maser position errors are $\approx0\rlap.{''}2$ (see Sect. \ref{sec:search}).
}
\label{fig:spots}
\end{center}
\end{figure*}

\begin{figure*}[ht]
\begin{center}
\includegraphics[width=0.40\textwidth,angle=0]{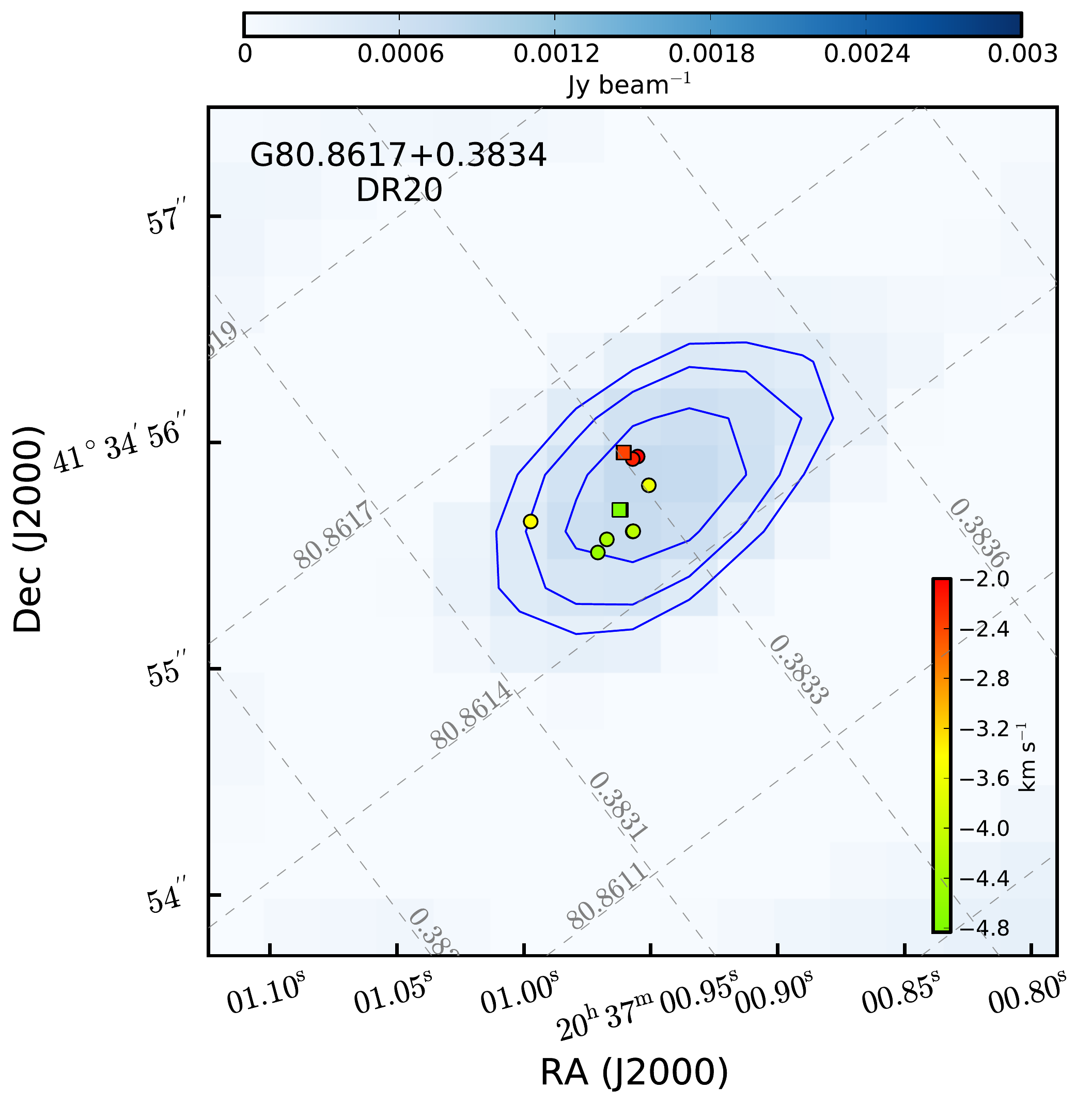} 
\includegraphics[width=0.40\textwidth,angle=0]{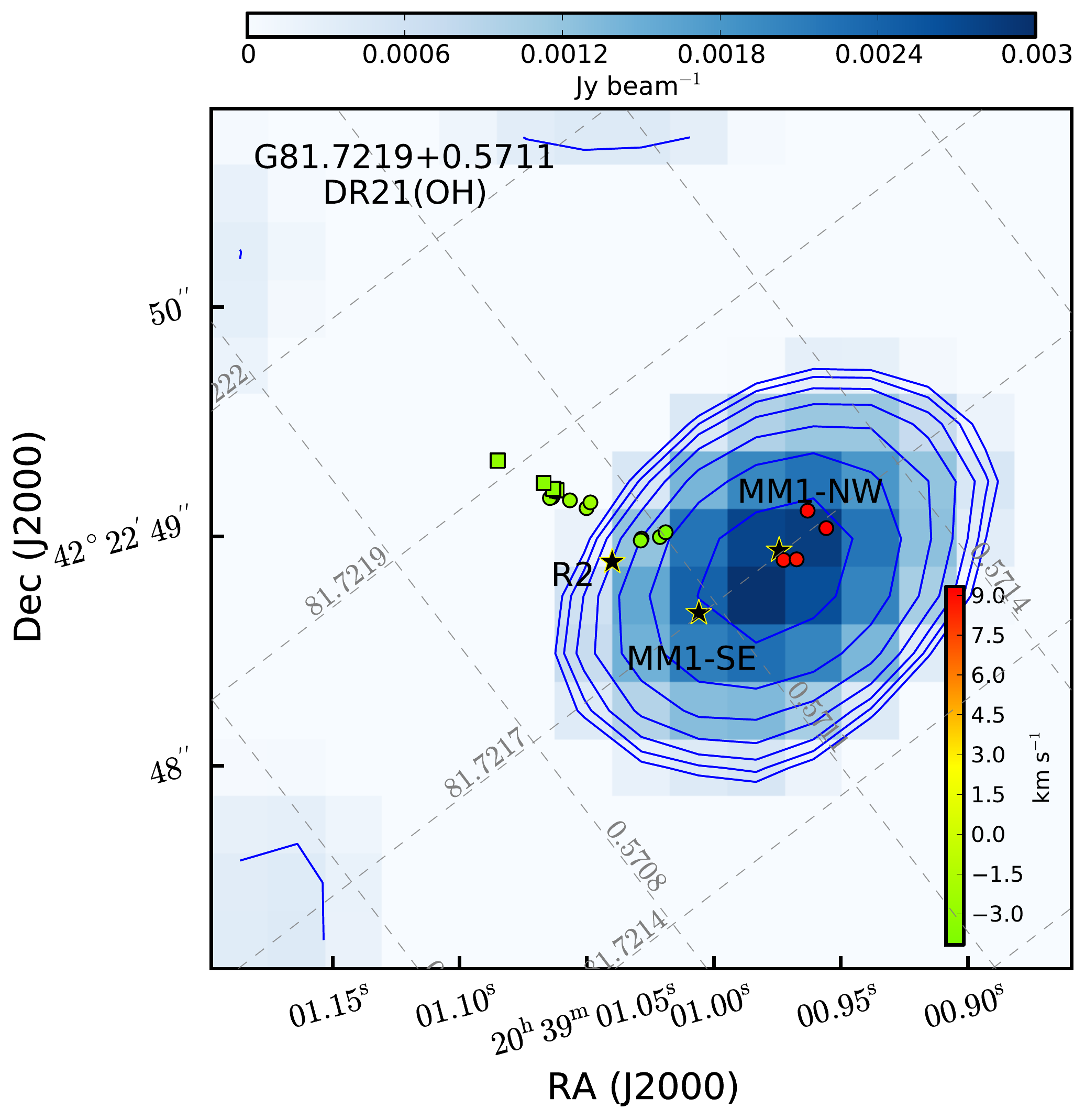}  
\includegraphics[width=0.40\textwidth,angle=0]{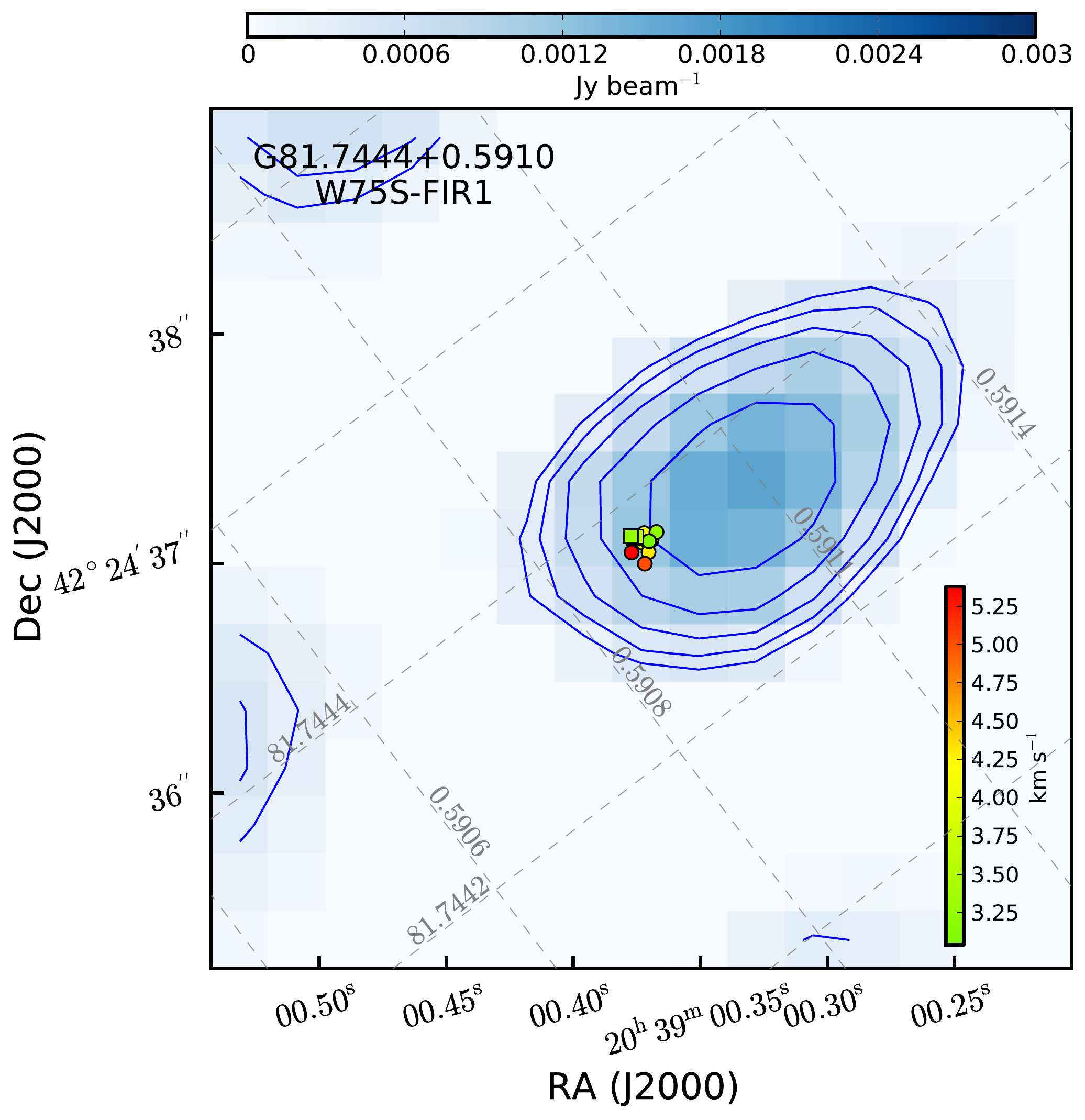}
\includegraphics[width=0.40\textwidth,angle=0]{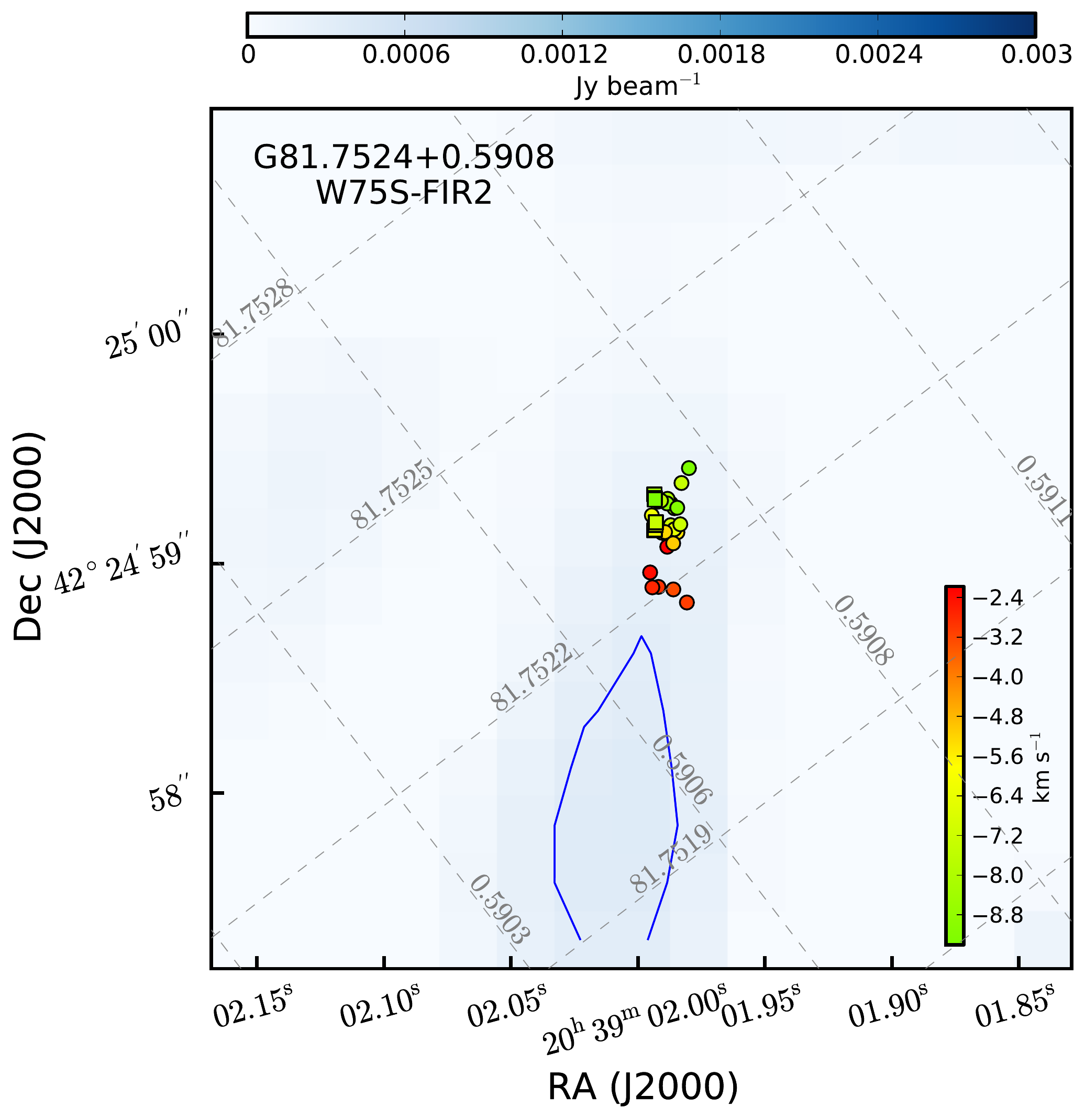}
\includegraphics[width=0.40\textwidth,angle=0]{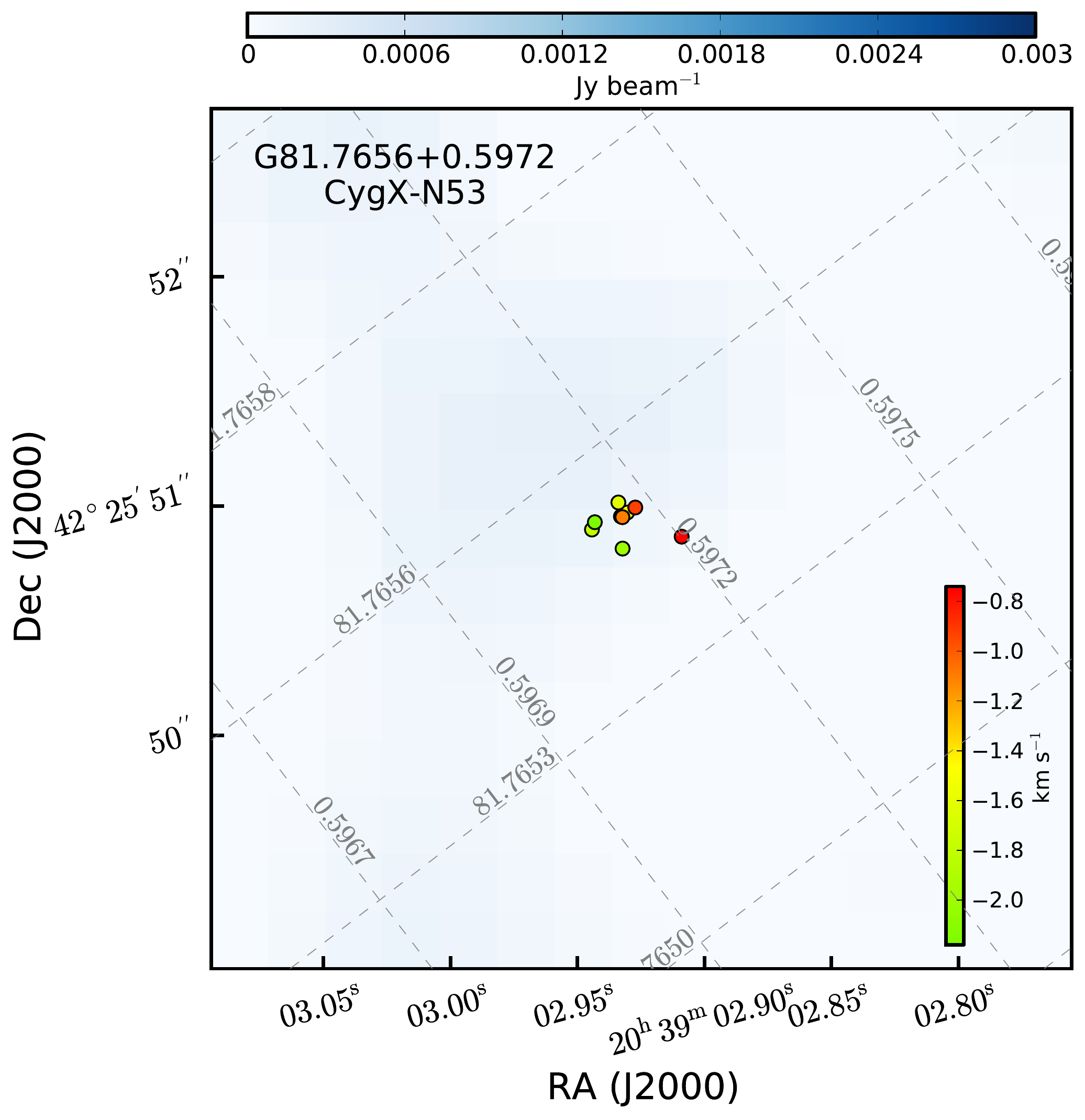} 
\includegraphics[width=0.40\textwidth,angle=0]{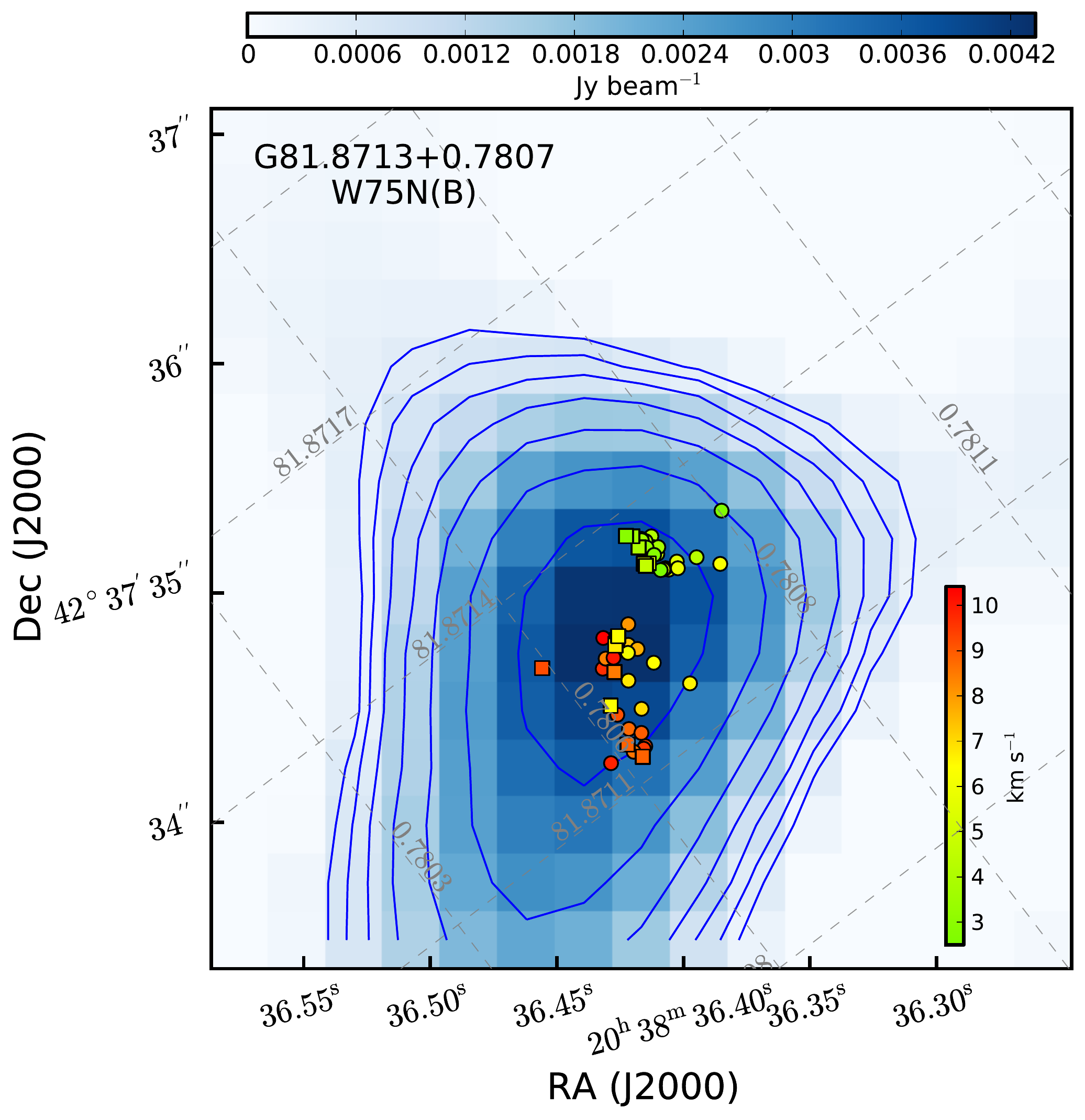} 
\caption{{\it Continued}. }
\end{center}
\end{figure*}

\begin{figure*}[ht]
\begin{center}
\includegraphics[width=0.40\textwidth,angle=0]{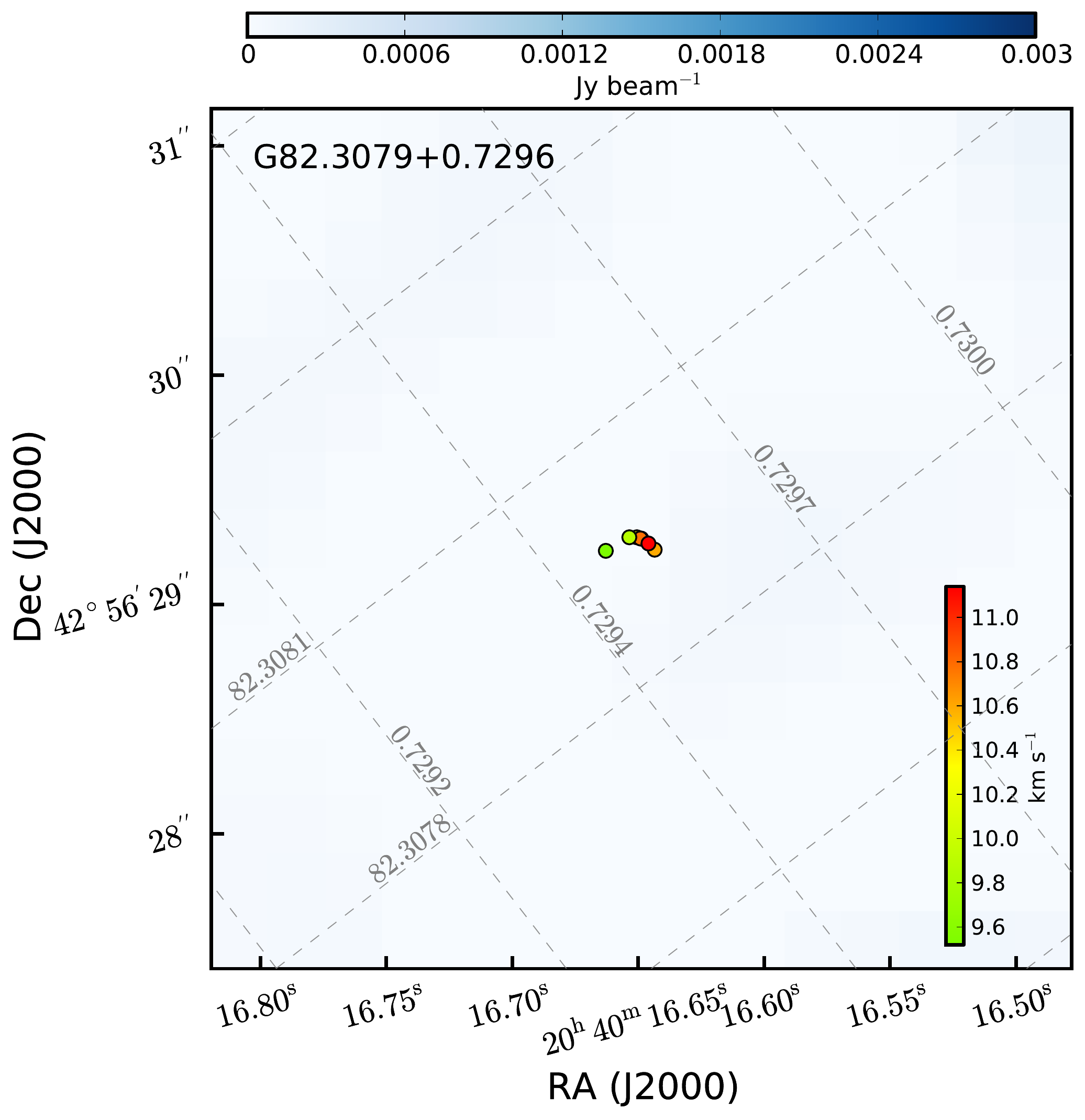} 
\caption{{\it Continued}. }
\end{center}
\end{figure*}

\clearpage

\section{SED fitting}\label{sec:sed-fitting}

Figure \ref{fig:sed-fit} displays the SED fitting of the dust cores that host maser emission. 

\renewcommand{\thefigure}{B.1}

\begin{figure*}[tbh]
\begin{center}
 \includegraphics[width=0.4\textwidth,angle=0]{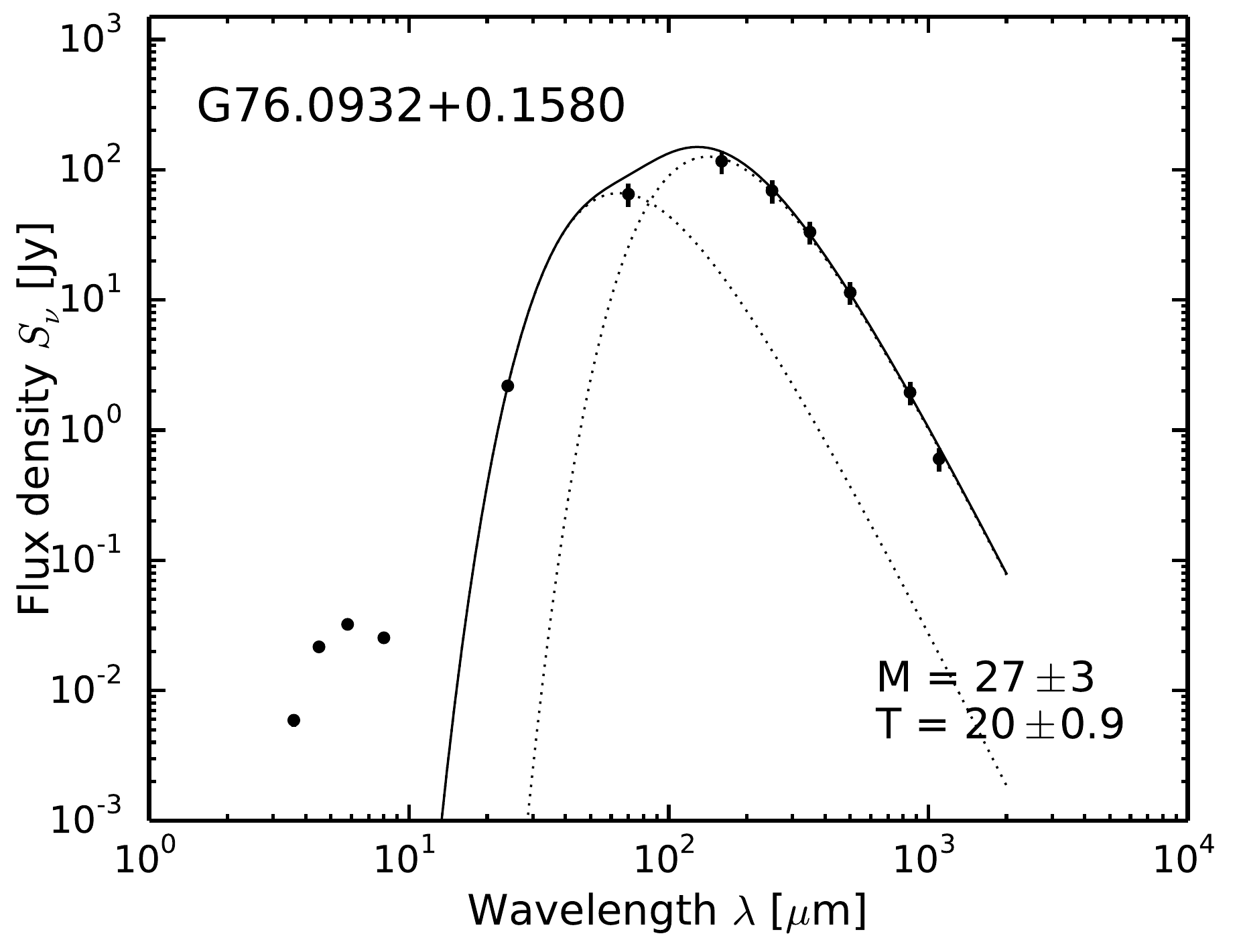}  
 \includegraphics[width=0.4\textwidth,angle=0]{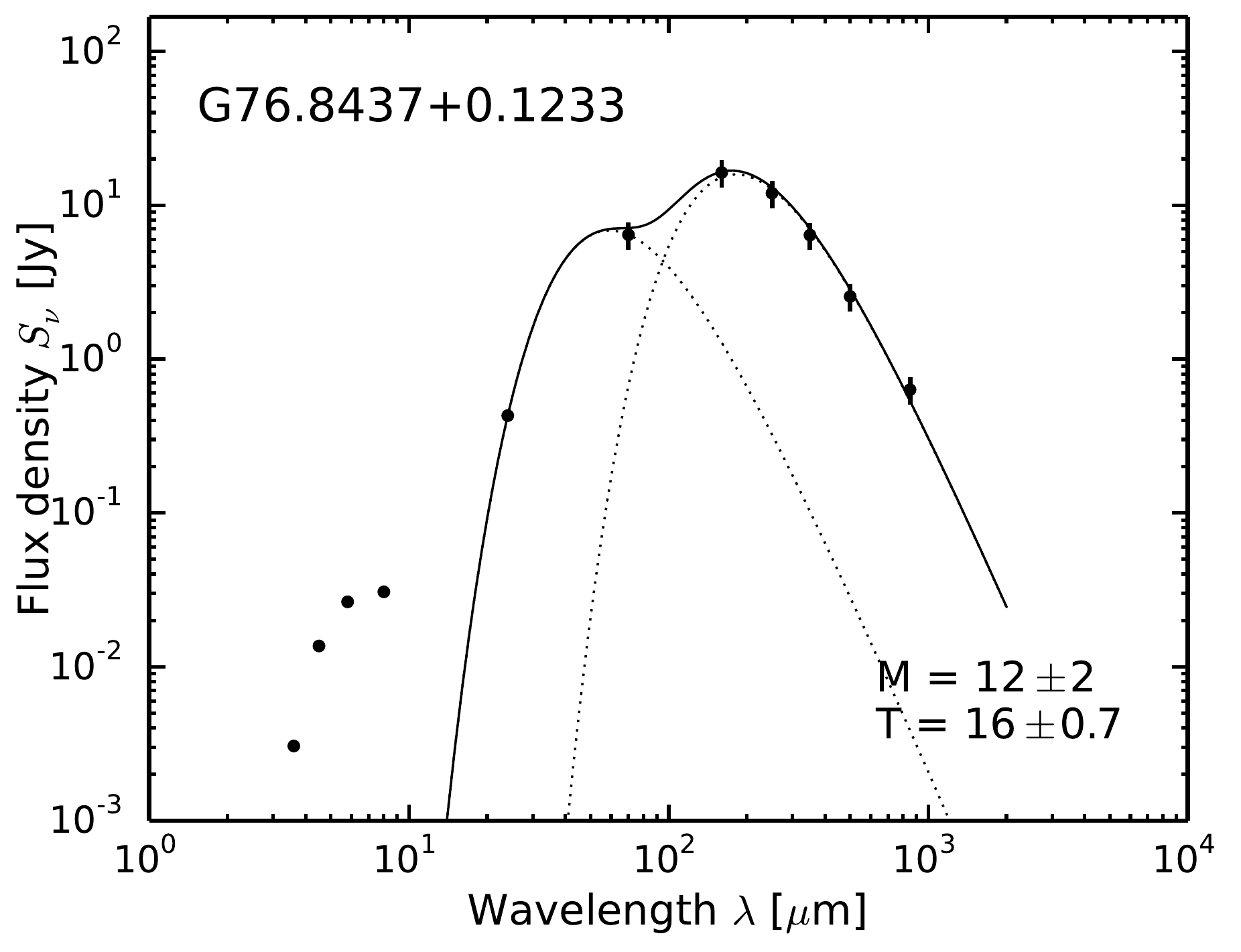}  
 \includegraphics[width=0.4\textwidth,angle=0]{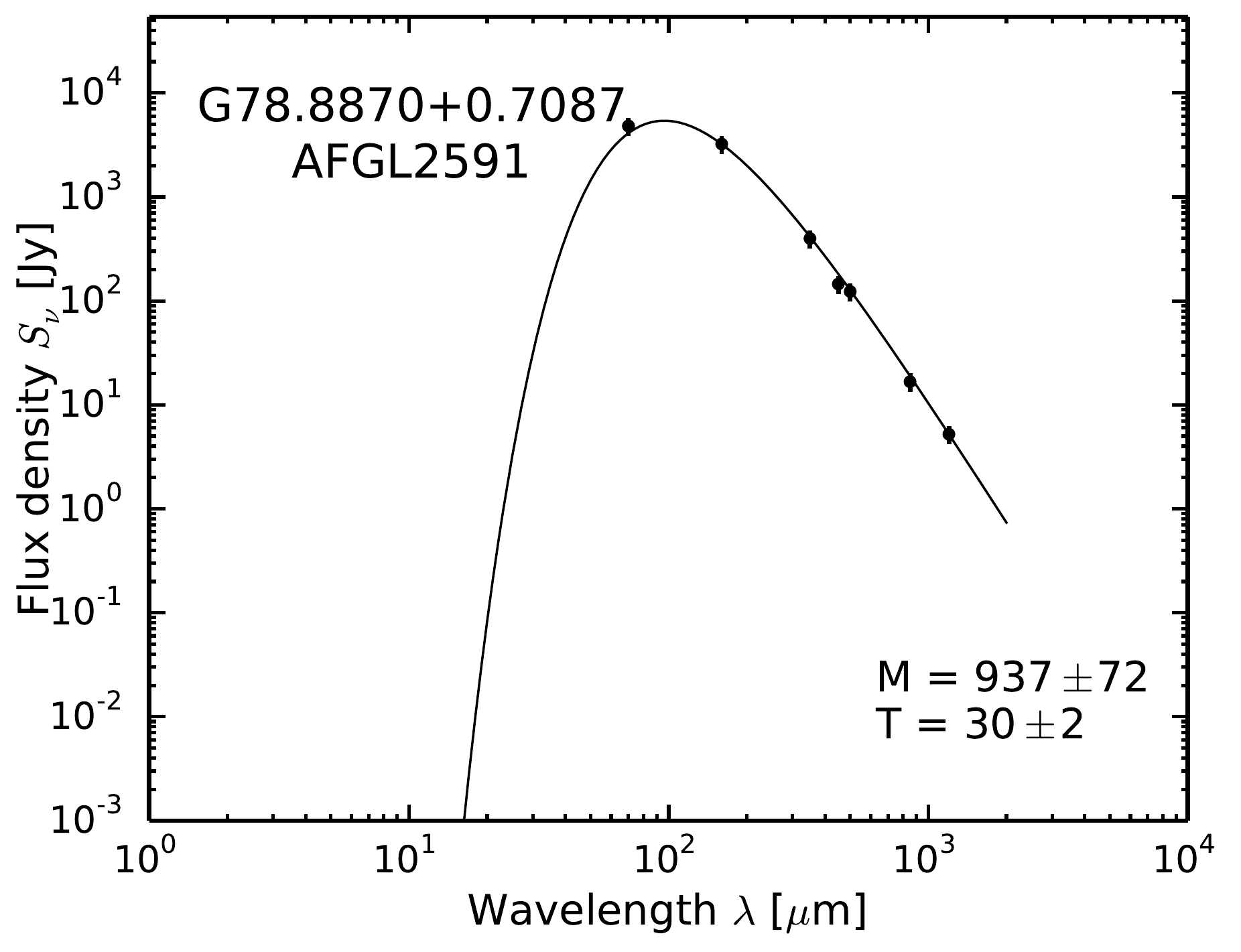}  
 \includegraphics[width=0.4\textwidth,angle=0]{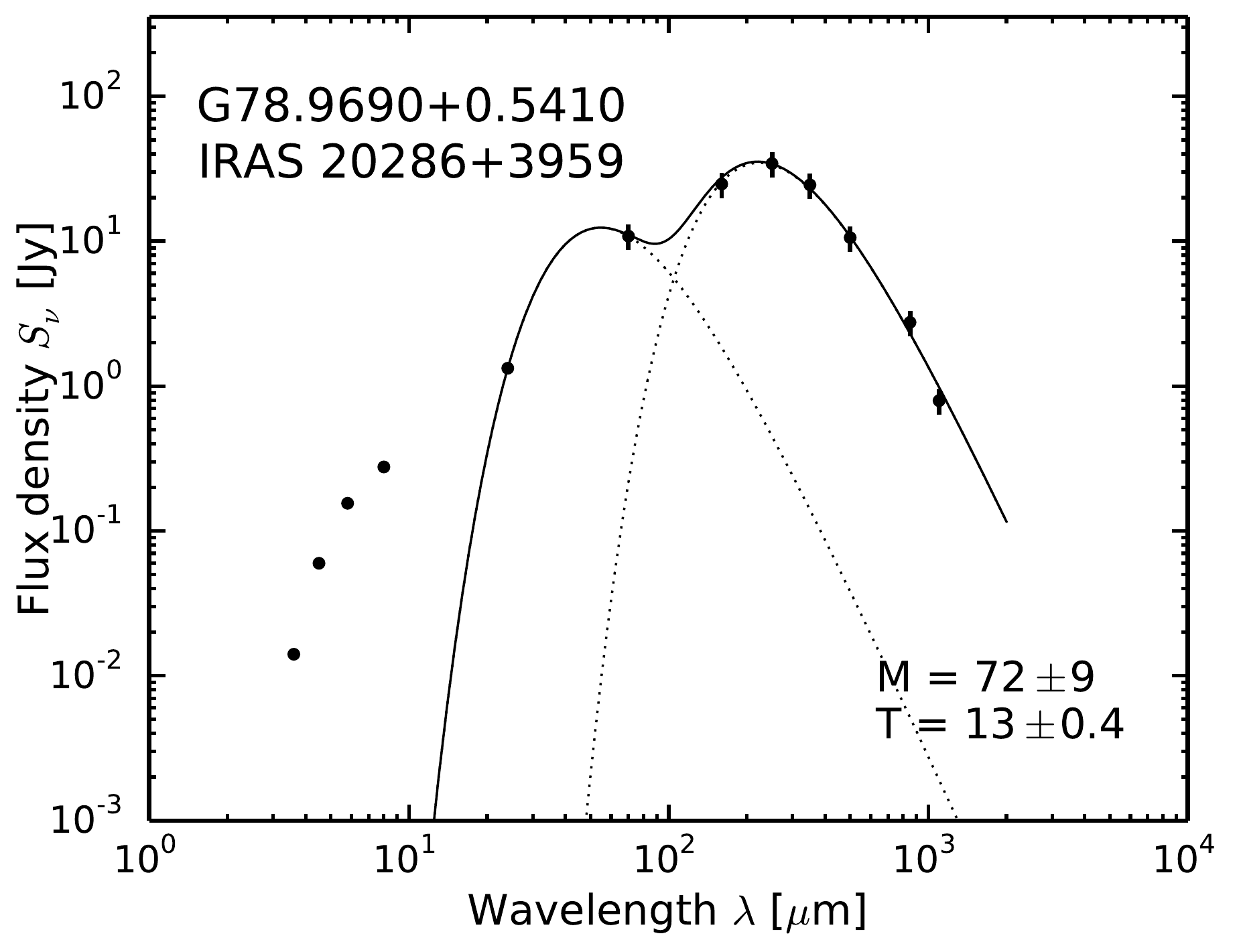}  
 \includegraphics[width=0.4\textwidth,angle=0]{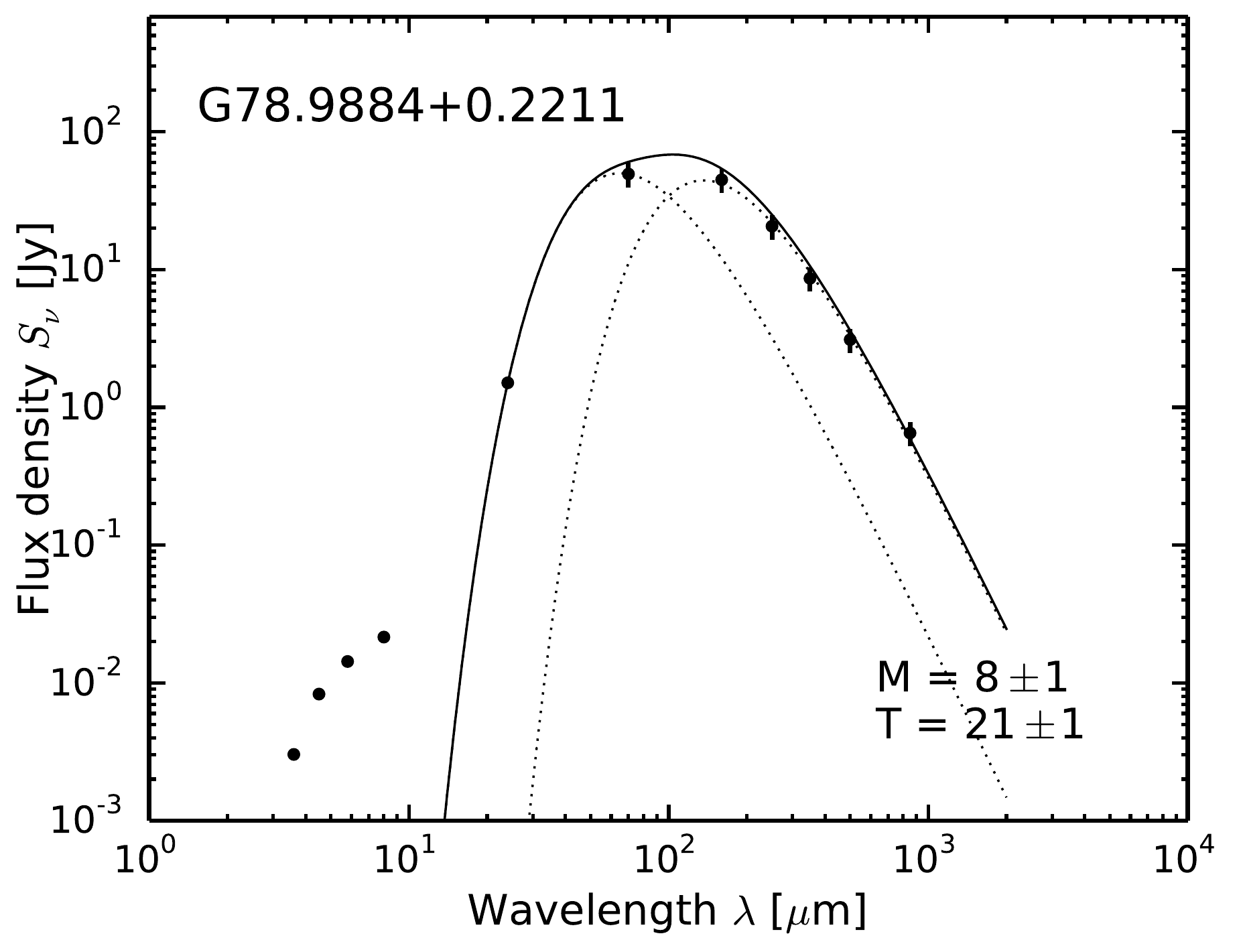}  
 \includegraphics[width=0.4\textwidth,angle=0]{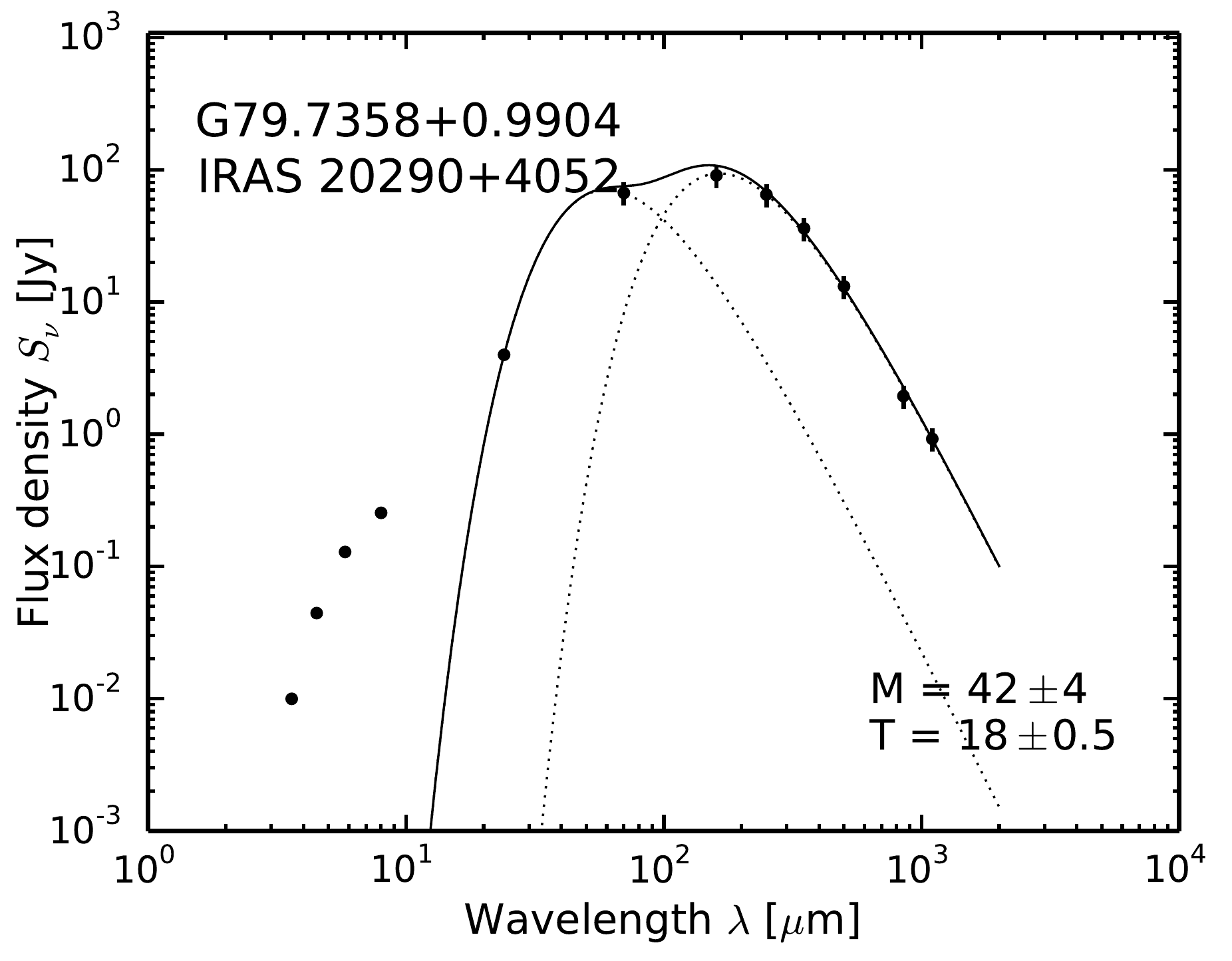}   
 \includegraphics[width=0.4\textwidth,angle=0]{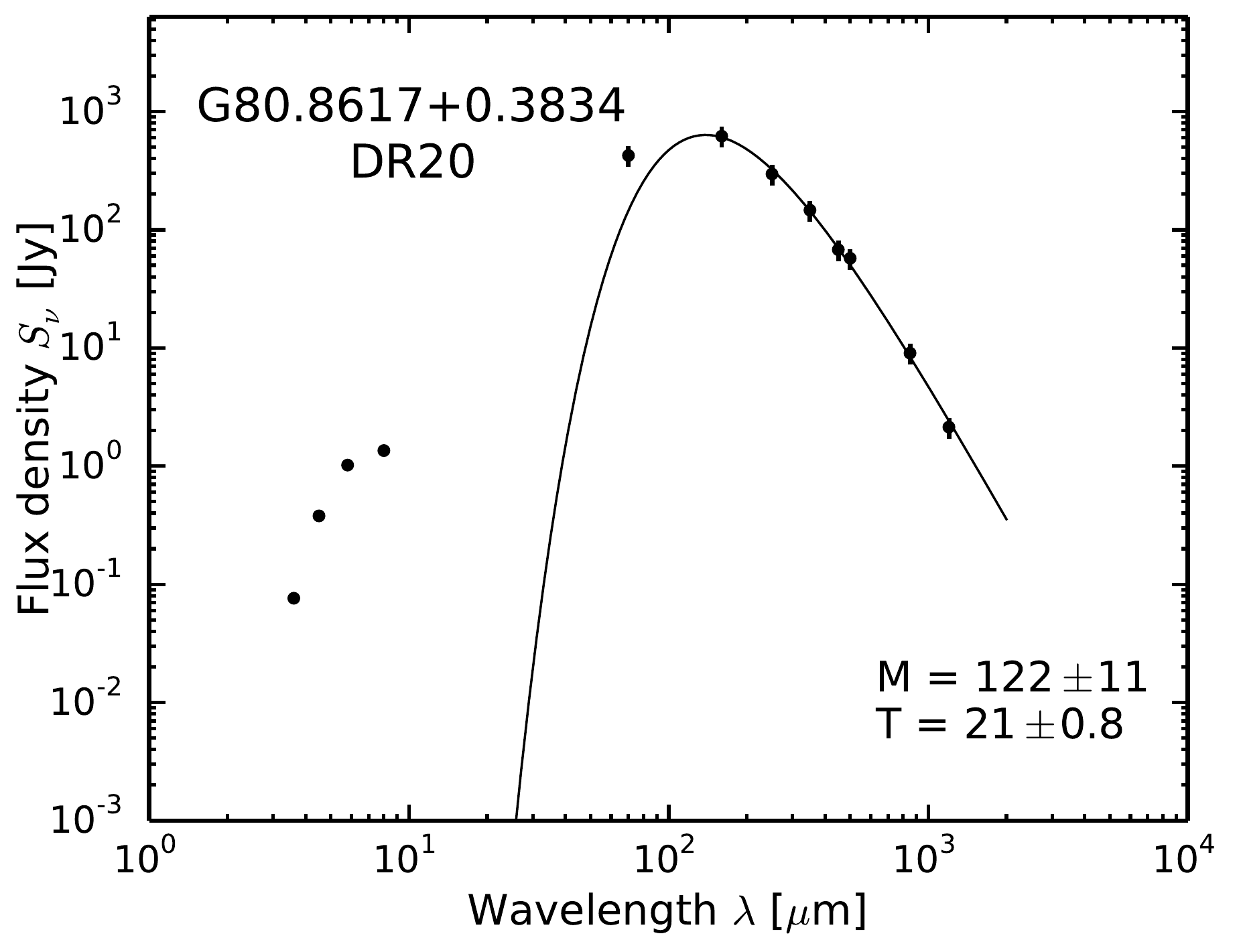}  
 \includegraphics[width=0.4\textwidth,angle=0]{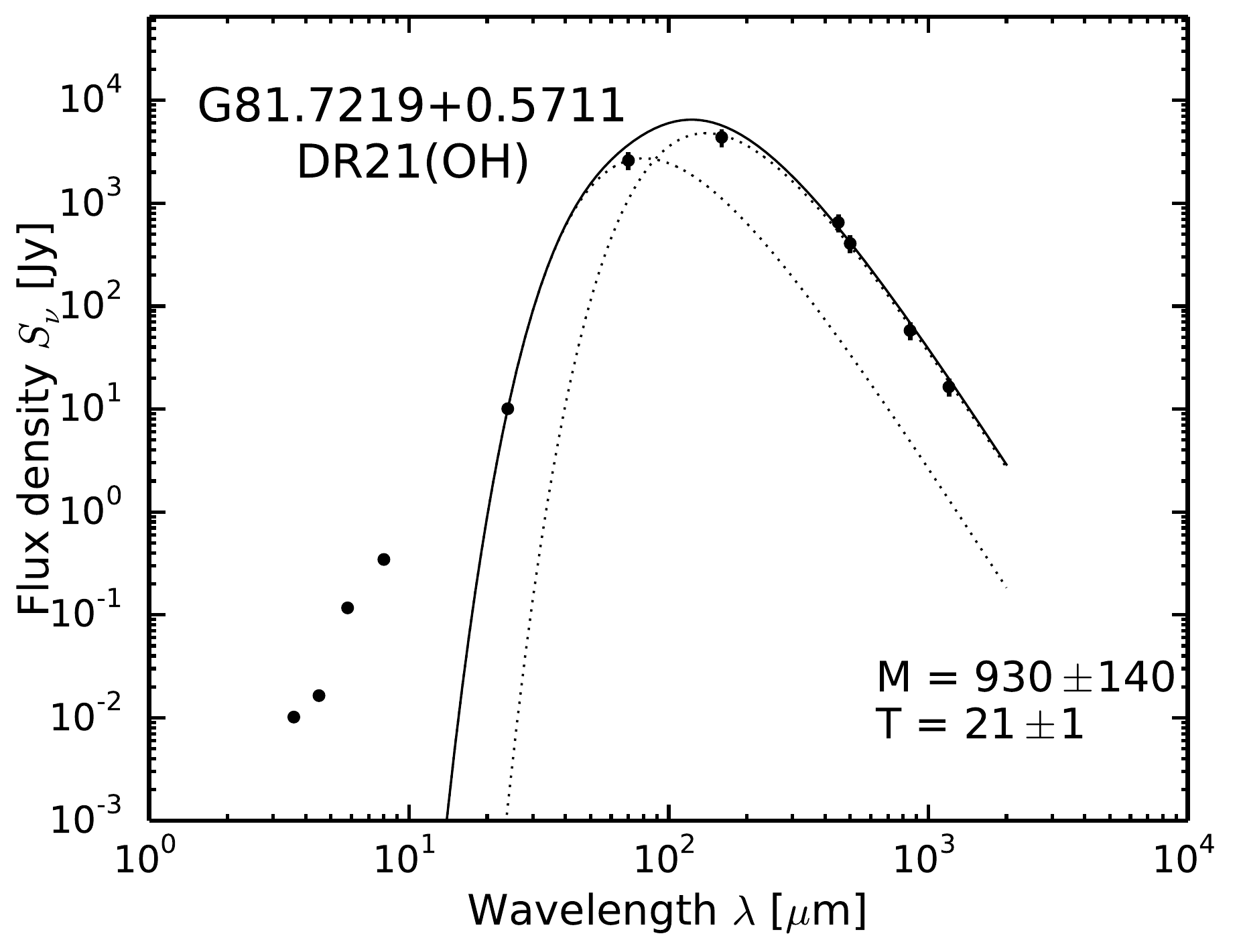} 
\caption{SED fitting for the mass and dust temperature derivation of the cores with associated maser emission. A model consisting of a warm and a cold component (solid line) is fit when the 24 and 70~$\mu$m points are available. Otherwise, the data are fit with a single gray-body model. 
The two dotted lines present the two components separately.
The IRAC 3.6--8~$\mu$m points -- likely originating from a third, hot and inner component -- were not used in the fitting.
}
\label{fig:sed-fit}
\end{center}
\end{figure*}

\setcounter{figure}{0}
\renewcommand{\thefigure}{B.1}

\begin{figure*}[tbh]
\begin{center}  
 \includegraphics[width=0.4\textwidth,angle=0]{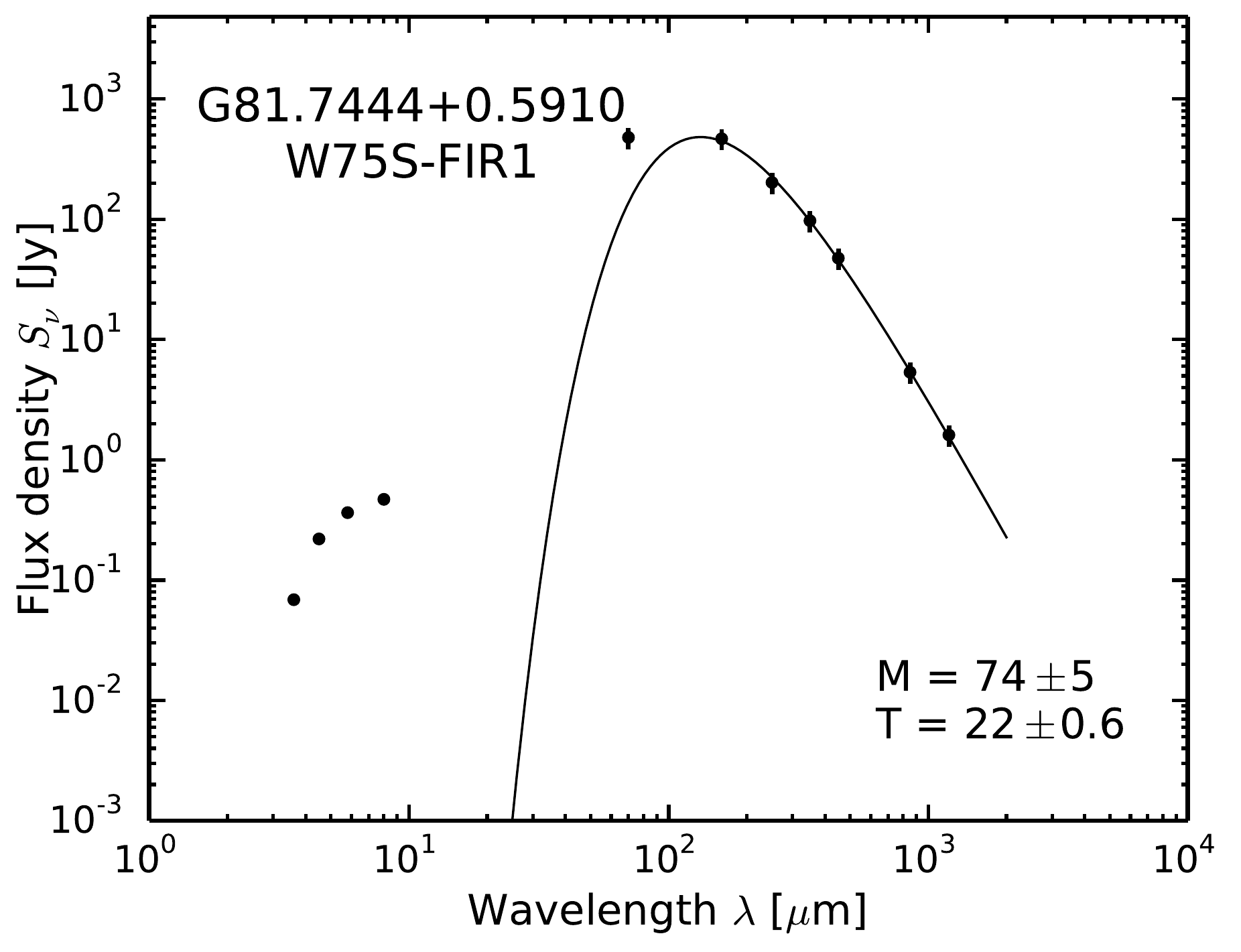}  
 \includegraphics[width=0.4\textwidth,angle=0]{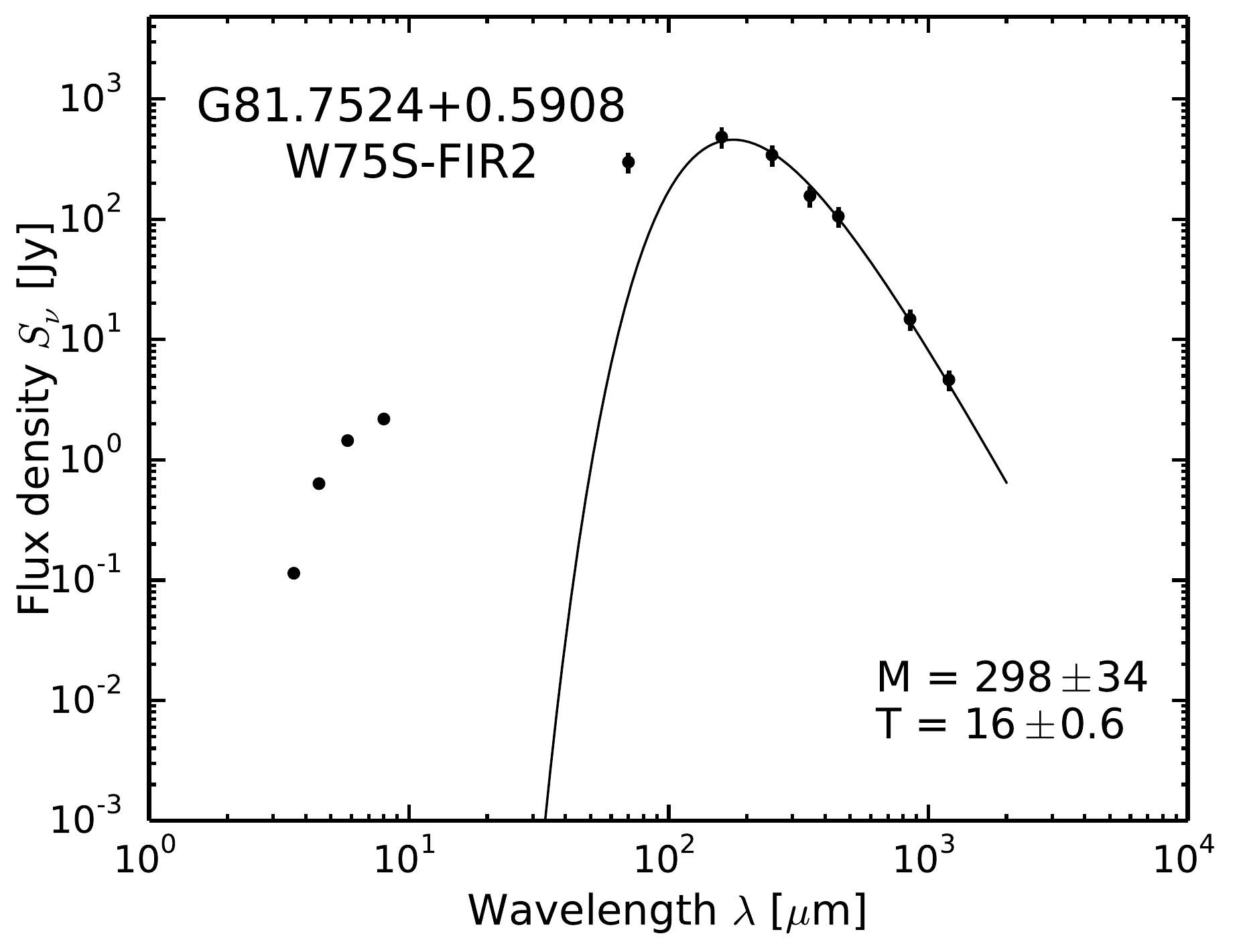}  
 \includegraphics[width=0.4\textwidth,angle=0]{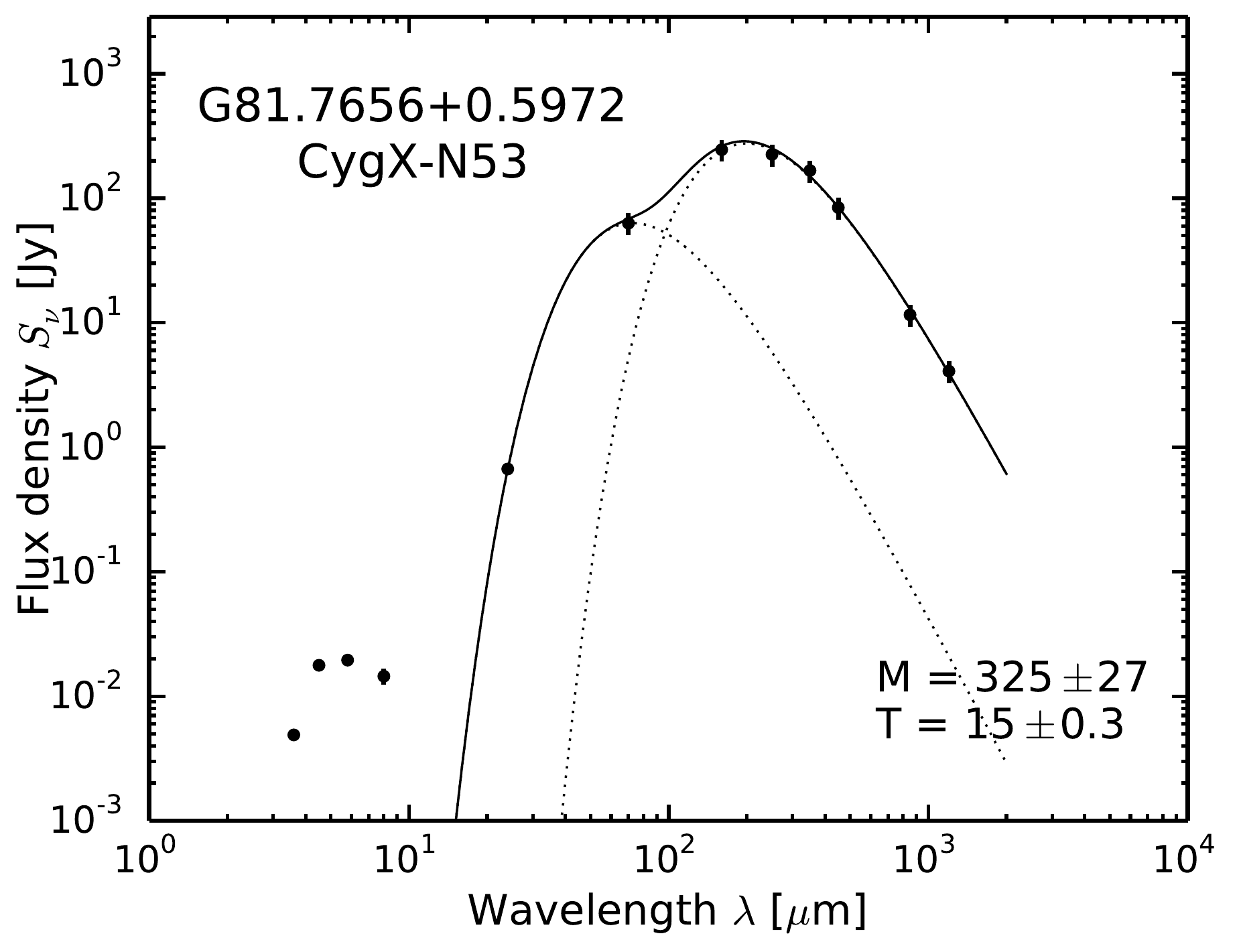}   
 \includegraphics[width=0.4\textwidth,angle=0]{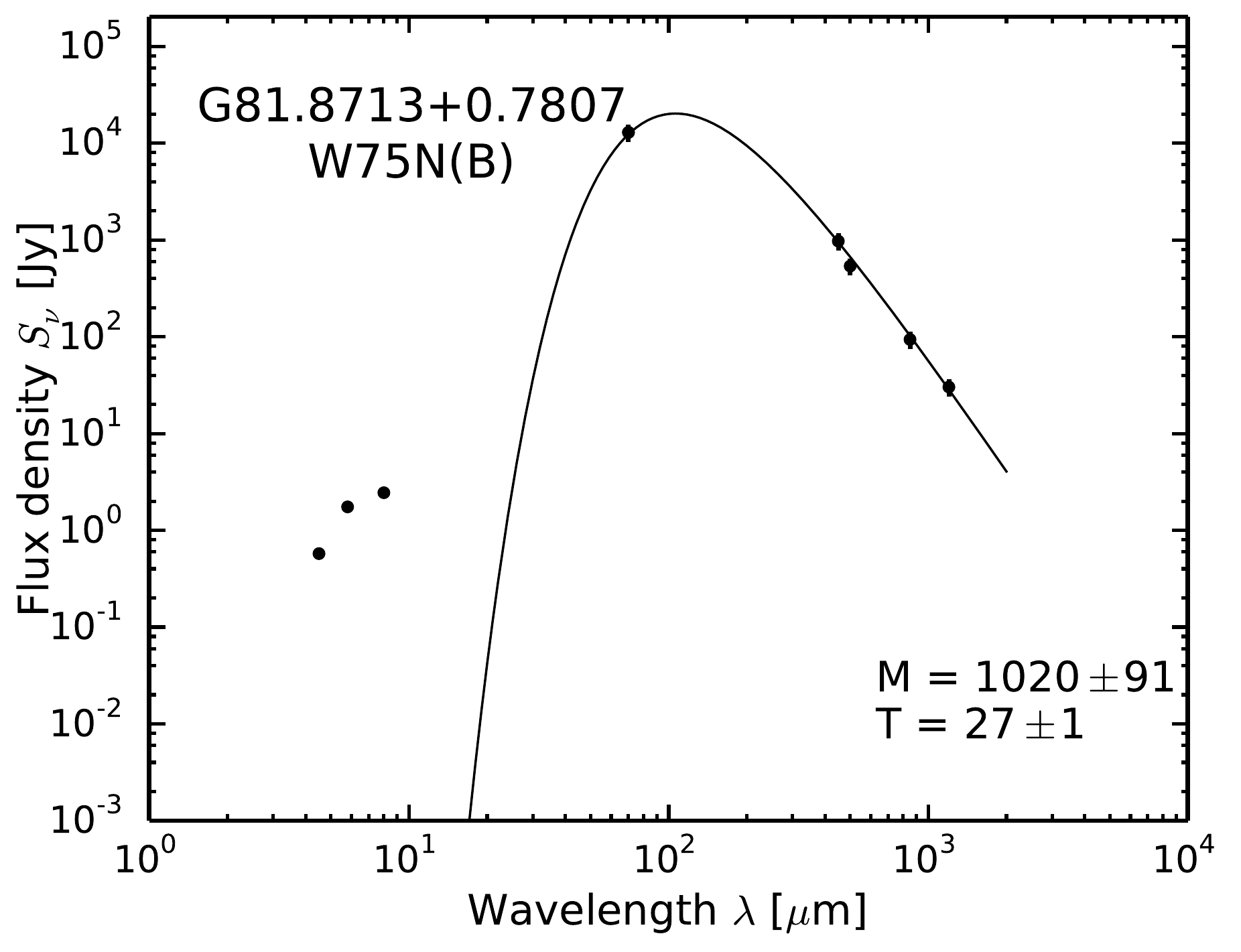}   
 \includegraphics[width=0.4\textwidth,angle=0]{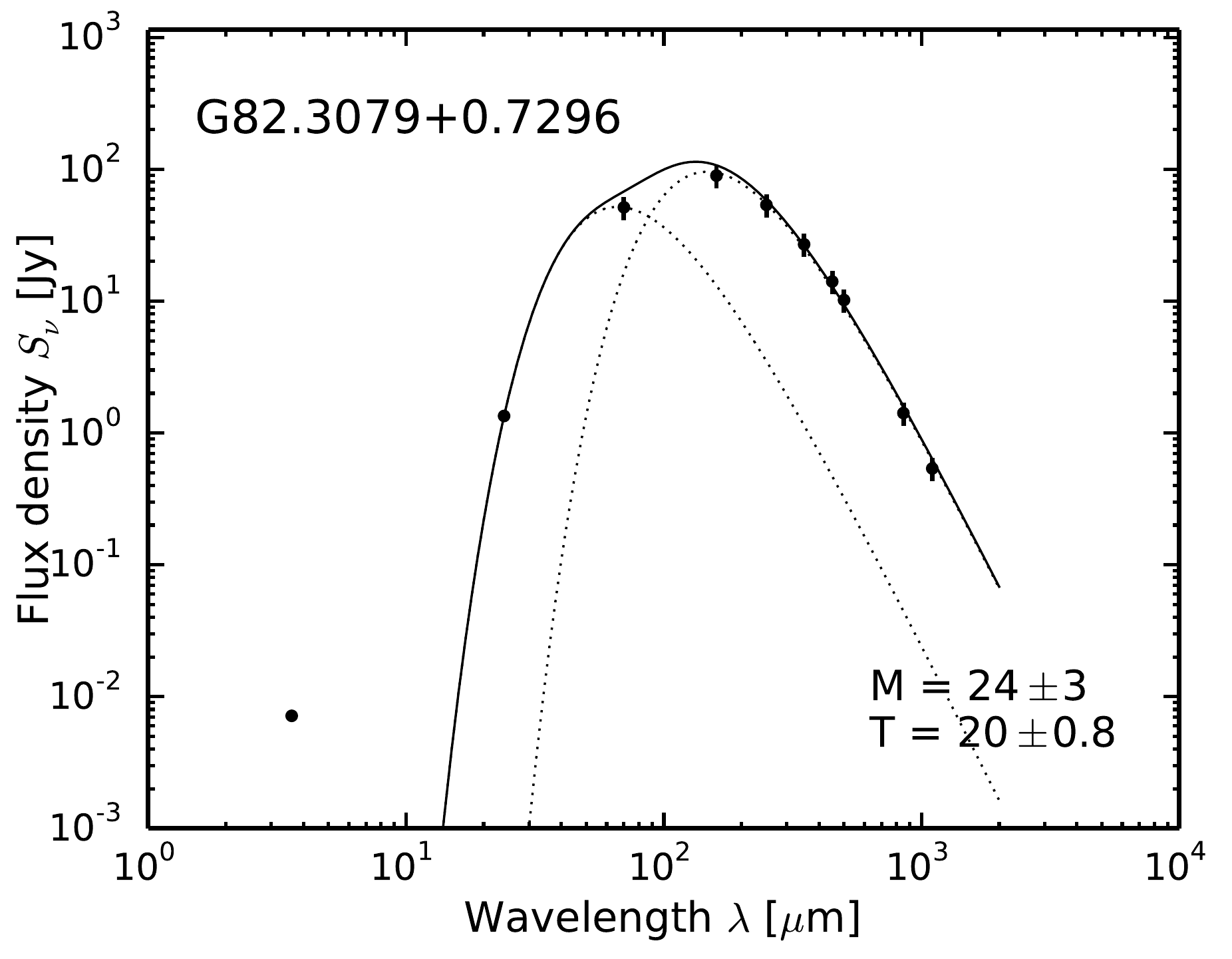}  
\caption{Continued. }
\end{center}
\end{figure*}

\clearpage

\section{Previously detected masers}\label{sec:maser-past}

Methanol masers at 6.7~GHz detected in previous surveys by \cite{Hu2016} and \cite{Yang2019} are listed in Table \ref{tab:hu16}.

\begin{table*}
\caption{Methanol masers toward Cygnus X reported in previous surveys. }
\label{tab:hu16} 
\centering 
\begin{tabular}{l l l r r}  
\hline\hline 
RA &  Dec. & Name & $V_{\rm LSR}$  &  $S_{\nu,~\rm Peak}$      \\ 
     &        &      &  (km~s$^{-1}$) & (Jy~beam$^{-1}$)     \\ 
\hline 
20 29 24.937    &       +40 11 19.29    &       G078.886+0.708  &       -7.02   &       0.54    \\
20 30 50.685    &       +41 02 27.50    &       G079.735+0.990  &       -5.44   &       23.17   \\
                        &                               &               &       -3.86   &       5.00    \\
                        &                               &               &       -3.16   &       1.91    \\
20 37 00.980    &       +41 34 55.05    &       G080.861+0.383  &       -11.06  &       2.53    \\
                        &                               &               &       -4.04   &       8.52    \\
                        &                               &               &       -1.93   &       1.17    \\
20 39 00.879    &       +42 22 48.93    &       G081.721+0.571  &       -3.51   &       0.86    \\
                        &                               &               &       -2.63   &       2.96    \\
20 39 00.378    &       +42 24 36.92    &       G081.744+0.590  &       3.86    &       3.82    \\
                        &                               &               &       4.57    &       5.53    \\
20 39 01.573    &       +42 24 59.09    &       G081.752+0.590  &       -8.60   &       8.35    \\
                        &                               &               &       -6.67   &       3.43    \\
                        &                               &               &       -5.80   &       3.87    \\
                        &                               &               &       -2.81   &       1.51    \\
                        &                               &               &       -2.28   &       0.90    \\
20 39 02.274    &       +42 25 50.81    &       G081.765+0.597  &       -1.23   &       2.54    \\
20 38 36.423    &       +42 37 35.01    &       G081.871+0.780  &       3.51    &       42.91   \\
                        &                               &               &       4.04    &       99.94   \\
                        &                               &               &       4.74    &       166.73  \\
                        &                               &               &       5.27    &       76.09   \\
                        &                               &               &       5.80    &       67.27   \\
                        &                               &               &       7.20    &       174.93  \\
                        &                               &               &       9.48    &       17.48   \\
\hline 
\hline 
     &        &      &  $V_{\rm LSR}$, $\Delta V$    &   $S_{\nu,~\rm Peak}$     \\ 
     &        &      &  (km~s$^{-1}$)    &         (Jy)         \\ 
\hline 
20  23  23.65   &       +37 35  34.3    &       G76.093+0.158   &       4.92 (4.27, 6.90)    &       0.49    \\
20      29      20.35   &       +40     11      37.5    &       G78.882+0.723   &       -6.94 (-7.41, -6.51)  &       0.95    \\
20  30  22.77   &       +40 09  23.2    &       G78.969+0.541   &       4.74 (4.01, 5.48)    &       1.24    \\
20      30      50.70   &       +41     02      28.8    &       G79.736+0.991   &       -5.51 (-6.63, -2.84) &        19.4    \\
20      37      01.02   &       +41     34      56.9    &       G80.862+0.383   &       -4.01 (-4.57, -1.77)  &       9.67    \\
20      39      02.01   &       +42     24      59.3    &       G81.752+0.591   &       -8.57 (-9.39, 5.04) & 15.4    \\
20  37  47.39   &       +42 38  39.0    &       G81.794+0.911\tablefootmark{a}  &       7.19 (-4.73, 7.54)   &       0.59    \\
20      38      36.67   &       +42     37      30.6    &       G81.871+0.779   &       4.53 (2.55, 9.87) &  225.7   \\
20  40  16.72   &       +42 56  28.6    &       G82.308+0.729   &       10.3 (10.0, 11.3) &  58.4    \\
\hline
\end{tabular}
\tablefoot{The top table lists maser components reported in the catalog of  \cite{Hu2016} and gives the LSR
velocity of the component and the flux density. The bottom table lists masers in the catalog of \cite{Yang2019} and gives 
LSR velocity of peak emission, the LSR velocity interval of the maser emission in parentheses, and peak flux density.
\tablefoottext{a}{Not detected in GLOSTAR.}
}
\end{table*}

\section{Methanol absorption toward DR21}\label{sec:dr21ch3oh}
Based on our absorption spectrum, we now constrain the methanol abundance in DR21~M.
In local thermodynamic equilibrium (LTE), the column density in a spectral line's lower energy level is given by the line's observed optical depth, $\tau$, integrated over the line profile, 

\begin{eqnarray}
N_l = {{8\pi~k}\over{c^3~h}}{g_l\over g_u}T_{ex}\int \tau dv,
\label{eq.Nmeth}
\end{eqnarray}from which the total column density of the molecule, 
$N({\rm CH}_3{\rm OH})$,  can be calculated:
\begin{eqnarray}
N = {N _l\over g_l} Q(T_{rot}) e^{E_l/{kT}}.
\label{eq.Ntot}
\end{eqnarray}Here, $k$, $h$, and $c$ are the Planck constant, the Boltzmann constant, and the speed of light, respectively, and $g_l$ and $g_u$ are the degeneracies of the line's lower and upper energy levels, respectively ($g_l = 13$ and $g_u = 11$); $Q$ is the partition function, and $E_l$ is the energy above ground state of the lower energy level (48.7 K).
Integrating our absorption spectrum from $-17$ to $+2$~ km~s$^{-1}$, we determine an integrated optical depth of 0.47~km~s$^{-1}$.

By definition, under LTE, the excitation temperature and rotation temperature of the line  are the same and are also equal to the kinetic temperature, $T_{ex} = T_{rot} \equiv T_{kin}$. A question then arises as to the appropriate temperature for the calculation of the 
methanol column density.

Modeling a variety of radio-wavelength OH absorption lines, \citet{Jones1994} derived an H$_2$ density of $(1.8 \pm 0.7) \times 10^7$~cm$^{-3}$ and a kinetic temperature of 175 K for the part of DR21~M's dense molecular core from which the OH-bearing gas originates, namely the dense PDR interface with the compact HII region 
(see Sect. \ref{sec:dr21}). 
These numbers might not be representative of the more extended condensation that gives rise to the methanol absorption.

\renewcommand{\thefigure}{D.1}

\begin{figure}[!tp]
        \centering
    \includegraphics[width=0.48\textwidth]{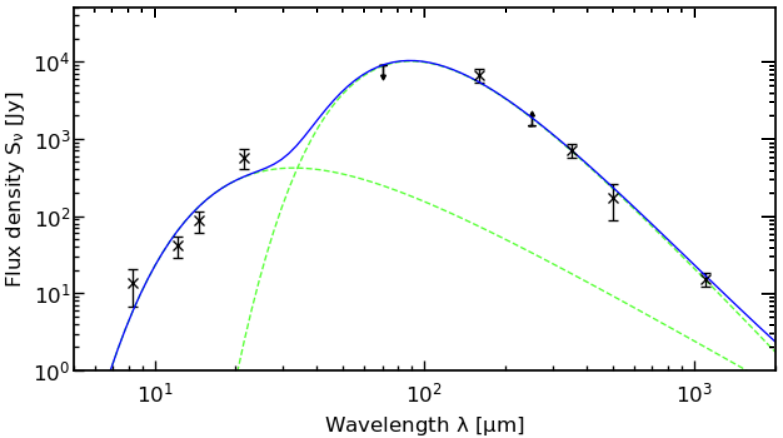}
        \caption[SED of DR21(M)]{Dust spectral energy distribution of DR21~M. The best fit SED is marked by the blue line and is the sum of the simultaneously fit gray body (right) and blackbody (left) components (green dashed lines).}
        \label{fig:outer:shell_hist_seds}
\end{figure}

Therefore, we characterized the molecular gas associated with DR21~M by fitting its dust SED (Fig. \ref{fig:outer:shell_hist_seds}) obtained through aperture photometry of mid-infrared to submillimeter continuum maps to derive its temperature, source size, H$_2$ column, and volume density.

Using the same procedures as described in \citet{koenig2021}, we determined the flux densities from Herschel/Hi-GAL \citep{Molinari2010} and MSX \citep{price2001} continuum maps in nine bands using an aperture-annulus approach. As the 160\,$\mu$m PACS band and the D-configuration maps of methanol have similar beam widths (i.e., $13\rlap.{''}5$ and 13$''$, respectively), we estimated the source size (i.e., the $\mathrm{FWHM}_\mathrm{160}$) through a two-dimensional Gaussian fit to the emission in the 160\,$\mu$m PACS band centered on the peak of the emission. The same position, aperture, and annulus sizes were subsequently used for all bands to measure the source and background flux, from which the background corrected source flux was calculated. The resulting SED was then fit with a two-component model consisting of a blackbody for the warm emission originating from the evolving star and a gray body representing the cold dust envelope, yielding the dust temperature, $T_\mathrm{dust}$.

With the ratio of the measured flux density, $F_{350}$, to the gray body's intensity, $B_{350}(T_\mathrm{dust})$, reflecting the dust emissivity, we calculated the H$_2$ column density according to

\begin{eqnarray}
N_\mathrm{H_2} = \frac{F_{350}}{B_{350}(T_\mathrm{dust})} \cdot \frac{\gamma}{\kappa_{350}} \cdot \frac{1}{\Omega_\mathrm{app} \cdot \mu_\mathrm{H_2}\cdot m_\mathrm{H}},
\label{eq.column_density}
\end{eqnarray}

\noindent where $\gamma=100$  (the gas-to-dust ratio), $\kappa_{350}=1.1$\,cm$^2$g$^{-1}$ (the dust opacity at 350\,$\mu$m calculated as the mean of the dust models from \citealt{Ossenkopf1994}), $\Omega_\mathrm{src}$ is the source solid angle, $\mu_\mathrm{H_2}=2.8$ is the mean molecular weight of the interstellar medium with respect to a hydrogen molecule \citep{Kauffmann2008}, and $m_\mathrm{H}$ is the mass of a hydrogen atom.

Using a distance of $d=1.4$\,kpc to DR21 and taking into account the fact that the measured source size $\mathrm{FWHM}$ is convolved with the beam, we determined the linear source size:

\begin{eqnarray}
r_\mathrm{src} = d \cdot \tan{\left(\sqrt{{\mathrm{FWHM}_{160}}^2 - {\theta_{160}}^2}\right)},
\label{eq.linear_source_size}
\end{eqnarray}

\noindent with $\theta_{70}=13\rlap.{''}51$ the beamwidth of the 160\,$\mu$m band.

Using the linear source size, $r_\mathrm{src}$, we determined the H$_2$ volume density: 
\begin{equation}
\begin{aligned}
n_\mathrm{H_2} &= \frac{M}{V} \cdot \frac{1}{\mu_\mathrm{H_2}\cdot m_\mathrm{H}}\\
 &= \frac{3}{4\pi \cdot r_\mathrm{src}^3}\cdot \frac{d^2\cdot F_{350}}{B_{350}(T_\mathrm{dust})}\cdot\frac{\gamma}{\kappa_{350}}\cdot \frac{1}{\mu_\mathrm{H_2}\cdot m_\mathrm{H}}.
\end{aligned}
\label{eq.mass}
\end{equation}We summarize the physical properties of DR21~M in Table\,\ref{tab:sed_dr21}.

\begin{table}[tp!]
\begin{center}
\caption[Physical properties of DR21(M)]{Summary of the physical properties of DR21~M.}\label{tab:sed_dr21}
\begin{tabular}{ll}
\hline
Parameter & Value \\
\hline\hline
Apparent source size $\mathrm{FWHM_{160}}$:      &  $37\rlap.{''}8$\\
Linear source radius $r_\mathrm{src}$:       &  0.128\,pc\\
Dust temperature $T_\mathrm{dust}$:&  $36.0 \pm 2.1$\,K\\
Average H$_2$ column density $N_\mathrm{H_2}$: &  $(1.7 \pm 0.4) \times 10^{23}$\,cm$^{-2}$\\
Average H$_2$ volume density $n_\mathrm{H_2}$: &  $(2.2 \pm 0.7) \times 10^5$\,cm$^{-3}$\\
\hline
\end{tabular}
\end{center}
\end{table}

At the derived density, the dust and gas temperatures, $T_{\rm dust}$ and $T_{\rm kin}$, respectively, are well coupled (i.e., equal to 36 K). If we assume that the CH$_3$OH energy levels
are in LTE at this temperature, we can derive the total CH$_3$OH column density using Eqs. \ref{eq.Nmeth} and \ref{eq.Ntot}. 
to calculate a total methanol column density, $N$(CH$_3$OH), of $2.8\times10^{17}$~cm$^{-2}$ and a CH$_3$OH abundance of $1.8\times10^{-6}$ relative to H$_2$ averaged over the DR21 M molecular core, for which we have interpolated the value for the partition function, $Q({\rm 36 K}),$ as 977 from the values tabulated on a CH$_3$OH data web page\footnote{{\url{https://cdms.astro.uni-koeln.de/cgi-bin/cdmsinfo?file=e032504.cat}}} of the Cologne Database for Molecular Spectroscopy \citep[CDMS;][]{Mueller2005}. 

While CH$_3$OH abundances of this magnitude (or even higher) have been determined for compact hot ($T \sim 150$~K) molecular cores that often surround  newly formed high-mass stars \citep[see, e.g., ][]{Menten1988, Bonfand2017, Molet2019}, $1.8\times10^{-6}$  appears to be a high value for the more extended, cooler, and more developed DR21 M molecular clump. We note that the much higher temperature (175 K) invoked from modeling the OH radio absorption lines characterizes a more limited hotter region that represents the dense PDR interface of the molecular clump with the compact HII region. 

As discussed in Sect. \ref{sec:dr21}, for certain ranges of density and temperature, the  $5_1\rightarrow 6_{0}~A^+$ transition can be ``overcooled'' (i.e., show deeper ``enhanced'' absorption) than expected under LTE, even against the cosmic microwave background radiation: \citet{Pandian2008} observed this line in absorption toward the hot corinos\footnote{Hot corinos are the equivalent of hot cores surrounding low-mass protostellar objects or low-mass young stellar objects.} NGC 1333-IRAS 4A and  4B, which do not show radio continuum emission. Were this the case, the CH$_3$OH abundance presented above would be an overestimate. To investigate this, we used the non-LTE radiative transfer program RADEX \citep{vanderTak2007} to study the line's excitation. For the gas densities and temperatures  determined by our modeling, we find that the line is indeed predicted to show enhanced absorption, that is, its excitation temperature is lower than the temperature of the cosmic microwave background. This means that the methanol abundance derived above represents an upper limit to the true value. Meaningful determinations of $N($CH$_3$OH) for DR21 M require constraints provided by data from thermally excited methanol lines. 

\section{Expected number of background sources}\label{sec:background}

Following \cite{Anglada1998}, we can estimate the number of expected background sources, $N_{\rm bg}$, inside a field of diameter $\theta_F$ as

\begin{equation}
\begin{split}
    N_{\rm bg} & = 1.4 \left\lbrace 1 -\exp \left[  -0.0066 \left( \frac{\theta_F}{\rm arcmin} \right)^2  \left( \frac{\nu}{\rm 5~GHz} \right)^2  \right] \right\rbrace  \\ & \times  \left( \frac{S_0}{\rm mJy} \right)^{-0.75} \left( \frac{\nu}{\rm 5~GHz} \right)^{-2.52},
\end{split}
\end{equation}

\noindent where $S_0$ is the detectable flux density threshold and $\nu$ the observing frequency. In our observations, $\nu=5.8$~GHz and $S_0=3\times{\rm rms}=0.18$~mJy. For field sizes of $\theta_F=1$, 0.5, and 0.017~arcmin, we expect $N_{\rm bg}=0.03$, 0.008, and 8.6$\times10^{-6}$. 

\end{document}